\documentclass[12pt]{article}
\usepackage{amsmath}
\usepackage{epsfig,graphics}

\newcommand{\be}{\begin{equation}}
\newcommand{\ee}{\end{equation}}

\newlength{\figsize}
\begin{document}

\begin{titlepage}

\vspace*{0.8in}
 
\begin{center}
{\large\bf Casimir scaling of domain wall tensions in the
\\ deconfined phase of D=3+1 SU(N) gauge theories.\\ }
\vspace*{0.75in}
{Francis Bursa and Michael Teper\\
\vspace*{.45in}
Rudolf Peierls Centre for Theoretical Physics, University of Oxford,\\
1 Keble Road, Oxford OX1 3NP, U.K.
}
\end{center}

\vspace*{0.75in}

\begin{center}
{\bf Abstract}
\end{center}

We perform lattice calculations of the spatial 't Hooft 
$k$-string tensions, $\tilde{\sigma}_k$, in the deconfined 
phase of SU($N$) gauge theories for $N=2,3,4,6$. These equal
(up to a factor of $T$) the surface tensions of the domain walls
between the corresponding (Euclidean) deconfined phases. 
For $T\gg T_c$ our results match on to the known perturbative result,
which exhibits Casimir Scaling, $\tilde{\sigma}_k \propto k(N-k)$.
At lower $T$ the coupling 
becomes stronger and, not surprisingly, our calculations show large 
deviations from the perturbative $T$-dependence. Despite this
we find that the behaviour $\tilde{\sigma}_k \propto k(N-k)$ 
persists very accurately down to temperatures very close to $T_c$.
Thus the Casimir Scaling of the
't Hooft tension appears to be a `universal' feature that
is more general than its appearance in the low order high-$T$
perturbative calculation. We observe the `wetting' of 
these $k$-walls at $T\simeq T_c$ and the (almost inevitable) 
`perfect wetting' of the $k=N/2$ domain wall. Our calculations
show that as $T\to T_c$ the magnitude of $\tilde{\sigma}_k(T)$ 
decreases rapidly. This suggests the existence of a (would-be) 
't Hooft string condensation transition at some temperature 
$T_{\tilde{H}}$ which is close to but below $T_c$. We speculate 
on the `dual' relationship between this and the 
(would-be) confining string condensation at 
the Hagedorn temperature $T_H$ that is close to but above $T_c$.

\end{titlepage}

\setcounter{page}{1}
\newpage
\pagestyle{plain}

\section{Introduction}
\label{section_intro}

In the Euclidean formulation of SU($N$) gauge theories at finite
temperature $T$, deconfinement at $T=T_c$ is associated with the 
spontaneous breaking of a $Z_N$ centre symmetry. This implies the
existence of $N$ degenerate phases for $T\geq T_c$ in our Euclidean 
box. When such phases co-exist they are separated by domain walls.
The tension of these domain walls is equal, up to a factor of $T$,
to the spatial 't Hooft string tension. This is an interesting
dynamical quantity that, for $T\leq T_c$, is in principle related 
to confinement.

As $T\to \infty$ one has $g^2(T)\to 0$ 
\cite{GPY} 
and so one can calculate
the  spatial 't Hooft tension in perturbation theory
\cite{CKA1,CKA2,DWd3}. 
These calculations   
\cite{CKA2}
show that this tension satisfies Casimir Scaling (to 2 loops).
This is an intriguing result given that the usual confining string 
tension between static sources is also known to satisfy Casimir 
Scaling to quite a good approximation
\cite{CS,oxsigk,pisasigk}.
Clearly it would be interesting to know whether the high $T$ Casimir 
Scaling that is characteristic of low order perturbation theory, 
persists to lower values of $T$ where the coupling becomes strong
and non-perturbative effects should be important. In particular 
it would be interesting to know what happens at $T \simeq T_c$
where one can imagine making contact with the confining phase.

In this paper we answer this question using numerical lattice
techniques applied to SU(2), SU(3), SU(4) and SU(6) gauge 
theories. The plan of the paper is as follows. In
Section~\ref{subsection_background} we discuss 
in more detail the theoretical background to this problem. 
We then provide a summary of our lattice setup in 
Section~\ref{subsection_lattice}. In Section~\ref{subsection_walls}
we describe how we can impose the existence of a domain wall
by using twisted boundary conditions and then, in 
Section~\ref{subsection_walltension}, we describe how we can calculate
(the derivative of) the surface tension from the average action.
In Section~\ref{subsection_wallprofile} we give the high-$T$
prediction for the profile of the domain wall and describe
how we obtain this profile in our lattice calculation. We
complete Section~\ref{section_prelim}
with a discussion of finite volume corrections in
Section~\ref{subsection_finiteV} and of `wetting' in 
Section~\ref{subsection_wetting}. We then move on in 
Section~\ref{section_results} 
to our calculations. Section~\ref{subsection_tension}
contains our results on the $k$-wall tension and
makes the comparison with perturbation theory. In 
Section~\ref{subsection_profile} we calculate the domain
wall profiles and again compare with perturbation theory.
In  Section~\ref{subsection_wettingres} we look for the 
appearance of a layer of confining phase in the centre
of the $k$-wall as $T\to T_c$, pointing out that one can use the 
pattern of `wetting' to tell us something about the
domain wall between confining and deconfining phases.
We finish in Section~\ref{subsection_stringcond}
with a conjecture about the existence of a dual to 
the usual string condensation `Hagedorn' transition.
Section~\ref{section_conclusions} gives our conclusions.

While this work was in progress a paper appeared 
\cite{pdf}
that addresses some of the same questions that we do,
using a novel lattice technique that allows the direct
calculation of the tension (rather than its derivative, as in
this paper). The calculations, for SU(4) at a fixed value of the
lattice spacing, observe consistency
with Casimir Scaling down to $T=1.2 T_c$ just as we do. 
However our calculations
test the Casimir scaling formula more tightly, because
we work with SU(6) as well as SU(4), and we vary the lattice
spacing so that we can address the continuum limit to some
extent. In addition our range of $T$ extends much closer to $T_c$.

\section{Preliminaries}
\label{section_prelim}

\subsection{Background}
\label{subsection_background}

As we remarked above, 
deconfinement in SU($N$) gauge theories is associated with the 
spontaneous breaking of a $Z_N$ centre symmetry. The symmetry arises from 
the fact that the partition function is unchanged if we impose that 
the time torus is periodic only up to a global gauge transformation
belonging to the centre of the group, $z_k = \exp\{2i \pi k/N\} \in Z_N$.  
(The group element $z_k$ contains a factor of the $N\times N$
unit matrix which we have suppressed. In fact it will be convenient
to use the same notation, $z_k$, both for the phase factor and for
the corresponding matrix. Which of the two we mean should always
be clear from the context.) The obvious order parameter is 
the trace of the Polyakov loop, $l_p$, which is a closed Wilson 
line that winds 
once around the Euclidean time torus and whose trace therefore acquires
a factor of $z_k$ under this symmetry transformation. In the
confined phase the effective potential for the Polyakov loop trace
averaged over the volume, $\bar{l}_p$, has its minimum at 
$\bar{l}_p = 0$ and so the symmetry is explicit. In the deconfined
phase the minimum is at $\bar{l}_p {\not =} 0$ and so the symmetry tells us
that there are $N$ degenerate minima: the symmetry is spontaneously
broken and there are $N$ possible deconfined phases. If 
$\bar{l}_p \propto z_k$ we label the deconfined phase by  $k$. When
two of these deconfined phases, $k_1$ and $k_2$ say, co-exist they 
will be separated by a domain wall whose surface tension 
$\sigma_W^k$ will depend on $k=k_1-k_2$, as well as on $N$ and $T$.  
At high $T$ the coupling on the relevant length scale, 
$g^2(T)$, becomes small
and one can calculate the domain wall surface tension in
perturbation theory. To two loops one finds
\cite{CKA1,CKA2}:
\begin{equation}
\sigma_W^k
=
k(N-k)
\frac{4\pi^2}{3\sqrt{3}}
\frac{T^3}{\sqrt{g^2(T) N}}
\{1 - \tilde{c}_2 g^2(T)N\}
\label{eqn_sigkPT}
\end{equation}
where $\tilde{c}_2 \simeq  0.09$. (While the precise value 
of  $\tilde{c}_2$ depends on the coupling scheme used
\cite{CKA1}, 
there appears to be only a $\sim 10\%$ variation between 
physically reasonable schemes.) The recently calculated
3-loop contribution
\cite{CKA2}
turns out to be negligible and so we do not include
it here. The spatial 't Hooft string tension (see below) is 
related to the wall tension by
\begin{equation}
\tilde{\sigma}_k
=
\frac{\sigma_W^k}{T}.
\label{eqn_sigdual}
\end{equation}
(We use this notation because $\sigma_k$ usually denotes the 
Wilson-loop $k$-string tension.) 

The factor of $T^3$ in eqn(\ref{eqn_sigkPT}) follows from 
dimensional analysis once $T$ is large enough for 
it to be the only relevant scale, and to two loops
$\sigma_W^k$ is only a function of the 't Hooft coupling 
$\lambda\equiv g^2N$ 
\cite{CKA1,CKA2}.
The leading dependence on the coupling is $1/g$ rather than $1/g^2$
because there are no semiclassical domain walls: they are
stabilised at one-loop. 
Although one might worry about the  infrared divergences
that eventually obstruct the perturbation expansion,
numerical tests in 2+1 dimensions
\cite{DWd3}
have shown that the lowest order perturbative calculations 
are remarkably accurate at all temperatures as long
as they are not very
close to $T_c$. Of course, as we reduce $T$ we increase $g^2(T)$
and at some point the perturbative expansion becomes unreliable.
For example if we use the 1-loop perturbative expression for the 
Debye screening mass, $m_D = T\sqrt{g^2(T)N/3}$
\cite{GPY},  
and equate it to the lattice values calculated in
\cite{oxT05}
then we would find
\begin{equation}
g^2(T)N \sim 7.5 \quad,\ T\sim 2T_c
\label{eqn_gN}
\end{equation}
at which point the 2-loop correction in eqn(\ref{eqn_sigkPT}) 
becomes $\sim 60\%$ of the 1-loop value and it is clear that we
are past the point where we can be confident in perturbation 
theory (and indeed in the reliability of the calculation
of $m_D$). At these values of $T$ numerical calculations are the
only unambiguous means of determining the physics.

The factor $k(N-k)$ in eqn(\ref{eqn_sigkPT}) is the $k$ dependence
of the Casimir, $Tr_{\cal{R}}T^aT^a$, when the represention
$\cal{R}$ is the totally antisymmetric representation of
a product of $k$ fundamentals of SU($N$). Thus it is the factor 
one obtains when calculating the Coulomb interaction
between sources in such a representation. In $D$=1+1 SU($N$) gauge
theories, the tension of the confining $k$-string that emanates
from  such a source has precisely this dependence. In fact there are 
old speculations
\cite{CS}
that this `Casimir Scaling' holds in $D$=3+1 and numerical
calculations show that it is in fact a good (but not exact) 
approximation to such $k$-string tensions 
\cite{oxsigk}
-- and an even better one in $D$=2+1 
\cite{oxsigk}.
A question we will try to answer in this paper is whether
the Casimir Scaling in eqn(\ref{eqn_sigkPT}), which is an
outcome of the low-order perturbative calculation, survives 
at very much lower $T$ where perturbation theory becomes a 
manifestly poor approximation. This non-perturbative question 
will be addressed using numerical lattice techniques.

What connects this question to the role of Casimir Scaling in 
confinement, is that these domain walls are closely related
to the centre vortices that provide a possible mechanism for 
confinement
\cite{tHooft,jgrev}.
Such a vortex will disorder any Wilson loop that encircles it, 
and the same for appropriate pairs of Polyakov loops.
(We temporarily locate our discussion in the in the more easily 
visualisable case of $D$=2+1
\cite{DWd3}.)
At low $T$, in the confined phase, one postulates a condensate
of such vortices, and a simple statistical calculation
(see e.g.
\cite{oxvor1})
shows how this leads to linear confinement. Since the Wilson loops 
are in a Euclidean space-time plane, the vortex cross-section
will be in the same plane.
As the Euclidean time extent is reduced so as to increase 
$T$, at some point the vortex core is squeezed by the short
temporal direction and becomes periodic in time. At this point 
one can no longer find a Wilson loop that encircle the vortex, 
but it will still disorder Polyakov loops that it separates.
This co-incides with the onset of the deconfined phase, which is 
a `Higgs' phase for these vortices, in which they become real, 
`massive' objects. They are now precisely the domain walls
which separate deconfined phases characterised by different Polyakov 
loop expectation values. 
The domain walls/vortices we discuss wind around a spatial torus.
However they could equally well be of finite length, ending on 
appropriate  $Z_N$ monopoles
\cite{susskind}.
Since the vortex cross-section is in a space-time plane, these 
monopoles will have world-lines in an orthogonal spatial
direction (returning to $D$=3+1). They form the boundary 
of a spatial 't Hooft loop,
and the domain wall tension is just the corresponding string
tension (multiplied by $T$). Where this tension is non-zero,
e.g. for $T>T_c$, these loops are not condensed. If however their
tensions exhibit Casimir Scaling all the way down to $T_c$,
where they might condense, then this may provide an extra
ingredient in our (hopefully) growing understanding of
confinement.  

As an aside, we remark that although these domain walls
certainly feature in the 4-dimensional Euclidean field theory, 
they almost certainly do not entail real domain walls
in the gauge theory at finite $T$. This has no implications
for the calculations described herein. (A physical manifestation
of these walls is described at the end of 
Section~\ref{subsection_walls}.)

\subsection{Lattice setup}
\label{subsection_lattice}

We discretise Euclidean space-time to a periodic hypercubic
lattice. The lattice spacing is $a$ and the lattice size,
in lattice units, is $L_\mu$ in the $\mu=0,1,2,3$ directions.
We choose the $\mu=0$ direction as the short Euclidean time
direction that generates our temperature:
\begin{equation}
T
=
\frac{1}{aL_0}.
\label{eqn_T}
\end{equation}
We will typically choose $L_1=L_2$ and place the domain wall 
to span the $L_1\times L_2$ torus. Typically $L_3 \gg L_1$
since it will need to contain 1 or 2 domain walls separating
regions in two phases. In addition, for thermodynamics to
be applicable, we need $aL_1,aL_2 \gg 1/T$.

The fields are SU($N$) matrices, $U_l$, on each link $l$ of the
lattice. (We sometimes label $U_l$ more explicitly as $U_\mu(n)$ where
$\mu$ is the direction of the link and $n=\{n_0,\vec{n}\}$ is the
site from which it emanates.) 
We label the elementary squares (plaquettes) of the
lattice by $p$ and the ordered product of link matrices around 
the boundary of $p$ by $U_p$. The partition function is then
\begin{equation}
Z
=
\int \prod_l dU_l e^{-\beta S}
\label{eqn_Z}
\end{equation}
and we use the plaquette action
\begin{equation}
S
=
\sum_p\{1-\frac{1}{N} \mathrm{ReTr} U_p\}.
\label{eqn_S}
\end{equation}
Analysing the continuum limit of this path integral, one
finds that
\begin{equation}
\beta
=
\frac{2N}{g^2}
\equiv
\frac{2N}{g_L^2(a)}
\label{eqn_g}
\end{equation}
where we have used the fact that $g^2$ is a lattice bare coupling 
to write it as some lattice scheme running coupling, $g_L$,
on the length scale $a$. To approach the $N\to\infty$ limit 
smoothly, we need to keep fixed the 't Hooft coupling 
\begin{equation}
\lambda(l)
\equiv 
g^2(l)N
\label{eqn_lambda}
\end{equation}
(where $l$ is fixed in units of some well behaved quantity such
as the mass gap) and so it is useful to define the 't Hooft 
inverse coupling
\begin{equation}
\gamma
\equiv
\frac{\beta}{2N^2}
=
\frac{1}{g_L^2(a)N}.
\label{eqn_gamma}
\end{equation}
If we keep $\gamma$ fixed as we take  $N\to\infty$ then we ensure
that $a$ remains fixed in physical units for large enough $N$. 
(See Fig.1 of 
\cite{oxT05} 
for an explicit lattice demonstration of this.)

\subsection{Imposing domain walls}
\label{subsection_walls}

Our order parameter will be based on the Polyakov loop
\begin{equation}
l_p(\vec{n})
=
\frac{1}{N}
\mathrm{Tr} \prod^{L_0}_{n_0=1} U_0(n_0,\vec{n}).
\label{eqn_lp}
\end{equation}

Consider the transformation where one multiplies all timelike 
link matrices at some fixed $n_0 = n_0^\prime$ by a non-trivial
element of the centre of SU(N), $z_k = e^{2\pi i\frac{k}{N}} \in Z_N$:
\begin{equation}
U_0(n_0=n_0^\prime,\vec{n})
\longrightarrow
z_k U_0(n_0=n_0^\prime,\vec{n})
\quad \forall \vec{n}.
\label{eqn_tw}
\end{equation}
A plaquette involving one such link matrix will necessarily 
involve the conjugate of a second and the centre factors
will cancel, $z_k {z_k}^\dagger =1$ so the plaquette remains
unchanged. Since the measure is
also unchanged, such a field has the same probability as
the original field and we have a corresponding $Z_N$ symmetry. 
A winding operator such as the Polyakov loop in eqn(\ref{eqn_lp}) 
will change under the symmetry transformation: $l_p \to z_k l_p$. 
It therefore provides an order parameter for the spontaneous
breaking of the symmetry. In the confined phase where
$\langle l_p \rangle = 0$ the symmetry is explicit, but in
the deconfined phase where $\langle l_p \rangle {\not=} 0$
it is spontaneously broken with $N$ degenerate phases in which
$\langle l_p \rangle = z_k c(\beta)$ where $c(\beta)$ is a 
real-valued renormalisation factor. Such phases can
co-exist at any $T\geq T_c$ and in that case they will be separated 
by domain walls. If the order parameter in the two phases differs
by a factor $z_k$ we shall refer to the domain wall as a $k$-wall.
The domain wall may, for example, extend right across
the 1-torus and 2-torus, and its world volume will wrap around
the time torus. In such a case 
periodicity in $x_3$ means there must be at least
two walls. Since the Polyakov loops on either side of the domain
wall differ by some factor $z_k\in Z_N$, 
we see that the wall disorders Polyakov
loops. Although we cannot encircle the wall with a Wilson loop in
the (3,0)-plane, since the wall wraps right around the time torus, 
if we could then the Wilson loop would acquire a factor $z_k$ as well. 
Thus, as remarked earlier, the domain wall is 
nothing but a spatial 't Hooft disordering loop in 
the (1,2)-plane, that has been  squashed in the short temporal 
direction. We describe it as squashed because the size in the
$x_3$-direction, where it is not squashed, is of the order 
of the inverse Debye mass
\begin{equation}
\frac{1}{m_D}
\stackrel{T\to\infty}{=}
 \frac{1}{T}\sqrt{\frac{3}{g^2(T)N}}
\gg  
\frac{1}{T} 
= 
a L_0
\label{eqn_mD}
\end{equation}
to leading order in perturbation theory
\cite{GPY}.

To study a $k$-wall we `twist' the action in
eqn(\ref{eqn_S}) so as to enforce the presence of a single domain
wall. The twisted action is defined by
\begin{equation}
S_{k}
=
\sum_p \Bigl(1-\frac{1}{N} \mathrm{ReTr} \{z(p) U_p\}\Bigr).
\label{eqn_Stw}
\end{equation}
where  $z(p)=1$ for all plaquettes except
\begin{equation}
z(p=\{\mu\nu,x_\mu\})
= z_k =
e^{2\pi i\frac{k}{N}}
 \  \  \quad \mu\nu=03;  \  x_0={x_0}^\prime,
 \  x_3={x_3}^\prime, \  x_{1,2}=1,...,L_{1,2}.
\label{eqn_Stw2}
\end{equation}
That is to say, the plaquettes in the (0,3)-plane that are at
the fixed values  $x_0={x_0}^\prime$ 
and  $x_3={x_3}^\prime$ and at all values
of $x_1$ and $x_2$ are multiplied by $z_k\in Z_N$. This is 
a plane of plaquettes that wraps around the 1 and 2-tori
at $x_0={x_0}^\prime$ and  $x_3={x_3}^\prime$. 
Although the plane appears to be in a
particular location this is an illusion. We can move it and
deform it by redefinitions $U_l\to z_k U_l$ of suitable subsets 
of links. It is clear from considering such redefinitions
that the Polyakov loops on either side of such a plane of
twisted plaquettes will differ by a factor of $z_k$. 
Periodicity in $x_3$ then demands that at some $x_3$ the
Polyakov loop must suffer a compensating factor of ${z_k}^{\dagger}$
so as to return to its original value. This requires an interface
-- a domain wall. Thus using such a twisted action we
ensure that in the deconfined phase each configuration possesses
at least one domain wall. This provides a convenient way to study
different $k$-walls corresponding to different choices of
$k$ in eqn(\ref{eqn_Stw}).

We will not discuss the interpretation of these walls in any detail, 
and instead refer the reader to
\cite{DWd3,oxvor2}
for a more explicit discussion in the readily visualisable context 
of $D$=2+1, and to 
\cite{monrev}
for a more general theoretical review. Here we content ourselves
with a few remarks for orientation. A finite line of twisted 
plaquettes enforces the presence of a $Z_N$ Dirac string so that 
we have $Z_N$ monopoles at the ends of the line
\cite{susskind}. 
The return flux between the monopoles is a centre vortex. 
A plane of such plaquettes thus corresponds to a `dual' Wilson
loop where a centre monopole propagates around the perimeter
of the loop. There is now a vortex sheet that also ends on
the perimeter of the loop. In our construction the loop is 
spacelike and we then extend the plane of twisted plaquettes right 
around the corresponding spatial two-torus, so that the boundary 
disappears and the vortex sheet, which now wraps around the
two-torus, becomes decoupled in position from the centre Dirac 
sheet. Nonetheless the presence of the latter enforces the presence 
of the former. One may regard the plane of twisted plaquettes as 
effectively constituting an ultraviolet $k$-wall at zero 
cost. In the picture where vortex sheets are degrees of freedom 
relevant to confinement in the Euclidean theory at $T=0$
\cite{monrev}, 
the idea 
is that a closed $k$-vortex sheet can `thread' a Wilson loop 
(just as a vortex loop would do in $D=2+1$) and the Wilson loop then
acquires a factor $z_k$. A condensate of such loops then leads 
to an area decay and linear confinement. These vortex sheets 
will in general close upon themselves rather than around 
the (infinite) torus.
Our periodic construction is a device to impose the presence of a
vortex sheet so that we can study it. 
In the deconfined phase such a spacelike vortex sheet is
squashed by the short Euclidean time direction so that it also extends
around the time torus. And simultaneously the vortex sheet becomes
real just as a 't Hooft-Polyakov monopole becomes real in the
Higgs phase of the Georgi-Glashow model. It is in this phase that we
can study it in the hope that what we learn will eventually have some
bearing on confinement. 

Finally we remark that this whole discussion 
becomes more transparent if we perform a cyclic relabelling of our 
co-ordinates,  $x_o \to x_1$ etc., so that we have a sytem at low 
$T=1/aL_3 \ll T_c$ but on a spatial three-torus where one of the tori
is very short. In this case our $Z_N$ monopole sources are timelike
and the dual 't Hooft loop is in a space-time plane. The 't Hooft
loop is imposed by the sheet of twisted plaquettes which generate
the monopole Dirac sheet. It is now 
the short spatial torus that squashes the return flux between
the monopoles.  If we make the sheet of twisted plaquettes
periodic in both directions then the squashed return flux
manifests itself as a domain wall separating two phases
in which the Polyakov loop around the short spatial torus
has non-zero vacuum expectation values that differ by a factor
of $z_k$ between the two phases. This relabelled description has 
the advantage that
it corresponds to real $T\simeq 0$ physics on such a spatial
volume, in contrast to the original description which (for subtle 
reasons) does not imply the existence of real domain walls in 
high-$T$ gauge theories.

\subsection{Domain wall tension}
\label{subsection_walltension}

Let us calculate the average action with and without a
$k$-twist as defined above. Then, on a lattice that is
of fixed size in lattice units,
\begin{equation}
\Delta S_k 
\equiv 
\langle S_k \rangle - \langle S_0 \rangle
=
\frac{\partial \ln Z_0}{\partial\beta}
-
\frac{\partial \ln Z_k}{\partial\beta}
=
\frac{\partial}{\partial\beta}\frac{F_k-F_0}{T}
=
\frac{\partial}{\partial\beta}\frac{\sigma_W^k A}{T}
\label{eqn_Swall}
\end{equation}
where $\sigma_W^k$ is the surface tension (energy per unit
area) of the domain wall and $A=a^2L_1L_2$ is its area.
We assume the spatial dimensions are large enough that
finite volume corrections are negligible. (We will of course
need to check that this is so for our particular calculations.)

We see from eqn(\ref{eqn_sigkPT}) that we expect 
\begin{equation}
\Delta S_k 
=
\frac{\partial}{\partial\beta}\frac{\sigma_W^k A}{T}
\stackrel{T\to \infty}{=}
 \alpha(L_0) \frac{k(N-k)}{\sqrt{N}} 
\frac{4\pi^2}{3\sqrt{3}}
\frac{L_1L_2}{L^2_0}
\frac{\partial}{\partial\beta}
\frac{1}{g(T)}
\label{eqn_SkPT1}
\end{equation}
when $T=1/a(\beta)L_0$ is large enough that leading order perturbation 
theory is accurate. The additional factor of  $\alpha(L_0)$ 
contains the $O(a^2T^2) = O(1/L^2_0)$ lattice correction to 
the continuum formula in eqn(\ref{eqn_sigkPT}). We have obtained 
it by a numerical evaluation of the (non-closed) analytic expression 
given in
\cite{DWd3}
and the values are listed in Table~\ref{table_alpha} for a
range of $aT$. The calculations in this paper will be for
$L_0= 4, 5$ where the correction is modest.

To compare the perturbative expression in eqn(\ref{eqn_SkPT1}) with 
our calculated values of $\Delta S_k$ we need to know the coupling
$g^2(T)$ in terms of $g_L^2(a)$ and hence $\beta$. 
The simplest choice is obtained by asserting that to leading
order the scale and distinction between running couplings is 
irrelevant, so we can use $g^2(T) \stackrel{lo}{=} g_L^2(a)$ 
and then
\begin{equation}
\frac{\Delta S_k}{L_1L_2} 
\stackrel{T\to \infty}{=}
 \alpha(L_0) 
\frac{2\pi^2}{3\sqrt{6}} 
\frac{k(N-k)}{N} 
\frac{1}{L^2_0}
\frac{1}{\sqrt{\beta}}.
\label{eqn_SkPT2}
\end{equation}
In practice this is too crude. It is well known that $g^2_L(a)$ 
is a bad running coupling in that it has very large higher 
order corrections, so that it is only at unattainably asymptotic 
$T$ that we could be confident of eqn(\ref{eqn_SkPT2}) holding. 
The simplest, if partial, remedy is mean field improvement
\cite{MFI}
\begin{equation}
\beta_I
=
\beta \times \langle \frac{1}{N}\mathrm{Tr}U_p \rangle
=
\frac{2N}{g^2_I(a)}.
\label{eqn_MFI}
\end{equation}
If we replace $g^2(T)$ in eqn(\ref{eqn_SkPT1}) by  $g^2_I(a)$
then we obtain
\begin{eqnarray}
\frac{\Delta S_k}{L_1L_2} 
& \stackrel{1-loop}{=} &
 \alpha(L_0) 
\frac{2\pi^2}{3\sqrt{6}} 
\frac{k(N-k)}{N} 
\frac{1}{L^2_0}
\frac{1}{\sqrt{\beta_I}}
\Bigl\{
1+\beta\frac{\partial}{\partial\beta}\Bigr\}
\langle \frac{1}{N}\mathrm{Tr}U_p \rangle 
 \nonumber\\
& \simeq &
 \alpha(L_0) 
\frac{2\pi^2}{3\sqrt{6}} 
\frac{k(N-k)}{N} 
\frac{1}{L^2_0}
\frac{1}{\sqrt{\beta_I}}
\label{eqn_Sk1loop}
\end{eqnarray}
where the difference between the second line and the first 
is $O(1/\beta^\frac{5}{2})$ and is therefore negligible
even to 2-loop order.  (If we use the expansion to $O(\beta^{-8})$ 
obtained in
\cite{plaq} 
for the average plaquette, we see that even at our maximum 
value of $a$, corresponding to $a=1/4T_c$, the numerical correction 
is no more than $\sim 17\%$.) In a similar way we obtain from
eqn(\ref{eqn_sigkPT})
\begin{eqnarray}
\Delta S_k
& \stackrel{2-loop}{=} &
\Delta S^{1-loop}_k
\times
\Bigl\{ 1 + \frac{2\tilde{c}_2 N^2}{\alpha(L_0)\beta_I}\Bigr\}.
\label{eqn_Sk2loop}
\end{eqnarray}
We do not have a calculation of the lattice correction for
the 2-loop term so we use no correction in that case.
The error from this should be small. The expressions in 
eqns(\ref{eqn_Sk1loop},\ref{eqn_Sk2loop}) are what we shall 
usually use when making comparisons between perturbation
theory and our lattice results. 

There still remains the ambiguity associated with the scale $T$ at 
which $g_I^2$ should be evaluated. Although, as we remarked above,
the scale is irrelevant at one loop, this is really a formal 
statement. The point is that with a reasonable coupling, using
an appropriate scale is the best way to ensure that higher
order corrections are likely to be small. Of course,
for $L_0 = 4,5$ plausible scales such as $\pi T$ and $2\pi T$ 
are numerically close to $1/a$, so it may be that using $g_I^2(a)$ 
for $g^2(T)$ is reasonable. However, it would be useful to have
some estimate of how eqn(\ref{eqn_Sk1loop}) changes if we use 
the scale $aL_0$, say, rather than $a$ in $g^2$. Near $T_c$ this
is a large spatial scale and so we need to use a lattice coupling 
scheme that we can have real confidence in. For this purpose we 
shall use the Schrodinger functional running coupling defined 
and calculated in
\cite{alpha_s}.
We shall leave the detailed discussion and calculation to the 
appropriate point in Section~\ref{section_results}.

\subsection{Domain wall profile}
\label{subsection_wallprofile}

In addition to calculating the surface tension of the 
$k$-wall as described above, we can also calculate how 
the average Polyakov loop varies as one passes from one vacuum 
to the other through the domain wall, just as was done for
the D=2+1 SU(2) case in 
\cite{DWd3}.   

Suppose the $k$-wall spans the $(1,2)$ torus and is centered at
$z=0$ where $z\in (-\infty,+\infty)$ and we use continuum 
variables. Suppose the Polyakov loop satisfies
\begin{equation}
\bar{l}_p(z=-\infty) = 1 \quad ; \quad
\bar{l}_p(z=+\infty) = e^{2\pi i \frac{k}{N}}.
\label{eqn_lpbc}
\end{equation}
Then to leading order in perturbation theory one finds
\cite{CKA1}
\begin{equation}
\bar{l}_p(z) 
= 
\frac{1}{N} \mathrm{Tr}
e^{2\pi i \frac{q(z)}{N}Y_k}
\label{eqn_lp1}
\end{equation}
where $Y_k$ is a diagonal matrix with 
$\{Y_k\}_{ii} = k$ for $i=1,...,N-k$ and
$\{Y_k\}_{ii} = k-N$ for $i=N-k+1,...,N$
and $q(z)$ is a function that is independent of $k$.
It is obtained from a one-loop effective potential calculation
for the Polyakov loop
\cite{CKA1,DWd3}
which leads to
\begin{equation}
q(z)
=
\frac{e^{\sqrt{\frac{g^2N}{3}}Tz}}
{1+e^{\sqrt{\frac{g^2N}{3}}Tz}}
\label{eqn_lp2}
\end{equation}
Eqns(\ref{eqn_lp1},\ref{eqn_lp2}) provide the 1-loop 
perturbative result for the domain wall profile for all
$k$ and $N$. 

To compare this with our lattice calculations we have to choose 
a coupling $g^2(T)$ and, just as for the 
surface tension, we choose to use the coupling $g^2_I(a)$ 
defined in eqn(\ref{eqn_MFI}). In addition the average
Polyakov loop in the deconfined phase does not have unit length,
but rather
\begin{equation}
|\langle \bar{l}_p\rangle|
=
c(\beta) .
\label{eqn_lpav}
\end{equation}
This can be interpreted as the self-energy correction to the static 
source and is dominated by ultra-violet fluctuations which are 
insensitive to which phase we are in. Thus if we want to compare
to the numerically calculated Polyakov loop we can simply renormalise  
the predicted Polyakov loop by a constant factor at all $z$
\begin{equation}
\bar{l}_p(z) 
\to
c(\beta) \bar{l}_p(z) 
\label{eqn_lp3}
\end{equation}
where $c(\beta)$ can be determined from field configurations
that are in a pure phase (and will depend on $N$ as well 
as on $\beta$). In principle we could obtain lattice corrections
to eqns(\ref{eqn_lp1},\ref{eqn_lp2}) analogous to the
factor $\alpha(L_0)$ for the surface tension. However we
expect these to be small and so we neglect them. Thus
eqns(\ref{eqn_lp1},\ref{eqn_lp2},\ref{eqn_MFI},\ref{eqn_lp3})
provide the 1-loop perturbative result for the domain wall 
profile against which we shall compare our numerical lattice results.

\subsection{Finite volume corrections}
\label{subsection_finiteV}

In principle the thermodynamic limit demands
that we take $L_i\to \infty \ ; i=1,2,3 $ at any fixed
$L_0$ and $\beta$. In practice what this means is that 
we need to make these
lengths large enough so that finite $V$ corrections
are small. We start with the most obvious.

\subsubsection{wall fluctuations}

The fluctuations of the domain wall lead to corrections
that are geometric and will not possess the $k,N$ 
dependence of the domain wall tension -- in exactly the
same way as the Luscher correction to the mass of
a confining string does not depend on the representation
of the sources. Thus we want to make the wall area, $L_1\times L_2$
large enough for these corrections to be negligible.
The leading corrections are typically
\cite{wallfluct}
\begin{eqnarray}
\frac{\delta \sigma^k_W}{\sigma^k_W}
& \sim &
\frac{c}{L_0L_1L_2} \frac{1}{\sigma^k_W} 
\nonumber \\
& \sim &
c
\frac{3\sqrt{3}}{4\pi^2}
\frac{\sqrt{g^2(T)N}}{k(N-k)}
\frac{L^2_0}{L_1L_2}
\nonumber \\
& < & 
0.01 
\label{eqn_fluct}
\end{eqnarray}
where we have used the fact that $c \sim 1$
\cite{wallfluct}, 
eqn(\ref{eqn_gN}) as some guide to
the largest value of $g^2(T)N$ and the fact that for
our calculations we have $L_0 =4,5$ and $L_{1,2} =20$.
Thus with our choice of lattice sizes any such corrections
will be very small.

\subsubsection{wall depth}

We must also make sure that $L_3$ is long enough to accommodate
the full depth of the domain wall. As pointed out in
\cite{DWd3} 
this requires a much large $L_3$ than one might naively expect.
If we define the full width of the $L_3=\infty$ wall, $d_W$, as 
extending between the points at which the Polyakov loop reaches
two-thirds of its vacuum value, then the domain wall
starts expanding once $a L_3 < 3 d_W$ and completely disappears
once $a L_3 \simeq 2 d_W$. We therefore explicitly monitor 
the domain wall profile in each calculation and make sure
that $L_3$ is large enough that there are no significant
finite-$L_3$ corrections.

\subsubsection{spatial symmetry breaking}

This correction is the most subtle and is typically relevant
to calculations at very high $T$. In our case the 
problem arose when we performed calculations in SU(4)
on $4\times 12^2\times 40$ and  $4\times 12^2\times 60$ lattices
at $\beta=18,20$ respectively. The latter $\beta$ corresponds, very
roughly, to $T\sim 1000T_c$ and we expected to obtain a good
agreement with perturbation theory. However, while our result for
the $k=2$ domain wall was in agreement with eqn(\ref{eqn_Sk1loop}), 
the $k=1$ result was about twice too large and even larger than the
$k=2$ result, which was clearly unphysical.

The cause of this phenomenon turns out to be a centre symmetry 
breaking in the spatial direction. That is to say, the 
average Polyakov loop in the $\mu=1$ or $\mu=2$ direction
acquires a non-zero vacuum expectation value. In the case under
discussion, the centre breaking occurred in the simulation
with $k=1$ twist but not for the untwisted $k=0$ case.
Thus the difference in actions was anomalous.

The practical cure, followed in the calculations in this
paper, is obvious -- calculate spatial Polyakov loops
and choose the spatial lattice large enough to ensure that the
symmetry breaking does not occur. For practical and theoretical
reasons it would be interesting to understand this phenomenon
sufficiently well to be able to estimate the spatial lattice
size needed. This we now turn to.

At high $T$, if we perform a Fourier decomposition in $x_0$ of the
SU($N$) fields, we can keep just the lowest constant mode since 
there is an effective mass gap to the higher modes of $O(\pi T)$. 
The Euclidean time integral in the action just gives $1/T$ and
the action becomes that of a $D=2+1$ gauge theory coupled to
an adjoint scalar field (the remnant of the $A_0$ field). 
The coefficient in
the action of the plaquette term is the corresponding D=2+1 
inverse coupling, which we call $g^2_3$, and which clearly
satisfies
\begin{equation}
\frac{1}{g^2_3}
=
\frac{1}{g^2(T)} \frac{1}{T}
\quad
\Longrightarrow
\quad
\beta_3 
\equiv
\frac{2N}{ag^2_3}
\simeq
\beta L_0
\label{eqn_beta3}
\end{equation}
at tree level. (An approximation that can be systematically 
improved upon). If we now neglect the massive adjoint scalar 
field (again an approximation that can be improved upon)
we obtain a  $D=2+1$ SU($N$) gauge theory on a 
$L_1\times L_2\times L_3$ lattice with the usual plaquette
action. All this is the well-known phenomenon of dimensional
reduction at high $T$. (See 
\cite{ML}
for a recent discussion.)

Suppose for a moment that $L_2,L_3 \gg L_1$. Then as we increase 
$\beta_3$ this $D=2+1$ SU($N$) gauge theory will at some point 
deconfine and at this point the average Polyakov loop in the
shorter $x_1$ direction (which plays the role of the Euclidean
time direction in the $D=2+1$ system) acquires a non-zero 
expectation value. This is precisely the `spatial symmetry breaking' 
in the parent $D=3+1$ gauge theory that we discussed above. 
When will it occur? This is easy to estimate. Using 
eqn(\ref{eqn_beta3}) we have a specific value of $\beta_3$
corresponding to given values of $\beta$ and $L_0$.
The string tension, $a^2\sigma$, can be obtained for any $N$ and 
for any  $\beta_3$ from the calculations in
\cite{d3}.
(It helps to know the average plaquette for the interpolation.)
We now note that in $D=2+1$ one finds $T_c \sim \sqrt{\sigma}$
\cite{tempd3}.
(This is based on SU(2) and SU(3); calculations for all $N$
are in progress
\cite{JLMTd3}.)
Thus in the case cited above, with $\beta=20$ and $L_0=4$ and
$N=4$, we have $\beta_3 \simeq 80$ which turns out to correspond 
to $a\sqrt{\sigma} \simeq 0.08$ 
\cite{tempd3}
so that the critical value of $L_1$ will be 
$L_1^{crit} = 1/aT_c \sim 1/a\sqrt{\sigma} \sim 12$. Since
in $D=2+1$ one has $a\sqrt{\sigma}\propto 1/\beta_3$ 
at large $\beta_3$,
the value of $L_1^{crit}$ corresponding to $\beta=18$ in the 
D=3+1 theory will not be very different. Finally, the fact
that $L_3$ is finite and
that $L_2{\not\gg}L_1$ but rather $L_2=L_1$ means that the
transition is smeared in $\beta_3$ (and hence in $\beta$) and
that it may occur for one of the two transverse
directions chosen at random. Thus
the observed scenario where on some lattices the transition
occurs and on others it does not, is just what one expects.
Using the above approach one can estimate the critical value
of $L_{1,2}$ for any particular $\beta$ and $L_0 \ll L_{1,2}$
and thus avoid the corresponding spatial symmetry breaking.

It is interesting to note that at high $T$ in $D=2+1$ there
should be no such spatial symmetry breaking because the
dimensionally reduced $D=1+1$ gauge theory confines at all
$T$. (The confining string cannot vibrate in one space dimension.) 
Indeed careful finite $V$ studies in $D$=2+1
\cite{DWd3} 
have not observed any unexpected behaviour of the domain wall
tension at very high $T$. It is perhaps over-familiarity with 
$D=2+1$ that led us to be surprised by what occurs in $D=3+1$.

\subsection{Wetting at $T\simeq T_c$}
\label{subsection_wetting}

Since the phase transition is first order for $N\geq 3$
one can have the confined phase as well as the various deconfined 
phases co-existing at $T=T_c$. We can always interpolate
between any two deconfined phases by passing through the
confined phase, and whether the system chooses to do so or not will
depend on the relevant wall free energies, i.e surface tensions.
A $k$-wall between two deconfined phases will be stable against
breaking up into two confined-deconfined domain walls, with 
surface tension $\sigma_{cd}$, provided that 
\begin{equation}
\sigma^k_W(T_c)
<
2\sigma_{cd}(T_c).
\label{eqn_wet1}
\end{equation}
In this case it is only at some $T < T_c$ that the $k$-wall 
would split. Another possibility is
\begin{equation}
\sigma^k_W(T_c)
=
2\sigma_{cd}(T_c).
\label{eqn_wet2}
\end{equation}
This corresponds to `perfect wetting'. It has been conjectured 
\cite{wet}
that this occurs in the SU(3) gauge theory, where there are 
only $k=1$ walls. (See 
\cite{pdf}
for a recent discussion.) A third possibility is
that the continuation from higher $T$ would give
\begin{equation}
\sigma^k_W(T_c)
>
2\sigma_{cd}(T_c).
\label{eqn_wet3}
\end{equation}
In that case there would be some critical value of $T$ above
$T_c$ where a pair of confined-deconfined domain walls would have
the same free energy and from there onwards, as $T\to T_c$,  
the $k$-wall will be on the $\sigma^k_W(T)=2\sigma_{cd}(T)$
branch. (So that in practice the condition in 
eqn(\ref{eqn_wet2}) will be satisfied.)

We recall that the confined and deconfined phases correspond
to the $N+1$ degenerate minima of the effective action,
calculated as a function of $\bar{l}_p$. The confined phase 
is at $\bar{l}_p = 0$ and the $N$ deconfined phases at
$\bar{l}_p = c(\beta) e^{2\pi i k/N} : k=0,...,N-1$.
A domain wall between two phases is a tunnelling path between 
the corresponding minima that minimises the effective action.
(A proper analysis, as in
\cite{CKA1,CKA2}
for very high $T$,
needs to express the effective potential not just in terms of the 
trace of the Polyakov loop but in terms of its eigenvalues.)
At high $T$ it is enough to keep the effective potential and
a term quadratic in derivatives
\cite{CKA1,CKA2}.
In this case one has the usual simple intuition about how
tunnelling paths are determined by the potential. 
As $T=1/aL_0$ decreases 
the coupling $g^2(T)$ increases and eqn(\ref{eqn_lp2}) tells
us that the domain wall will become thinner, $\propto 1/g(T)$,
on the scale of $1/T$.  This could imply 
that higher order derivative terms will become more
important in the effective action. In reality however, as we shall 
see, the walls become thicker rather than thinner as $T\to T_c$,
suggesting that the simple picture of a potential plus
quadratic derivatives might remain reliable beyond its naive domain 
of validity.

If we increase $T$ slightly away from $T_c$ then the minimum at 
$\bar{l}_p = 0$ rises and the confining phase becomes metastable.  
If we keep increasing $T$ then eventually this minimum ceases
to play any role and we need only consider the $N$ deconfined
minima. At high $T$ one can calculate these tunnelling paths
in perturbation theory 
\cite{CKA1,CKA2}
as summarised in Section~\ref{subsection_wallprofile}. 
For example, one finds
that the $k=N/2$ domain wall that runs between the $k=0$ and 
$k=N/2$ phases passes through $\bar{l}_p = 0$.  
In practice, as we shall see below, this behaviour persists 
all the way down to $T=T_c$. Given this
we expect $\sigma^{k=N/2}_W(T_c)= 2\sigma_{cd}(T_c)$;
i.e `perfect wetting'. Indeed we expect this equality to hold
at any $T$ close enough to $T_c$ for $\sigma_{cd}(T)$ to make 
sense. Thus we expect that at $T=T_c$ the $k=N/2$ wall will  
break up into two confined-deconfined walls. Since at $T_c$
the confined and deconfined free energies are equal,
these two walls can separate and rejoin.

As an aside we remark that the above scenario allows us to
calculate $T_c$ on arbitrarily large volumes, in contrast
to the standard calculations such as in
\cite{oxT05}.
For example, direct tunnelling at $T=T_c$ from the confined to the
deconfined phase on a lattice larger than $16^35$ will simply 
not occur in a SU(6) Monte Carlo calculation of feasible length
\cite{oxT05}.
The reason is the thermodynamic suppression of the large
domain walls that occur as a necessary intermediate step
in the tunnelling. Here, by contrast, our twisted boundary
conditions enforce the presence of a $k$-wall that can have
enough free energy to split into two confined-deconfined
walls irrespective of how large is the volume. In this context 
$\beta_c$ can be identified by the fact that the growth
and shrinkage of the confined and deconfined sub-volumes
follows an unbiased random walk. In principle this 
provides a practical technique for directly locating $T_c$ on
arbitrarily large volumes.  

A $k{\not =}N/2$ domain wall, on the other hand,
does not pass through  $\bar{l}_p = 0$ at high $T$. Thus an
alternative tunnelling path from the $k=0$ to $K=N/2$ minima
is not via  $\bar{l}_p = 0$, but rather via the paths joining
the sequence of minima $K=0 \to 1 \to ... \to N/2$. While this is
clearly a local minimum of the effective action, this minimum
turns out to be sufficiently higher than the direct 
path through  $\bar{l}_p = 0$ that it does not play a role 
in practice. More interesting is the reverse possibility:
for tunnelling between the $k$ and $k^\prime$ phases, there is
a minimal path that first runs from the $k$ minimum to
$\bar{l}_p = 0$ (just like the first half of the $k=N/2$
domain wall) and then runs out from $\bar{l}_p = 0$ to the
$k^\prime$ minimum. We would expect this to have an effective
action equal (approximately) to that of the $k=N/2$
tunnelling path and thus to be too large to contribute
at high $T$ (where we know that we have paths that satisfy
Casimir Scaling). However as $T$ drops to and below $T_c$ there 
will certainly be a point at which the path through
$\bar{l}_p = 0$ comes to have a lower action than the
path not through $\bar{l}_p = 0$. It could of course be
that the shape of the effective potential also changes so
that the latter path smoothly deforms into the former
at $T\simeq T_c$. However our calculations (see below) suggest 
that what actually occurs is that there is a tunnelling from 
one kind of path to the other. So if Casimir Scaling persists
all the way down to $T_c$, we expect that it will only
be at some $T$ below $T_c$ that the $k=1$ wall breaks
up into two $cd$ walls. In this sense we do not expect
`perfect wetting' for domain walls in SU(3), since these
are necessarily $k=1$.

The above remarks suggest a possible way of determining
whether  $\sigma_{cd}$ is $O(N)$ or $O(N^2)$, something
which direct calculations
\cite{oxT05} 
have so far been unable to determine unambiguously. 
Since we have found that
$\sigma^k_W(T)$ satifies Casimir Scaling very accurately
down to $T\simeq 1.02 T_c$, it is plausible to assume
that it will continue to do so down to $T=T_c$. In that case
the $N$-dependence is $\sigma^k_W(T_c) \propto k(N-k)$
and so $2\sigma_{cd} = \sigma^{k=N/2}_W(T_c) \propto N^2$.
More generally by observing for which values
of $k$ and $N$ the domain walls wet at $T=T_c$,
one can imagine being able to pin down $\sigma_{cd}$
quite accurately. The calculation in this paper does not 
have the range of $k$ and $N$ to make this a realistic
goal, but we can at least attempt to gauge the
potential of such a strategy.

\section{Results}
\label{section_results}

The strategy of our calculation is as follows. We first perform
calculations of $k=1$ and $k=2$ domain walls in SU(4)
at a very high temperature, $T \sim 1000T_c$, where we can 
expect our results to match perturbation theory quite precisely.
We then repeat the calculations much closer to $T_c$,
where perturbation theory is {\it{\`a priori}} no longer useful.
We do so for  $L_0=4$ and for $L_0=5$, so that
at each $T$ we have results at two values of $a$. i.e.
$a=1/4T$ and $a=1/5T$. This allows us to make some statement about
the continuum limit. We then perform calculations of $k=1,2,3$
domain walls in SU(6), for $L_t=5$, which together with the
SU(4) results provide us accurate information on the full
$k$ and $N$ dependence of the surface tension. As a check on
the $N$-dependence of the low $T$ behaviour of the
$k=1$ surface tension we also perform some SU(2)
and SU(3) calculations.

We simultaneously calculate the domain wall profiles and
compare them with perturbation theory. As $T\to T_c$
we look for the breakup of the $k$-wall into two 
confined-deconfined  domain walls separated by a region of 
confining phase (`wetting' and `perfect wetting'). 
In principle this provides a way to obtain information
about the $N$-dependence of the corresponding surface 
tension, $\sigma_{cd}$.

Finally we list the values of $\beta$ that correspond to
$T=T_c$ for the calculations in this paper:
\begin{equation}
\beta_c
= \begin{cases}
10.4865(3) - 0.067(7)/VT^3
&:\quad L_0=4 \ ; \ SU(4) \\ 
10.6373(6) - 0.090(17)/VT^3
&:\quad L_0=5 \ ; \ SU(4) \\ 
24.5139(24) - 0.112(19)/VT^3
&:\quad L_0=5 \ ; \ SU(6) \\ 
\end{cases}
\label{eqn_bc}
\end{equation}
The last two values are taken from 
\cite{oxT05,blmtuw-T03}
while the first is a calculation performed in this paper,
Exhibiting the volume dependence allows us to correct for
a finite volume shift, although in practice this will
be completely negligible for us.

\subsection{Surface tension}
\label{subsection_tension}

The parameters of our high statistics calculations are listed 
in Table~\ref{table_runs} and the corresponding values of
the average plaquette are listed in Table~\ref{table_aplaq}.
We have deliberately chosen the values of $\beta$ so that
the values of $T/T_c$ are closely matched across the various
SU($N$) groups. (For $T_c$ we use the values in eqn(\ref{eqn_bc}) 
and for interpolations we use the the string tension values in 
\cite{oxglue04}
as well as the extra calculations listed in Table~\ref{table_sigN4}.
One can instead interpolate using the values of $\beta_c(L_0)$ in
\cite{oxT05,blmtuw-T03}
and this will give values of $T/T_c$ that differ at the $1\%$
level from the ones we list.)
These values, $T\simeq 1.88T_c$ and $T\simeq 1.02T_c$, take us very
close to the deconfinement transition itself. In addition we have 
performed a calculation for SU(4) at a very large value of $\beta$,
which corresponds to $T \sim 1000T_c$, where we expect to
find good agreement with the perturbative result. 

In Table~\ref{table_result} we list the results of our calculations 
of $\Delta S^k_W$. We also list the 2-loop perturbative expectation
obtained from eqn(\ref{eqn_Sk2loop}), using for $g^2(T)$ the 
mean-field improved coupling $g^2_I(a)$ defined in eqn(\ref{eqn_MFI}).

Consider first the values of $\Delta S^k_W$ that we obtain in SU(4) 
at $\beta=20$, which corresponds to $T \sim 1000T_c$. We observe 
excellent agreement with perturbation theory for both $k=2$
and $k=1$ walls. The ratio of the $k=2$ and $k=1$ values therefore
exhibits Casimir Scaling, i.e. is $\propto k(N-k)$.  At such a very 
high $T$ all this is to be expected -- but it provides a reassuring 
check on our calculations.

Moving on to the other SU(4) calculations with $aT=1/L_0=1/4$,
we observe that as $T\to T_c$ the value of $\Delta S^k_W$ grows
much more rapidly than the perturbative prediction. Indeed
at $T\simeq 1.02 T_c$ the discrepancy is very large, a factor 
of $\sim 18$, and  even at  $T\simeq 1.88 T_c$ it is a substantial 
factor $\sim 3$. This is essentially telling us that
$\partial\sigma^k_W/\partial T$ becomes much larger than the 
low-order perturbative expectation as $T\to T_c$. Since at large 
$T$ the domain wall tension tends to its perturbative value,
what this implies is that $\sigma^k_W(T)$ is increasingly
suppressed relative to its perturbative value, 
as we approach $T_c$. Later on in this Section we 
shall provide an estimate of this suppression.

Although we find that $\Delta S^k_W$ deviates strongly from
low-order perturbation theory as $T$ approaches $T_c$, we also see
from Table~\ref{table_result} that the
ratio of $\Delta S^{k=2}_W$ to $\Delta S^{k=1}_W$ continues 
to satisfy Casimir Scaling, to a high precision, even down
to $T\simeq 1.02 T_c$. This is a remarkable and
unexpected result.

All these calculations are at a fixed value of the lattice
spacing in units of $T$, i.e. $aT=1/4$. At this value of $a$
we expect typical lattice spacing corrections to both the
non-perturbative and perturbative
calculations to be small but not negligible (see e.g.
Table~\ref{table_alpha}). Nonetheless it would be useful
to check that the effects we are seeing do not change
significantly in the continuum limit.
To obtain some control over that limit we repeat
our calculations for $L_0=5$, i.e. $aT=1/5$. The leading lattice
corrections should be $O(a^2T^2)$ and so should be reduced
by a factor $\sim 2/3$ when we go from $L_0=4$ to $L_0=5$. 
We see from Table~\ref{table_result} that at $T\simeq 1.88 T_c$ 
we observe a discrepancy
with perturbation theory that is essentially identical to
the one we saw at the coarser lattice spacing. Thus we can infer
that this discrepancy 
is a continuum effect. For $T\simeq 1.02 T_c$ the ratio to the
perturbative result is  $\sim 15$ for $L_0=5$ rather than 
the $\sim 18$ we saw for $L_0=4$, but since this ratio
is varying very rapidly with $T$ when we are this close to $T_c$,
it is not clear whether this apparent variation with $a$  is 
due to lattice 
spacing corrections or is due to a slight mismatch between the 
values of $T/T_c$ at the two values of $a$. In any case it is 
clear that a large and
growing mismatch with perturbation theory as $T\to T_c$ is a 
feature of the continuum theory. Nonetheless we see 
that the ratio of the $\Delta S^k_W$ continues to accurately
satisfy Casimir Scaling when we decrease $a$, so that is 
also a property of the continuum theory.

We now turn to our SU(6) calculations. These have been performed
for $L_0=5$ rather than $L_0=4$ so as to avoid the first order
bulk transition
\cite{oxT05}
that separates the lattice strong and weak coupling regimes
for $N\geq 5$. 
This and the fact that our statistics are smaller 
(the computational cost grows roughly $\propto N^3$) means
that our results are significantly less precise than for SU(4).
Nonetheless it is clear from Table~\ref{table_result} 
that we observe at both values of $T$
precisely the same discrepancy with perturbation theory as we 
saw for SU(4) at the same value of $L_0$. In addition the
$\Delta S^k_W$ ratios continue to satisfy Casimir Scaling
(albeit only at the 2$\sigma$ level at $T\simeq 1.02 T_c$). 
This and the fact that the discrepancy with perturbation
theory is the same as for SU(4) tells us that the domain
wall tension has no factors of $k$ and $N$ except for
the Casimir scaling factor $k(N-k)$ and its dependence on
the 't Hooft coupling,  $g^2 N$. (The latter follows from the
fact that everything is only a function of $T/T_c$ and that 
fixed $T_c$ is obtained by keeping $g^2 N$ fixed at large $N$
\cite{oxT05}.)

Finally we note in Table~\ref{table_result} that our calculations 
of the SU(2) and SU(3) $k=1$ domain wall tensions
show a very similar discrepancy with perturbation theory.
(The SU(2) transition is second order and this prevents us from 
performing reliable domain wall calculations at 
$T\simeq 1.02 T_c$.) Thus the suppression of  $\sigma^k_W(T)$
as $T\to T_c$ is largely independent of $N$.

To illustrate the above, we plot in Fig.~\ref{fig_wallPT} 
the values of $\Delta S^{k=1}_W$ for our $L_0=4$ SU(4) 
calculations, and compare to the two-loop perturbative 
expectation, eqn(\ref{eqn_SkPT2}), using the mean-field
improved coupling, $\beta_I$, as well as the unimproved 
lattice bare coupling $\beta$. (The fact that both the lattice 
and perturbative calculations exhibit Casimir Scaling
means that we would obtain a very similar comparison in SU(6).)
We see the growing discrepancy as $T\to T_c$ that was discussed 
above. We also see a substantial difference between the
two perturbative predictions. It is therefore relevant to ask 
how much of the observed discrepancy
with perturbation theory might be due to the fact that our
choice of coupling or scale is becoming inappropriate on 
the large distance scales associated with the limit $T \sim T_c$. 
To address this question we shall choose some standard `good'
coupling whose running has been evaluated
down to low enough energy scales that we can
extrapolate with some confidence to scales $\sim T_c$.
(This approach is at best qualitative: a proper calculation 
would require us to redo every part of the calculation in 
whatever coupling scheme we use, and this (more formidable)
calculation we do not attempt.) An example of such a coupling
is the Schrodinger functional coupling whose running
has been calculated non-perturbatively in 
\cite{alpha_s}.  
In Appendix A, we describe the evaluation of this coupling
on the length scale $T^{-1} = a L_0$ and we show the result 
in Fig.\ref{fig_alphaSF}. Note that the direct calculation 
\cite{alpha_s} 
goes down to about $T\simeq 2 T_c$ and from there onwards
we extrapolate using the `3-loop' beta function calculated in 
\cite{alpha_s}.  
We use this coupling to obtain $\Delta S^k_W$ and we show the
result in  Fig.~\ref{fig_wallPT}. We see that while using this 
coupling reduces the discrepancy at $T\simeq 1.88T_c$, it
it is still far from large enough to bridge the observed gap
at $T\simeq 1.02T_c$. This suggests that for $T\to T_c$ the 
domain wall tension becomes increasingly dominated by 
non-perturbative physics.

Returning to the evidence for the Casimir Scaling of the 
$k$-wall tensions, we average our accurate SU(4) ratios for 
$aT=0.20, 0.25$ giving 
\begin{equation}
\frac{\Delta S^{k=2}_W}{\Delta S^{k=1}_W}
= \begin{cases}
1.327(26) &:\quad T\simeq 1.02 T_c \\ 
1.345(27) &:\quad T\simeq 1.88 T_c \\
1.341(49) &:\quad T\sim 1000 T_c \\
1.333...  &:\quad \mathrm{Casimir\ Scaling}.
\end{cases}
\label{eqn_CSsu4}
\end{equation}
This (together with the results at $T/T_c=2.3,1.5,1.2$ of
\cite{pdf})
provides impressively accurate evidence for the Casimir
scaling of domain wall tensions (or 't Hooft string tensions)
from $T=\infty$ all the way down to  $T\simeq 1.02T_c$.

The fact that, when $T$ decreases towards $T_c$,
$\partial \sigma^k_W/\partial T$ increases more strongly 
than expected from perturbation theory, tells us that 
$\sigma^k_W$ must decrease more rapidly than expected as 
$T\to T_c$. In principle we can calculate $\sigma^k_W$ as a
function of $T$ by interpolating $\partial \sigma^k_W/\partial\beta$
in $\beta$ and then integrating from large $\beta$, where it
asymptotes to the perturbative value, down to the desired
value of $T$. In practice our calculations are 
not dense enough in $\beta$ to determine the interpolation
unambiguously. (They were not intended for this purpose).
Therefore the best we can do is to obtain  
a qualitative picture by assuming
some functional form for $\partial \sigma^k_W/\partial\beta$ and 
fitting it to the few calculated values we have. 
Choosing the $L_0=4$ SU(4) calculation, we display in 
Fig.~\ref{fig_wallPT} our best fit to $\Delta S_W$ based 
on a simple modification of the one-loop formula to the
$k$-wall free energy: 
\begin{equation}
\frac{F^k_W}{L_1L_2T}
=
[\mathrm{1 \ loop}]
+
a\exp(-b\sqrt{\beta_I-\beta_{Ic}})
\label{eqn_Ffit}
\end{equation}
(The fit is also constrained to go through the very large $\beta$
value which is not shown in the Figure.)
Using this fit we then obtain the the $T$-dependence of $\sigma^k_W$ 
shown in Fig.~\ref{fig_wallT}.  We display the uncertainty
that arises from the errors in the fitted parameters, 
but it is clear that we cannot reliably estimate the systematic 
error inherent in the choice of fitting function. However we
can see from Fig.~\ref{fig_wallPT} that our interpolation is 
conservative in that, if anything, it errs on the side of minimising 
the value of $\partial \sigma_W/\partial T$ in the region 
$T\geq 1.02 T_c$. We also show Fig.~\ref{fig_wallT} in the two-loop
perturbative prediction (using the improved lattice coupling).
The qualitative picture is that $\sigma^k_W$ is strongly
suppressed at $T\leq 2T_c$, well below the perturbative
expectations, to the point of being close to zero
at $T=T_c$. 
This is not inconsistent with our picture of these
domain walls as being squashed but `real' $Z_N$ vortices that 
condense into the vacuum in the confined phase.

\subsection{Profile}
\label{subsection_profile}

As described above, imposing suitable twisted boundary conditions
in the deconfined phase ensures that each lattice field contains 
a single domain wall. (Assuming, as is always the case for us,
that the thermal production of extra pairs of domain walls is
completely negligible.) If we average the Polyakov loop
over the transverse co-ordinates we obtain the profile
$\bar{l}_p(x_3)$. This profile maps the interval $x_3\in [0,L-1]$ 
to a line in the complex plane corresponding to the values that
$\bar{l}_p(x_3)$ takes as $x_3$ runs over the $x_3$-torus. 
This line can be directly compared to the perturbative 
one loop prediction in eqns(\ref{eqn_lpbc}-\ref{eqn_lp3}).
Note that this comparison is independent of the choice
of $g^2(T)$. (As we see from eqn(\ref{eqn_lp2}) the value of the 
coupling determines the rate at which $\bar{l}_p$ runs along the 
line in the complex plane, as $x_3$ is varied, but not the actual
locus of the line.) Thus any disagreement between calculated
and perturbative loci cannot be put down to the use of an
inappropriate coupling or scale. 

In Fig.~\ref{fig_avprofc1} and  Fig.~\ref{fig_avprofc2} we compare 
the one-loop perturbative prediction for the SU(4) $k=1$
locus, to what we obtain at $T\simeq 1.88 T_c$ and at
$T\simeq 1.02 T_c$ respectively, in our lattice calculations. 
Plotting the fluctuating locus for each lattice field 
would clutter the plot so what we show is an average line. 
These results are also characteristic of what we observe for $L_0=5$
and for SU(6) and other $k$. What we observe is not only excellent 
agreement at the very highest $T$ (not shown here), but also,
from Fig.~\ref{fig_avprofc1}, that the 
agreement remains remarkably good down to $T\simeq 1.88 T_c$.
This implies that the shape of the relevant region of
the effective potential remains close to the one-loop 
perturbative form, even at these low $T$. By contrast, once we descend
very close to $T_c$, as in  Fig.~\ref{fig_avprofc2} at 
$T\simeq 1.02 T_c$, the average locus has
moved far from the one-loop curve suggesting that the effective action
is now far from its high-$T$ perturbative form. This observation
makes it all the more remarkable that we continue to see
Casimir Scaling at such very low $T$.

In SU(4) the average $k=2$ wall profile between the $k=0$ and $k=2$
phases, must lie along the
real axis by symmetry, just like the perturbative prediction.
It is interesting to ask whether this is achieved by 
averaging over, say, two sets of tunnelling paths fluctuating
around two degenerate minimum action paths, each away from the 
real axis, but which are related by 
${\mathrm{Im}}{\bar{l}_p} \to -{\mathrm{Im}}{\bar{l}_p}$, 
or if there is only one such minimum tunnelling path, as in 
perturbation
theory. In Fig.~\ref{fig_profc1} and  in Fig.~\ref{fig_profc2}
we show the loci for individual lattice fields containing
a $k=2$ wall, to illustrate
that the latter is the case, both at $T\simeq 1.88 T_c$ and at
$T\simeq 1.02 T_c$. Thus we can expect, as discussed in
Section~\ref{subsection_wetting}, that the $k=N/2$ wall
will indeed split into two confined-deconfined domain walls
precisely at $T=T_c$.

It is also relevant to look at the corresponding plots for
the $k=1$ SU(4) wall. In Fig.~\ref{fig_profc3} we show
what happens at $T\simeq 1.88 T_c$. More interesting
is $T\simeq 1.02 T_c$. We have seen in Fig.~\ref{fig_avprofc2} 
that at this very low $T$ the average locus is deformed so 
as to pass not very far from ${\bar{l}_p}=0$. As we pointed 
out in  Section~\ref{subsection_wetting} we expect 
that as $T\to T_c$ there will be a second minimum action
tunnelling path that passes along the real and imaginary
axes near ${\bar{l}_p}=0$, and which becomes the global
minimum near $T_c$. Does the average in Fig.~\ref{fig_avprofc2}
include some paths that are tracing this alternative trajectory? 
In fact this appears not to be the case, at $T\simeq 1.02 T_c$, 
as we see from Fig.~\ref{fig_profc4}. In the next Section
we shall see that at $T=T_c$ things do in fact become different.

As we have just seen, in practice the fluctuations are small 
enough that we can 
identify quite accurately the profile and hence the centre 
of the domain wall in each lattice field, at any of the 
above values of $T$. We can 
then translate that centre to a common value of $x_3$ and take 
an average over all the lattice fields so as to obtain the averaged
domain wall profile. Let us relabel the position in the 3-direction
by $z$, with the domain wall centered at $z=(L-1)/2$ and 
$z\in [0,L-1]$. In  Fig.~\ref{fig_profz1} we plot the average 
of the real part of the Polyakov loop as a function of $z$
for the $L_0=4$ SU(4) calculation at $T \simeq 1.88T_c$.
Note that for the $k=1$ SU(4) wall the real part is not 
symmetric about the centre:
one must exchange the real and imaginary parts.
We do not show the imaginary part, to avoid cluttering
the plot, but the same analysis and results follow in that case.
We also show curves obtained using the one-loop formula
in eqns(\ref{eqn_lpbc}-\ref{eqn_lp3}) with four different choices 
of coupling. Simply using the bare coupling $g^2=2N/\beta$
does not reproduce the shape at all. The mean-field improved
coupling defined in eqn(\ref{eqn_MFI}) works better but not 
very well. The Schrodinger functional coupling discussed 
in Appendix A works very well indeed.
The last coupling, which is simply defined so as to minimise the
discrepancy between the one-loop formula and the lattice
result, works only slightly better than this. 

At $T\simeq 1.02 T_c$ we already know from  Fig.~\ref{fig_avprofc2}
that no coupling will work well. We illustrate this in
Fig.~\ref{fig_profz2} where we plot the modulus of the Polyakov
loop against $z$ and compare it to the various perturbative
predictions. Here we see the interesting feature that the
`better' the coupling, the worse is its description of the profile.
This is in contrast to what we find in Fig.~\ref{fig_profz3} in
a similar plot at $T\simeq 1.88 T_c$. The reason for this
is that the `better' couplings are larger at lower $T$
than the lattice bare coupling, and eqn(\ref{eqn_lp2}) then
predicts a thinner domain wall in lattice units (recalling that
$T=1/aL_0$ is fixed in lattice units). 
Here however the observed thickness of the wall actually grows 
as we lower $T/T_c$ from 1.88 to 1.02. This would appear to contradict
asymptotic freedom which tells us that $g^2(T)$ grows as 
$T\downarrow$ and hence, by eqn(\ref{eqn_lp2}), that the wall
narrows. This points to the fact that near $T_c$ low-order
perturbation theory is qualitatively, and not just quantitatively,
unreliable. It is interesting to note that precisely the same 
phenomenon is found
\cite{oxT05}
to occur for the lightest mass that couples (strongly) to the 
Polyakov loop. (At high $T$ this is just twice the electric 
screening mass $m_D$.) As we see from eqn(\ref{eqn_mD}), we 
would expect  $m_D$ to approach its high-$T$ linear behaviour
from above since $g^2(T)$ decreases with $T$; however lattice 
calculations show that the approach is from below
\cite{oxT05}.
Since $1/m_D$ is the natural length scale for the thickness 
of the domain wall, this observation ties in with what we 
have found here.

\subsection{Wetting}
\label{subsection_wettingres}

Motivated by the discussion in Section~\ref{subsection_wetting}, 
we have performed a set of calculations
extremely close to $T=T_c$. They are not intended to be
accurate enough to provide a useful estimate of $\Delta S$;
rather we wish to see if the $k$-wall separates into two
$cd$ walls at $T\simeq T_c$ in a Monte Carlo sequence of
several thousand field configurations. We perform such 
calculations for $k=1,2$ walls in SU(4) on a 
$4\times 100 \times 20^2$ lattice and for $k=1,2,3$ walls 
in SU(6) on a $5\times 100 \times 20^2$ lattice. We
use long lattices so that if a $k$-wall breaks up into two 
$cd$ walls there is enough space for two well separated walls.

Our observations are summarised in Table~\ref{table_wetsu4}
and Table~\ref{table_wetsu6}. The first thing we note is that 
the transition from no sign of wetting to the complete splitting 
of the domain walls takes place over a very small range
of $T$ which is precisely around $T=T_c$. (See eqn(\ref{eqn_bc}).)
For $k{\not =}N/2$ we first have a profile
that is well away from the origin (`no'). Then at lower $T$ we 
typically see fluctuations of the wall profile which deform
it so that it sometimes passes through the 
origin (`via origin') but does not visibly break up. If we
lower $T$ further the wall fluctuates more strongly, so that the 
profile not only goes through the origin, but the wall splits, 
reforms, and again reassumes a profile well away from the origin
(`mixed'). Finally, at lower $T$ still, the $k$-wall
splits completely (`splits'). All this takes place over
an interval $\delta T/T_c \sim 0.005 $. 
For the $k=N/2$ walls not all these
stages are identifiable because the profile normally
passes through the origin. But a precurser to the
separation, a central `flattening', can be identified
and presumably overlaps with the `via origin' and
`mixed' stages for $k{\not =}N/2$.

It is interesting to ask what happens to the $k=1$ SU(4) profile,
shown in Fig.~\ref{fig_profc4} for $T\simeq 1.02 T_c$, when
we lower the temperature to $T\simeq T_c$. To illustrate
this we show in Fig.~\ref{fig_profc5} a set of 10 profiles obtained
over a sequence of 1000 Monte Carlo sweeps and in Fig.~\ref{fig_profc6}
over a later sequence of 1000 sweeps, all at $T\simeq 1.006 T_c$. 
Each locus has been obtained by averaging over 
subsequence of 100 field configurations. We observe
that the domain walls in Fig.~\ref{fig_profc5} have profiles
that fluctuate around an average locus that is similar to the 
one we saw in Fig.~\ref{fig_profc3}. The profiles in
Fig.~\ref{fig_profc6}, on the other hand, fluctuate around
a quite different locus that passes through $\bar{l}_p =0$
and which is identical to the minimum $k=2$ path
up to the $\pi/2$ rotation at the origin. What we appear to see is
a tunnelling between these two paths and this strongly suggests
that the effective potential has two minimal paths separated
by a ridge through which the system has to tunnel. Once it has
tunnelled to the set of paths passing through $\bar{l}_p =0$
it will readily split into two $cd$ walls at $T=T_c$. 
One would expect that the surface tension of the $k=1$ and
$k=2$ walls that pass through  $\bar{l}_p =0$ will be approximately
the same rather than being related by Casimir Scaling. The
statistics of our calculations near $T=T_c$ do not allow a
direct test of this reasonable conjecture.
  
We observe that, within our resolution, the wetting of the  
SU(4) $k=2$ and $k=1$ walls appears to occur simultaneously
and it appears to occur at $T=T_c$. This would appear to tell 
us that $\sigma^k_W(T_c) {\not <} 2\sigma_{cd}(T_c)$ for both
$k=1$ and $k=2$ at $N=4$. For SU(6) there is perhaps
some evidence that the $k=2$ and $k=3$ walls begin to split slightly 
before the $k=1$ wall. If this hint were to survive a more careful
calculation, then we would have the beginnings of a
calculation of the $N$-dependence of $\sigma_{cd}$ as
oulined earlier. 

The fact is that all the $k$-walls appear 
to break up at almost the same value of $T$ again raises
the  question whether this can be consistent with the $k$-wall 
tensions being different and related by Casimir Scaling
as $T\to T_c$, since in that case it is clear that one would 
expect the higher-$k$ walls to split at a higher $T$ than
the lower-$k$ walls.

To address this question and also to
understand why everything occurs in such a very narrow range
of $T$ around $T_c$, we need to look at the magnitudes of
the quantities involved. First consider the `natural' magnitude 
of $\sigma^k_W$ near $T=T_c$ as given by the one loop part 
of eqn(\ref{eqn_sigkPT}) and using a realistic coupling. Plugging in
the value that the Schrodinger functional coupling takes at $T_c$ 
in SU(3), $g(T_c) \simeq 2.2$ and assuming $g^2(T_c)N$ is constant,
we obtain 
\begin{equation}
\sigma_W^k(\mathrm{1-loop})
=
k(N-k)
\frac{4\pi^2}{3\sqrt{3}}
\frac{T_c^3}{\sqrt{g^2(T)N}}
\simeq
2 k(N-k) T_c^3 .
\label{eqn_sigknat}
\end{equation}
We now compare it to the latent heat
\cite{oxT05}
\begin{equation}
\frac{L_h}{T_c}
\simeq
0.35 N^2 T_c^3 + O(\frac{1}{N^2}).
\label{eqn_Lhnat}
\end{equation}
It is natural to compare $L_h$ divided by $T_c$ to $\sigma_W^k$
because the latter is a free energy per unit area that is
effectively integrated over the thickness of the wall, which
is $O(1/T_c)$. We observe from eqn(\ref{eqn_sigknat})
that for the maximal wall, $k=N/2$,
$\sigma_W^k(\mathrm{1-loop})\simeq 0.5 N^2 T_c^2$ and 
this is of the same order as the value of $L_h/T_c$. This seems 
reasonable, given that there is one scale in the problem. 
Now consider $\sigma_{cd}$. Although, as we have emphasised,
it is not known if it increases as $N^2$, such a fit is
certainly possible, and one then finds
\cite{oxT05}
\begin{equation}
\sigma_{cd}
\simeq
(0.014 N^2 - 0.104) T_c^3.
\label{eqn_sigcdnat}
\end{equation}
This is very small compared to the previous two quantities;
e.g. for SU(4) it is about a factor of 50 smaller than $L_h/T_c$.
The reason for this mismatch between these two energy scales is 
not entirely clear (but see below). Nonetheless, as we shall now argue,
it is what leads to the onset of wetting occuring over such a 
narrow range of $T$ around $T_c$.

To be explicit we consider the SU(4) calculation with $L_0=4$. As we 
have seen in Fig.~\ref{fig_wallT} the actual value of $\sigma_W^k$
drops well below the one-loop value in eqn(\ref{eqn_sigknat})
once $T \leq 2T_c$ and it collapses to a value consistent
with zero as $T\to T_c$. Nonetheless we have seen that 
neither of the $k$-walls shows any sign of splitting into two 
$cd$ walls even at $T\simeq 1.02 T_c$. This is also true
even closer to $T_c$, as shown in Table~\ref{table_wetsu4},
but the relevant runs are much shorter there. Consider first
the $k=2$ wall. Consider a fluctuation just above $T_c$
in which it breaks up into two $cd$ walls separated by a
region of confining phase. For this region of confining phase
to be clearly distinguishable, it needs to be of some minimal 
size, presumably $O(1/T_c)$ long. Assuming (as we have argued
above) that  $\sigma_W^{k=N/2} \simeq 2\sigma_{cd}$ at and near
$T_c$, the probability of such a fluctuation will simply
be due to the free energy cost, $\Delta F_{cd}$, of the region 
of confining phase. Approximating  this by the latent heat, $L_h$,
multiplied by the area $A$ and the length $1/T_c$,
one has a suppression
\begin{equation}
\propto
\exp\biggl\{-\frac{L_h A}{T T_c}\Bigl\{1-\frac{T_c}{T}\Bigr\}\biggr\}
\stackrel{T\simeq T_c}{\simeq}
\exp\biggl\{-\frac{L_h}{T^4_c}\frac{L_1L_2}{L^2_0}
\Bigl\{1-\frac{T_c}{T}\Bigr\}\biggr\}
\label{eqn_pk2wetT}
\end{equation}
Plugging in the value for $L_h/T^4_c$ in eqn(\ref{eqn_Lhnat}) 
and $L_1=L_2=5L_0$ we 
see that the range of $T/T_c$ in which such wetting fluctuations have 
a significant probability is very small and on the order of
\begin{equation}
\frac{T-T_c}{T_c}
\leq
\frac{1}{140}.
\label{eqn_rTk2wet}
\end{equation}
Consider now a $k=1$ wall in our SU(4) calculation. We do not 
know enough about its dynamics to analyse the situation precisely,
but we can outline a plausible scenario within which we can
see why the splitting takes place very close to $T_c$.
(The splitting essentially refers to the tunnelling
exemplified in Fig.\ref{fig_profc5} and Fig.\ref{fig_profc6}.)
If the $k=2$ wall splits into two $cd$ walls at $T=T_c$
we have $\sigma_W^{k=N/2}(T_c) = 2\sigma_{cd}(T_c)$ and 
and if we also have Casimir Scaling then 
$\sigma_W^{k=1}(T_c) = 4/3\times \sigma_{cd}(T_c)$.
and the $k=1$ wall will not split at $T=T_c$ but only at a lower
$T$ where the free energy advantage of a layer of confining phase
provides the required extra $2/3\times \sigma_{cd}(T_c)$
needed for splitting. Assuming a region of confining phase of 
length $O(1/T_c)$ and neglecting the variation with $T$ of
both $\sigma_W^{k=1}$ and $\sigma_{cd}$ leads to an estimate
for the $T$ where the $k=1$ wall splits of
\begin{equation}
\frac{T_c-T}{T_c}
\sim
\frac{2}{3}\frac{\sigma_{cd}}{L_h/T_c}
\sim
0.013
\label{eqn_k2dt}
\end{equation}
using eqn(\ref{eqn_sigcdnat}) and eqn(\ref{eqn_Lhnat}).
This is a very crude estimate, of course, but it serves to
illustrate how the large ratio of the energy scales associated
with $L_h/T_c$ and $\sigma_{cd}$ mean that the difference
between the values of $T$ at which the $k=2$ and
$k=1$ walls break up is, assuming Casimir Scaling, very small
and on the order of the resolution of our calculations.
That is to say, our calculations do not in fact provide
evidence against Casimir Scaling persisting at all $T$
(until the walls break up). A similar argument will explain 
why the process of breaking up occurs over a very narrow
range of $T$ around $T_c$. It also implies that any attempt
to verify dynamical `perfect wetting' must have an extremely
fine resolution in $T$ near $T_c$, and particularly so for
SU(3) where the ratio $L_h/T_c\sigma_{cd}$ is even larger.
 
We have ignored the fact that both $\sigma_{cd}$ and $\sigma_W^{k=1}$ 
will vary with $T$. This variation, as $T$ moves away from $T_c$, 
is likely to be driven by the growing free energy difference
between the two phases and so is likely again to be on the
order of $L_h/T_c\times (T-T_c)/T_c$, in which case we will
obtain a result similar to the one above. Of course, as we
move away from $T_c$ either way, we will eventually encounter
the spinodal point at which one or both types of wall
ceases to exist.

We briefly return to the question why $\sigma_{cd}(T_c)$ is so 
small compared to $L_h/T_c$. At $T=T_c$ there is a cancellation
of differences in entropy and internal energies between the
confined and deconfined phases. Nonetheless one might have expected
the `hills' and `passes' in the effective potential at$T=T_c$ to 
have a natural scale of $O(L_h)$. This would imply a value
for $\sigma_W^{k}(T_c)$ and $\sigma_{cd}(T_c)$ that is
$O(L_h/T_c)$, which is consistent with low-order perturbation theory, 
but not with what we actually find to be the case. Clearly our
expectation is mistaken: the `passes' are also strongly
suppressed. As we move a little away
from $T_c$ the free energy density difference between the phases
increases roughly like $\sim L_h |1-T/T_c|$. The behaviour of the
domain walls suggests that the whole effective potential, not just
the wrong vacuum, scales something like this: $\sigma_{cd}(T_c)$ and
$\sigma_W^{k}(T_c)$ are small because the height of the `pass'
which the tunnelling path traverses is also small -- presumably
the same dynamics as the free energy cancellation. As we move
away from $T_c$ the height of the pass also grows roughly as  
$\sim L_h |1-T/T_c|$. This suggests that  $\sigma_W^{k}(T)$
will grow in this way as $T$ increases above $T_c$ -- qualitatively
in accord with what we see in Fig.\ref{fig_wallT}. 

This might also suggest a scenario in which
both $\sigma_{cd}(T)$ and $\sigma_W^{k}(T)$ have a minimum at 
$T=T_c$ but this requires  the effective potential to
be very similar in shape above and below $T_c$ (with just the
depth of the minima reversed) which hardly seems likely.

The fact that the $k$-walls do break up only at $T\simeq T_c$
does tell us that $\sigma^k_W$ cannot be very much larger than
$2\sigma_{cd}$ once we are very near to $T=T_c$. This confirms   
the strong suppression of $\sigma^k_W$ indicated in 
Fig.~\ref{fig_wallPT}. We note that this appears to be an effect 
that survives the $N\to\infty$ limit and provides further evidence
that even in that limit the deconfined phase is far from being 
a weakly coupled gluon plasma in the region from $T_c$ to several 
times $T_c$. This fits in with the observation that the
SU(3) pressure anomaly is also a large-$N$ effect
\cite{bbmt1}.

\subsection{'t Hooft string condensation}
\label{subsection_stringcond}

The fact that $\sigma_W^k(T)$ decreases rapidly towards
a value that is close to zero, as we decrease $T$ towards $T_c$,
tells us that the spatial 't Hooft string tension, 
$\tilde{\sigma}_k(T) = \sigma_W^k(T)/T$, also decreases.

Since the deconfining transition is robustly first order 
(for larger $N$) we can use its metastability to discuss 
what happens in the deconfined phase for at least some
range of temperatures $T<T_c$. 
Doing so, the observed rapid decrease suggests that 
$\tilde{\sigma}_k = 0$
at some temperature $T_{\tilde{H}}$ that is close to and below $T_c$.
This suggests that as we decrease $T$ in the deconfined phase,
then at $T=T_{\tilde{H}}$ we would encounter a second-order 
phase transition driven by the condensation of spatial 't Hooft 
strings -- if it were not for the intervening presence of the first
order transition to the (more-or-less) separate confining ensemble 
of energy eigenstates of the theory. 

This is reminiscent of 
the expectation that as one increases $T$ within the confining phase,
there would be a second order Hagedorn-like string condensation 
transition at $T=T_H$, if it were not, once again, for the
intervening presence 
of the first order transition to the (more-or-less) separate 
`gluon gas' ensemble of energy eigenstates of the theory. Indeed
one can use the metastability of this first order transition
to obtain direct evidence for this would-be second order transition
and to estimate its location at $T= T_H > T_c$, as has been done in 
\cite{bbmt2}. 

The condensation of confining strings as $T\to T^-_H$, leads to
deconfinement, but without reference to the `gluon plasma'. Similarly
one can see how the condensation of (spatial) 't\nolinebreak Hooft strings 
as $T\to T^-_{\tilde{H}}$, leads to confinement, but without 
reference to the usual confining phase. As the domain wall
tension goes to zero, the vacuum will break down into a
gas of bubbles containing different $Z_N$ phases so that the
Polyakov loop, $\bar{l}_p$, is zero.

All this hints at some kind of a duality
between the confined and deconfined phases.

\section{Conclusions}
\label{section_conclusions}

We have shown in this paper that the surface tensions of 
the domain walls separating different deconfined phases
in the Euclidean formulation of finite-$T$ SU($N$) satisfy
Casimir Scaling, i.e.
\begin{equation}
\sigma^k_W
=
k(N-k)\, f(g^2(T)N,T/T_c)\, T^3 ,
\label{eqn_factor1}
\end{equation}
to a good approximation. We have shown this to be the case
not only at very high $T$ where our calculations merely reproduce 
the perturbative predictions
\cite{CKA1,CKA2}, 
but all the way down to $T\simeq 1.02 T_c$ where, not surprisingly,
both the wall profile and the magnitude of the surface tension 
are very far from perturbation theory. It appears that the
Casimir Scaling predicted by low-order perturbation theory is
in fact a much more general property of these domain walls and,
equivalently, the spatial 't Hooft string tension.

Our numerical results are for $\partial \sigma^k_W/ \partial T$
rather than for $\sigma^k_W$ itself, but we can interpolate the
former and integrate it so as to estimate the $T$ dependence 
of the latter. We have too few values of 
$\partial \sigma^k_W/ \partial T$ to be able to do so with
any precision, however the  
striking qualitative feature is that as we reduce $T$ 
below, say, $2T_c$ the surface tension becomes strongly
suppressed relative to its perturbative value, reaching values 
close to zero at $T\simeq T_c$.  

We have also performed calculations near and 
passing through $T_c$. Here, in a very narrow range of 
temperatures around $T_c$, we see all the $k$-walls splitting
into confined-deconfined walls. The fact that the Polyakov loop 
profiles of $k=N/2$ domain walls are found to pass through
zero, at any $T$, (almost) inevitably leads to the `perfect wetting'
at $T=T_c$ of such walls. The reason that all the $k$-walls `wet'
at values of $T$ very close to $T_c$, is intimately linked
with the fact that the latent heat, which provides a measure of
how the confining-deconfining free energy difference increase with 
$T-T_c$, is very much larger than the surface tension, $\sigma_{cd}$, 
of the confined-deconfined domain wall. This means that to test
ideas of `perfect wetting' for $k{\not =} N/2$ domain walls,
e.g. the usual domain walls in SU(3), will require calculations 
with a very fine resolution in $T$ -- certainly finer than ours,
and quite possibly finer than any previous calculations.

There are a number of ways in which our calculations
could be usefully improved upon. Firstly, the comparison
with perturbation theory would be under much better control
if we were to have two-loop calculations of $\sigma^k_W$ in 
the various coupling schemes that we have tried.
Turning to our lattice calculations, while these 
do allow us some control of the
continuum limit and the large-$N$ limit, it is clear that
in both respects our calculations could be significantly,
and usefully, improved upon. Calculations at more values of
$T$ would transform our very crude reconstruction of
the value of $\sigma^k_W(T)$ into something much more
controlled and credible. In addition it would be
very useful to analyse the Polyakov loop profiles, and hence 
the effective action, not just in terms
of the trace, but in terms of the fuller range of gauge invariant
variables provided by the eigenvalues of the Polyakov loop.
This would allow us to transform our plausible inferences
into strict derivations (or the opposite, as the case may be).

We have noted that the rapid decrease of both 
$\sigma_W^k(T)$ and the  spatial 't Hooft 
string tension, $\tilde{\sigma}_k(T) = \sigma_W^k(T)/T$,
as we decrease $T$ towards $T_c$, suggests the presence of 
what would be a second order phase transition a little below 
$T_c$ if  it were not for the intervening presence of the first
order transition. This is reminiscent of    
the conjecture that as one increases $T$ within the confining phase,
the condensation of confining strings would occur, a little above
$T_c$,  if it were not, 
once again, for the intervening presence of the first order
transition. Just as has been done in the latter case 
\cite{bbmt2}
it would be interesting to use the metastability of the first 
order transition to go beyond $T_c$ for direct evidence of the 
second order transition. It would also be very interesting
to perform the sort of calculations we have done, but more densely in
$T$, so as to determine the functional form of  $\tilde{\sigma}_k(T)$
much more reliably and precisely. This would help us to
explore the possibility of a duality between the behaviours
as $T\to T^-_H$ and $T\to T^+_{\tilde{H}}$.

\section*{Acknowledgements}

Our lattice calculations were carried out on PPARC 
and EPSRC funded computers in Oxford Theoretical 
Physics. FB acknowledges the support of a PPARC
graduate studentship.

\section*{Appendix A}

When $T\to T_c$ the coupling $g^2(T)$ becomes large and our 
choice of the mean-field improved coupling evaluated at $a=1/L_0T$, 
i.e. $g^2_I(a)$, may become inadequate. To estimate the range of
uncertainty introduced by this choice, we calculate the 
perturbative expression for the wall tension in 
Section~\ref{subsection_walltension} with a `good' coupling
at the scale $T^{-1} = aL_0$.

We need a coupling that has been explicitly calculated to
length scales that are quite large. For this purpose we shall
use the calculation in
\cite{alpha_s}.
This calculation is for SU(3) and we shall therefore perform
our analysis for $N=3$. The coupling in
\cite{alpha_s}
has been extrapolated to the continuum limit, but we shall
apply it to a finite $a$, assuming the corrections are
insignificant for our purposes.

The running of the coupling $\bar{g}$ in
\cite{alpha_s}
is given by
\begin{equation}
- l \frac{\partial \bar{g}}{\partial l}
=
\beta(\bar{g})
=
- b_o \bar{g}^3 - b_1 \bar{g}^5- b_2^{eff} \bar{g}^7 
\label{eqn_alpha1}
\end{equation}
where
\begin{equation}
b_0 = 11(4\pi)^{-2}, \quad 
b_1 = 102(4\pi)^{-4}, \quad 
b_2^{eff} = 1.5(8) (4\pi)^{-3}.
\label{eqn_alpha2}
\end{equation}
(Here $b_2^{eff}$ is a fitted effective coefficient that
subsumes higher order corrections and is not intended to
be an estimate of the coefficient $b_2$ of the 
$\beta$-function in this coupling scheme.)  
The normalisation of the coupling is then fixed by
its value at the largest length scale at which it is
calculated
\begin{equation}
\bar{g}^2(l_m) 
=
3.48
\quad; \quad 
l_m = 4 a(\beta_0=5.904).
\label{eqn_alpha3}
\end{equation}
This length scale is $l_m \simeq 2/3T_c$ which is not 
quite as large as we would like but is large enough 
that extrapolating it to  $l=1/T_c$ might be qualitatively
reliable.

Defining the reduced length 
\begin{equation}
\hat{l} = \frac{l}{l_m}
\label{eqn_alpha4}
\end{equation}
we solve eqns(\ref{eqn_alpha1},\ref{eqn_alpha2}) numerically,
with the boundary condition of eqn(\ref{eqn_alpha3}), so
obtaining $\bar{g}^2(\hat{l})$ for any desired value
of $\hat{l}$. The quantity that appears in eqn(\ref{eqn_SkPT1}) and
which we now need to evaluate is 
\begin{equation}
\frac{\partial}{\partial\beta}
\frac{1}{\bar{g}(\hat{l})} 
=
- \frac{1}{\bar{g}^2(\hat{l})}
\frac{\partial\bar{g}(\hat{l})}{\partial\hat{l}}
\frac{\partial\hat{l}}{\partial\beta}.
\label{eqn_alpha5}
\end{equation}
We will perform the calculation for a lattice with $L_0=4$ 
and with varying $T=1/a(\beta)L_0$. So we evaluate the coupling 
at 
\begin{equation}
\hat{l} 
= 
\frac{l}{l_m}
=
\frac{L_0 a(\beta)}{L_0 a(\beta_0)}
=
\frac{a\sqrt{\sigma}(\beta)}{a\sqrt{\sigma}(\beta_0)}.
\label{eqn_alpha6}
\end{equation}
We now use the expression for $a\sqrt{\sigma}(\beta)$
as a function of $\beta$ that is provided in Table~5 of 
\cite{oxT05}
for the range $5.6925 \leq \beta \leq 6.338$, to evaluate
both $\hat{l}$ in eqn(\ref{eqn_alpha6}) and 
$\partial\hat{l}/\partial\beta$ in eqn(\ref{eqn_alpha5})
in this range of $\beta$. (Like any multi-parameter
interpolation, this cannot be assumed to be useful
outside the range in which it was fitted.) 

We now have all the ingredients to calculate each of the 
factors on the rhs of  eqn(\ref{eqn_alpha5}). The first
factor comes by plugging in the calculated value of $\hat{l}$ 
into the numerically determined $\bar{g}^2(\hat{l})$.
The second factor is just the $\beta$-function in
eqn(\ref{eqn_alpha1}) evaluated with this  $\bar{g}^2(\hat{l})$.
To obtain the third factor one differentiates the 
interpolating formula in Table~5 of 
\cite{oxT05}
and evaluates it at the desired value of $\beta$, and
normalises it to the formula at $\beta_0 = 5.904$.
Since $T_c = 1/4a(\beta_c)$ for $\beta_c = 5.6924(2)$
\cite{blmtuw-T03}
and this lies within our interpolation range, we are able 
to perform this calculation down to $T=T_c$. To transform
this coupling to one for SU(4) we simply assume that
$g^2N$ is fixed at fixed $T$.

\begin	{figure}[p]
\begin	{center}
\leavevmode
\begingroup%
  \makeatletter%
  \newcommand{\GNUPLOTspecial}{%
    \@sanitize\catcode`\%=14\relax\special}%
  \setlength{\unitlength}{0.1bp}%
{\GNUPLOTspecial{!
/gnudict 256 dict def
gnudict begin
/Color false def
/Solid false def
/gnulinewidth 5.000 def
/userlinewidth gnulinewidth def
/vshift -33 def
/dl {10 mul} def
/hpt_ 31.5 def
/vpt_ 31.5 def
/hpt hpt_ def
/vpt vpt_ def
/M {moveto} bind def
/L {lineto} bind def
/R {rmoveto} bind def
/V {rlineto} bind def
/vpt2 vpt 2 mul def
/hpt2 hpt 2 mul def
/Lshow { currentpoint stroke M
  0 vshift R show } def
/Rshow { currentpoint stroke M
  dup stringwidth pop neg vshift R show } def
/Cshow { currentpoint stroke M
  dup stringwidth pop -2 div vshift R show } def
/UP { dup vpt_ mul /vpt exch def hpt_ mul /hpt exch def
  /hpt2 hpt 2 mul def /vpt2 vpt 2 mul def } def
/DL { Color {setrgbcolor Solid {pop []} if 0 setdash }
 {pop pop pop Solid {pop []} if 0 setdash} ifelse } def
/BL { stroke userlinewidth 2 mul setlinewidth } def
/AL { stroke userlinewidth 2 div setlinewidth } def
/UL { dup gnulinewidth mul /userlinewidth exch def
      10 mul /udl exch def } def
/PL { stroke userlinewidth setlinewidth } def
/LTb { BL [] 0 0 0 DL } def
/LTa { AL [1 udl mul 2 udl mul] 0 setdash 0 0 0 setrgbcolor } def
/LT0 { PL [] 1 0 0 DL } def
/LT1 { PL [4 dl 2 dl] 0 1 0 DL } def
/LT2 { PL [2 dl 3 dl] 0 0 1 DL } def
/LT3 { PL [1 dl 1.5 dl] 1 0 1 DL } def
/LT4 { PL [5 dl 2 dl 1 dl 2 dl] 0 1 1 DL } def
/LT5 { PL [4 dl 3 dl 1 dl 3 dl] 1 1 0 DL } def
/LT6 { PL [2 dl 2 dl 2 dl 4 dl] 0 0 0 DL } def
/LT7 { PL [2 dl 2 dl 2 dl 2 dl 2 dl 4 dl] 1 0.3 0 DL } def
/LT8 { PL [2 dl 2 dl 2 dl 2 dl 2 dl 2 dl 2 dl 4 dl] 0.5 0.5 0.5 DL } def
/Pnt { stroke [] 0 setdash
   gsave 1 setlinecap M 0 0 V stroke grestore } def
/Dia { stroke [] 0 setdash 2 copy vpt add M
  hpt neg vpt neg V hpt vpt neg V
  hpt vpt V hpt neg vpt V closepath stroke
  Pnt } def
/Pls { stroke [] 0 setdash vpt sub M 0 vpt2 V
  currentpoint stroke M
  hpt neg vpt neg R hpt2 0 V stroke
  } def
/Box { stroke [] 0 setdash 2 copy exch hpt sub exch vpt add M
  0 vpt2 neg V hpt2 0 V 0 vpt2 V
  hpt2 neg 0 V closepath stroke
  Pnt } def
/Crs { stroke [] 0 setdash exch hpt sub exch vpt add M
  hpt2 vpt2 neg V currentpoint stroke M
  hpt2 neg 0 R hpt2 vpt2 V stroke } def
/TriU { stroke [] 0 setdash 2 copy vpt 1.12 mul add M
  hpt neg vpt -1.62 mul V
  hpt 2 mul 0 V
  hpt neg vpt 1.62 mul V closepath stroke
  Pnt  } def
/Star { 2 copy Pls Crs } def
/BoxF { stroke [] 0 setdash exch hpt sub exch vpt add M
  0 vpt2 neg V  hpt2 0 V  0 vpt2 V
  hpt2 neg 0 V  closepath fill } def
/TriUF { stroke [] 0 setdash vpt 1.12 mul add M
  hpt neg vpt -1.62 mul V
  hpt 2 mul 0 V
  hpt neg vpt 1.62 mul V closepath fill } def
/TriD { stroke [] 0 setdash 2 copy vpt 1.12 mul sub M
  hpt neg vpt 1.62 mul V
  hpt 2 mul 0 V
  hpt neg vpt -1.62 mul V closepath stroke
  Pnt  } def
/TriDF { stroke [] 0 setdash vpt 1.12 mul sub M
  hpt neg vpt 1.62 mul V
  hpt 2 mul 0 V
  hpt neg vpt -1.62 mul V closepath fill} def
/DiaF { stroke [] 0 setdash vpt add M
  hpt neg vpt neg V hpt vpt neg V
  hpt vpt V hpt neg vpt V closepath fill } def
/Pent { stroke [] 0 setdash 2 copy gsave
  translate 0 hpt M 4 {72 rotate 0 hpt L} repeat
  closepath stroke grestore Pnt } def
/PentF { stroke [] 0 setdash gsave
  translate 0 hpt M 4 {72 rotate 0 hpt L} repeat
  closepath fill grestore } def
/Circle { stroke [] 0 setdash 2 copy
  hpt 0 360 arc stroke Pnt } def
/CircleF { stroke [] 0 setdash hpt 0 360 arc fill } def
/C0 { BL [] 0 setdash 2 copy moveto vpt 90 450  arc } bind def
/C1 { BL [] 0 setdash 2 copy        moveto
       2 copy  vpt 0 90 arc closepath fill
               vpt 0 360 arc closepath } bind def
/C2 { BL [] 0 setdash 2 copy moveto
       2 copy  vpt 90 180 arc closepath fill
               vpt 0 360 arc closepath } bind def
/C3 { BL [] 0 setdash 2 copy moveto
       2 copy  vpt 0 180 arc closepath fill
               vpt 0 360 arc closepath } bind def
/C4 { BL [] 0 setdash 2 copy moveto
       2 copy  vpt 180 270 arc closepath fill
               vpt 0 360 arc closepath } bind def
/C5 { BL [] 0 setdash 2 copy moveto
       2 copy  vpt 0 90 arc
       2 copy moveto
       2 copy  vpt 180 270 arc closepath fill
               vpt 0 360 arc } bind def
/C6 { BL [] 0 setdash 2 copy moveto
      2 copy  vpt 90 270 arc closepath fill
              vpt 0 360 arc closepath } bind def
/C7 { BL [] 0 setdash 2 copy moveto
      2 copy  vpt 0 270 arc closepath fill
              vpt 0 360 arc closepath } bind def
/C8 { BL [] 0 setdash 2 copy moveto
      2 copy vpt 270 360 arc closepath fill
              vpt 0 360 arc closepath } bind def
/C9 { BL [] 0 setdash 2 copy moveto
      2 copy  vpt 270 450 arc closepath fill
              vpt 0 360 arc closepath } bind def
/C10 { BL [] 0 setdash 2 copy 2 copy moveto vpt 270 360 arc closepath fill
       2 copy moveto
       2 copy vpt 90 180 arc closepath fill
               vpt 0 360 arc closepath } bind def
/C11 { BL [] 0 setdash 2 copy moveto
       2 copy  vpt 0 180 arc closepath fill
       2 copy moveto
       2 copy  vpt 270 360 arc closepath fill
               vpt 0 360 arc closepath } bind def
/C12 { BL [] 0 setdash 2 copy moveto
       2 copy  vpt 180 360 arc closepath fill
               vpt 0 360 arc closepath } bind def
/C13 { BL [] 0 setdash  2 copy moveto
       2 copy  vpt 0 90 arc closepath fill
       2 copy moveto
       2 copy  vpt 180 360 arc closepath fill
               vpt 0 360 arc closepath } bind def
/C14 { BL [] 0 setdash 2 copy moveto
       2 copy  vpt 90 360 arc closepath fill
               vpt 0 360 arc } bind def
/C15 { BL [] 0 setdash 2 copy vpt 0 360 arc closepath fill
               vpt 0 360 arc closepath } bind def
/Rec   { newpath 4 2 roll moveto 1 index 0 rlineto 0 exch rlineto
       neg 0 rlineto closepath } bind def
/Square { dup Rec } bind def
/Bsquare { vpt sub exch vpt sub exch vpt2 Square } bind def
/S0 { BL [] 0 setdash 2 copy moveto 0 vpt rlineto BL Bsquare } bind def
/S1 { BL [] 0 setdash 2 copy vpt Square fill Bsquare } bind def
/S2 { BL [] 0 setdash 2 copy exch vpt sub exch vpt Square fill Bsquare } bind def
/S3 { BL [] 0 setdash 2 copy exch vpt sub exch vpt2 vpt Rec fill Bsquare } bind def
/S4 { BL [] 0 setdash 2 copy exch vpt sub exch vpt sub vpt Square fill Bsquare } bind def
/S5 { BL [] 0 setdash 2 copy 2 copy vpt Square fill
       exch vpt sub exch vpt sub vpt Square fill Bsquare } bind def
/S6 { BL [] 0 setdash 2 copy exch vpt sub exch vpt sub vpt vpt2 Rec fill Bsquare } bind def
/S7 { BL [] 0 setdash 2 copy exch vpt sub exch vpt sub vpt vpt2 Rec fill
       2 copy vpt Square fill
       Bsquare } bind def
/S8 { BL [] 0 setdash 2 copy vpt sub vpt Square fill Bsquare } bind def
/S9 { BL [] 0 setdash 2 copy vpt sub vpt vpt2 Rec fill Bsquare } bind def
/S10 { BL [] 0 setdash 2 copy vpt sub vpt Square fill 2 copy exch vpt sub exch vpt Square fill
       Bsquare } bind def
/S11 { BL [] 0 setdash 2 copy vpt sub vpt Square fill 2 copy exch vpt sub exch vpt2 vpt Rec fill
       Bsquare } bind def
/S12 { BL [] 0 setdash 2 copy exch vpt sub exch vpt sub vpt2 vpt Rec fill Bsquare } bind def
/S13 { BL [] 0 setdash 2 copy exch vpt sub exch vpt sub vpt2 vpt Rec fill
       2 copy vpt Square fill Bsquare } bind def
/S14 { BL [] 0 setdash 2 copy exch vpt sub exch vpt sub vpt2 vpt Rec fill
       2 copy exch vpt sub exch vpt Square fill Bsquare } bind def
/S15 { BL [] 0 setdash 2 copy Bsquare fill Bsquare } bind def
/D0 { gsave translate 45 rotate 0 0 S0 stroke grestore } bind def
/D1 { gsave translate 45 rotate 0 0 S1 stroke grestore } bind def
/D2 { gsave translate 45 rotate 0 0 S2 stroke grestore } bind def
/D3 { gsave translate 45 rotate 0 0 S3 stroke grestore } bind def
/D4 { gsave translate 45 rotate 0 0 S4 stroke grestore } bind def
/D5 { gsave translate 45 rotate 0 0 S5 stroke grestore } bind def
/D6 { gsave translate 45 rotate 0 0 S6 stroke grestore } bind def
/D7 { gsave translate 45 rotate 0 0 S7 stroke grestore } bind def
/D8 { gsave translate 45 rotate 0 0 S8 stroke grestore } bind def
/D9 { gsave translate 45 rotate 0 0 S9 stroke grestore } bind def
/D10 { gsave translate 45 rotate 0 0 S10 stroke grestore } bind def
/D11 { gsave translate 45 rotate 0 0 S11 stroke grestore } bind def
/D12 { gsave translate 45 rotate 0 0 S12 stroke grestore } bind def
/D13 { gsave translate 45 rotate 0 0 S13 stroke grestore } bind def
/D14 { gsave translate 45 rotate 0 0 S14 stroke grestore } bind def
/D15 { gsave translate 45 rotate 0 0 S15 stroke grestore } bind def
/DiaE { stroke [] 0 setdash vpt add M
  hpt neg vpt neg V hpt vpt neg V
  hpt vpt V hpt neg vpt V closepath stroke } def
/BoxE { stroke [] 0 setdash exch hpt sub exch vpt add M
  0 vpt2 neg V hpt2 0 V 0 vpt2 V
  hpt2 neg 0 V closepath stroke } def
/TriUE { stroke [] 0 setdash vpt 1.12 mul add M
  hpt neg vpt -1.62 mul V
  hpt 2 mul 0 V
  hpt neg vpt 1.62 mul V closepath stroke } def
/TriDE { stroke [] 0 setdash vpt 1.12 mul sub M
  hpt neg vpt 1.62 mul V
  hpt 2 mul 0 V
  hpt neg vpt -1.62 mul V closepath stroke } def
/PentE { stroke [] 0 setdash gsave
  translate 0 hpt M 4 {72 rotate 0 hpt L} repeat
  closepath stroke grestore } def
/CircE { stroke [] 0 setdash 
  hpt 0 360 arc stroke } def
/Opaque { gsave closepath 1 setgray fill grestore 0 setgray closepath } def
/DiaW { stroke [] 0 setdash vpt add M
  hpt neg vpt neg V hpt vpt neg V
  hpt vpt V hpt neg vpt V Opaque stroke } def
/BoxW { stroke [] 0 setdash exch hpt sub exch vpt add M
  0 vpt2 neg V hpt2 0 V 0 vpt2 V
  hpt2 neg 0 V Opaque stroke } def
/TriUW { stroke [] 0 setdash vpt 1.12 mul add M
  hpt neg vpt -1.62 mul V
  hpt 2 mul 0 V
  hpt neg vpt 1.62 mul V Opaque stroke } def
/TriDW { stroke [] 0 setdash vpt 1.12 mul sub M
  hpt neg vpt 1.62 mul V
  hpt 2 mul 0 V
  hpt neg vpt -1.62 mul V Opaque stroke } def
/PentW { stroke [] 0 setdash gsave
  translate 0 hpt M 4 {72 rotate 0 hpt L} repeat
  Opaque stroke grestore } def
/CircW { stroke [] 0 setdash 
  hpt 0 360 arc Opaque stroke } def
/BoxFill { gsave Rec 1 setgray fill grestore } def
end
}}%
\begin{picture}(3600,2160)(0,0)%
{\GNUPLOTspecial{"
gnudict begin
gsave
0 0 translate
0.100 0.100 scale
0 setgray
newpath
1.000 UL
LTb
400 300 M
63 0 V
2987 0 R
-63 0 V
400 496 M
63 0 V
2987 0 R
-63 0 V
400 691 M
63 0 V
2987 0 R
-63 0 V
400 887 M
63 0 V
2987 0 R
-63 0 V
400 1082 M
63 0 V
2987 0 R
-63 0 V
400 1278 M
63 0 V
2987 0 R
-63 0 V
400 1473 M
63 0 V
2987 0 R
-63 0 V
400 1669 M
63 0 V
2987 0 R
-63 0 V
400 1864 M
63 0 V
2987 0 R
-63 0 V
400 2060 M
63 0 V
2987 0 R
-63 0 V
400 300 M
0 63 V
0 1697 R
0 -63 V
908 300 M
0 63 V
0 1697 R
0 -63 V
1417 300 M
0 63 V
0 1697 R
0 -63 V
1925 300 M
0 63 V
0 1697 R
0 -63 V
2433 300 M
0 63 V
0 1697 R
0 -63 V
2942 300 M
0 63 V
0 1697 R
0 -63 V
3450 300 M
0 63 V
0 1697 R
0 -63 V
1.000 UL
LTb
400 300 M
3050 0 V
0 1760 V
-3050 0 V
400 300 L
1.000 UL
LT0
400 356 M
14 0 V
36 0 V
37 0 V
37 0 V
37 -1 V
37 0 V
38 0 V
38 0 V
38 0 V
39 0 V
39 0 V
39 0 V
39 0 V
40 0 V
39 0 V
40 0 V
41 0 V
40 0 V
41 0 V
41 0 V
41 0 V
42 0 V
41 0 V
42 0 V
42 0 V
43 0 V
42 0 V
43 0 V
43 0 V
43 0 V
44 0 V
44 0 V
43 -1 V
44 0 V
45 0 V
44 0 V
45 0 V
44 0 V
45 0 V
46 0 V
45 0 V
45 0 V
46 0 V
46 0 V
46 0 V
46 0 V
46 0 V
47 0 V
46 0 V
47 0 V
47 0 V
47 0 V
47 0 V
48 0 V
47 0 V
48 0 V
47 0 V
48 0 V
48 0 V
48 0 V
48 0 V
48 -1 V
48 0 V
49 0 V
48 0 V
49 0 V
48 0 V
49 0 V
49 0 V
49 0 V
14 0 V
1.000 UL
LT1
400 389 M
31 0 V
31 0 V
30 0 V
31 -1 V
31 0 V
31 0 V
31 0 V
30 0 V
31 0 V
31 -1 V
31 0 V
31 0 V
31 0 V
30 0 V
31 0 V
31 0 V
31 -1 V
31 0 V
30 0 V
31 0 V
31 0 V
31 0 V
31 0 V
30 -1 V
31 0 V
31 0 V
31 0 V
31 0 V
30 0 V
31 0 V
31 0 V
31 -1 V
31 0 V
30 0 V
31 0 V
31 0 V
31 0 V
31 0 V
31 0 V
30 0 V
31 -1 V
31 0 V
31 0 V
31 0 V
30 0 V
31 0 V
31 0 V
31 0 V
31 0 V
30 0 V
31 -1 V
31 0 V
31 0 V
31 0 V
30 0 V
31 0 V
31 0 V
31 0 V
31 0 V
30 0 V
31 0 V
31 -1 V
31 0 V
31 0 V
31 0 V
30 0 V
31 0 V
31 0 V
31 0 V
31 0 V
30 0 V
31 0 V
31 0 V
31 0 V
31 -1 V
30 0 V
31 0 V
31 0 V
31 0 V
31 0 V
30 0 V
31 0 V
31 0 V
31 0 V
31 0 V
30 0 V
31 0 V
31 -1 V
31 0 V
31 0 V
31 0 V
30 0 V
31 0 V
31 0 V
31 0 V
31 0 V
30 0 V
31 0 V
31 0 V
1.000 UL
LT2
400 505 M
14 0 V
36 -2 V
37 -1 V
37 -1 V
37 -2 V
37 -1 V
38 -1 V
38 -2 V
38 -1 V
39 -1 V
39 -1 V
39 -1 V
39 -1 V
40 -2 V
39 -1 V
40 -1 V
41 -1 V
40 -1 V
41 -1 V
41 -1 V
41 -1 V
42 -1 V
41 -1 V
42 -1 V
42 0 V
43 -1 V
42 -1 V
43 -1 V
43 -1 V
43 -1 V
44 -1 V
44 0 V
43 -1 V
44 -1 V
45 -1 V
44 0 V
45 -1 V
44 -1 V
45 -1 V
46 0 V
45 -1 V
45 -1 V
46 0 V
46 -1 V
46 -1 V
46 0 V
46 -1 V
47 -1 V
46 0 V
47 -1 V
47 0 V
47 -1 V
47 -1 V
48 0 V
47 -1 V
48 0 V
47 -1 V
48 -1 V
48 0 V
48 -1 V
48 0 V
48 -1 V
48 0 V
49 -1 V
48 0 V
49 -1 V
48 0 V
49 -1 V
49 0 V
49 -1 V
14 0 V
1.000 UL
LT3
430 2060 M
1 -24 V
31 -396 V
30 -217 V
31 -142 V
31 -102 V
31 -77 V
31 -61 V
30 -50 V
31 -42 V
31 -36 V
31 -30 V
31 -27 V
31 -24 V
30 -21 V
31 -18 V
31 -18 V
31 -15 V
31 -14 V
30 -13 V
31 -12 V
31 -11 V
31 -10 V
31 -10 V
30 -9 V
31 -8 V
31 -8 V
31 -7 V
31 -7 V
30 -7 V
31 -6 V
31 -6 V
31 -5 V
31 -6 V
30 -5 V
31 -4 V
31 -5 V
31 -4 V
31 -5 V
31 -4 V
30 -3 V
31 -4 V
31 -4 V
31 -3 V
31 -3 V
30 -3 V
31 -3 V
31 -3 V
31 -3 V
31 -3 V
30 -3 V
31 -2 V
31 -3 V
31 -2 V
31 -2 V
30 -3 V
31 -2 V
31 -2 V
31 -2 V
31 -2 V
30 -2 V
31 -2 V
31 -2 V
31 -2 V
31 -1 V
31 -2 V
30 -2 V
31 -2 V
31 -1 V
31 -2 V
31 -1 V
30 -2 V
31 -1 V
31 -2 V
31 -1 V
31 -2 V
30 -1 V
31 -2 V
31 -1 V
31 -1 V
31 -2 V
30 -1 V
31 -1 V
31 -1 V
31 -2 V
31 -1 V
30 -1 V
31 -1 V
31 -1 V
31 -2 V
31 -1 V
31 -1 V
30 -1 V
31 -1 V
31 -1 V
31 -1 V
31 -2 V
30 -1 V
31 -1 V
31 -1 V
1.000 UP
1.000 UL
LT4
450 1850 M
0 58 V
-31 -58 R
62 0 V
-62 58 R
62 0 V
2570 510 M
0 11 V
-31 -11 R
62 0 V
-62 11 R
62 0 V
450 1879 Pls
2570 515 Pls
stroke
grestore
end
showpage
}}%
\put(1925,50){\makebox(0,0){$\frac{T}{T_c}$}}%
\put(100,1180){%
\makebox(0,0)[b]{\shortstack{$\frac{S_w}{L_xL_y}$}}%
}%
\put(3450,200){\makebox(0,0){2.2}}%
\put(2942,200){\makebox(0,0){2}}%
\put(2433,200){\makebox(0,0){1.8}}%
\put(1925,200){\makebox(0,0){1.6}}%
\put(1417,200){\makebox(0,0){1.4}}%
\put(908,200){\makebox(0,0){1.2}}%
\put(400,200){\makebox(0,0){1}}%
\put(350,2060){\makebox(0,0)[r]{1.8}}%
\put(350,1864){\makebox(0,0)[r]{1.6}}%
\put(350,1669){\makebox(0,0)[r]{1.4}}%
\put(350,1473){\makebox(0,0)[r]{1.2}}%
\put(350,1278){\makebox(0,0)[r]{1}}%
\put(350,1082){\makebox(0,0)[r]{0.8}}%
\put(350,887){\makebox(0,0)[r]{0.6}}%
\put(350,691){\makebox(0,0)[r]{0.4}}%
\put(350,496){\makebox(0,0)[r]{0.2}}%
\put(350,300){\makebox(0,0)[r]{0}}%
\end{picture}%
\endgroup

\end	{center}
\vskip 0.15in
\caption{Action per unit area of the $k=1$ domain wall in SU(4)
with $aT=0.25$. Monte Carlo values, $+$, compared with 
perturbation theory based on verious couplings:
$g^2(a)$, solid line, $g^2_I(a)$, long dashed line, $g^2_{SF}(T)$,
short  dashed line. The dotted line is the interpolation in
eqn(\ref{eqn_Ffit}).}
\label{fig_wallPT}
\end 	{figure}
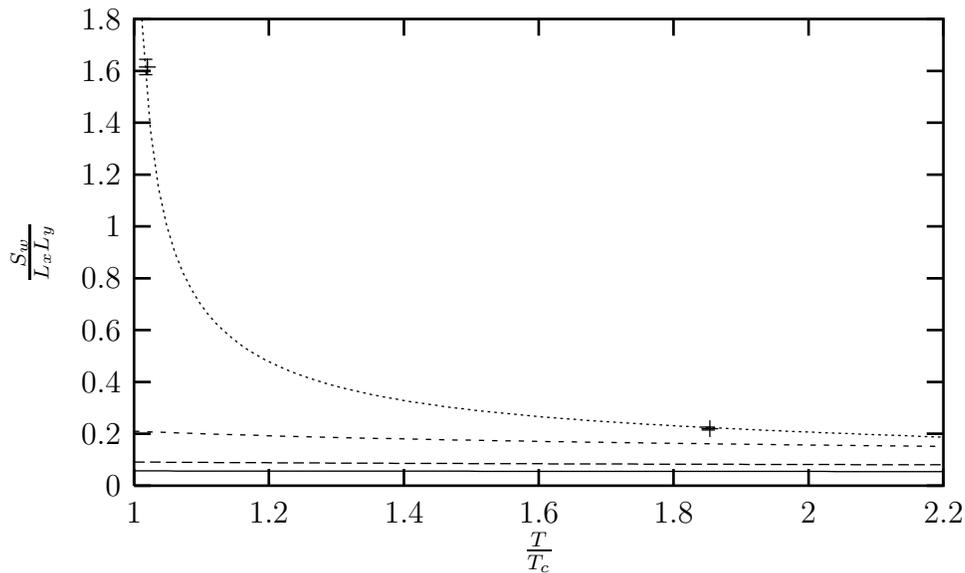

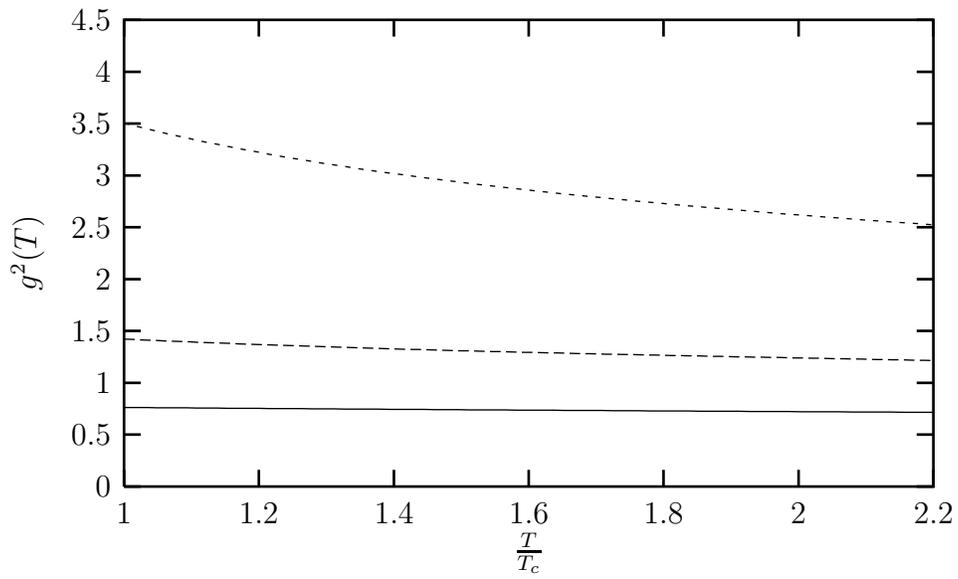
\begin	{figure}[p]
\begin	{center}
\leavevmode
\begingroup%
  \makeatletter%
  \newcommand{\GNUPLOTspecial}{%
    \@sanitize\catcode`\%=14\relax\special}%
  \setlength{\unitlength}{0.1bp}%
{\GNUPLOTspecial{!
/gnudict 256 dict def
gnudict begin
/Color false def
/Solid false def
/gnulinewidth 5.000 def
/userlinewidth gnulinewidth def
/vshift -33 def
/dl {10 mul} def
/hpt_ 31.5 def
/vpt_ 31.5 def
/hpt hpt_ def
/vpt vpt_ def
/M {moveto} bind def
/L {lineto} bind def
/R {rmoveto} bind def
/V {rlineto} bind def
/vpt2 vpt 2 mul def
/hpt2 hpt 2 mul def
/Lshow { currentpoint stroke M
  0 vshift R show } def
/Rshow { currentpoint stroke M
  dup stringwidth pop neg vshift R show } def
/Cshow { currentpoint stroke M
  dup stringwidth pop -2 div vshift R show } def
/UP { dup vpt_ mul /vpt exch def hpt_ mul /hpt exch def
  /hpt2 hpt 2 mul def /vpt2 vpt 2 mul def } def
/DL { Color {setrgbcolor Solid {pop []} if 0 setdash }
 {pop pop pop Solid {pop []} if 0 setdash} ifelse } def
/BL { stroke userlinewidth 2 mul setlinewidth } def
/AL { stroke userlinewidth 2 div setlinewidth } def
/UL { dup gnulinewidth mul /userlinewidth exch def
      10 mul /udl exch def } def
/PL { stroke userlinewidth setlinewidth } def
/LTb { BL [] 0 0 0 DL } def
/LTa { AL [1 udl mul 2 udl mul] 0 setdash 0 0 0 setrgbcolor } def
/LT0 { PL [] 1 0 0 DL } def
/LT1 { PL [4 dl 2 dl] 0 1 0 DL } def
/LT2 { PL [2 dl 3 dl] 0 0 1 DL } def
/LT3 { PL [1 dl 1.5 dl] 1 0 1 DL } def
/LT4 { PL [5 dl 2 dl 1 dl 2 dl] 0 1 1 DL } def
/LT5 { PL [4 dl 3 dl 1 dl 3 dl] 1 1 0 DL } def
/LT6 { PL [2 dl 2 dl 2 dl 4 dl] 0 0 0 DL } def
/LT7 { PL [2 dl 2 dl 2 dl 2 dl 2 dl 4 dl] 1 0.3 0 DL } def
/LT8 { PL [2 dl 2 dl 2 dl 2 dl 2 dl 2 dl 2 dl 4 dl] 0.5 0.5 0.5 DL } def
/Pnt { stroke [] 0 setdash
   gsave 1 setlinecap M 0 0 V stroke grestore } def
/Dia { stroke [] 0 setdash 2 copy vpt add M
  hpt neg vpt neg V hpt vpt neg V
  hpt vpt V hpt neg vpt V closepath stroke
  Pnt } def
/Pls { stroke [] 0 setdash vpt sub M 0 vpt2 V
  currentpoint stroke M
  hpt neg vpt neg R hpt2 0 V stroke
  } def
/Box { stroke [] 0 setdash 2 copy exch hpt sub exch vpt add M
  0 vpt2 neg V hpt2 0 V 0 vpt2 V
  hpt2 neg 0 V closepath stroke
  Pnt } def
/Crs { stroke [] 0 setdash exch hpt sub exch vpt add M
  hpt2 vpt2 neg V currentpoint stroke M
  hpt2 neg 0 R hpt2 vpt2 V stroke } def
/TriU { stroke [] 0 setdash 2 copy vpt 1.12 mul add M
  hpt neg vpt -1.62 mul V
  hpt 2 mul 0 V
  hpt neg vpt 1.62 mul V closepath stroke
  Pnt  } def
/Star { 2 copy Pls Crs } def
/BoxF { stroke [] 0 setdash exch hpt sub exch vpt add M
  0 vpt2 neg V  hpt2 0 V  0 vpt2 V
  hpt2 neg 0 V  closepath fill } def
/TriUF { stroke [] 0 setdash vpt 1.12 mul add M
  hpt neg vpt -1.62 mul V
  hpt 2 mul 0 V
  hpt neg vpt 1.62 mul V closepath fill } def
/TriD { stroke [] 0 setdash 2 copy vpt 1.12 mul sub M
  hpt neg vpt 1.62 mul V
  hpt 2 mul 0 V
  hpt neg vpt -1.62 mul V closepath stroke
  Pnt  } def
/TriDF { stroke [] 0 setdash vpt 1.12 mul sub M
  hpt neg vpt 1.62 mul V
  hpt 2 mul 0 V
  hpt neg vpt -1.62 mul V closepath fill} def
/DiaF { stroke [] 0 setdash vpt add M
  hpt neg vpt neg V hpt vpt neg V
  hpt vpt V hpt neg vpt V closepath fill } def
/Pent { stroke [] 0 setdash 2 copy gsave
  translate 0 hpt M 4 {72 rotate 0 hpt L} repeat
  closepath stroke grestore Pnt } def
/PentF { stroke [] 0 setdash gsave
  translate 0 hpt M 4 {72 rotate 0 hpt L} repeat
  closepath fill grestore } def
/Circle { stroke [] 0 setdash 2 copy
  hpt 0 360 arc stroke Pnt } def
/CircleF { stroke [] 0 setdash hpt 0 360 arc fill } def
/C0 { BL [] 0 setdash 2 copy moveto vpt 90 450  arc } bind def
/C1 { BL [] 0 setdash 2 copy        moveto
       2 copy  vpt 0 90 arc closepath fill
               vpt 0 360 arc closepath } bind def
/C2 { BL [] 0 setdash 2 copy moveto
       2 copy  vpt 90 180 arc closepath fill
               vpt 0 360 arc closepath } bind def
/C3 { BL [] 0 setdash 2 copy moveto
       2 copy  vpt 0 180 arc closepath fill
               vpt 0 360 arc closepath } bind def
/C4 { BL [] 0 setdash 2 copy moveto
       2 copy  vpt 180 270 arc closepath fill
               vpt 0 360 arc closepath } bind def
/C5 { BL [] 0 setdash 2 copy moveto
       2 copy  vpt 0 90 arc
       2 copy moveto
       2 copy  vpt 180 270 arc closepath fill
               vpt 0 360 arc } bind def
/C6 { BL [] 0 setdash 2 copy moveto
      2 copy  vpt 90 270 arc closepath fill
              vpt 0 360 arc closepath } bind def
/C7 { BL [] 0 setdash 2 copy moveto
      2 copy  vpt 0 270 arc closepath fill
              vpt 0 360 arc closepath } bind def
/C8 { BL [] 0 setdash 2 copy moveto
      2 copy vpt 270 360 arc closepath fill
              vpt 0 360 arc closepath } bind def
/C9 { BL [] 0 setdash 2 copy moveto
      2 copy  vpt 270 450 arc closepath fill
              vpt 0 360 arc closepath } bind def
/C10 { BL [] 0 setdash 2 copy 2 copy moveto vpt 270 360 arc closepath fill
       2 copy moveto
       2 copy vpt 90 180 arc closepath fill
               vpt 0 360 arc closepath } bind def
/C11 { BL [] 0 setdash 2 copy moveto
       2 copy  vpt 0 180 arc closepath fill
       2 copy moveto
       2 copy  vpt 270 360 arc closepath fill
               vpt 0 360 arc closepath } bind def
/C12 { BL [] 0 setdash 2 copy moveto
       2 copy  vpt 180 360 arc closepath fill
               vpt 0 360 arc closepath } bind def
/C13 { BL [] 0 setdash  2 copy moveto
       2 copy  vpt 0 90 arc closepath fill
       2 copy moveto
       2 copy  vpt 180 360 arc closepath fill
               vpt 0 360 arc closepath } bind def
/C14 { BL [] 0 setdash 2 copy moveto
       2 copy  vpt 90 360 arc closepath fill
               vpt 0 360 arc } bind def
/C15 { BL [] 0 setdash 2 copy vpt 0 360 arc closepath fill
               vpt 0 360 arc closepath } bind def
/Rec   { newpath 4 2 roll moveto 1 index 0 rlineto 0 exch rlineto
       neg 0 rlineto closepath } bind def
/Square { dup Rec } bind def
/Bsquare { vpt sub exch vpt sub exch vpt2 Square } bind def
/S0 { BL [] 0 setdash 2 copy moveto 0 vpt rlineto BL Bsquare } bind def
/S1 { BL [] 0 setdash 2 copy vpt Square fill Bsquare } bind def
/S2 { BL [] 0 setdash 2 copy exch vpt sub exch vpt Square fill Bsquare } bind def
/S3 { BL [] 0 setdash 2 copy exch vpt sub exch vpt2 vpt Rec fill Bsquare } bind def
/S4 { BL [] 0 setdash 2 copy exch vpt sub exch vpt sub vpt Square fill Bsquare } bind def
/S5 { BL [] 0 setdash 2 copy 2 copy vpt Square fill
       exch vpt sub exch vpt sub vpt Square fill Bsquare } bind def
/S6 { BL [] 0 setdash 2 copy exch vpt sub exch vpt sub vpt vpt2 Rec fill Bsquare } bind def
/S7 { BL [] 0 setdash 2 copy exch vpt sub exch vpt sub vpt vpt2 Rec fill
       2 copy vpt Square fill
       Bsquare } bind def
/S8 { BL [] 0 setdash 2 copy vpt sub vpt Square fill Bsquare } bind def
/S9 { BL [] 0 setdash 2 copy vpt sub vpt vpt2 Rec fill Bsquare } bind def
/S10 { BL [] 0 setdash 2 copy vpt sub vpt Square fill 2 copy exch vpt sub exch vpt Square fill
       Bsquare } bind def
/S11 { BL [] 0 setdash 2 copy vpt sub vpt Square fill 2 copy exch vpt sub exch vpt2 vpt Rec fill
       Bsquare } bind def
/S12 { BL [] 0 setdash 2 copy exch vpt sub exch vpt sub vpt2 vpt Rec fill Bsquare } bind def
/S13 { BL [] 0 setdash 2 copy exch vpt sub exch vpt sub vpt2 vpt Rec fill
       2 copy vpt Square fill Bsquare } bind def
/S14 { BL [] 0 setdash 2 copy exch vpt sub exch vpt sub vpt2 vpt Rec fill
       2 copy exch vpt sub exch vpt Square fill Bsquare } bind def
/S15 { BL [] 0 setdash 2 copy Bsquare fill Bsquare } bind def
/D0 { gsave translate 45 rotate 0 0 S0 stroke grestore } bind def
/D1 { gsave translate 45 rotate 0 0 S1 stroke grestore } bind def
/D2 { gsave translate 45 rotate 0 0 S2 stroke grestore } bind def
/D3 { gsave translate 45 rotate 0 0 S3 stroke grestore } bind def
/D4 { gsave translate 45 rotate 0 0 S4 stroke grestore } bind def
/D5 { gsave translate 45 rotate 0 0 S5 stroke grestore } bind def
/D6 { gsave translate 45 rotate 0 0 S6 stroke grestore } bind def
/D7 { gsave translate 45 rotate 0 0 S7 stroke grestore } bind def
/D8 { gsave translate 45 rotate 0 0 S8 stroke grestore } bind def
/D9 { gsave translate 45 rotate 0 0 S9 stroke grestore } bind def
/D10 { gsave translate 45 rotate 0 0 S10 stroke grestore } bind def
/D11 { gsave translate 45 rotate 0 0 S11 stroke grestore } bind def
/D12 { gsave translate 45 rotate 0 0 S12 stroke grestore } bind def
/D13 { gsave translate 45 rotate 0 0 S13 stroke grestore } bind def
/D14 { gsave translate 45 rotate 0 0 S14 stroke grestore } bind def
/D15 { gsave translate 45 rotate 0 0 S15 stroke grestore } bind def
/DiaE { stroke [] 0 setdash vpt add M
  hpt neg vpt neg V hpt vpt neg V
  hpt vpt V hpt neg vpt V closepath stroke } def
/BoxE { stroke [] 0 setdash exch hpt sub exch vpt add M
  0 vpt2 neg V hpt2 0 V 0 vpt2 V
  hpt2 neg 0 V closepath stroke } def
/TriUE { stroke [] 0 setdash vpt 1.12 mul add M
  hpt neg vpt -1.62 mul V
  hpt 2 mul 0 V
  hpt neg vpt 1.62 mul V closepath stroke } def
/TriDE { stroke [] 0 setdash vpt 1.12 mul sub M
  hpt neg vpt 1.62 mul V
  hpt 2 mul 0 V
  hpt neg vpt -1.62 mul V closepath stroke } def
/PentE { stroke [] 0 setdash gsave
  translate 0 hpt M 4 {72 rotate 0 hpt L} repeat
  closepath stroke grestore } def
/CircE { stroke [] 0 setdash 
  hpt 0 360 arc stroke } def
/Opaque { gsave closepath 1 setgray fill grestore 0 setgray closepath } def
/DiaW { stroke [] 0 setdash vpt add M
  hpt neg vpt neg V hpt vpt neg V
  hpt vpt V hpt neg vpt V Opaque stroke } def
/BoxW { stroke [] 0 setdash exch hpt sub exch vpt add M
  0 vpt2 neg V hpt2 0 V 0 vpt2 V
  hpt2 neg 0 V Opaque stroke } def
/TriUW { stroke [] 0 setdash vpt 1.12 mul add M
  hpt neg vpt -1.62 mul V
  hpt 2 mul 0 V
  hpt neg vpt 1.62 mul V Opaque stroke } def
/TriDW { stroke [] 0 setdash vpt 1.12 mul sub M
  hpt neg vpt 1.62 mul V
  hpt 2 mul 0 V
  hpt neg vpt -1.62 mul V Opaque stroke } def
/PentW { stroke [] 0 setdash gsave
  translate 0 hpt M 4 {72 rotate 0 hpt L} repeat
  Opaque stroke grestore } def
/CircW { stroke [] 0 setdash 
  hpt 0 360 arc Opaque stroke } def
/BoxFill { gsave Rec 1 setgray fill grestore } def
end
}}%
\begin{picture}(3600,2160)(0,0)%
{\GNUPLOTspecial{"
gnudict begin
gsave
0 0 translate
0.100 0.100 scale
0 setgray
newpath
1.000 UL
LTb
400 300 M
63 0 V
2987 0 R
-63 0 V
400 496 M
63 0 V
2987 0 R
-63 0 V
400 691 M
63 0 V
2987 0 R
-63 0 V
400 887 M
63 0 V
2987 0 R
-63 0 V
400 1082 M
63 0 V
2987 0 R
-63 0 V
400 1278 M
63 0 V
2987 0 R
-63 0 V
400 1473 M
63 0 V
2987 0 R
-63 0 V
400 1669 M
63 0 V
2987 0 R
-63 0 V
400 1864 M
63 0 V
2987 0 R
-63 0 V
400 2060 M
63 0 V
2987 0 R
-63 0 V
400 300 M
0 63 V
0 1697 R
0 -63 V
908 300 M
0 63 V
0 1697 R
0 -63 V
1417 300 M
0 63 V
0 1697 R
0 -63 V
1925 300 M
0 63 V
0 1697 R
0 -63 V
2433 300 M
0 63 V
0 1697 R
0 -63 V
2942 300 M
0 63 V
0 1697 R
0 -63 V
3450 300 M
0 63 V
0 1697 R
0 -63 V
1.000 UL
LTb
400 300 M
3050 0 V
0 1760 V
-3050 0 V
400 300 L
1.000 UL
LT0
400 598 M
14 0 V
36 0 V
37 0 V
37 -1 V
37 0 V
37 0 V
38 0 V
38 -1 V
38 0 V
39 0 V
39 -1 V
39 0 V
39 0 V
40 0 V
39 -1 V
40 0 V
41 0 V
40 0 V
41 -1 V
41 0 V
41 0 V
42 -1 V
41 0 V
42 0 V
42 0 V
43 -1 V
42 0 V
43 0 V
43 0 V
43 -1 V
44 0 V
44 0 V
43 -1 V
44 0 V
45 0 V
44 0 V
45 -1 V
44 0 V
45 0 V
46 0 V
45 -1 V
45 0 V
46 0 V
46 0 V
46 -1 V
46 0 V
46 0 V
47 -1 V
46 0 V
47 0 V
47 0 V
47 -1 V
47 0 V
48 0 V
47 0 V
48 -1 V
47 0 V
48 0 V
48 0 V
48 -1 V
48 0 V
48 0 V
48 0 V
49 -1 V
48 0 V
49 0 V
48 0 V
49 -1 V
49 0 V
49 0 V
14 0 V
1.000 UL
LT1
400 856 M
31 -1 V
31 -2 V
30 -1 V
31 -1 V
31 -2 V
31 -1 V
31 -1 V
30 -1 V
31 -2 V
31 -1 V
31 -1 V
31 -1 V
31 -1 V
30 -2 V
31 -1 V
31 -1 V
31 -1 V
31 -1 V
30 -1 V
31 -1 V
31 -1 V
31 -1 V
31 -1 V
30 -1 V
31 -1 V
31 -1 V
31 -1 V
31 -1 V
30 -1 V
31 -1 V
31 -1 V
31 -1 V
31 -1 V
30 -1 V
31 -1 V
31 0 V
31 -1 V
31 -1 V
31 -1 V
30 -1 V
31 -1 V
31 0 V
31 -1 V
31 -1 V
30 -1 V
31 0 V
31 -1 V
31 -1 V
31 -1 V
30 0 V
31 -1 V
31 -1 V
31 0 V
31 -1 V
30 -1 V
31 -1 V
31 0 V
31 -1 V
31 -1 V
30 0 V
31 -1 V
31 0 V
31 -1 V
31 -1 V
31 0 V
30 -1 V
31 -1 V
31 0 V
31 -1 V
31 0 V
30 -1 V
31 -1 V
31 0 V
31 -1 V
31 -1 V
30 0 V
31 -1 V
31 0 V
31 -1 V
31 0 V
30 -1 V
31 -1 V
31 0 V
31 -1 V
31 0 V
30 -1 V
31 -1 V
31 0 V
31 -1 V
31 0 V
31 -1 V
30 -1 V
31 0 V
31 -1 V
31 0 V
31 -1 V
30 -1 V
31 0 V
31 -1 V
1.000 UL
LT2
400 1671 M
14 -4 V
36 -9 V
37 -9 V
37 -9 V
37 -9 V
37 -8 V
38 -8 V
38 -9 V
38 -8 V
39 -7 V
39 -8 V
39 -8 V
39 -7 V
40 -7 V
39 -8 V
40 -7 V
41 -7 V
40 -6 V
41 -7 V
41 -7 V
41 -6 V
42 -6 V
41 -7 V
42 -6 V
42 -6 V
43 -6 V
42 -5 V
43 -6 V
43 -6 V
43 -5 V
44 -6 V
44 -5 V
43 -6 V
44 -5 V
45 -5 V
44 -5 V
45 -5 V
44 -5 V
45 -5 V
46 -5 V
45 -5 V
45 -4 V
46 -5 V
46 -4 V
46 -5 V
46 -4 V
46 -5 V
47 -4 V
46 -4 V
47 -5 V
47 -4 V
47 -4 V
47 -4 V
48 -4 V
47 -4 V
48 -4 V
47 -4 V
48 -4 V
48 -4 V
48 -3 V
48 -4 V
48 -4 V
48 -3 V
49 -4 V
48 -4 V
49 -3 V
48 -4 V
49 -3 V
49 -3 V
49 -4 V
14 -1 V
stroke
grestore
end
showpage
}}%
\put(1925,50){\makebox(0,0){$\frac{T}{T_c}$}}%
\put(100,1180){%
\makebox(0,0)[b]{\shortstack{$g^2(T)$}}%
}%
\put(3450,200){\makebox(0,0){2.2}}%
\put(2942,200){\makebox(0,0){2}}%
\put(2433,200){\makebox(0,0){1.8}}%
\put(1925,200){\makebox(0,0){1.6}}%
\put(1417,200){\makebox(0,0){1.4}}%
\put(908,200){\makebox(0,0){1.2}}%
\put(400,200){\makebox(0,0){1}}%
\put(350,2060){\makebox(0,0)[r]{4.5}}%
\put(350,1864){\makebox(0,0)[r]{4}}%
\put(350,1669){\makebox(0,0)[r]{3.5}}%
\put(350,1473){\makebox(0,0)[r]{3}}%
\put(350,1278){\makebox(0,0)[r]{2.5}}%
\put(350,1082){\makebox(0,0)[r]{2}}%
\put(350,887){\makebox(0,0)[r]{1.5}}%
\put(350,691){\makebox(0,0)[r]{1}}%
\put(350,496){\makebox(0,0)[r]{0.5}}%
\put(350,300){\makebox(0,0)[r]{0}}%
\end{picture}%
\endgroup

\end	{center}
\vskip 0.15in
\caption{The running coupling $g^2(T)$ obtained from
the lattice bare coupling (solid line), the mean field improved
bare coupling (dashed line), and the Schrodinger functional
coupling (short dashed line). All for SU(4) and  $aT=0.25$.}
\label{fig_alphaSF}
\end 	{figure}

We use the above in the formulae of Section~\ref{subsection_walltension} 
and show in Fig.\ref{fig_wallPT} the resulting value for 
$\Delta S_W/L_1L_2$ of the 2-loop perturbative wall tension using  
the coupling $\bar{g}^2(T^{-1}=aL_0)$.
We also show the values obtained with the mean-field improved
coupling, $g^2_I(a)$, and the unimproved bare coupling, $g^2(a)$.
Finally we show the values obtained from the lattice simulation.
We observe that the more realistic the coupling, the closer
the perturbative value moves towards the lattice values.
However a substantial discrepancy remains, strongly suggesting 
at $T\simeq T_c$ the domain wall can no longer be usefully
described within perturbation theory.

\vfill\eject

\begin{table}
\begin{center}
\begin{tabular}{|c|c|}\hline
$L_0$ & $\alpha(L_0)$ \\ \hline
$\infty$ & 1 \\
6 & 1.060 \\
5 & 1.098 \\
4 & 1.170 \\
3 & 1.294 \\
2 & 1.423 \\  \hline
\end{tabular}
\caption{\label{table_alpha} }
\end{center}
\end{table}

\begin{table}
\begin{center}
\begin{tabular}{|c|c|r|r|r|r|c|c|} \hline
N & $\beta$ & $L_0$ & $L_3$ & $L_{1,2}$ & Sweeps & $\beta_I$  
& $T/T_c$  \\ \hline
2 & 2.4843 & 4 & 30 & 20 & 100000 & 1.6144 & 1.88\\ \hline
3 & 5.702 & 4 & 30 & 20 & 100000 & 3.159 & 1.02 \\
3 & 6.013 & 4 & 30 & 20 & 100000 & 3.585 & 1.88 \\ \hline
4 & 10.5 & 4 & 30 & 20 & 100000 & 5.644 & 1.02 \\
4 & 11 & 4 & 30 & 20 & 100000 & 6.360 & 1.88 \\
4 & 20 & 4 & 30 & 20 & 100000 & 15.954 & $\sim1000$ \\
4 & 10.654 & 5 & 30 & 20 & 68000 & 5.878 & 1.02 \\
4 & 11.239 & 5 & 30 & 20 & 100000 & 6.654 & 1.88 \\ \hline
6 & 24.545 & 5 & 30 & 20 & 20000 & 13.190 & 1.02 \\
6 & 25.863 & 5 & 30 & 20 & 20000 & 15.019 & 1.88 \\ \hline
\end{tabular}
\caption{\label{table_runs} }
\end{center}
\end{table}

\begin{table}
\begin{center}
\begin{tabular}{|c|c|l|l|l|l|} \hline
N & $\beta$ & $\overline{u}_p$ & $\overline{u}_p^{k=1}$ & $\overline{u}_p^{k=2}$ &  $\overline{u}_p^{k=3}$ \\ \hline
2 & 2.4843 & 0.6498239(90) & 0.6493949(74) && \\ \hline
3 & 5.702 & 0.554069(23) & 0.552153(51) && \\
3 & 6.013 & 0.5962242(67) & 0.5958470(65) && \\ \hline
4 & 10.5  & 0.537515(16) & 0.535272(39) & 0.534541(39) & \\
4 & 11 & 0.5781191(54) & 0.5778135(56) & 0.5777084(57) & \\
4 & 20 & 0.7977065(13) & 0.7976478(18) & 0.7976278(15) & \\
4 & 10.654 & 0.551683(12) & 0.550862(27) & 0.550592(29) & \\
4 & 11.239 & 0.5920068(40) & 0.5918633(45) & 0.5918131(54) & \\ \hline
6 & 24.545 & 0.537364(17) & 0.536822(22) & 0.536423(25) & 0.536236(27) \\
6 & 25.863 & 0.5807296(78) & 0.5806203(71) & 0.5805524(80) & 0.5805399(85) \\ \hline
\end{tabular}
\caption{\label{table_aplaq} }
\end{center}
\end{table}

\begin{table}
\begin{center}
\begin{tabular}{|c|c|c|c|c|c|}\hline
$\beta$ & $L^4$ & $a\surd\sigma_{k=1}$ & $a\surd\sigma_{k=2}$ &
 $\sigma_{k=2}/\sigma_{k=1}$ & plaq \\ \hline
10.48 & $8^4$ & 0.4300(21) & 0.496(7) & 1.331(20) & 0.525301(54) \\
10.50 & $8^4$ & 0.4170(16) & 0.481(4) & 1.331(12) & 0.529210(50) \\ \hline
\end{tabular}
\caption{\label{table_sigN4} }
\end{center}
\end{table}

\begin{table}
\begin{center}
\begin{tabular}{|c|c|c|c|c|c|c|c|} \hline
N & $aT$ & k & Prediction & $\Delta S^k_w/L_xL_y$ & 
$\Delta S^k_w/\Delta S^1_w$ & CS & $T/T_c$ \\ \hline
2 & 0.25 & 1 & 0.1042 & 0.3089(84) & & & 1.88 \\ \hline
3 & 0.25 & 1 & 0.1032 & 1.380(40)  & & & 1.02 \\ 
3 & 0.25 & 1 & 0.0936 & 0.2716(67) & & & 1.88 \\ \hline
4 & 0.25 & 1 & 0.0867 & 1.615(30)  & & & 1.02 \\
  &      & 2 & 0.1157 & 2.141(30)  & 1.326(29) & 1.333 & 1.02 \\
4 & 0.25 & 1 & 0.0791 & 0.2200(56) & & & 1.88 \\
  &      & 2 & 0.1055 & 0.2957(57) & 1.344(31) & 1.333 & 1.88 \\ 
4 & 0.25 &  1 & 0.0421 & 0.0423(16) & & & $\sim1000$ \\
  &      & 2 & 0.0561 & 0.0567(14) & 1.341(49) & 1.333 & $\sim1000$  \\
4 & 0.20 & 1 & 0.0514 & 0.739(27)  & & & 1.02 \\
  &      & 2 & 0.0685 & 0.982(28)  & 1.329(56) & 1.333 & 1.02 \\
4 & 0.20 & 1 & 0.0467 & 0.1292(54) & & & 1.88 \\
  &      & 2 & 0.0622 & 0.1743(60) & 1.350(57) & 1.333 & 1.88 \\ \hline
6 & 0.20 & 1 & 0.0381 & 0.488(25)  & & & 1.02 \\
  &      & 2 & 0.0610 & 0.847(27)  & 1.74(9)   & 1.60 & 1.02 \\
  &      & 3 & 0.0687 & 1.015(29)  & 2.08(10)  & 1.80 & 1.02 \\ 
6 & 0.20 & 1 & 0.0345 & 0.098(9)   & & & 1.88 \\
  &      & 2  & 0.0552 & 0.159(10)  & 1.62(14) & 1.60 & 1.88 \\
  &      & 3  & 0.0621 & 0.171(10)  & 1.74(15) & 1.80 & 1.88 \\ \hline
\end{tabular}
\caption{\label{table_result} }
\end{center}
\end{table}

\begin{table}
\begin{center}
\begin{tabular}{|c|c|c|c|} \hline
\multicolumn{4}{|c|}{SU(4)} \\ \hline
$\beta$ & $T/T_c$ & $k=1$ & $k=2$ \\ \hline
10.497  & 1.016  & no  & no  \\
10.494  & 1.012 & no  & no  \\
10.491  & 1.007 & mixed & flattens \\
10.488  & 1.002 & mixed & mixed   \\
10.485  & 0.998 & splits  & splits \\
10.482  & 0.993 & splits  & splits \\ \hline
\end{tabular}
\caption{\label{table_wetsu4} }
\end{center}
\end{table}

\begin{table}
\begin{center}
\begin{tabular}{|c|c|c|c|c|} \hline
\multicolumn{5}{|c|}{SU(6)} \\ \hline
$\beta$ & $T/T_c$ & $k=1$ & $k=2$ & $k=3$ \\ \hline
24.532  & 1.013 & no  & no & no  \\
24.526  & 1.009 & no  & no & no  \\
24.520  & 1.004 & no  & via origin & flattens  \\
24.514  & 1.000 & no  & via origin & flattens  \\
24.508  & 0.996 & splits   & splits    & splits  \\
24.502  & 0.991 & splits   & splits    & splits  \\ \hline
\end{tabular}
\caption{\label{table_wetsu6} }
\end{center}
\end{table}

\clearpage

\begin	{figure}[p]
\begin	{center}
\leavevmode
\begingroup%
  \makeatletter%
  \newcommand{\GNUPLOTspecial}{%
    \@sanitize\catcode`\%=14\relax\special}%
  \setlength{\unitlength}{0.1bp}%
{\GNUPLOTspecial{!
/gnudict 256 dict def
gnudict begin
/Color false def
/Solid false def
/gnulinewidth 5.000 def
/userlinewidth gnulinewidth def
/vshift -33 def
/dl {10 mul} def
/hpt_ 31.5 def
/vpt_ 31.5 def
/hpt hpt_ def
/vpt vpt_ def
/M {moveto} bind def
/L {lineto} bind def
/R {rmoveto} bind def
/V {rlineto} bind def
/vpt2 vpt 2 mul def
/hpt2 hpt 2 mul def
/Lshow { currentpoint stroke M
  0 vshift R show } def
/Rshow { currentpoint stroke M
  dup stringwidth pop neg vshift R show } def
/Cshow { currentpoint stroke M
  dup stringwidth pop -2 div vshift R show } def
/UP { dup vpt_ mul /vpt exch def hpt_ mul /hpt exch def
  /hpt2 hpt 2 mul def /vpt2 vpt 2 mul def } def
/DL { Color {setrgbcolor Solid {pop []} if 0 setdash }
 {pop pop pop Solid {pop []} if 0 setdash} ifelse } def
/BL { stroke userlinewidth 2 mul setlinewidth } def
/AL { stroke userlinewidth 2 div setlinewidth } def
/UL { dup gnulinewidth mul /userlinewidth exch def
      10 mul /udl exch def } def
/PL { stroke userlinewidth setlinewidth } def
/LTb { BL [] 0 0 0 DL } def
/LTa { AL [1 udl mul 2 udl mul] 0 setdash 0 0 0 setrgbcolor } def
/LT0 { PL [] 1 0 0 DL } def
/LT1 { PL [4 dl 2 dl] 0 1 0 DL } def
/LT2 { PL [2 dl 3 dl] 0 0 1 DL } def
/LT3 { PL [1 dl 1.5 dl] 1 0 1 DL } def
/LT4 { PL [5 dl 2 dl 1 dl 2 dl] 0 1 1 DL } def
/LT5 { PL [4 dl 3 dl 1 dl 3 dl] 1 1 0 DL } def
/LT6 { PL [2 dl 2 dl 2 dl 4 dl] 0 0 0 DL } def
/LT7 { PL [2 dl 2 dl 2 dl 2 dl 2 dl 4 dl] 1 0.3 0 DL } def
/LT8 { PL [2 dl 2 dl 2 dl 2 dl 2 dl 2 dl 2 dl 4 dl] 0.5 0.5 0.5 DL } def
/Pnt { stroke [] 0 setdash
   gsave 1 setlinecap M 0 0 V stroke grestore } def
/Dia { stroke [] 0 setdash 2 copy vpt add M
  hpt neg vpt neg V hpt vpt neg V
  hpt vpt V hpt neg vpt V closepath stroke
  Pnt } def
/Pls { stroke [] 0 setdash vpt sub M 0 vpt2 V
  currentpoint stroke M
  hpt neg vpt neg R hpt2 0 V stroke
  } def
/Box { stroke [] 0 setdash 2 copy exch hpt sub exch vpt add M
  0 vpt2 neg V hpt2 0 V 0 vpt2 V
  hpt2 neg 0 V closepath stroke
  Pnt } def
/Crs { stroke [] 0 setdash exch hpt sub exch vpt add M
  hpt2 vpt2 neg V currentpoint stroke M
  hpt2 neg 0 R hpt2 vpt2 V stroke } def
/TriU { stroke [] 0 setdash 2 copy vpt 1.12 mul add M
  hpt neg vpt -1.62 mul V
  hpt 2 mul 0 V
  hpt neg vpt 1.62 mul V closepath stroke
  Pnt  } def
/Star { 2 copy Pls Crs } def
/BoxF { stroke [] 0 setdash exch hpt sub exch vpt add M
  0 vpt2 neg V  hpt2 0 V  0 vpt2 V
  hpt2 neg 0 V  closepath fill } def
/TriUF { stroke [] 0 setdash vpt 1.12 mul add M
  hpt neg vpt -1.62 mul V
  hpt 2 mul 0 V
  hpt neg vpt 1.62 mul V closepath fill } def
/TriD { stroke [] 0 setdash 2 copy vpt 1.12 mul sub M
  hpt neg vpt 1.62 mul V
  hpt 2 mul 0 V
  hpt neg vpt -1.62 mul V closepath stroke
  Pnt  } def
/TriDF { stroke [] 0 setdash vpt 1.12 mul sub M
  hpt neg vpt 1.62 mul V
  hpt 2 mul 0 V
  hpt neg vpt -1.62 mul V closepath fill} def
/DiaF { stroke [] 0 setdash vpt add M
  hpt neg vpt neg V hpt vpt neg V
  hpt vpt V hpt neg vpt V closepath fill } def
/Pent { stroke [] 0 setdash 2 copy gsave
  translate 0 hpt M 4 {72 rotate 0 hpt L} repeat
  closepath stroke grestore Pnt } def
/PentF { stroke [] 0 setdash gsave
  translate 0 hpt M 4 {72 rotate 0 hpt L} repeat
  closepath fill grestore } def
/Circle { stroke [] 0 setdash 2 copy
  hpt 0 360 arc stroke Pnt } def
/CircleF { stroke [] 0 setdash hpt 0 360 arc fill } def
/C0 { BL [] 0 setdash 2 copy moveto vpt 90 450  arc } bind def
/C1 { BL [] 0 setdash 2 copy        moveto
       2 copy  vpt 0 90 arc closepath fill
               vpt 0 360 arc closepath } bind def
/C2 { BL [] 0 setdash 2 copy moveto
       2 copy  vpt 90 180 arc closepath fill
               vpt 0 360 arc closepath } bind def
/C3 { BL [] 0 setdash 2 copy moveto
       2 copy  vpt 0 180 arc closepath fill
               vpt 0 360 arc closepath } bind def
/C4 { BL [] 0 setdash 2 copy moveto
       2 copy  vpt 180 270 arc closepath fill
               vpt 0 360 arc closepath } bind def
/C5 { BL [] 0 setdash 2 copy moveto
       2 copy  vpt 0 90 arc
       2 copy moveto
       2 copy  vpt 180 270 arc closepath fill
               vpt 0 360 arc } bind def
/C6 { BL [] 0 setdash 2 copy moveto
      2 copy  vpt 90 270 arc closepath fill
              vpt 0 360 arc closepath } bind def
/C7 { BL [] 0 setdash 2 copy moveto
      2 copy  vpt 0 270 arc closepath fill
              vpt 0 360 arc closepath } bind def
/C8 { BL [] 0 setdash 2 copy moveto
      2 copy vpt 270 360 arc closepath fill
              vpt 0 360 arc closepath } bind def
/C9 { BL [] 0 setdash 2 copy moveto
      2 copy  vpt 270 450 arc closepath fill
              vpt 0 360 arc closepath } bind def
/C10 { BL [] 0 setdash 2 copy 2 copy moveto vpt 270 360 arc closepath fill
       2 copy moveto
       2 copy vpt 90 180 arc closepath fill
               vpt 0 360 arc closepath } bind def
/C11 { BL [] 0 setdash 2 copy moveto
       2 copy  vpt 0 180 arc closepath fill
       2 copy moveto
       2 copy  vpt 270 360 arc closepath fill
               vpt 0 360 arc closepath } bind def
/C12 { BL [] 0 setdash 2 copy moveto
       2 copy  vpt 180 360 arc closepath fill
               vpt 0 360 arc closepath } bind def
/C13 { BL [] 0 setdash  2 copy moveto
       2 copy  vpt 0 90 arc closepath fill
       2 copy moveto
       2 copy  vpt 180 360 arc closepath fill
               vpt 0 360 arc closepath } bind def
/C14 { BL [] 0 setdash 2 copy moveto
       2 copy  vpt 90 360 arc closepath fill
               vpt 0 360 arc } bind def
/C15 { BL [] 0 setdash 2 copy vpt 0 360 arc closepath fill
               vpt 0 360 arc closepath } bind def
/Rec   { newpath 4 2 roll moveto 1 index 0 rlineto 0 exch rlineto
       neg 0 rlineto closepath } bind def
/Square { dup Rec } bind def
/Bsquare { vpt sub exch vpt sub exch vpt2 Square } bind def
/S0 { BL [] 0 setdash 2 copy moveto 0 vpt rlineto BL Bsquare } bind def
/S1 { BL [] 0 setdash 2 copy vpt Square fill Bsquare } bind def
/S2 { BL [] 0 setdash 2 copy exch vpt sub exch vpt Square fill Bsquare } bind def
/S3 { BL [] 0 setdash 2 copy exch vpt sub exch vpt2 vpt Rec fill Bsquare } bind def
/S4 { BL [] 0 setdash 2 copy exch vpt sub exch vpt sub vpt Square fill Bsquare } bind def
/S5 { BL [] 0 setdash 2 copy 2 copy vpt Square fill
       exch vpt sub exch vpt sub vpt Square fill Bsquare } bind def
/S6 { BL [] 0 setdash 2 copy exch vpt sub exch vpt sub vpt vpt2 Rec fill Bsquare } bind def
/S7 { BL [] 0 setdash 2 copy exch vpt sub exch vpt sub vpt vpt2 Rec fill
       2 copy vpt Square fill
       Bsquare } bind def
/S8 { BL [] 0 setdash 2 copy vpt sub vpt Square fill Bsquare } bind def
/S9 { BL [] 0 setdash 2 copy vpt sub vpt vpt2 Rec fill Bsquare } bind def
/S10 { BL [] 0 setdash 2 copy vpt sub vpt Square fill 2 copy exch vpt sub exch vpt Square fill
       Bsquare } bind def
/S11 { BL [] 0 setdash 2 copy vpt sub vpt Square fill 2 copy exch vpt sub exch vpt2 vpt Rec fill
       Bsquare } bind def
/S12 { BL [] 0 setdash 2 copy exch vpt sub exch vpt sub vpt2 vpt Rec fill Bsquare } bind def
/S13 { BL [] 0 setdash 2 copy exch vpt sub exch vpt sub vpt2 vpt Rec fill
       2 copy vpt Square fill Bsquare } bind def
/S14 { BL [] 0 setdash 2 copy exch vpt sub exch vpt sub vpt2 vpt Rec fill
       2 copy exch vpt sub exch vpt Square fill Bsquare } bind def
/S15 { BL [] 0 setdash 2 copy Bsquare fill Bsquare } bind def
/D0 { gsave translate 45 rotate 0 0 S0 stroke grestore } bind def
/D1 { gsave translate 45 rotate 0 0 S1 stroke grestore } bind def
/D2 { gsave translate 45 rotate 0 0 S2 stroke grestore } bind def
/D3 { gsave translate 45 rotate 0 0 S3 stroke grestore } bind def
/D4 { gsave translate 45 rotate 0 0 S4 stroke grestore } bind def
/D5 { gsave translate 45 rotate 0 0 S5 stroke grestore } bind def
/D6 { gsave translate 45 rotate 0 0 S6 stroke grestore } bind def
/D7 { gsave translate 45 rotate 0 0 S7 stroke grestore } bind def
/D8 { gsave translate 45 rotate 0 0 S8 stroke grestore } bind def
/D9 { gsave translate 45 rotate 0 0 S9 stroke grestore } bind def
/D10 { gsave translate 45 rotate 0 0 S10 stroke grestore } bind def
/D11 { gsave translate 45 rotate 0 0 S11 stroke grestore } bind def
/D12 { gsave translate 45 rotate 0 0 S12 stroke grestore } bind def
/D13 { gsave translate 45 rotate 0 0 S13 stroke grestore } bind def
/D14 { gsave translate 45 rotate 0 0 S14 stroke grestore } bind def
/D15 { gsave translate 45 rotate 0 0 S15 stroke grestore } bind def
/DiaE { stroke [] 0 setdash vpt add M
  hpt neg vpt neg V hpt vpt neg V
  hpt vpt V hpt neg vpt V closepath stroke } def
/BoxE { stroke [] 0 setdash exch hpt sub exch vpt add M
  0 vpt2 neg V hpt2 0 V 0 vpt2 V
  hpt2 neg 0 V closepath stroke } def
/TriUE { stroke [] 0 setdash vpt 1.12 mul add M
  hpt neg vpt -1.62 mul V
  hpt 2 mul 0 V
  hpt neg vpt 1.62 mul V closepath stroke } def
/TriDE { stroke [] 0 setdash vpt 1.12 mul sub M
  hpt neg vpt 1.62 mul V
  hpt 2 mul 0 V
  hpt neg vpt -1.62 mul V closepath stroke } def
/PentE { stroke [] 0 setdash gsave
  translate 0 hpt M 4 {72 rotate 0 hpt L} repeat
  closepath stroke grestore } def
/CircE { stroke [] 0 setdash 
  hpt 0 360 arc stroke } def
/Opaque { gsave closepath 1 setgray fill grestore 0 setgray closepath } def
/DiaW { stroke [] 0 setdash vpt add M
  hpt neg vpt neg V hpt vpt neg V
  hpt vpt V hpt neg vpt V Opaque stroke } def
/BoxW { stroke [] 0 setdash exch hpt sub exch vpt add M
  0 vpt2 neg V hpt2 0 V 0 vpt2 V
  hpt2 neg 0 V Opaque stroke } def
/TriUW { stroke [] 0 setdash vpt 1.12 mul add M
  hpt neg vpt -1.62 mul V
  hpt 2 mul 0 V
  hpt neg vpt 1.62 mul V Opaque stroke } def
/TriDW { stroke [] 0 setdash vpt 1.12 mul sub M
  hpt neg vpt 1.62 mul V
  hpt 2 mul 0 V
  hpt neg vpt -1.62 mul V Opaque stroke } def
/PentW { stroke [] 0 setdash gsave
  translate 0 hpt M 4 {72 rotate 0 hpt L} repeat
  Opaque stroke grestore } def
/CircW { stroke [] 0 setdash 
  hpt 0 360 arc Opaque stroke } def
/BoxFill { gsave Rec 1 setgray fill grestore } def
end
}}%
\begin{picture}(3600,2160)(0,0)%
{\GNUPLOTspecial{"
gnudict begin
gsave
0 0 translate
0.100 0.100 scale
0 setgray
newpath
1.000 UL
LTb
300 300 M
63 0 V
3087 0 R
-63 0 V
300 520 M
63 0 V
3087 0 R
-63 0 V
300 740 M
63 0 V
3087 0 R
-63 0 V
300 960 M
63 0 V
3087 0 R
-63 0 V
300 1180 M
63 0 V
3087 0 R
-63 0 V
300 1400 M
63 0 V
3087 0 R
-63 0 V
300 1620 M
63 0 V
3087 0 R
-63 0 V
300 1840 M
63 0 V
3087 0 R
-63 0 V
300 2060 M
63 0 V
3087 0 R
-63 0 V
300 300 M
0 63 V
0 1697 R
0 -63 V
825 300 M
0 63 V
0 1697 R
0 -63 V
1350 300 M
0 63 V
0 1697 R
0 -63 V
1875 300 M
0 63 V
0 1697 R
0 -63 V
2400 300 M
0 63 V
0 1697 R
0 -63 V
2925 300 M
0 63 V
0 1697 R
0 -63 V
3450 300 M
0 63 V
0 1697 R
0 -63 V
1.000 UL
LTb
300 300 M
3150 0 V
0 1760 V
-3150 0 V
300 300 L
1.000 UL
LT0
300 1608 M
32 5 V
32 4 V
31 5 V
32 4 V
32 4 V
32 5 V
32 4 V
32 4 V
31 4 V
32 4 V
32 4 V
32 4 V
32 4 V
31 4 V
32 4 V
32 4 V
32 4 V
32 3 V
32 4 V
31 4 V
32 3 V
32 4 V
32 3 V
32 4 V
31 3 V
32 4 V
32 3 V
32 3 V
32 4 V
32 3 V
31 3 V
32 3 V
32 3 V
32 3 V
32 4 V
31 3 V
32 3 V
32 3 V
32 2 V
32 3 V
32 3 V
31 3 V
32 3 V
32 3 V
32 2 V
32 3 V
31 3 V
32 3 V
32 2 V
32 3 V
32 2 V
32 3 V
31 3 V
32 2 V
32 3 V
32 2 V
32 3 V
31 2 V
32 3 V
32 2 V
32 2 V
32 3 V
32 2 V
31 3 V
32 2 V
32 2 V
32 3 V
32 2 V
31 2 V
32 2 V
32 3 V
32 2 V
32 2 V
32 3 V
31 2 V
32 2 V
32 2 V
32 3 V
32 2 V
31 2 V
32 2 V
32 3 V
32 2 V
32 2 V
32 3 V
31 2 V
32 2 V
32 2 V
32 3 V
32 2 V
31 2 V
32 3 V
32 2 V
32 2 V
32 3 V
32 2 V
31 3 V
32 2 V
32 2 V
1.000 UL
LT1
300 510 M
32 111 V
32 76 V
31 61 V
32 52 V
32 45 V
32 41 V
32 37 V
32 34 V
31 31 V
32 30 V
32 27 V
32 26 V
32 25 V
31 24 V
32 22 V
32 21 V
32 21 V
32 19 V
32 19 V
31 18 V
32 17 V
32 17 V
32 16 V
32 16 V
31 15 V
32 15 V
32 14 V
32 14 V
32 13 V
32 13 V
31 13 V
32 13 V
32 12 V
32 11 V
32 12 V
31 11 V
32 11 V
32 10 V
32 11 V
32 10 V
32 10 V
31 10 V
32 9 V
32 9 V
32 10 V
32 9 V
31 8 V
32 9 V
32 8 V
32 9 V
32 8 V
32 8 V
31 8 V
32 7 V
32 8 V
32 8 V
32 7 V
31 7 V
32 7 V
32 7 V
32 7 V
32 7 V
32 7 V
31 7 V
32 6 V
32 7 V
32 6 V
32 7 V
31 6 V
32 6 V
32 7 V
32 6 V
32 6 V
32 6 V
31 6 V
32 6 V
32 6 V
32 6 V
32 6 V
31 5 V
32 6 V
32 6 V
32 6 V
32 5 V
32 6 V
31 6 V
32 6 V
32 5 V
32 6 V
32 5 V
31 6 V
32 6 V
32 5 V
32 6 V
32 6 V
32 5 V
31 6 V
32 6 V
32 5 V
0.500 UL
LT3
300 618 M
32 111 V
32 76 V
31 61 V
32 52 V
32 46 V
32 40 V
32 37 V
32 34 V
31 31 V
32 30 V
32 27 V
32 26 V
32 24 V
31 24 V
32 22 V
32 21 V
32 20 V
32 19 V
32 19 V
31 18 V
32 17 V
32 17 V
32 16 V
32 15 V
31 15 V
32 15 V
32 14 V
32 13 V
32 14 V
32 12 V
31 13 V
32 12 V
32 12 V
32 11 V
32 12 V
31 11 V
32 10 V
32 11 V
32 10 V
32 10 V
32 10 V
31 9 V
32 10 V
32 9 V
32 9 V
32 9 V
31 8 V
32 9 V
32 8 V
32 8 V
32 8 V
32 8 V
31 8 V
32 7 V
32 8 V
32 7 V
32 7 V
31 7 V
32 7 V
32 7 V
32 7 V
32 7 V
32 6 V
31 7 V
32 6 V
32 7 V
32 6 V
32 6 V
31 7 V
32 6 V
32 6 V
32 6 V
32 6 V
32 6 V
31 6 V
32 6 V
32 5 V
32 6 V
32 6 V
31 5 V
32 6 V
32 6 V
32 5 V
32 6 V
32 6 V
31 5 V
32 6 V
32 5 V
32 6 V
32 5 V
31 6 V
32 5 V
32 6 V
32 5 V
32 6 V
32 5 V
31 6 V
32 5 V
32 6 V
0.500 UL
LT3
300 402 M
32 109 V
32 75 V
31 61 V
32 52 V
32 45 V
32 40 V
32 37 V
32 34 V
31 32 V
32 29 V
32 28 V
32 26 V
32 25 V
31 23 V
32 23 V
32 21 V
32 21 V
32 19 V
32 19 V
31 18 V
32 18 V
32 17 V
32 16 V
32 16 V
31 15 V
32 15 V
32 15 V
32 14 V
32 13 V
32 13 V
31 13 V
32 13 V
32 12 V
32 12 V
32 11 V
31 12 V
32 11 V
32 10 V
32 11 V
32 10 V
32 10 V
31 10 V
32 10 V
32 9 V
32 10 V
32 9 V
31 9 V
32 8 V
32 9 V
32 8 V
32 9 V
32 8 V
31 8 V
32 8 V
32 7 V
32 8 V
32 7 V
31 8 V
32 7 V
32 7 V
32 7 V
32 7 V
32 7 V
31 7 V
32 7 V
32 7 V
32 6 V
32 7 V
31 6 V
32 7 V
32 6 V
32 6 V
32 6 V
32 7 V
31 6 V
32 6 V
32 6 V
32 6 V
32 6 V
31 6 V
32 6 V
32 6 V
32 5 V
32 6 V
32 6 V
31 6 V
32 6 V
32 5 V
32 6 V
32 6 V
31 6 V
32 6 V
32 5 V
32 6 V
32 6 V
32 6 V
31 5 V
32 6 V
32 6 V
stroke
grestore
end
showpage
}}%
\put(1875,50){\makebox(0,0){$\frac{T}{T_c}$}}%
\put(100,1180){%
\makebox(0,0)[b]{\shortstack{$\frac{\sigma}{T^3}$}}%
}%
\put(3450,200){\makebox(0,0){2.2}}%
\put(2925,200){\makebox(0,0){2}}%
\put(2400,200){\makebox(0,0){1.8}}%
\put(1875,200){\makebox(0,0){1.6}}%
\put(1350,200){\makebox(0,0){1.4}}%
\put(825,200){\makebox(0,0){1.2}}%
\put(300,200){\makebox(0,0){1}}%
\put(250,2060){\makebox(0,0)[r]{8}}%
\put(250,1840){\makebox(0,0)[r]{7}}%
\put(250,1620){\makebox(0,0)[r]{6}}%
\put(250,1400){\makebox(0,0)[r]{5}}%
\put(250,1180){\makebox(0,0)[r]{4}}%
\put(250,960){\makebox(0,0)[r]{3}}%
\put(250,740){\makebox(0,0)[r]{2}}%
\put(250,520){\makebox(0,0)[r]{1}}%
\put(250,300){\makebox(0,0)[r]{0}}%
\end{picture}%
\endgroup

\end	{center}
\vskip 0.15in
\caption{Surface tension in units of $T$, using the interpolation
shown in Fig.~\ref{fig_wallPT}. For comparison we show the
2-loop perturbative result using the mean field improved coupling.
All for the $k=1$ wall in SU(4).}
\label{fig_wallT}
\end 	{figure}
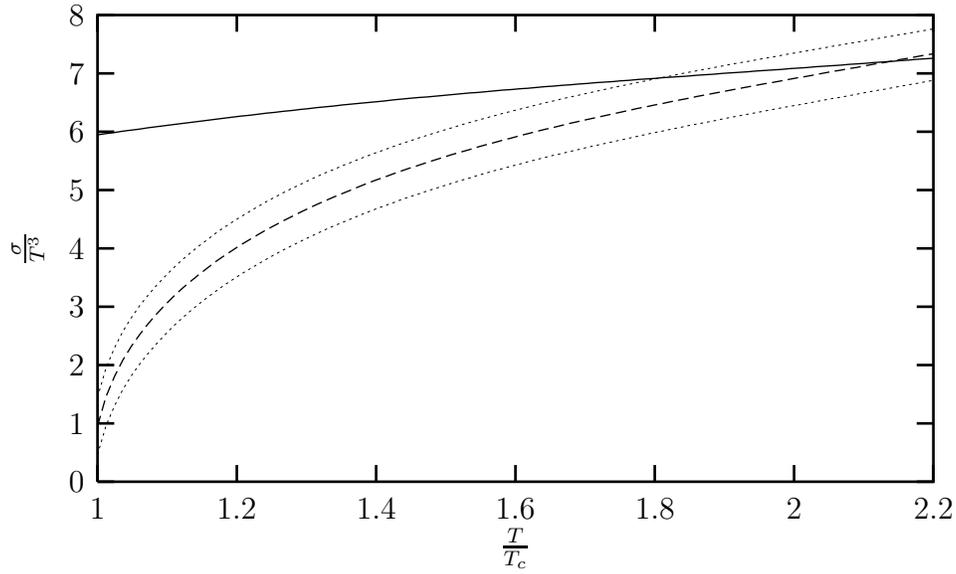

\begin	{figure}[p]
\begin	{center}
\leavevmode
\begingroup%
  \makeatletter%
  \newcommand{\GNUPLOTspecial}{%
    \@sanitize\catcode`\%=14\relax\special}%
  \setlength{\unitlength}{0.1bp}%
{\GNUPLOTspecial{!
/gnudict 256 dict def
gnudict begin
/Color false def
/Solid false def
/gnulinewidth 5.000 def
/userlinewidth gnulinewidth def
/vshift -33 def
/dl {10 mul} def
/hpt_ 31.5 def
/vpt_ 31.5 def
/hpt hpt_ def
/vpt vpt_ def
/M {moveto} bind def
/L {lineto} bind def
/R {rmoveto} bind def
/V {rlineto} bind def
/vpt2 vpt 2 mul def
/hpt2 hpt 2 mul def
/Lshow { currentpoint stroke M
  0 vshift R show } def
/Rshow { currentpoint stroke M
  dup stringwidth pop neg vshift R show } def
/Cshow { currentpoint stroke M
  dup stringwidth pop -2 div vshift R show } def
/UP { dup vpt_ mul /vpt exch def hpt_ mul /hpt exch def
  /hpt2 hpt 2 mul def /vpt2 vpt 2 mul def } def
/DL { Color {setrgbcolor Solid {pop []} if 0 setdash }
 {pop pop pop Solid {pop []} if 0 setdash} ifelse } def
/BL { stroke userlinewidth 2 mul setlinewidth } def
/AL { stroke userlinewidth 2 div setlinewidth } def
/UL { dup gnulinewidth mul /userlinewidth exch def
      10 mul /udl exch def } def
/PL { stroke userlinewidth setlinewidth } def
/LTb { BL [] 0 0 0 DL } def
/LTa { AL [1 udl mul 2 udl mul] 0 setdash 0 0 0 setrgbcolor } def
/LT0 { PL [] 1 0 0 DL } def
/LT1 { PL [4 dl 2 dl] 0 1 0 DL } def
/LT2 { PL [2 dl 3 dl] 0 0 1 DL } def
/LT3 { PL [1 dl 1.5 dl] 1 0 1 DL } def
/LT4 { PL [5 dl 2 dl 1 dl 2 dl] 0 1 1 DL } def
/LT5 { PL [4 dl 3 dl 1 dl 3 dl] 1 1 0 DL } def
/LT6 { PL [2 dl 2 dl 2 dl 4 dl] 0 0 0 DL } def
/LT7 { PL [2 dl 2 dl 2 dl 2 dl 2 dl 4 dl] 1 0.3 0 DL } def
/LT8 { PL [2 dl 2 dl 2 dl 2 dl 2 dl 2 dl 2 dl 4 dl] 0.5 0.5 0.5 DL } def
/Pnt { stroke [] 0 setdash
   gsave 1 setlinecap M 0 0 V stroke grestore } def
/Dia { stroke [] 0 setdash 2 copy vpt add M
  hpt neg vpt neg V hpt vpt neg V
  hpt vpt V hpt neg vpt V closepath stroke
  Pnt } def
/Pls { stroke [] 0 setdash vpt sub M 0 vpt2 V
  currentpoint stroke M
  hpt neg vpt neg R hpt2 0 V stroke
  } def
/Box { stroke [] 0 setdash 2 copy exch hpt sub exch vpt add M
  0 vpt2 neg V hpt2 0 V 0 vpt2 V
  hpt2 neg 0 V closepath stroke
  Pnt } def
/Crs { stroke [] 0 setdash exch hpt sub exch vpt add M
  hpt2 vpt2 neg V currentpoint stroke M
  hpt2 neg 0 R hpt2 vpt2 V stroke } def
/TriU { stroke [] 0 setdash 2 copy vpt 1.12 mul add M
  hpt neg vpt -1.62 mul V
  hpt 2 mul 0 V
  hpt neg vpt 1.62 mul V closepath stroke
  Pnt  } def
/Star { 2 copy Pls Crs } def
/BoxF { stroke [] 0 setdash exch hpt sub exch vpt add M
  0 vpt2 neg V  hpt2 0 V  0 vpt2 V
  hpt2 neg 0 V  closepath fill } def
/TriUF { stroke [] 0 setdash vpt 1.12 mul add M
  hpt neg vpt -1.62 mul V
  hpt 2 mul 0 V
  hpt neg vpt 1.62 mul V closepath fill } def
/TriD { stroke [] 0 setdash 2 copy vpt 1.12 mul sub M
  hpt neg vpt 1.62 mul V
  hpt 2 mul 0 V
  hpt neg vpt -1.62 mul V closepath stroke
  Pnt  } def
/TriDF { stroke [] 0 setdash vpt 1.12 mul sub M
  hpt neg vpt 1.62 mul V
  hpt 2 mul 0 V
  hpt neg vpt -1.62 mul V closepath fill} def
/DiaF { stroke [] 0 setdash vpt add M
  hpt neg vpt neg V hpt vpt neg V
  hpt vpt V hpt neg vpt V closepath fill } def
/Pent { stroke [] 0 setdash 2 copy gsave
  translate 0 hpt M 4 {72 rotate 0 hpt L} repeat
  closepath stroke grestore Pnt } def
/PentF { stroke [] 0 setdash gsave
  translate 0 hpt M 4 {72 rotate 0 hpt L} repeat
  closepath fill grestore } def
/Circle { stroke [] 0 setdash 2 copy
  hpt 0 360 arc stroke Pnt } def
/CircleF { stroke [] 0 setdash hpt 0 360 arc fill } def
/C0 { BL [] 0 setdash 2 copy moveto vpt 90 450  arc } bind def
/C1 { BL [] 0 setdash 2 copy        moveto
       2 copy  vpt 0 90 arc closepath fill
               vpt 0 360 arc closepath } bind def
/C2 { BL [] 0 setdash 2 copy moveto
       2 copy  vpt 90 180 arc closepath fill
               vpt 0 360 arc closepath } bind def
/C3 { BL [] 0 setdash 2 copy moveto
       2 copy  vpt 0 180 arc closepath fill
               vpt 0 360 arc closepath } bind def
/C4 { BL [] 0 setdash 2 copy moveto
       2 copy  vpt 180 270 arc closepath fill
               vpt 0 360 arc closepath } bind def
/C5 { BL [] 0 setdash 2 copy moveto
       2 copy  vpt 0 90 arc
       2 copy moveto
       2 copy  vpt 180 270 arc closepath fill
               vpt 0 360 arc } bind def
/C6 { BL [] 0 setdash 2 copy moveto
      2 copy  vpt 90 270 arc closepath fill
              vpt 0 360 arc closepath } bind def
/C7 { BL [] 0 setdash 2 copy moveto
      2 copy  vpt 0 270 arc closepath fill
              vpt 0 360 arc closepath } bind def
/C8 { BL [] 0 setdash 2 copy moveto
      2 copy vpt 270 360 arc closepath fill
              vpt 0 360 arc closepath } bind def
/C9 { BL [] 0 setdash 2 copy moveto
      2 copy  vpt 270 450 arc closepath fill
              vpt 0 360 arc closepath } bind def
/C10 { BL [] 0 setdash 2 copy 2 copy moveto vpt 270 360 arc closepath fill
       2 copy moveto
       2 copy vpt 90 180 arc closepath fill
               vpt 0 360 arc closepath } bind def
/C11 { BL [] 0 setdash 2 copy moveto
       2 copy  vpt 0 180 arc closepath fill
       2 copy moveto
       2 copy  vpt 270 360 arc closepath fill
               vpt 0 360 arc closepath } bind def
/C12 { BL [] 0 setdash 2 copy moveto
       2 copy  vpt 180 360 arc closepath fill
               vpt 0 360 arc closepath } bind def
/C13 { BL [] 0 setdash  2 copy moveto
       2 copy  vpt 0 90 arc closepath fill
       2 copy moveto
       2 copy  vpt 180 360 arc closepath fill
               vpt 0 360 arc closepath } bind def
/C14 { BL [] 0 setdash 2 copy moveto
       2 copy  vpt 90 360 arc closepath fill
               vpt 0 360 arc } bind def
/C15 { BL [] 0 setdash 2 copy vpt 0 360 arc closepath fill
               vpt 0 360 arc closepath } bind def
/Rec   { newpath 4 2 roll moveto 1 index 0 rlineto 0 exch rlineto
       neg 0 rlineto closepath } bind def
/Square { dup Rec } bind def
/Bsquare { vpt sub exch vpt sub exch vpt2 Square } bind def
/S0 { BL [] 0 setdash 2 copy moveto 0 vpt rlineto BL Bsquare } bind def
/S1 { BL [] 0 setdash 2 copy vpt Square fill Bsquare } bind def
/S2 { BL [] 0 setdash 2 copy exch vpt sub exch vpt Square fill Bsquare } bind def
/S3 { BL [] 0 setdash 2 copy exch vpt sub exch vpt2 vpt Rec fill Bsquare } bind def
/S4 { BL [] 0 setdash 2 copy exch vpt sub exch vpt sub vpt Square fill Bsquare } bind def
/S5 { BL [] 0 setdash 2 copy 2 copy vpt Square fill
       exch vpt sub exch vpt sub vpt Square fill Bsquare } bind def
/S6 { BL [] 0 setdash 2 copy exch vpt sub exch vpt sub vpt vpt2 Rec fill Bsquare } bind def
/S7 { BL [] 0 setdash 2 copy exch vpt sub exch vpt sub vpt vpt2 Rec fill
       2 copy vpt Square fill
       Bsquare } bind def
/S8 { BL [] 0 setdash 2 copy vpt sub vpt Square fill Bsquare } bind def
/S9 { BL [] 0 setdash 2 copy vpt sub vpt vpt2 Rec fill Bsquare } bind def
/S10 { BL [] 0 setdash 2 copy vpt sub vpt Square fill 2 copy exch vpt sub exch vpt Square fill
       Bsquare } bind def
/S11 { BL [] 0 setdash 2 copy vpt sub vpt Square fill 2 copy exch vpt sub exch vpt2 vpt Rec fill
       Bsquare } bind def
/S12 { BL [] 0 setdash 2 copy exch vpt sub exch vpt sub vpt2 vpt Rec fill Bsquare } bind def
/S13 { BL [] 0 setdash 2 copy exch vpt sub exch vpt sub vpt2 vpt Rec fill
       2 copy vpt Square fill Bsquare } bind def
/S14 { BL [] 0 setdash 2 copy exch vpt sub exch vpt sub vpt2 vpt Rec fill
       2 copy exch vpt sub exch vpt Square fill Bsquare } bind def
/S15 { BL [] 0 setdash 2 copy Bsquare fill Bsquare } bind def
/D0 { gsave translate 45 rotate 0 0 S0 stroke grestore } bind def
/D1 { gsave translate 45 rotate 0 0 S1 stroke grestore } bind def
/D2 { gsave translate 45 rotate 0 0 S2 stroke grestore } bind def
/D3 { gsave translate 45 rotate 0 0 S3 stroke grestore } bind def
/D4 { gsave translate 45 rotate 0 0 S4 stroke grestore } bind def
/D5 { gsave translate 45 rotate 0 0 S5 stroke grestore } bind def
/D6 { gsave translate 45 rotate 0 0 S6 stroke grestore } bind def
/D7 { gsave translate 45 rotate 0 0 S7 stroke grestore } bind def
/D8 { gsave translate 45 rotate 0 0 S8 stroke grestore } bind def
/D9 { gsave translate 45 rotate 0 0 S9 stroke grestore } bind def
/D10 { gsave translate 45 rotate 0 0 S10 stroke grestore } bind def
/D11 { gsave translate 45 rotate 0 0 S11 stroke grestore } bind def
/D12 { gsave translate 45 rotate 0 0 S12 stroke grestore } bind def
/D13 { gsave translate 45 rotate 0 0 S13 stroke grestore } bind def
/D14 { gsave translate 45 rotate 0 0 S14 stroke grestore } bind def
/D15 { gsave translate 45 rotate 0 0 S15 stroke grestore } bind def
/DiaE { stroke [] 0 setdash vpt add M
  hpt neg vpt neg V hpt vpt neg V
  hpt vpt V hpt neg vpt V closepath stroke } def
/BoxE { stroke [] 0 setdash exch hpt sub exch vpt add M
  0 vpt2 neg V hpt2 0 V 0 vpt2 V
  hpt2 neg 0 V closepath stroke } def
/TriUE { stroke [] 0 setdash vpt 1.12 mul add M
  hpt neg vpt -1.62 mul V
  hpt 2 mul 0 V
  hpt neg vpt 1.62 mul V closepath stroke } def
/TriDE { stroke [] 0 setdash vpt 1.12 mul sub M
  hpt neg vpt 1.62 mul V
  hpt 2 mul 0 V
  hpt neg vpt -1.62 mul V closepath stroke } def
/PentE { stroke [] 0 setdash gsave
  translate 0 hpt M 4 {72 rotate 0 hpt L} repeat
  closepath stroke grestore } def
/CircE { stroke [] 0 setdash 
  hpt 0 360 arc stroke } def
/Opaque { gsave closepath 1 setgray fill grestore 0 setgray closepath } def
/DiaW { stroke [] 0 setdash vpt add M
  hpt neg vpt neg V hpt vpt neg V
  hpt vpt V hpt neg vpt V Opaque stroke } def
/BoxW { stroke [] 0 setdash exch hpt sub exch vpt add M
  0 vpt2 neg V hpt2 0 V 0 vpt2 V
  hpt2 neg 0 V Opaque stroke } def
/TriUW { stroke [] 0 setdash vpt 1.12 mul add M
  hpt neg vpt -1.62 mul V
  hpt 2 mul 0 V
  hpt neg vpt 1.62 mul V Opaque stroke } def
/TriDW { stroke [] 0 setdash vpt 1.12 mul sub M
  hpt neg vpt 1.62 mul V
  hpt 2 mul 0 V
  hpt neg vpt -1.62 mul V Opaque stroke } def
/PentW { stroke [] 0 setdash gsave
  translate 0 hpt M 4 {72 rotate 0 hpt L} repeat
  Opaque stroke grestore } def
/CircW { stroke [] 0 setdash 
  hpt 0 360 arc Opaque stroke } def
/BoxFill { gsave Rec 1 setgray fill grestore } def
end
}}%
\begin{picture}(3600,2160)(0,0)%
{\GNUPLOTspecial{"
gnudict begin
gsave
0 0 translate
0.100 0.100 scale
0 setgray
newpath
1.000 UL
LTb
450 300 M
63 0 V
2937 0 R
-63 0 V
450 652 M
63 0 V
2937 0 R
-63 0 V
450 1004 M
63 0 V
2937 0 R
-63 0 V
450 1356 M
63 0 V
2937 0 R
-63 0 V
450 1708 M
63 0 V
2937 0 R
-63 0 V
450 2060 M
63 0 V
2937 0 R
-63 0 V
450 300 M
0 63 V
0 1697 R
0 -63 V
1050 300 M
0 63 V
0 1697 R
0 -63 V
1650 300 M
0 63 V
0 1697 R
0 -63 V
2250 300 M
0 63 V
0 1697 R
0 -63 V
2850 300 M
0 63 V
0 1697 R
0 -63 V
3450 300 M
0 63 V
0 1697 R
0 -63 V
1.000 UL
LTb
450 300 M
3000 0 V
0 1760 V
-3000 0 V
450 300 L
1.000 UP
1.000 UL
LT0
3263 300 M
-1 0 V
-1 0 V
-1 0 V
-1 0 V
-6 0 V
-12 0 V
-22 1 V
-47 2 V
-91 6 V
-168 20 V
-291 51 V
2191 497 L
1671 702 L
-533 312 V
787 1320 L
587 1573 L
-88 170 V
-33 100 V
-12 54 V
-3 27 V
-1 13 V
0 5 V
0 6 R
0 1 V
0 1 V
0 -1 V
3263 300 Pls
3262 300 Pls
3262 300 Pls
3261 300 Pls
3260 300 Pls
3259 300 Pls
3253 300 Pls
3241 300 Pls
3219 301 Pls
3172 303 Pls
3081 309 Pls
2913 329 Pls
2622 380 Pls
2191 497 Pls
1671 702 Pls
1138 1014 Pls
787 1320 Pls
587 1573 Pls
499 1743 Pls
466 1843 Pls
454 1897 Pls
451 1924 Pls
450 1937 Pls
450 1949 Pls
450 1950 Pls
450 1950 Pls
1.000 UL
LT1
3264 300 M
-1 0 V
-1 0 V
-2 0 V
-3 0 V
-5 0 V
-8 1 V
-15 0 V
-26 2 V
-42 3 V
-68 8 V
-106 15 V
-158 28 V
-222 51 V
-286 84 V
1981 618 L
1621 786 L
1279 987 L
992 1198 L
777 1397 L
634 1566 L
-87 129 V
-48 93 V
-26 63 V
-12 40 V
-6 24 V
-3 15 V
-1 9 V
-1 5 V
0 3 V
0 2 V
0 1 V
0 1 V
stroke
grestore
end
showpage
}}%
\put(1950,50){\makebox(0,0){$\mathrm{Re}\, \overline{l_{p}}$}}%
\put(100,1180){%
\makebox(0,0)[b]{\shortstack{$\mathrm{Im}\, \overline{l_{p}}$}}%
}%
\put(3450,200){\makebox(0,0){0.25}}%
\put(2850,200){\makebox(0,0){0.2}}%
\put(2250,200){\makebox(0,0){0.15}}%
\put(1650,200){\makebox(0,0){0.1}}%
\put(1050,200){\makebox(0,0){0.05}}%
\put(450,200){\makebox(0,0){0}}%
\put(400,2060){\makebox(0,0)[r]{0.25}}%
\put(400,1708){\makebox(0,0)[r]{0.2}}%
\put(400,1356){\makebox(0,0)[r]{0.15}}%
\put(400,1004){\makebox(0,0)[r]{0.1}}%
\put(400,652){\makebox(0,0)[r]{0.05}}%
\put(400,300){\makebox(0,0)[r]{0}}%
\end{picture}%
\endgroup

\end	{center}
\vskip 0.15in
\caption{The values in the complex plane taken by the Polyakov 
loop for $x_3\in [0,L-1]$, for the $k=1$ wall in SU(4) with
$aT=0.25$ and $T\simeq 1.88 T_c$. Dashed line is one-loop
perturbation theory.}
\label{fig_avprofc1}
\end 	{figure}

\begin	{figure}[p]
\begin	{center}
\leavevmode
\begingroup%
  \makeatletter%
  \newcommand{\GNUPLOTspecial}{%
    \@sanitize\catcode`\%=14\relax\special}%
  \setlength{\unitlength}{0.1bp}%
{\GNUPLOTspecial{!
/gnudict 256 dict def
gnudict begin
/Color false def
/Solid false def
/gnulinewidth 5.000 def
/userlinewidth gnulinewidth def
/vshift -33 def
/dl {10 mul} def
/hpt_ 31.5 def
/vpt_ 31.5 def
/hpt hpt_ def
/vpt vpt_ def
/M {moveto} bind def
/L {lineto} bind def
/R {rmoveto} bind def
/V {rlineto} bind def
/vpt2 vpt 2 mul def
/hpt2 hpt 2 mul def
/Lshow { currentpoint stroke M
  0 vshift R show } def
/Rshow { currentpoint stroke M
  dup stringwidth pop neg vshift R show } def
/Cshow { currentpoint stroke M
  dup stringwidth pop -2 div vshift R show } def
/UP { dup vpt_ mul /vpt exch def hpt_ mul /hpt exch def
  /hpt2 hpt 2 mul def /vpt2 vpt 2 mul def } def
/DL { Color {setrgbcolor Solid {pop []} if 0 setdash }
 {pop pop pop Solid {pop []} if 0 setdash} ifelse } def
/BL { stroke userlinewidth 2 mul setlinewidth } def
/AL { stroke userlinewidth 2 div setlinewidth } def
/UL { dup gnulinewidth mul /userlinewidth exch def
      10 mul /udl exch def } def
/PL { stroke userlinewidth setlinewidth } def
/LTb { BL [] 0 0 0 DL } def
/LTa { AL [1 udl mul 2 udl mul] 0 setdash 0 0 0 setrgbcolor } def
/LT0 { PL [] 1 0 0 DL } def
/LT1 { PL [4 dl 2 dl] 0 1 0 DL } def
/LT2 { PL [2 dl 3 dl] 0 0 1 DL } def
/LT3 { PL [1 dl 1.5 dl] 1 0 1 DL } def
/LT4 { PL [5 dl 2 dl 1 dl 2 dl] 0 1 1 DL } def
/LT5 { PL [4 dl 3 dl 1 dl 3 dl] 1 1 0 DL } def
/LT6 { PL [2 dl 2 dl 2 dl 4 dl] 0 0 0 DL } def
/LT7 { PL [2 dl 2 dl 2 dl 2 dl 2 dl 4 dl] 1 0.3 0 DL } def
/LT8 { PL [2 dl 2 dl 2 dl 2 dl 2 dl 2 dl 2 dl 4 dl] 0.5 0.5 0.5 DL } def
/Pnt { stroke [] 0 setdash
   gsave 1 setlinecap M 0 0 V stroke grestore } def
/Dia { stroke [] 0 setdash 2 copy vpt add M
  hpt neg vpt neg V hpt vpt neg V
  hpt vpt V hpt neg vpt V closepath stroke
  Pnt } def
/Pls { stroke [] 0 setdash vpt sub M 0 vpt2 V
  currentpoint stroke M
  hpt neg vpt neg R hpt2 0 V stroke
  } def
/Box { stroke [] 0 setdash 2 copy exch hpt sub exch vpt add M
  0 vpt2 neg V hpt2 0 V 0 vpt2 V
  hpt2 neg 0 V closepath stroke
  Pnt } def
/Crs { stroke [] 0 setdash exch hpt sub exch vpt add M
  hpt2 vpt2 neg V currentpoint stroke M
  hpt2 neg 0 R hpt2 vpt2 V stroke } def
/TriU { stroke [] 0 setdash 2 copy vpt 1.12 mul add M
  hpt neg vpt -1.62 mul V
  hpt 2 mul 0 V
  hpt neg vpt 1.62 mul V closepath stroke
  Pnt  } def
/Star { 2 copy Pls Crs } def
/BoxF { stroke [] 0 setdash exch hpt sub exch vpt add M
  0 vpt2 neg V  hpt2 0 V  0 vpt2 V
  hpt2 neg 0 V  closepath fill } def
/TriUF { stroke [] 0 setdash vpt 1.12 mul add M
  hpt neg vpt -1.62 mul V
  hpt 2 mul 0 V
  hpt neg vpt 1.62 mul V closepath fill } def
/TriD { stroke [] 0 setdash 2 copy vpt 1.12 mul sub M
  hpt neg vpt 1.62 mul V
  hpt 2 mul 0 V
  hpt neg vpt -1.62 mul V closepath stroke
  Pnt  } def
/TriDF { stroke [] 0 setdash vpt 1.12 mul sub M
  hpt neg vpt 1.62 mul V
  hpt 2 mul 0 V
  hpt neg vpt -1.62 mul V closepath fill} def
/DiaF { stroke [] 0 setdash vpt add M
  hpt neg vpt neg V hpt vpt neg V
  hpt vpt V hpt neg vpt V closepath fill } def
/Pent { stroke [] 0 setdash 2 copy gsave
  translate 0 hpt M 4 {72 rotate 0 hpt L} repeat
  closepath stroke grestore Pnt } def
/PentF { stroke [] 0 setdash gsave
  translate 0 hpt M 4 {72 rotate 0 hpt L} repeat
  closepath fill grestore } def
/Circle { stroke [] 0 setdash 2 copy
  hpt 0 360 arc stroke Pnt } def
/CircleF { stroke [] 0 setdash hpt 0 360 arc fill } def
/C0 { BL [] 0 setdash 2 copy moveto vpt 90 450  arc } bind def
/C1 { BL [] 0 setdash 2 copy        moveto
       2 copy  vpt 0 90 arc closepath fill
               vpt 0 360 arc closepath } bind def
/C2 { BL [] 0 setdash 2 copy moveto
       2 copy  vpt 90 180 arc closepath fill
               vpt 0 360 arc closepath } bind def
/C3 { BL [] 0 setdash 2 copy moveto
       2 copy  vpt 0 180 arc closepath fill
               vpt 0 360 arc closepath } bind def
/C4 { BL [] 0 setdash 2 copy moveto
       2 copy  vpt 180 270 arc closepath fill
               vpt 0 360 arc closepath } bind def
/C5 { BL [] 0 setdash 2 copy moveto
       2 copy  vpt 0 90 arc
       2 copy moveto
       2 copy  vpt 180 270 arc closepath fill
               vpt 0 360 arc } bind def
/C6 { BL [] 0 setdash 2 copy moveto
      2 copy  vpt 90 270 arc closepath fill
              vpt 0 360 arc closepath } bind def
/C7 { BL [] 0 setdash 2 copy moveto
      2 copy  vpt 0 270 arc closepath fill
              vpt 0 360 arc closepath } bind def
/C8 { BL [] 0 setdash 2 copy moveto
      2 copy vpt 270 360 arc closepath fill
              vpt 0 360 arc closepath } bind def
/C9 { BL [] 0 setdash 2 copy moveto
      2 copy  vpt 270 450 arc closepath fill
              vpt 0 360 arc closepath } bind def
/C10 { BL [] 0 setdash 2 copy 2 copy moveto vpt 270 360 arc closepath fill
       2 copy moveto
       2 copy vpt 90 180 arc closepath fill
               vpt 0 360 arc closepath } bind def
/C11 { BL [] 0 setdash 2 copy moveto
       2 copy  vpt 0 180 arc closepath fill
       2 copy moveto
       2 copy  vpt 270 360 arc closepath fill
               vpt 0 360 arc closepath } bind def
/C12 { BL [] 0 setdash 2 copy moveto
       2 copy  vpt 180 360 arc closepath fill
               vpt 0 360 arc closepath } bind def
/C13 { BL [] 0 setdash  2 copy moveto
       2 copy  vpt 0 90 arc closepath fill
       2 copy moveto
       2 copy  vpt 180 360 arc closepath fill
               vpt 0 360 arc closepath } bind def
/C14 { BL [] 0 setdash 2 copy moveto
       2 copy  vpt 90 360 arc closepath fill
               vpt 0 360 arc } bind def
/C15 { BL [] 0 setdash 2 copy vpt 0 360 arc closepath fill
               vpt 0 360 arc closepath } bind def
/Rec   { newpath 4 2 roll moveto 1 index 0 rlineto 0 exch rlineto
       neg 0 rlineto closepath } bind def
/Square { dup Rec } bind def
/Bsquare { vpt sub exch vpt sub exch vpt2 Square } bind def
/S0 { BL [] 0 setdash 2 copy moveto 0 vpt rlineto BL Bsquare } bind def
/S1 { BL [] 0 setdash 2 copy vpt Square fill Bsquare } bind def
/S2 { BL [] 0 setdash 2 copy exch vpt sub exch vpt Square fill Bsquare } bind def
/S3 { BL [] 0 setdash 2 copy exch vpt sub exch vpt2 vpt Rec fill Bsquare } bind def
/S4 { BL [] 0 setdash 2 copy exch vpt sub exch vpt sub vpt Square fill Bsquare } bind def
/S5 { BL [] 0 setdash 2 copy 2 copy vpt Square fill
       exch vpt sub exch vpt sub vpt Square fill Bsquare } bind def
/S6 { BL [] 0 setdash 2 copy exch vpt sub exch vpt sub vpt vpt2 Rec fill Bsquare } bind def
/S7 { BL [] 0 setdash 2 copy exch vpt sub exch vpt sub vpt vpt2 Rec fill
       2 copy vpt Square fill
       Bsquare } bind def
/S8 { BL [] 0 setdash 2 copy vpt sub vpt Square fill Bsquare } bind def
/S9 { BL [] 0 setdash 2 copy vpt sub vpt vpt2 Rec fill Bsquare } bind def
/S10 { BL [] 0 setdash 2 copy vpt sub vpt Square fill 2 copy exch vpt sub exch vpt Square fill
       Bsquare } bind def
/S11 { BL [] 0 setdash 2 copy vpt sub vpt Square fill 2 copy exch vpt sub exch vpt2 vpt Rec fill
       Bsquare } bind def
/S12 { BL [] 0 setdash 2 copy exch vpt sub exch vpt sub vpt2 vpt Rec fill Bsquare } bind def
/S13 { BL [] 0 setdash 2 copy exch vpt sub exch vpt sub vpt2 vpt Rec fill
       2 copy vpt Square fill Bsquare } bind def
/S14 { BL [] 0 setdash 2 copy exch vpt sub exch vpt sub vpt2 vpt Rec fill
       2 copy exch vpt sub exch vpt Square fill Bsquare } bind def
/S15 { BL [] 0 setdash 2 copy Bsquare fill Bsquare } bind def
/D0 { gsave translate 45 rotate 0 0 S0 stroke grestore } bind def
/D1 { gsave translate 45 rotate 0 0 S1 stroke grestore } bind def
/D2 { gsave translate 45 rotate 0 0 S2 stroke grestore } bind def
/D3 { gsave translate 45 rotate 0 0 S3 stroke grestore } bind def
/D4 { gsave translate 45 rotate 0 0 S4 stroke grestore } bind def
/D5 { gsave translate 45 rotate 0 0 S5 stroke grestore } bind def
/D6 { gsave translate 45 rotate 0 0 S6 stroke grestore } bind def
/D7 { gsave translate 45 rotate 0 0 S7 stroke grestore } bind def
/D8 { gsave translate 45 rotate 0 0 S8 stroke grestore } bind def
/D9 { gsave translate 45 rotate 0 0 S9 stroke grestore } bind def
/D10 { gsave translate 45 rotate 0 0 S10 stroke grestore } bind def
/D11 { gsave translate 45 rotate 0 0 S11 stroke grestore } bind def
/D12 { gsave translate 45 rotate 0 0 S12 stroke grestore } bind def
/D13 { gsave translate 45 rotate 0 0 S13 stroke grestore } bind def
/D14 { gsave translate 45 rotate 0 0 S14 stroke grestore } bind def
/D15 { gsave translate 45 rotate 0 0 S15 stroke grestore } bind def
/DiaE { stroke [] 0 setdash vpt add M
  hpt neg vpt neg V hpt vpt neg V
  hpt vpt V hpt neg vpt V closepath stroke } def
/BoxE { stroke [] 0 setdash exch hpt sub exch vpt add M
  0 vpt2 neg V hpt2 0 V 0 vpt2 V
  hpt2 neg 0 V closepath stroke } def
/TriUE { stroke [] 0 setdash vpt 1.12 mul add M
  hpt neg vpt -1.62 mul V
  hpt 2 mul 0 V
  hpt neg vpt 1.62 mul V closepath stroke } def
/TriDE { stroke [] 0 setdash vpt 1.12 mul sub M
  hpt neg vpt 1.62 mul V
  hpt 2 mul 0 V
  hpt neg vpt -1.62 mul V closepath stroke } def
/PentE { stroke [] 0 setdash gsave
  translate 0 hpt M 4 {72 rotate 0 hpt L} repeat
  closepath stroke grestore } def
/CircE { stroke [] 0 setdash 
  hpt 0 360 arc stroke } def
/Opaque { gsave closepath 1 setgray fill grestore 0 setgray closepath } def
/DiaW { stroke [] 0 setdash vpt add M
  hpt neg vpt neg V hpt vpt neg V
  hpt vpt V hpt neg vpt V Opaque stroke } def
/BoxW { stroke [] 0 setdash exch hpt sub exch vpt add M
  0 vpt2 neg V hpt2 0 V 0 vpt2 V
  hpt2 neg 0 V Opaque stroke } def
/TriUW { stroke [] 0 setdash vpt 1.12 mul add M
  hpt neg vpt -1.62 mul V
  hpt 2 mul 0 V
  hpt neg vpt 1.62 mul V Opaque stroke } def
/TriDW { stroke [] 0 setdash vpt 1.12 mul sub M
  hpt neg vpt 1.62 mul V
  hpt 2 mul 0 V
  hpt neg vpt -1.62 mul V Opaque stroke } def
/PentW { stroke [] 0 setdash gsave
  translate 0 hpt M 4 {72 rotate 0 hpt L} repeat
  Opaque stroke grestore } def
/CircW { stroke [] 0 setdash 
  hpt 0 360 arc Opaque stroke } def
/BoxFill { gsave Rec 1 setgray fill grestore } def
end
}}%
\begin{picture}(3600,2160)(0,0)%
{\GNUPLOTspecial{"
gnudict begin
gsave
0 0 translate
0.100 0.100 scale
0 setgray
newpath
1.000 UL
LTb
450 302 M
63 0 V
2937 0 R
-63 0 V
450 522 M
63 0 V
2937 0 R
-63 0 V
450 742 M
63 0 V
2937 0 R
-63 0 V
450 961 M
63 0 V
2937 0 R
-63 0 V
450 1181 M
63 0 V
2937 0 R
-63 0 V
450 1401 M
63 0 V
2937 0 R
-63 0 V
450 1621 M
63 0 V
2937 0 R
-63 0 V
450 1840 M
63 0 V
2937 0 R
-63 0 V
450 2060 M
63 0 V
2937 0 R
-63 0 V
454 300 M
0 63 V
0 1697 R
0 -63 V
828 300 M
0 63 V
0 1697 R
0 -63 V
1203 300 M
0 63 V
0 1697 R
0 -63 V
1577 300 M
0 63 V
0 1697 R
0 -63 V
1952 300 M
0 63 V
0 1697 R
0 -63 V
2326 300 M
0 63 V
0 1697 R
0 -63 V
2701 300 M
0 63 V
0 1697 R
0 -63 V
3075 300 M
0 63 V
0 1697 R
0 -63 V
3450 300 M
0 63 V
0 1697 R
0 -63 V
1.000 UL
LTb
450 300 M
3000 0 V
0 1760 V
-3000 0 V
450 300 L
1.000 UP
1.000 UL
LT0
3050 302 M
-9 1 V
-9 1 V
-16 -1 V
-32 1 V
-45 -3 V
-58 1 V
-97 0 V
-135 3 V
-181 3 V
-225 10 V
-267 18 V
-289 33 V
-291 54 V
-265 84 V
798 695 L
650 844 L
-96 163 V
-56 162 V
-26 150 V
-13 132 V
-3 108 V
-3 84 V
-1 62 V
0 43 V
0 27 V
-1 19 V
2 12 V
-1 7 V
1 5 V
3050 302 Pls
3041 303 Pls
3032 304 Pls
3016 303 Pls
2984 304 Pls
2939 301 Pls
2881 302 Pls
2784 302 Pls
2649 305 Pls
2468 308 Pls
2243 318 Pls
1976 336 Pls
1687 369 Pls
1396 423 Pls
1131 507 Pls
798 695 Pls
650 844 Pls
554 1007 Pls
498 1169 Pls
472 1319 Pls
459 1451 Pls
456 1559 Pls
453 1643 Pls
452 1705 Pls
452 1748 Pls
452 1775 Pls
451 1794 Pls
453 1806 Pls
452 1813 Pls
453 1818 Pls
1.000 UL
LT1
3085 302 M
-1 0 V
-1 0 V
-2 0 V
-2 0 V
-5 0 V
-8 1 V
-14 0 V
-24 2 V
-39 3 V
-64 7 V
-99 14 V
-148 26 V
-207 48 V
-268 79 V
1886 599 L
1548 757 L
1229 944 L
960 1142 L
760 1328 L
626 1486 L
-81 121 V
-46 87 V
-23 58 V
-12 37 V
-6 24 V
-2 13 V
-1 9 V
-1 4 V
0 3 V
0 2 V
0 1 V
0 1 V
stroke
grestore
end
showpage
}}%
\put(1950,50){\makebox(0,0){$\mathrm{Re}\, \overline{l_{p}}$}}%
\put(100,1180){%
\makebox(0,0)[b]{\shortstack{$\mathrm{Im}\, \overline{l_{p}}$}}%
}%
\put(3450,200){\makebox(0,0){0.16}}%
\put(3075,200){\makebox(0,0){0.14}}%
\put(2701,200){\makebox(0,0){0.12}}%
\put(2326,200){\makebox(0,0){0.1}}%
\put(1952,200){\makebox(0,0){0.08}}%
\put(1577,200){\makebox(0,0){0.06}}%
\put(1203,200){\makebox(0,0){0.04}}%
\put(828,200){\makebox(0,0){0.02}}%
\put(454,200){\makebox(0,0){0}}%
\put(400,2060){\makebox(0,0)[r]{0.16}}%
\put(400,1840){\makebox(0,0)[r]{0.14}}%
\put(400,1621){\makebox(0,0)[r]{0.12}}%
\put(400,1401){\makebox(0,0)[r]{0.1}}%
\put(400,1181){\makebox(0,0)[r]{0.08}}%
\put(400,961){\makebox(0,0)[r]{0.06}}%
\put(400,742){\makebox(0,0)[r]{0.04}}%
\put(400,522){\makebox(0,0)[r]{0.02}}%
\put(400,302){\makebox(0,0)[r]{0}}%
\end{picture}%
\endgroup

\end	{center}
\vskip 0.15in
\caption{The values in the complex plane taken by the Polyakov 
loop for $x_3\in [0,L-1]$, for the $k=1$ wall in SU(4) with
$aT=0.25$ and $T\simeq 1.02 T_c$. Dashed line is one-loop
perturbation theory.}
\label{fig_avprofc2}
\end 	{figure}

\begin	{figure}[p]
\begin	{center}
\leavevmode
\begingroup%
  \makeatletter%
  \newcommand{\GNUPLOTspecial}{%
    \@sanitize\catcode`\%=14\relax\special}%
  \setlength{\unitlength}{0.1bp}%
{\GNUPLOTspecial{!
/gnudict 256 dict def
gnudict begin
/Color false def
/Solid false def
/gnulinewidth 5.000 def
/userlinewidth gnulinewidth def
/vshift -33 def
/dl {10 mul} def
/hpt_ 31.5 def
/vpt_ 31.5 def
/hpt hpt_ def
/vpt vpt_ def
/M {moveto} bind def
/L {lineto} bind def
/R {rmoveto} bind def
/V {rlineto} bind def
/vpt2 vpt 2 mul def
/hpt2 hpt 2 mul def
/Lshow { currentpoint stroke M
  0 vshift R show } def
/Rshow { currentpoint stroke M
  dup stringwidth pop neg vshift R show } def
/Cshow { currentpoint stroke M
  dup stringwidth pop -2 div vshift R show } def
/UP { dup vpt_ mul /vpt exch def hpt_ mul /hpt exch def
  /hpt2 hpt 2 mul def /vpt2 vpt 2 mul def } def
/DL { Color {setrgbcolor Solid {pop []} if 0 setdash }
 {pop pop pop Solid {pop []} if 0 setdash} ifelse } def
/BL { stroke userlinewidth 2 mul setlinewidth } def
/AL { stroke userlinewidth 2 div setlinewidth } def
/UL { dup gnulinewidth mul /userlinewidth exch def
      10 mul /udl exch def } def
/PL { stroke userlinewidth setlinewidth } def
/LTb { BL [] 0 0 0 DL } def
/LTa { AL [1 udl mul 2 udl mul] 0 setdash 0 0 0 setrgbcolor } def
/LT0 { PL [] 1 0 0 DL } def
/LT1 { PL [4 dl 2 dl] 0 1 0 DL } def
/LT2 { PL [2 dl 3 dl] 0 0 1 DL } def
/LT3 { PL [1 dl 1.5 dl] 1 0 1 DL } def
/LT4 { PL [5 dl 2 dl 1 dl 2 dl] 0 1 1 DL } def
/LT5 { PL [4 dl 3 dl 1 dl 3 dl] 1 1 0 DL } def
/LT6 { PL [2 dl 2 dl 2 dl 4 dl] 0 0 0 DL } def
/LT7 { PL [2 dl 2 dl 2 dl 2 dl 2 dl 4 dl] 1 0.3 0 DL } def
/LT8 { PL [2 dl 2 dl 2 dl 2 dl 2 dl 2 dl 2 dl 4 dl] 0.5 0.5 0.5 DL } def
/Pnt { stroke [] 0 setdash
   gsave 1 setlinecap M 0 0 V stroke grestore } def
/Dia { stroke [] 0 setdash 2 copy vpt add M
  hpt neg vpt neg V hpt vpt neg V
  hpt vpt V hpt neg vpt V closepath stroke
  Pnt } def
/Pls { stroke [] 0 setdash vpt sub M 0 vpt2 V
  currentpoint stroke M
  hpt neg vpt neg R hpt2 0 V stroke
  } def
/Box { stroke [] 0 setdash 2 copy exch hpt sub exch vpt add M
  0 vpt2 neg V hpt2 0 V 0 vpt2 V
  hpt2 neg 0 V closepath stroke
  Pnt } def
/Crs { stroke [] 0 setdash exch hpt sub exch vpt add M
  hpt2 vpt2 neg V currentpoint stroke M
  hpt2 neg 0 R hpt2 vpt2 V stroke } def
/TriU { stroke [] 0 setdash 2 copy vpt 1.12 mul add M
  hpt neg vpt -1.62 mul V
  hpt 2 mul 0 V
  hpt neg vpt 1.62 mul V closepath stroke
  Pnt  } def
/Star { 2 copy Pls Crs } def
/BoxF { stroke [] 0 setdash exch hpt sub exch vpt add M
  0 vpt2 neg V  hpt2 0 V  0 vpt2 V
  hpt2 neg 0 V  closepath fill } def
/TriUF { stroke [] 0 setdash vpt 1.12 mul add M
  hpt neg vpt -1.62 mul V
  hpt 2 mul 0 V
  hpt neg vpt 1.62 mul V closepath fill } def
/TriD { stroke [] 0 setdash 2 copy vpt 1.12 mul sub M
  hpt neg vpt 1.62 mul V
  hpt 2 mul 0 V
  hpt neg vpt -1.62 mul V closepath stroke
  Pnt  } def
/TriDF { stroke [] 0 setdash vpt 1.12 mul sub M
  hpt neg vpt 1.62 mul V
  hpt 2 mul 0 V
  hpt neg vpt -1.62 mul V closepath fill} def
/DiaF { stroke [] 0 setdash vpt add M
  hpt neg vpt neg V hpt vpt neg V
  hpt vpt V hpt neg vpt V closepath fill } def
/Pent { stroke [] 0 setdash 2 copy gsave
  translate 0 hpt M 4 {72 rotate 0 hpt L} repeat
  closepath stroke grestore Pnt } def
/PentF { stroke [] 0 setdash gsave
  translate 0 hpt M 4 {72 rotate 0 hpt L} repeat
  closepath fill grestore } def
/Circle { stroke [] 0 setdash 2 copy
  hpt 0 360 arc stroke Pnt } def
/CircleF { stroke [] 0 setdash hpt 0 360 arc fill } def
/C0 { BL [] 0 setdash 2 copy moveto vpt 90 450  arc } bind def
/C1 { BL [] 0 setdash 2 copy        moveto
       2 copy  vpt 0 90 arc closepath fill
               vpt 0 360 arc closepath } bind def
/C2 { BL [] 0 setdash 2 copy moveto
       2 copy  vpt 90 180 arc closepath fill
               vpt 0 360 arc closepath } bind def
/C3 { BL [] 0 setdash 2 copy moveto
       2 copy  vpt 0 180 arc closepath fill
               vpt 0 360 arc closepath } bind def
/C4 { BL [] 0 setdash 2 copy moveto
       2 copy  vpt 180 270 arc closepath fill
               vpt 0 360 arc closepath } bind def
/C5 { BL [] 0 setdash 2 copy moveto
       2 copy  vpt 0 90 arc
       2 copy moveto
       2 copy  vpt 180 270 arc closepath fill
               vpt 0 360 arc } bind def
/C6 { BL [] 0 setdash 2 copy moveto
      2 copy  vpt 90 270 arc closepath fill
              vpt 0 360 arc closepath } bind def
/C7 { BL [] 0 setdash 2 copy moveto
      2 copy  vpt 0 270 arc closepath fill
              vpt 0 360 arc closepath } bind def
/C8 { BL [] 0 setdash 2 copy moveto
      2 copy vpt 270 360 arc closepath fill
              vpt 0 360 arc closepath } bind def
/C9 { BL [] 0 setdash 2 copy moveto
      2 copy  vpt 270 450 arc closepath fill
              vpt 0 360 arc closepath } bind def
/C10 { BL [] 0 setdash 2 copy 2 copy moveto vpt 270 360 arc closepath fill
       2 copy moveto
       2 copy vpt 90 180 arc closepath fill
               vpt 0 360 arc closepath } bind def
/C11 { BL [] 0 setdash 2 copy moveto
       2 copy  vpt 0 180 arc closepath fill
       2 copy moveto
       2 copy  vpt 270 360 arc closepath fill
               vpt 0 360 arc closepath } bind def
/C12 { BL [] 0 setdash 2 copy moveto
       2 copy  vpt 180 360 arc closepath fill
               vpt 0 360 arc closepath } bind def
/C13 { BL [] 0 setdash  2 copy moveto
       2 copy  vpt 0 90 arc closepath fill
       2 copy moveto
       2 copy  vpt 180 360 arc closepath fill
               vpt 0 360 arc closepath } bind def
/C14 { BL [] 0 setdash 2 copy moveto
       2 copy  vpt 90 360 arc closepath fill
               vpt 0 360 arc } bind def
/C15 { BL [] 0 setdash 2 copy vpt 0 360 arc closepath fill
               vpt 0 360 arc closepath } bind def
/Rec   { newpath 4 2 roll moveto 1 index 0 rlineto 0 exch rlineto
       neg 0 rlineto closepath } bind def
/Square { dup Rec } bind def
/Bsquare { vpt sub exch vpt sub exch vpt2 Square } bind def
/S0 { BL [] 0 setdash 2 copy moveto 0 vpt rlineto BL Bsquare } bind def
/S1 { BL [] 0 setdash 2 copy vpt Square fill Bsquare } bind def
/S2 { BL [] 0 setdash 2 copy exch vpt sub exch vpt Square fill Bsquare } bind def
/S3 { BL [] 0 setdash 2 copy exch vpt sub exch vpt2 vpt Rec fill Bsquare } bind def
/S4 { BL [] 0 setdash 2 copy exch vpt sub exch vpt sub vpt Square fill Bsquare } bind def
/S5 { BL [] 0 setdash 2 copy 2 copy vpt Square fill
       exch vpt sub exch vpt sub vpt Square fill Bsquare } bind def
/S6 { BL [] 0 setdash 2 copy exch vpt sub exch vpt sub vpt vpt2 Rec fill Bsquare } bind def
/S7 { BL [] 0 setdash 2 copy exch vpt sub exch vpt sub vpt vpt2 Rec fill
       2 copy vpt Square fill
       Bsquare } bind def
/S8 { BL [] 0 setdash 2 copy vpt sub vpt Square fill Bsquare } bind def
/S9 { BL [] 0 setdash 2 copy vpt sub vpt vpt2 Rec fill Bsquare } bind def
/S10 { BL [] 0 setdash 2 copy vpt sub vpt Square fill 2 copy exch vpt sub exch vpt Square fill
       Bsquare } bind def
/S11 { BL [] 0 setdash 2 copy vpt sub vpt Square fill 2 copy exch vpt sub exch vpt2 vpt Rec fill
       Bsquare } bind def
/S12 { BL [] 0 setdash 2 copy exch vpt sub exch vpt sub vpt2 vpt Rec fill Bsquare } bind def
/S13 { BL [] 0 setdash 2 copy exch vpt sub exch vpt sub vpt2 vpt Rec fill
       2 copy vpt Square fill Bsquare } bind def
/S14 { BL [] 0 setdash 2 copy exch vpt sub exch vpt sub vpt2 vpt Rec fill
       2 copy exch vpt sub exch vpt Square fill Bsquare } bind def
/S15 { BL [] 0 setdash 2 copy Bsquare fill Bsquare } bind def
/D0 { gsave translate 45 rotate 0 0 S0 stroke grestore } bind def
/D1 { gsave translate 45 rotate 0 0 S1 stroke grestore } bind def
/D2 { gsave translate 45 rotate 0 0 S2 stroke grestore } bind def
/D3 { gsave translate 45 rotate 0 0 S3 stroke grestore } bind def
/D4 { gsave translate 45 rotate 0 0 S4 stroke grestore } bind def
/D5 { gsave translate 45 rotate 0 0 S5 stroke grestore } bind def
/D6 { gsave translate 45 rotate 0 0 S6 stroke grestore } bind def
/D7 { gsave translate 45 rotate 0 0 S7 stroke grestore } bind def
/D8 { gsave translate 45 rotate 0 0 S8 stroke grestore } bind def
/D9 { gsave translate 45 rotate 0 0 S9 stroke grestore } bind def
/D10 { gsave translate 45 rotate 0 0 S10 stroke grestore } bind def
/D11 { gsave translate 45 rotate 0 0 S11 stroke grestore } bind def
/D12 { gsave translate 45 rotate 0 0 S12 stroke grestore } bind def
/D13 { gsave translate 45 rotate 0 0 S13 stroke grestore } bind def
/D14 { gsave translate 45 rotate 0 0 S14 stroke grestore } bind def
/D15 { gsave translate 45 rotate 0 0 S15 stroke grestore } bind def
/DiaE { stroke [] 0 setdash vpt add M
  hpt neg vpt neg V hpt vpt neg V
  hpt vpt V hpt neg vpt V closepath stroke } def
/BoxE { stroke [] 0 setdash exch hpt sub exch vpt add M
  0 vpt2 neg V hpt2 0 V 0 vpt2 V
  hpt2 neg 0 V closepath stroke } def
/TriUE { stroke [] 0 setdash vpt 1.12 mul add M
  hpt neg vpt -1.62 mul V
  hpt 2 mul 0 V
  hpt neg vpt 1.62 mul V closepath stroke } def
/TriDE { stroke [] 0 setdash vpt 1.12 mul sub M
  hpt neg vpt 1.62 mul V
  hpt 2 mul 0 V
  hpt neg vpt -1.62 mul V closepath stroke } def
/PentE { stroke [] 0 setdash gsave
  translate 0 hpt M 4 {72 rotate 0 hpt L} repeat
  closepath stroke grestore } def
/CircE { stroke [] 0 setdash 
  hpt 0 360 arc stroke } def
/Opaque { gsave closepath 1 setgray fill grestore 0 setgray closepath } def
/DiaW { stroke [] 0 setdash vpt add M
  hpt neg vpt neg V hpt vpt neg V
  hpt vpt V hpt neg vpt V Opaque stroke } def
/BoxW { stroke [] 0 setdash exch hpt sub exch vpt add M
  0 vpt2 neg V hpt2 0 V 0 vpt2 V
  hpt2 neg 0 V Opaque stroke } def
/TriUW { stroke [] 0 setdash vpt 1.12 mul add M
  hpt neg vpt -1.62 mul V
  hpt 2 mul 0 V
  hpt neg vpt 1.62 mul V Opaque stroke } def
/TriDW { stroke [] 0 setdash vpt 1.12 mul sub M
  hpt neg vpt 1.62 mul V
  hpt 2 mul 0 V
  hpt neg vpt -1.62 mul V Opaque stroke } def
/PentW { stroke [] 0 setdash gsave
  translate 0 hpt M 4 {72 rotate 0 hpt L} repeat
  Opaque stroke grestore } def
/CircW { stroke [] 0 setdash 
  hpt 0 360 arc Opaque stroke } def
/BoxFill { gsave Rec 1 setgray fill grestore } def
end
}}%
\begin{picture}(3600,2160)(0,0)%
{\GNUPLOTspecial{"
gnudict begin
gsave
0 0 translate
0.100 0.100 scale
0 setgray
newpath
1.000 UL
LTb
500 300 M
63 0 V
2887 0 R
-63 0 V
500 740 M
63 0 V
2887 0 R
-63 0 V
500 1180 M
63 0 V
2887 0 R
-63 0 V
500 1620 M
63 0 V
2887 0 R
-63 0 V
500 2060 M
63 0 V
2887 0 R
-63 0 V
500 300 M
0 63 V
0 1697 R
0 -63 V
992 300 M
0 63 V
0 1697 R
0 -63 V
1483 300 M
0 63 V
0 1697 R
0 -63 V
1975 300 M
0 63 V
0 1697 R
0 -63 V
2467 300 M
0 63 V
0 1697 R
0 -63 V
2958 300 M
0 63 V
0 1697 R
0 -63 V
3450 300 M
0 63 V
0 1697 R
0 -63 V
1.000 UL
LTb
500 300 M
2950 0 V
0 1760 V
-2950 0 V
500 300 L
0.300 UL
LT3
3144 1180 M
51 19 V
-5 95 V
-143 -40 V
28 -74 V
17 98 V
-21 31 V
-29 11 V
56 -6 V
-13 -38 V
-21 -53 V
48 -32 V
-217 44 V
-226 -63 V
-274 -24 V
-464 -85 V
-300 176 V
-360 -4 V
995 1109 L
-186 87 V
-4 -111 V
-2 13 V
10 72 V
119 -80 V
822 1074 L
-41 37 V
68 -26 V
727 1188 L
879 1069 L
28 24 V
2271 87 R
-4 -45 V
-55 8 V
-78 47 V
121 -70 V
-45 -3 V
69 35 V
-107 -16 V
16 -16 V
52 157 V
-55 -33 V
-85 -131 V
-55 116 V
2787 1080 L
-374 176 V
-444 82 V
-428 78 V
1217 1188 L
-133 79 V
893 1198 L
11 -83 V
-43 66 V
-8 125 V
-45 -1 V
-6 2 V
23 -199 V
-52 72 V
129 123 V
842 1119 L
-25 82 V
2297 -21 R
-31 -47 V
-38 -37 V
-50 -9 V
54 90 V
88 -6 V
42 -76 V
-87 71 V
85 35 V
-71 -26 V
-47 -55 V
-98 135 V
-128 -50 V
-167 46 V
-252 54 V
-483 -28 V
-482 -31 V
-259 56 V
972 1251 L
-88 -1 V
-34 -4 V
42 -42 V
-72 -54 V
30 84 V
-34 58 V
33 -114 V
-205 97 V
277 -75 V
749 1328 L
886 1211 L
2323 -31 R
-80 -97 V
-74 -54 V
-2 143 V
35 -119 V
-26 52 V
-34 -91 V
62 58 V
-35 -89 V
98 41 V
-115 77 V
-60 -18 V
-125 5 V
-224 27 V
-414 83 V
-406 -44 V
-533 4 V
-11 71 V
-271 85 V
-47 -25 V
-81 -9 V
-24 -75 V
-9 109 V
11 18 V
-24 32 V
43 -29 V
-40 -43 V
-23 -34 V
61 13 V
-148 -2 V
2329 -89 R
100 22 V
1 -32 V
81 -44 V
-170 145 V
117 4 V
-196 -96 V
-6 28 V
126 76 V
34 -87 V
-39 -93 V
-126 60 V
-52 49 V
-239 20 V
-501 -55 V
-399 88 V
1330 1143 L
-282 -68 V
-91 90 V
-116 1 V
102 -78 V
770 1035 L
43 94 V
772 1011 L
20 59 V
743 994 L
94 128 V
-36 98 V
53 -139 V
-69 1 V
2299 98 R
63 -8 V
10 13 V
14 -5 V
-43 -29 V
-9 57 V
-28 -36 V
-24 -7 V
-82 61 V
145 -99 V
-36 179 V
-11 -76 V
-116 -56 V
-379 -98 V
-359 26 V
-330 37 V
-455 -8 V
-330 -3 V
974 1290 L
-65 -21 V
769 1255 L
-6 0 V
-49 -81 V
20 34 V
-12 20 V
93 -3 V
704 1173 L
104 -93 V
-25 190 V
87 -107 V
2311 17 R
-97 -19 V
89 91 V
-147 -92 V
2 -43 V
64 87 V
10 68 V
-1 45 V
198 -43 V
-182 -3 V
2996 1062 L
61 70 V
-201 -32 V
-288 30 V
-280 86 V
1899 1040 L
-337 14 V
-359 -6 V
-49 -99 V
928 1113 L
-87 106 V
5 8 V
80 -92 V
-34 159 V
25 -23 V
-16 -94 V
-46 -22 V
-2 34 V
-99 -51 V
70 1 V
2234 41 R
108 -129 V
-15 -31 V
15 109 V
-28 -114 V
-40 9 V
-55 31 V
-3 2 V
64 167 V
23 -50 V
5 -65 V
-66 -92 V
-111 55 V
-418 61 V
-399 11 V
1825 1022 L
-397 147 V
-205 -17 V
-160 54 V
852 1111 L
36 129 V
-21 10 V
67 -23 V
-21 -48 V
35 34 V
-48 75 V
-48 32 V
15 -42 V
-142 22 V
875 1165 L
2183 15 R
-66 -111 V
141 58 V
71 5 V
-187 7 V
53 -18 V
25 25 V
2 72 V
-6 39 V
-72 -205 V
-14 138 V
-56 -16 V
-157 84 V
2480 1060 L
2171 933 L
-351 141 V
-363 179 V
-188 -59 V
-191 -24 V
-147 48 V
-75 -82 V
45 37 V
767 1134 L
76 -37 V
-61 22 V
65 130 V
-98 -10 V
96 -83 V
-28 94 V
705 1160 L
2404 20 R
66 -31 V
-32 -96 V
-34 45 V
-19 22 V
15 9 V
25 37 V
-7 -20 V
-47 -62 V
-20 67 V
-52 19 V
2935 967 L
-50 38 V
-170 69 V
-369 152 V
-392 -7 V
-374 44 V
-356 -77 V
-228 34 V
-63 8 V
-117 23 V
-80 23 V
76 -6 V
921 1128 L
-87 99 V
-5 45 V
-22 -85 V
-78 -13 V
1 94 V
23 -49 V
2444 -39 R
-32 9 V
-25 66 V
27 -40 V
55 66 V
-60 -86 V
-43 12 V
-39 101 V
18 -136 V
17 43 V
-61 85 V
-93 -129 V
-135 -13 V
-220 14 V
-322 -70 V
-519 89 V
-355 -75 V
-375 54 V
-24 33 V
854 1147 L
-132 98 V
114 13 V
15 -128 V
56 135 V
0 -174 V
-97 -3 V
-28 23 V
11 15 V
-55 3 V
61 -68 V
2359 119 R
-11 -51 V
-52 -91 V
-5 57 V
44 -69 V
24 21 V
-14 -62 V
-54 10 V
150 -1 V
-69 190 V
-170 16 V
64 -220 V
-232 168 V
-199 86 V
-490 29 V
-428 5 V
-159 10 V
-335 -36 V
-156 58 V
-92 42 V
961 1240 L
-58 95 V
23 -23 V
760 1294 L
57 -59 V
79 35 V
-33 9 V
-89 -55 V
52 5 V
-176 -8 V
2528 -41 R
-75 -31 V
140 41 V
-123 4 V
-68 0 V
141 46 V
-71 -213 V
-61 113 V
24 86 V
108 -121 V
-140 80 V
64 -90 V
-180 32 V
-279 21 V
2295 1047 L
-448 133 V
-361 0 V
-308 -64 V
-122 162 V
980 1103 L
877 1001 L
76 172 V
-114 74 V
736 1116 L
131 54 V
-110 61 V
-30 -30 V
128 93 V
721 1242 L
101 50 V
3201 1180 M
-73 214 V
-14 -176 V
-57 105 V
-29 80 V
148 -83 V
-128 107 V
130 -134 V
-55 -65 V
90 44 V
currentpoint stroke M
-120 124 V
-50 -88 V
2821 1113 L
-202 80 V
-322 87 V
-424 -57 V
-533 -80 V
-264 -82 V
-9 127 V
-75 -25 V
-81 -73 V
-17 -72 V
-48 84 V
45 -92 V
861 979 L
-218 99 V
142 -32 V
710 972 L
193 84 V
-1 58 V
2166 66 R
38 61 V
-59 -95 V
-8 25 V
46 111 V
24 -81 V
-69 65 V
121 29 V
-22 -176 V
-39 13 V
-4 -70 V
2933 985 L
-82 186 V
-194 -96 V
-441 44 V
-468 18 V
-321 162 V
1130 1108 L
-98 -50 V
-85 63 V
741 1062 L
43 104 V
18 8 V
57 16 V
-84 2 V
80 -59 V
-31 -27 V
-43 93 V
1 -41 V
92 43 V
2185 -21 R
42 -43 V
-57 3 V
101 44 V
15 -144 V
-103 -20 V
-121 51 V
148 61 V
-13 0 V
-24 7 V
-7 -35 V
-50 100 V
-38 -97 V
-310 21 V
-293 108 V
-486 51 V
-323 -16 V
-393 90 V
-82 -182 V
-294 79 V
-10 -35 V
97 69 V
27 -206 V
-57 124 V
20 -65 V
-88 127 V
-10 20 V
88 -91 V
88 -81 V
-67 196 V
3189 1180 M
-46 93 V
-10 32 V
-16 -185 V
73 163 V
3086 1175 L
19 -6 V
-61 -151 V
-13 205 V
50 90 V
-18 -50 V
-67 -42 V
-70 79 V
-258 49 V
-371 -53 V
-509 -37 V
1398 1111 L
-208 -6 V
927 1090 L
-50 7 V
-17 1 V
88 56 V
-32 16 V
-92 -66 V
-98 122 V
42 -54 V
88 -82 V
-78 186 V
746 1134 L
59 -96 V
2296 142 R
-40 20 V
-30 28 V
3 -6 V
9 -36 V
12 -80 V
71 171 V
11 28 V
3017 1175 L
69 21 V
-37 111 V
-147 -58 V
-105 -13 V
-114 -99 V
-484 40 V
-426 44 V
-391 -98 V
-263 91 V
-88 -99 V
-63 48 V
-73 103 V
17 -144 V
5 24 V
-18 116 V
9 -141 V
-125 96 V
-29 -18 V
-43 36 V
65 -110 V
1 23 V
2261 32 R
32 -29 V
26 29 V
19 30 V
-49 1 V
81 0 V
3 56 V
18 -1 V
-62 -181 V
-3 125 V
-51 32 V
-85 -103 V
-282 141 V
-291 -15 V
-228 130 V
1677 1205 L
-412 -8 V
-159 -41 V
1 -24 V
-177 32 V
6 32 V
-15 -90 V
-136 20 V
6 -16 V
-68 17 V
-4 6 V
188 137 V
-96 -13 V
685 1145 L
146 51 V
2372 -16 R
22 35 V
-43 70 V
-98 3 V
147 -96 V
-57 -42 V
-115 133 V
51 -79 V
47 74 V
39 -71 V
-55 -21 V
-102 15 V
-159 29 V
-66 -75 V
-455 40 V
1916 1064 L
-268 16 V
1434 867 L
1172 995 L
-83 203 V
897 1108 L
10 174 V
808 1155 L
-33 37 V
63 -24 V
-44 31 V
-23 -64 V
43 -65 V
-85 97 V
146 -29 V
stroke
grestore
end
showpage
}}%
\put(1975,50){\makebox(0,0){$\mathrm{Re}\, \overline{l_{p}}$}}%
\put(100,1180){%
\makebox(0,0)[b]{\shortstack{$\mathrm{Im}\, \overline{l_{p}}$}}%
}%
\put(3450,200){\makebox(0,0){0.3}}%
\put(2958,200){\makebox(0,0){0.2}}%
\put(2467,200){\makebox(0,0){0.1}}%
\put(1975,200){\makebox(0,0){0}}%
\put(1483,200){\makebox(0,0){-0.1}}%
\put(992,200){\makebox(0,0){-0.2}}%
\put(500,200){\makebox(0,0){-0.3}}%
\put(450,2060){\makebox(0,0)[r]{0.1}}%
\put(450,1620){\makebox(0,0)[r]{0.05}}%
\put(450,1180){\makebox(0,0)[r]{0}}%
\put(450,740){\makebox(0,0)[r]{-0.05}}%
\put(450,300){\makebox(0,0)[r]{-0.1}}%
\end{picture}%
\endgroup

\end	{center}
\vskip 0.15in
\caption{The values in the complex plane taken by the Polyakov 
loop for $x_3\in [0,L-1]$ for a sample of individual $k=2$
walls in SU(4) with $aT=0.25$ and $T\simeq 1.88 T_c$.}
\label{fig_profc1}
\end 	{figure}

\begin	{figure}[p]
\begin	{center}
\leavevmode
\begingroup%
  \makeatletter%
  \newcommand{\GNUPLOTspecial}{%
    \@sanitize\catcode`\%=14\relax\special}%
  \setlength{\unitlength}{0.1bp}%
{\GNUPLOTspecial{!
/gnudict 256 dict def
gnudict begin
/Color false def
/Solid false def
/gnulinewidth 5.000 def
/userlinewidth gnulinewidth def
/vshift -33 def
/dl {10 mul} def
/hpt_ 31.5 def
/vpt_ 31.5 def
/hpt hpt_ def
/vpt vpt_ def
/M {moveto} bind def
/L {lineto} bind def
/R {rmoveto} bind def
/V {rlineto} bind def
/vpt2 vpt 2 mul def
/hpt2 hpt 2 mul def
/Lshow { currentpoint stroke M
  0 vshift R show } def
/Rshow { currentpoint stroke M
  dup stringwidth pop neg vshift R show } def
/Cshow { currentpoint stroke M
  dup stringwidth pop -2 div vshift R show } def
/UP { dup vpt_ mul /vpt exch def hpt_ mul /hpt exch def
  /hpt2 hpt 2 mul def /vpt2 vpt 2 mul def } def
/DL { Color {setrgbcolor Solid {pop []} if 0 setdash }
 {pop pop pop Solid {pop []} if 0 setdash} ifelse } def
/BL { stroke userlinewidth 2 mul setlinewidth } def
/AL { stroke userlinewidth 2 div setlinewidth } def
/UL { dup gnulinewidth mul /userlinewidth exch def
      10 mul /udl exch def } def
/PL { stroke userlinewidth setlinewidth } def
/LTb { BL [] 0 0 0 DL } def
/LTa { AL [1 udl mul 2 udl mul] 0 setdash 0 0 0 setrgbcolor } def
/LT0 { PL [] 1 0 0 DL } def
/LT1 { PL [4 dl 2 dl] 0 1 0 DL } def
/LT2 { PL [2 dl 3 dl] 0 0 1 DL } def
/LT3 { PL [1 dl 1.5 dl] 1 0 1 DL } def
/LT4 { PL [5 dl 2 dl 1 dl 2 dl] 0 1 1 DL } def
/LT5 { PL [4 dl 3 dl 1 dl 3 dl] 1 1 0 DL } def
/LT6 { PL [2 dl 2 dl 2 dl 4 dl] 0 0 0 DL } def
/LT7 { PL [2 dl 2 dl 2 dl 2 dl 2 dl 4 dl] 1 0.3 0 DL } def
/LT8 { PL [2 dl 2 dl 2 dl 2 dl 2 dl 2 dl 2 dl 4 dl] 0.5 0.5 0.5 DL } def
/Pnt { stroke [] 0 setdash
   gsave 1 setlinecap M 0 0 V stroke grestore } def
/Dia { stroke [] 0 setdash 2 copy vpt add M
  hpt neg vpt neg V hpt vpt neg V
  hpt vpt V hpt neg vpt V closepath stroke
  Pnt } def
/Pls { stroke [] 0 setdash vpt sub M 0 vpt2 V
  currentpoint stroke M
  hpt neg vpt neg R hpt2 0 V stroke
  } def
/Box { stroke [] 0 setdash 2 copy exch hpt sub exch vpt add M
  0 vpt2 neg V hpt2 0 V 0 vpt2 V
  hpt2 neg 0 V closepath stroke
  Pnt } def
/Crs { stroke [] 0 setdash exch hpt sub exch vpt add M
  hpt2 vpt2 neg V currentpoint stroke M
  hpt2 neg 0 R hpt2 vpt2 V stroke } def
/TriU { stroke [] 0 setdash 2 copy vpt 1.12 mul add M
  hpt neg vpt -1.62 mul V
  hpt 2 mul 0 V
  hpt neg vpt 1.62 mul V closepath stroke
  Pnt  } def
/Star { 2 copy Pls Crs } def
/BoxF { stroke [] 0 setdash exch hpt sub exch vpt add M
  0 vpt2 neg V  hpt2 0 V  0 vpt2 V
  hpt2 neg 0 V  closepath fill } def
/TriUF { stroke [] 0 setdash vpt 1.12 mul add M
  hpt neg vpt -1.62 mul V
  hpt 2 mul 0 V
  hpt neg vpt 1.62 mul V closepath fill } def
/TriD { stroke [] 0 setdash 2 copy vpt 1.12 mul sub M
  hpt neg vpt 1.62 mul V
  hpt 2 mul 0 V
  hpt neg vpt -1.62 mul V closepath stroke
  Pnt  } def
/TriDF { stroke [] 0 setdash vpt 1.12 mul sub M
  hpt neg vpt 1.62 mul V
  hpt 2 mul 0 V
  hpt neg vpt -1.62 mul V closepath fill} def
/DiaF { stroke [] 0 setdash vpt add M
  hpt neg vpt neg V hpt vpt neg V
  hpt vpt V hpt neg vpt V closepath fill } def
/Pent { stroke [] 0 setdash 2 copy gsave
  translate 0 hpt M 4 {72 rotate 0 hpt L} repeat
  closepath stroke grestore Pnt } def
/PentF { stroke [] 0 setdash gsave
  translate 0 hpt M 4 {72 rotate 0 hpt L} repeat
  closepath fill grestore } def
/Circle { stroke [] 0 setdash 2 copy
  hpt 0 360 arc stroke Pnt } def
/CircleF { stroke [] 0 setdash hpt 0 360 arc fill } def
/C0 { BL [] 0 setdash 2 copy moveto vpt 90 450  arc } bind def
/C1 { BL [] 0 setdash 2 copy        moveto
       2 copy  vpt 0 90 arc closepath fill
               vpt 0 360 arc closepath } bind def
/C2 { BL [] 0 setdash 2 copy moveto
       2 copy  vpt 90 180 arc closepath fill
               vpt 0 360 arc closepath } bind def
/C3 { BL [] 0 setdash 2 copy moveto
       2 copy  vpt 0 180 arc closepath fill
               vpt 0 360 arc closepath } bind def
/C4 { BL [] 0 setdash 2 copy moveto
       2 copy  vpt 180 270 arc closepath fill
               vpt 0 360 arc closepath } bind def
/C5 { BL [] 0 setdash 2 copy moveto
       2 copy  vpt 0 90 arc
       2 copy moveto
       2 copy  vpt 180 270 arc closepath fill
               vpt 0 360 arc } bind def
/C6 { BL [] 0 setdash 2 copy moveto
      2 copy  vpt 90 270 arc closepath fill
              vpt 0 360 arc closepath } bind def
/C7 { BL [] 0 setdash 2 copy moveto
      2 copy  vpt 0 270 arc closepath fill
              vpt 0 360 arc closepath } bind def
/C8 { BL [] 0 setdash 2 copy moveto
      2 copy vpt 270 360 arc closepath fill
              vpt 0 360 arc closepath } bind def
/C9 { BL [] 0 setdash 2 copy moveto
      2 copy  vpt 270 450 arc closepath fill
              vpt 0 360 arc closepath } bind def
/C10 { BL [] 0 setdash 2 copy 2 copy moveto vpt 270 360 arc closepath fill
       2 copy moveto
       2 copy vpt 90 180 arc closepath fill
               vpt 0 360 arc closepath } bind def
/C11 { BL [] 0 setdash 2 copy moveto
       2 copy  vpt 0 180 arc closepath fill
       2 copy moveto
       2 copy  vpt 270 360 arc closepath fill
               vpt 0 360 arc closepath } bind def
/C12 { BL [] 0 setdash 2 copy moveto
       2 copy  vpt 180 360 arc closepath fill
               vpt 0 360 arc closepath } bind def
/C13 { BL [] 0 setdash  2 copy moveto
       2 copy  vpt 0 90 arc closepath fill
       2 copy moveto
       2 copy  vpt 180 360 arc closepath fill
               vpt 0 360 arc closepath } bind def
/C14 { BL [] 0 setdash 2 copy moveto
       2 copy  vpt 90 360 arc closepath fill
               vpt 0 360 arc } bind def
/C15 { BL [] 0 setdash 2 copy vpt 0 360 arc closepath fill
               vpt 0 360 arc closepath } bind def
/Rec   { newpath 4 2 roll moveto 1 index 0 rlineto 0 exch rlineto
       neg 0 rlineto closepath } bind def
/Square { dup Rec } bind def
/Bsquare { vpt sub exch vpt sub exch vpt2 Square } bind def
/S0 { BL [] 0 setdash 2 copy moveto 0 vpt rlineto BL Bsquare } bind def
/S1 { BL [] 0 setdash 2 copy vpt Square fill Bsquare } bind def
/S2 { BL [] 0 setdash 2 copy exch vpt sub exch vpt Square fill Bsquare } bind def
/S3 { BL [] 0 setdash 2 copy exch vpt sub exch vpt2 vpt Rec fill Bsquare } bind def
/S4 { BL [] 0 setdash 2 copy exch vpt sub exch vpt sub vpt Square fill Bsquare } bind def
/S5 { BL [] 0 setdash 2 copy 2 copy vpt Square fill
       exch vpt sub exch vpt sub vpt Square fill Bsquare } bind def
/S6 { BL [] 0 setdash 2 copy exch vpt sub exch vpt sub vpt vpt2 Rec fill Bsquare } bind def
/S7 { BL [] 0 setdash 2 copy exch vpt sub exch vpt sub vpt vpt2 Rec fill
       2 copy vpt Square fill
       Bsquare } bind def
/S8 { BL [] 0 setdash 2 copy vpt sub vpt Square fill Bsquare } bind def
/S9 { BL [] 0 setdash 2 copy vpt sub vpt vpt2 Rec fill Bsquare } bind def
/S10 { BL [] 0 setdash 2 copy vpt sub vpt Square fill 2 copy exch vpt sub exch vpt Square fill
       Bsquare } bind def
/S11 { BL [] 0 setdash 2 copy vpt sub vpt Square fill 2 copy exch vpt sub exch vpt2 vpt Rec fill
       Bsquare } bind def
/S12 { BL [] 0 setdash 2 copy exch vpt sub exch vpt sub vpt2 vpt Rec fill Bsquare } bind def
/S13 { BL [] 0 setdash 2 copy exch vpt sub exch vpt sub vpt2 vpt Rec fill
       2 copy vpt Square fill Bsquare } bind def
/S14 { BL [] 0 setdash 2 copy exch vpt sub exch vpt sub vpt2 vpt Rec fill
       2 copy exch vpt sub exch vpt Square fill Bsquare } bind def
/S15 { BL [] 0 setdash 2 copy Bsquare fill Bsquare } bind def
/D0 { gsave translate 45 rotate 0 0 S0 stroke grestore } bind def
/D1 { gsave translate 45 rotate 0 0 S1 stroke grestore } bind def
/D2 { gsave translate 45 rotate 0 0 S2 stroke grestore } bind def
/D3 { gsave translate 45 rotate 0 0 S3 stroke grestore } bind def
/D4 { gsave translate 45 rotate 0 0 S4 stroke grestore } bind def
/D5 { gsave translate 45 rotate 0 0 S5 stroke grestore } bind def
/D6 { gsave translate 45 rotate 0 0 S6 stroke grestore } bind def
/D7 { gsave translate 45 rotate 0 0 S7 stroke grestore } bind def
/D8 { gsave translate 45 rotate 0 0 S8 stroke grestore } bind def
/D9 { gsave translate 45 rotate 0 0 S9 stroke grestore } bind def
/D10 { gsave translate 45 rotate 0 0 S10 stroke grestore } bind def
/D11 { gsave translate 45 rotate 0 0 S11 stroke grestore } bind def
/D12 { gsave translate 45 rotate 0 0 S12 stroke grestore } bind def
/D13 { gsave translate 45 rotate 0 0 S13 stroke grestore } bind def
/D14 { gsave translate 45 rotate 0 0 S14 stroke grestore } bind def
/D15 { gsave translate 45 rotate 0 0 S15 stroke grestore } bind def
/DiaE { stroke [] 0 setdash vpt add M
  hpt neg vpt neg V hpt vpt neg V
  hpt vpt V hpt neg vpt V closepath stroke } def
/BoxE { stroke [] 0 setdash exch hpt sub exch vpt add M
  0 vpt2 neg V hpt2 0 V 0 vpt2 V
  hpt2 neg 0 V closepath stroke } def
/TriUE { stroke [] 0 setdash vpt 1.12 mul add M
  hpt neg vpt -1.62 mul V
  hpt 2 mul 0 V
  hpt neg vpt 1.62 mul V closepath stroke } def
/TriDE { stroke [] 0 setdash vpt 1.12 mul sub M
  hpt neg vpt 1.62 mul V
  hpt 2 mul 0 V
  hpt neg vpt -1.62 mul V closepath stroke } def
/PentE { stroke [] 0 setdash gsave
  translate 0 hpt M 4 {72 rotate 0 hpt L} repeat
  closepath stroke grestore } def
/CircE { stroke [] 0 setdash 
  hpt 0 360 arc stroke } def
/Opaque { gsave closepath 1 setgray fill grestore 0 setgray closepath } def
/DiaW { stroke [] 0 setdash vpt add M
  hpt neg vpt neg V hpt vpt neg V
  hpt vpt V hpt neg vpt V Opaque stroke } def
/BoxW { stroke [] 0 setdash exch hpt sub exch vpt add M
  0 vpt2 neg V hpt2 0 V 0 vpt2 V
  hpt2 neg 0 V Opaque stroke } def
/TriUW { stroke [] 0 setdash vpt 1.12 mul add M
  hpt neg vpt -1.62 mul V
  hpt 2 mul 0 V
  hpt neg vpt 1.62 mul V Opaque stroke } def
/TriDW { stroke [] 0 setdash vpt 1.12 mul sub M
  hpt neg vpt 1.62 mul V
  hpt 2 mul 0 V
  hpt neg vpt -1.62 mul V Opaque stroke } def
/PentW { stroke [] 0 setdash gsave
  translate 0 hpt M 4 {72 rotate 0 hpt L} repeat
  Opaque stroke grestore } def
/CircW { stroke [] 0 setdash 
  hpt 0 360 arc Opaque stroke } def
/BoxFill { gsave Rec 1 setgray fill grestore } def
end
}}%
\begin{picture}(3600,2160)(0,0)%
{\GNUPLOTspecial{"
gnudict begin
gsave
0 0 translate
0.100 0.100 scale
0 setgray
newpath
1.000 UL
LTb
500 300 M
63 0 V
2887 0 R
-63 0 V
500 740 M
63 0 V
2887 0 R
-63 0 V
500 1180 M
63 0 V
2887 0 R
-63 0 V
500 1620 M
63 0 V
2887 0 R
-63 0 V
500 2060 M
63 0 V
2887 0 R
-63 0 V
500 300 M
0 63 V
0 1697 R
0 -63 V
869 300 M
0 63 V
0 1697 R
0 -63 V
1238 300 M
0 63 V
0 1697 R
0 -63 V
1606 300 M
0 63 V
0 1697 R
0 -63 V
1975 300 M
0 63 V
0 1697 R
0 -63 V
2344 300 M
0 63 V
0 1697 R
0 -63 V
2713 300 M
0 63 V
0 1697 R
0 -63 V
3081 300 M
0 63 V
0 1697 R
0 -63 V
3450 300 M
0 63 V
0 1697 R
0 -63 V
1.000 UL
LTb
500 300 M
2950 0 V
0 1760 V
-2950 0 V
500 300 L
0.300 UL
LT3
2970 1180 M
3 -33 V
-52 -110 V
37 91 V
19 84 V
128 -131 V
-192 12 V
-42 80 V
-214 -13 V
2264 856 L
82 269 V
-54 -12 V
-151 -86 V
-74 220 V
-32 -87 V
-369 85 V
-155 58 V
-82 -73 V
-76 -4 V
-287 -69 V
88 -39 V
60 224 V
946 1217 L
20 55 V
-45 5 V
37 -232 V
-21 51 V
756 1238 L
46 14 V
-25 47 V
3047 1180 M
-111 -70 V
104 21 V
-198 64 V
-72 -118 V
6 77 V
118 -55 V
152 51 V
-158 44 V
-246 -68 V
2594 994 L
-205 9 V
-101 20 V
-109 193 V
-36 63 V
1778 1147 L
-216 312 V
1328 1298 L
26 -56 V
-109 19 V
-34 -16 V
-62 -33 V
111 63 V
-332 10 V
144 -10 V
927 1499 L
800 1407 L
73 -166 V
670 1210 L
19 -22 V
2287 -8 R
-8 -69 V
-42 -82 V
56 103 V
-119 -34 V
3044 913 L
-37 -16 V
-39 17 V
-171 123 V
-98 30 V
2422 918 L
-121 46 V
-87 163 V
2094 931 L
-147 178 V
-155 77 V
-142 62 V
-241 -85 V
-123 13 V
24 51 V
-252 98 V
127 87 V
996 1329 L
-12 -6 V
91 86 V
-83 -31 V
951 1256 L
-94 208 V
-83 22 V
54 -100 V
3172 1180 M
-93 64 V
66 -122 V
-200 144 V
-69 -197 V
20 101 V
-23 213 V
184 -97 V
-277 -75 V
-160 4 V
-72 5 V
-97 -157 V
-114 79 V
-178 57 V
-124 223 V
1834 1305 L
1672 1132 L
-252 50 V
-70 78 V
58 -150 V
1129 979 L
5 106 V
-230 49 V
-90 -16 V
305 -48 V
-28 42 V
791 1102 L
19 193 V
786 1192 L
111 14 V
2257 -26 R
-169 125 V
23 -152 V
-160 6 V
106 124 V
48 -147 V
13 126 V
-69 -31 V
-96 -54 V
-52 96 V
-100 -41 V
-11 -121 V
-152 46 V
-290 80 V
-186 113 V
1925 1216 L
-200 -19 V
-185 -75 V
-143 -85 V
-145 32 V
41 -49 V
-188 105 V
-89 -49 V
119 2 V
-109 4 V
1009 922 L
-77 109 V
736 952 L
141 162 V
722 1044 L
2297 136 R
93 -102 V
-51 283 V
56 -192 V
3059 987 L
-237 177 V
136 -42 V
-102 -61 V
-59 80 V
51 194 V
-258 51 V
148 -192 V
2347 1020 L
-54 78 V
-202 4 V
-371 -50 V
-61 182 V
1514 1114 L
-104 181 V
1196 1152 L
-108 -13 V
-171 88 V
52 -153 V
-15 39 V
156 172 V
869 1252 L
-23 -63 V
-22 51 V
12 13 V
160 -72 V
2245 -1 R
3126 1056 L
-50 -2 V
2 0 V
33 56 V
-51 187 V
-40 -128 V
-20 -115 V
-142 35 V
-59 38 V
50 79 V
-156 -43 V
-86 -127 V
-235 28 V
-38 8 V
-366 37 V
-176 28 V
-121 64 V
-234 92 V
-22 47 V
-339 29 V
981 1172 L
-15 129 V
6 -90 V
-43 102 V
-52 87 V
28 -203 V
63 47 V
-238 79 V
-52 62 V
3177 1180 M
-89 -94 V
120 20 V
-28 57 V
-160 50 V
108 -209 V
-278 205 V
-125 -85 V
21 104 V
199 -40 V
-198 47 V
-216 -34 V
-118 53 V
-135 78 V
2145 1153 L
-186 -47 V
-348 -25 V
-32 102 V
-284 78 V
968 1233 L
118 -21 V
-19 58 V
-78 -60 V
91 2 V
43 91 V
725 1263 L
854 1148 L
4 -24 V
5 -46 V
40 225 V
3084 1180 M
16 69 V
72 -84 V
-31 59 V
35 59 V
2983 1114 L
-50 61 V
-35 121 V
-125 -24 V
-80 -156 V
-147 88 V
74 -199 V
-192 149 V
-390 -8 V
38 -30 V
-230 80 V
-118 5 V
-269 -64 V
-301 13 V
-31 -123 V
-117 67 V
-127 38 V
84 35 V
53 -20 V
777 1098 L
-62 103 V
127 28 V
88 -76 V
-40 -48 V
726 1065 L
2271 115 R
77 139 V
5 -47 V
188 66 V
3164 1228 L
-115 61 V
-145 117 V
84 -62 V
-3 -6 V
-77 -198 V
-197 -18 V
28 -2 V
-233 -24 V
-144 -23 V
-41 48 V
-396 92 V
-285 -80 V
-278 -34 V
-6 -23 V
-322 -16 V
879 1024 L
-37 117 V
-31 -43 V
219 -67 V
-28 53 V
-30 88 V
3 -52 V
-69 37 V
39 31 V
22 -94 V
1872 86 R
180 303 V
27 -16 V
35 -11 V
155 -194 V
-139 103 V
-39 62 V
-60 -173 V
-34 -22 V
-42 111 V
-177 29 V
-23 -32 V
-176 -17 V
-99 -100 V
-231 -91 V
-283 199 V
1546 1114 L
1330 988 L
1227 896 L
-214 34 V
874 852 L
70 149 V
-12 21 V
1166 891 L
854 1043 L
988 866 L
36 244 V
-80 -47 V
819 917 L
250 147 V
2060 116 R
-120 -79 V
128 39 V
42 -9 V
-4 -19 V
-135 -80 V
-153 179 V
-50 59 V
2676 1166 L
218 15 V
-82 15 V
-157 2 V
-145 -39 V
-234 24 V
-76 -171 V
-245 110 V
-437 139 V
-12 -144 V
-314 62 V
-141 6 V
-8 18 V
25 50 V
868 1161 L
28 131 V
-46 -89 V
-145 40 V
101 45 V
87 -53 V
-84 -36 V
181 73 V
2046 -92 R
-33 -116 V
-40 -19 V
93 108 V
36 -307 V
-111 26 V
-132 44 V
-95 136 V
-21 -5 V
166 -61 V
-148 46 V
-12 158 V
2520 1003 L
-240 204 V
-125 10 V
-362 15 V
-218 1 V
-325 61 V
35 -110 V
-169 174 V
-19 -3 V
76 -60 V
-222 3 V
53 -143 V
-22 158 V
736 1475 L
895 1315 L
5 4 V
-6 17 V
-19 58 V
2991 1180 M
155 34 V
-26 -82 V
-69 120 V
-105 18 V
-89 -30 V
-10 102 V
-115 -14 V
71 24 V
18 -24 V
currentpoint stroke M
2665 1200 L
-79 48 V
-404 39 V
-62 -188 V
-98 44 V
-284 107 V
-209 -98 V
-373 9 V
37 24 V
-74 58 V
132 -62 V
881 1137 L
151 84 V
49 -123 V
827 1065 L
714 1044 L
28 177 V
898 1058 L
774 1214 L
112 -7 V
2216 -27 R
55 3 V
-90 -131 V
62 67 V
-104 73 V
-108 90 V
67 -57 V
-113 52 V
2 -37 V
-85 29 V
2670 1168 L
-287 213 V
2237 1196 L
-13 -80 V
-192 134 V
1703 1090 L
-345 -59 V
-70 131 V
-183 45 V
-64 -69 V
-109 45 V
17 1 V
267 22 V
-144 -14 V
-263 18 V
112 125 V
145 -98 V
-88 -57 V
-94 21 V
37 84 V
3214 1180 M
-41 -123 V
-242 -25 V
11 111 V
-18 -78 V
-79 12 V
32 101 V
155 -40 V
-43 -43 V
-14 26 V
-70 -34 V
-167 -6 V
-183 113 V
2433 1020 L
-304 160 V
-271 -75 V
-174 116 V
1 41 V
-291 7 V
-163 48 V
-73 1 V
869 1246 L
153 21 V
-58 -9 V
127 -11 V
47 1 V
-131 142 V
63 -113 V
-138 23 V
27 -95 V
2161 -25 R
-64 71 V
-22 -77 V
-11 24 V
-106 58 V
116 88 V
-105 -65 V
47 -112 V
40 25 V
-120 162 V
2717 1174 L
-313 173 V
-193 179 V
1934 1328 L
216 -170 V
1862 1033 L
-239 147 V
1483 1039 L
-291 105 V
-128 -51 V
1016 915 L
863 936 L
129 263 V
54 21 V
-89 26 V
5 -69 V
49 -48 V
-138 -3 V
160 -31 V
13 24 V
2005 61 R
-141 59 V
273 52 V
-137 55 V
175 -111 V
-188 17 V
-53 170 V
-183 -57 V
89 19 V
-59 21 V
-100 -91 V
-248 150 V
2329 1306 L
-215 110 V
-71 -162 V
-287 -58 V
-113 134 V
1521 1096 L
1220 996 L
0 66 V
-44 41 V
-85 -73 V
-42 -90 V
952 853 L
66 234 V
56 -35 V
1003 892 L
968 992 L
31 69 V
-12 -59 V
2058 178 R
47 -119 V
-3 375 V
2954 1291 L
8 -10 V
-10 -96 V
-60 -72 V
-39 107 V
318 -183 V
-88 153 V
-175 64 V
-127 -38 V
-290 112 V
-273 9 V
-14 -121 V
-304 -56 V
-91 182 V
1669 1140 L
75 3 V
-324 -11 V
-139 -61 V
-179 -47 V
25 15 V
996 1154 L
-6 -5 V
116 91 V
-19 -198 V
-42 17 V
-40 77 V
-42 -6 V
2055 50 R
104 -18 V
-34 83 V
10 -70 V
-6 -99 V
-202 12 V
110 -51 V
165 57 V
-82 62 V
-131 -88 V
-88 97 V
-300 19 V
-201 264 V
2199 1029 L
-233 340 V
-70 -297 V
-81 35 V
-87 69 V
-190 -74 V
-212 47 V
-240 25 V
127 63 V
-32 -107 V
-61 74 V
13 22 V
-123 99 V
10 -149 V
-32 48 V
-54 -77 V
63 165 V
stroke
grestore
end
showpage
}}%
\put(1975,50){\makebox(0,0){$\mathrm{Re}\, \overline{l_{p}}$}}%
\put(100,1180){%
\makebox(0,0)[b]{\shortstack{$\mathrm{Im}\, \overline{l_{p}}$}}%
}%
\put(3450,200){\makebox(0,0){0.2}}%
\put(3081,200){\makebox(0,0){0.15}}%
\put(2713,200){\makebox(0,0){0.1}}%
\put(2344,200){\makebox(0,0){0.05}}%
\put(1975,200){\makebox(0,0){0}}%
\put(1606,200){\makebox(0,0){-0.05}}%
\put(1238,200){\makebox(0,0){-0.1}}%
\put(869,200){\makebox(0,0){-0.15}}%
\put(500,200){\makebox(0,0){-0.2}}%
\put(450,2060){\makebox(0,0)[r]{0.1}}%
\put(450,1620){\makebox(0,0)[r]{0.05}}%
\put(450,1180){\makebox(0,0)[r]{0}}%
\put(450,740){\makebox(0,0)[r]{-0.05}}%
\put(450,300){\makebox(0,0)[r]{-0.1}}%
\end{picture}%
\endgroup

\end	{center}
\vskip 0.15in
\caption{The values in the complex plane taken by the Polyakov 
loop for $x_3\in [0,L-1]$ for a sample of  individual $k=2$ walls 
in SU(4) with $aT=0.25$ and $T\simeq 1.02 T_c$.}
\label{fig_profc2}
\end 	{figure}

\begin	{figure}[p]
\begin	{center}
\leavevmode
\begingroup%
  \makeatletter%
  \newcommand{\GNUPLOTspecial}{%
    \@sanitize\catcode`\%=14\relax\special}%
  \setlength{\unitlength}{0.1bp}%
{\GNUPLOTspecial{!
/gnudict 256 dict def
gnudict begin
/Color false def
/Solid false def
/gnulinewidth 5.000 def
/userlinewidth gnulinewidth def
/vshift -33 def
/dl {10 mul} def
/hpt_ 31.5 def
/vpt_ 31.5 def
/hpt hpt_ def
/vpt vpt_ def
/M {moveto} bind def
/L {lineto} bind def
/R {rmoveto} bind def
/V {rlineto} bind def
/vpt2 vpt 2 mul def
/hpt2 hpt 2 mul def
/Lshow { currentpoint stroke M
  0 vshift R show } def
/Rshow { currentpoint stroke M
  dup stringwidth pop neg vshift R show } def
/Cshow { currentpoint stroke M
  dup stringwidth pop -2 div vshift R show } def
/UP { dup vpt_ mul /vpt exch def hpt_ mul /hpt exch def
  /hpt2 hpt 2 mul def /vpt2 vpt 2 mul def } def
/DL { Color {setrgbcolor Solid {pop []} if 0 setdash }
 {pop pop pop Solid {pop []} if 0 setdash} ifelse } def
/BL { stroke userlinewidth 2 mul setlinewidth } def
/AL { stroke userlinewidth 2 div setlinewidth } def
/UL { dup gnulinewidth mul /userlinewidth exch def
      10 mul /udl exch def } def
/PL { stroke userlinewidth setlinewidth } def
/LTb { BL [] 0 0 0 DL } def
/LTa { AL [1 udl mul 2 udl mul] 0 setdash 0 0 0 setrgbcolor } def
/LT0 { PL [] 1 0 0 DL } def
/LT1 { PL [4 dl 2 dl] 0 1 0 DL } def
/LT2 { PL [2 dl 3 dl] 0 0 1 DL } def
/LT3 { PL [1 dl 1.5 dl] 1 0 1 DL } def
/LT4 { PL [5 dl 2 dl 1 dl 2 dl] 0 1 1 DL } def
/LT5 { PL [4 dl 3 dl 1 dl 3 dl] 1 1 0 DL } def
/LT6 { PL [2 dl 2 dl 2 dl 4 dl] 0 0 0 DL } def
/LT7 { PL [2 dl 2 dl 2 dl 2 dl 2 dl 4 dl] 1 0.3 0 DL } def
/LT8 { PL [2 dl 2 dl 2 dl 2 dl 2 dl 2 dl 2 dl 4 dl] 0.5 0.5 0.5 DL } def
/Pnt { stroke [] 0 setdash
   gsave 1 setlinecap M 0 0 V stroke grestore } def
/Dia { stroke [] 0 setdash 2 copy vpt add M
  hpt neg vpt neg V hpt vpt neg V
  hpt vpt V hpt neg vpt V closepath stroke
  Pnt } def
/Pls { stroke [] 0 setdash vpt sub M 0 vpt2 V
  currentpoint stroke M
  hpt neg vpt neg R hpt2 0 V stroke
  } def
/Box { stroke [] 0 setdash 2 copy exch hpt sub exch vpt add M
  0 vpt2 neg V hpt2 0 V 0 vpt2 V
  hpt2 neg 0 V closepath stroke
  Pnt } def
/Crs { stroke [] 0 setdash exch hpt sub exch vpt add M
  hpt2 vpt2 neg V currentpoint stroke M
  hpt2 neg 0 R hpt2 vpt2 V stroke } def
/TriU { stroke [] 0 setdash 2 copy vpt 1.12 mul add M
  hpt neg vpt -1.62 mul V
  hpt 2 mul 0 V
  hpt neg vpt 1.62 mul V closepath stroke
  Pnt  } def
/Star { 2 copy Pls Crs } def
/BoxF { stroke [] 0 setdash exch hpt sub exch vpt add M
  0 vpt2 neg V  hpt2 0 V  0 vpt2 V
  hpt2 neg 0 V  closepath fill } def
/TriUF { stroke [] 0 setdash vpt 1.12 mul add M
  hpt neg vpt -1.62 mul V
  hpt 2 mul 0 V
  hpt neg vpt 1.62 mul V closepath fill } def
/TriD { stroke [] 0 setdash 2 copy vpt 1.12 mul sub M
  hpt neg vpt 1.62 mul V
  hpt 2 mul 0 V
  hpt neg vpt -1.62 mul V closepath stroke
  Pnt  } def
/TriDF { stroke [] 0 setdash vpt 1.12 mul sub M
  hpt neg vpt 1.62 mul V
  hpt 2 mul 0 V
  hpt neg vpt -1.62 mul V closepath fill} def
/DiaF { stroke [] 0 setdash vpt add M
  hpt neg vpt neg V hpt vpt neg V
  hpt vpt V hpt neg vpt V closepath fill } def
/Pent { stroke [] 0 setdash 2 copy gsave
  translate 0 hpt M 4 {72 rotate 0 hpt L} repeat
  closepath stroke grestore Pnt } def
/PentF { stroke [] 0 setdash gsave
  translate 0 hpt M 4 {72 rotate 0 hpt L} repeat
  closepath fill grestore } def
/Circle { stroke [] 0 setdash 2 copy
  hpt 0 360 arc stroke Pnt } def
/CircleF { stroke [] 0 setdash hpt 0 360 arc fill } def
/C0 { BL [] 0 setdash 2 copy moveto vpt 90 450  arc } bind def
/C1 { BL [] 0 setdash 2 copy        moveto
       2 copy  vpt 0 90 arc closepath fill
               vpt 0 360 arc closepath } bind def
/C2 { BL [] 0 setdash 2 copy moveto
       2 copy  vpt 90 180 arc closepath fill
               vpt 0 360 arc closepath } bind def
/C3 { BL [] 0 setdash 2 copy moveto
       2 copy  vpt 0 180 arc closepath fill
               vpt 0 360 arc closepath } bind def
/C4 { BL [] 0 setdash 2 copy moveto
       2 copy  vpt 180 270 arc closepath fill
               vpt 0 360 arc closepath } bind def
/C5 { BL [] 0 setdash 2 copy moveto
       2 copy  vpt 0 90 arc
       2 copy moveto
       2 copy  vpt 180 270 arc closepath fill
               vpt 0 360 arc } bind def
/C6 { BL [] 0 setdash 2 copy moveto
      2 copy  vpt 90 270 arc closepath fill
              vpt 0 360 arc closepath } bind def
/C7 { BL [] 0 setdash 2 copy moveto
      2 copy  vpt 0 270 arc closepath fill
              vpt 0 360 arc closepath } bind def
/C8 { BL [] 0 setdash 2 copy moveto
      2 copy vpt 270 360 arc closepath fill
              vpt 0 360 arc closepath } bind def
/C9 { BL [] 0 setdash 2 copy moveto
      2 copy  vpt 270 450 arc closepath fill
              vpt 0 360 arc closepath } bind def
/C10 { BL [] 0 setdash 2 copy 2 copy moveto vpt 270 360 arc closepath fill
       2 copy moveto
       2 copy vpt 90 180 arc closepath fill
               vpt 0 360 arc closepath } bind def
/C11 { BL [] 0 setdash 2 copy moveto
       2 copy  vpt 0 180 arc closepath fill
       2 copy moveto
       2 copy  vpt 270 360 arc closepath fill
               vpt 0 360 arc closepath } bind def
/C12 { BL [] 0 setdash 2 copy moveto
       2 copy  vpt 180 360 arc closepath fill
               vpt 0 360 arc closepath } bind def
/C13 { BL [] 0 setdash  2 copy moveto
       2 copy  vpt 0 90 arc closepath fill
       2 copy moveto
       2 copy  vpt 180 360 arc closepath fill
               vpt 0 360 arc closepath } bind def
/C14 { BL [] 0 setdash 2 copy moveto
       2 copy  vpt 90 360 arc closepath fill
               vpt 0 360 arc } bind def
/C15 { BL [] 0 setdash 2 copy vpt 0 360 arc closepath fill
               vpt 0 360 arc closepath } bind def
/Rec   { newpath 4 2 roll moveto 1 index 0 rlineto 0 exch rlineto
       neg 0 rlineto closepath } bind def
/Square { dup Rec } bind def
/Bsquare { vpt sub exch vpt sub exch vpt2 Square } bind def
/S0 { BL [] 0 setdash 2 copy moveto 0 vpt rlineto BL Bsquare } bind def
/S1 { BL [] 0 setdash 2 copy vpt Square fill Bsquare } bind def
/S2 { BL [] 0 setdash 2 copy exch vpt sub exch vpt Square fill Bsquare } bind def
/S3 { BL [] 0 setdash 2 copy exch vpt sub exch vpt2 vpt Rec fill Bsquare } bind def
/S4 { BL [] 0 setdash 2 copy exch vpt sub exch vpt sub vpt Square fill Bsquare } bind def
/S5 { BL [] 0 setdash 2 copy 2 copy vpt Square fill
       exch vpt sub exch vpt sub vpt Square fill Bsquare } bind def
/S6 { BL [] 0 setdash 2 copy exch vpt sub exch vpt sub vpt vpt2 Rec fill Bsquare } bind def
/S7 { BL [] 0 setdash 2 copy exch vpt sub exch vpt sub vpt vpt2 Rec fill
       2 copy vpt Square fill
       Bsquare } bind def
/S8 { BL [] 0 setdash 2 copy vpt sub vpt Square fill Bsquare } bind def
/S9 { BL [] 0 setdash 2 copy vpt sub vpt vpt2 Rec fill Bsquare } bind def
/S10 { BL [] 0 setdash 2 copy vpt sub vpt Square fill 2 copy exch vpt sub exch vpt Square fill
       Bsquare } bind def
/S11 { BL [] 0 setdash 2 copy vpt sub vpt Square fill 2 copy exch vpt sub exch vpt2 vpt Rec fill
       Bsquare } bind def
/S12 { BL [] 0 setdash 2 copy exch vpt sub exch vpt sub vpt2 vpt Rec fill Bsquare } bind def
/S13 { BL [] 0 setdash 2 copy exch vpt sub exch vpt sub vpt2 vpt Rec fill
       2 copy vpt Square fill Bsquare } bind def
/S14 { BL [] 0 setdash 2 copy exch vpt sub exch vpt sub vpt2 vpt Rec fill
       2 copy exch vpt sub exch vpt Square fill Bsquare } bind def
/S15 { BL [] 0 setdash 2 copy Bsquare fill Bsquare } bind def
/D0 { gsave translate 45 rotate 0 0 S0 stroke grestore } bind def
/D1 { gsave translate 45 rotate 0 0 S1 stroke grestore } bind def
/D2 { gsave translate 45 rotate 0 0 S2 stroke grestore } bind def
/D3 { gsave translate 45 rotate 0 0 S3 stroke grestore } bind def
/D4 { gsave translate 45 rotate 0 0 S4 stroke grestore } bind def
/D5 { gsave translate 45 rotate 0 0 S5 stroke grestore } bind def
/D6 { gsave translate 45 rotate 0 0 S6 stroke grestore } bind def
/D7 { gsave translate 45 rotate 0 0 S7 stroke grestore } bind def
/D8 { gsave translate 45 rotate 0 0 S8 stroke grestore } bind def
/D9 { gsave translate 45 rotate 0 0 S9 stroke grestore } bind def
/D10 { gsave translate 45 rotate 0 0 S10 stroke grestore } bind def
/D11 { gsave translate 45 rotate 0 0 S11 stroke grestore } bind def
/D12 { gsave translate 45 rotate 0 0 S12 stroke grestore } bind def
/D13 { gsave translate 45 rotate 0 0 S13 stroke grestore } bind def
/D14 { gsave translate 45 rotate 0 0 S14 stroke grestore } bind def
/D15 { gsave translate 45 rotate 0 0 S15 stroke grestore } bind def
/DiaE { stroke [] 0 setdash vpt add M
  hpt neg vpt neg V hpt vpt neg V
  hpt vpt V hpt neg vpt V closepath stroke } def
/BoxE { stroke [] 0 setdash exch hpt sub exch vpt add M
  0 vpt2 neg V hpt2 0 V 0 vpt2 V
  hpt2 neg 0 V closepath stroke } def
/TriUE { stroke [] 0 setdash vpt 1.12 mul add M
  hpt neg vpt -1.62 mul V
  hpt 2 mul 0 V
  hpt neg vpt 1.62 mul V closepath stroke } def
/TriDE { stroke [] 0 setdash vpt 1.12 mul sub M
  hpt neg vpt 1.62 mul V
  hpt 2 mul 0 V
  hpt neg vpt -1.62 mul V closepath stroke } def
/PentE { stroke [] 0 setdash gsave
  translate 0 hpt M 4 {72 rotate 0 hpt L} repeat
  closepath stroke grestore } def
/CircE { stroke [] 0 setdash 
  hpt 0 360 arc stroke } def
/Opaque { gsave closepath 1 setgray fill grestore 0 setgray closepath } def
/DiaW { stroke [] 0 setdash vpt add M
  hpt neg vpt neg V hpt vpt neg V
  hpt vpt V hpt neg vpt V Opaque stroke } def
/BoxW { stroke [] 0 setdash exch hpt sub exch vpt add M
  0 vpt2 neg V hpt2 0 V 0 vpt2 V
  hpt2 neg 0 V Opaque stroke } def
/TriUW { stroke [] 0 setdash vpt 1.12 mul add M
  hpt neg vpt -1.62 mul V
  hpt 2 mul 0 V
  hpt neg vpt 1.62 mul V Opaque stroke } def
/TriDW { stroke [] 0 setdash vpt 1.12 mul sub M
  hpt neg vpt 1.62 mul V
  hpt 2 mul 0 V
  hpt neg vpt -1.62 mul V Opaque stroke } def
/PentW { stroke [] 0 setdash gsave
  translate 0 hpt M 4 {72 rotate 0 hpt L} repeat
  Opaque stroke grestore } def
/CircW { stroke [] 0 setdash 
  hpt 0 360 arc Opaque stroke } def
/BoxFill { gsave Rec 1 setgray fill grestore } def
end
}}%
\begin{picture}(3600,2160)(0,0)%
{\GNUPLOTspecial{"
gnudict begin
gsave
0 0 translate
0.100 0.100 scale
0 setgray
newpath
1.000 UL
LTb
500 300 M
63 0 V
2887 0 R
-63 0 V
500 551 M
63 0 V
2887 0 R
-63 0 V
500 803 M
63 0 V
2887 0 R
-63 0 V
500 1054 M
63 0 V
2887 0 R
-63 0 V
500 1306 M
63 0 V
2887 0 R
-63 0 V
500 1557 M
63 0 V
2887 0 R
-63 0 V
500 1809 M
63 0 V
2887 0 R
-63 0 V
500 2060 M
63 0 V
2887 0 R
-63 0 V
500 300 M
0 63 V
0 1697 R
0 -63 V
921 300 M
0 63 V
0 1697 R
0 -63 V
1343 300 M
0 63 V
0 1697 R
0 -63 V
1764 300 M
0 63 V
0 1697 R
0 -63 V
2186 300 M
0 63 V
0 1697 R
0 -63 V
2607 300 M
0 63 V
0 1697 R
0 -63 V
3029 300 M
0 63 V
0 1697 R
0 -63 V
3450 300 M
0 63 V
0 1697 R
0 -63 V
1.000 UL
LTb
500 300 M
2950 0 V
0 1760 V
-2950 0 V
500 300 L
1.000 UP
1.500 UL
LT0
2898 551 M
-2 1 V
-1 0 V
-1 0 V
-4 -1 V
-8 0 V
-16 1 V
-33 1 V
-64 5 V
-118 14 V
-204 37 V
-302 83 V
1779 839 L
-374 223 V
-247 218 V
-140 180 V
-62 122 V
-24 72 V
-8 38 V
-2 19 V
0 10 V
-1 5 V
0 2 V
1 1 V
0 1 V
-1 0 V
2898 551 Pls
2896 552 Pls
2896 552 Pls
2896 552 Pls
2895 552 Pls
2894 552 Pls
2890 551 Pls
2882 551 Pls
2866 552 Pls
2833 553 Pls
2769 558 Pls
2651 572 Pls
2447 609 Pls
2145 692 Pls
1779 839 Pls
1405 1062 Pls
1158 1280 Pls
1018 1460 Pls
956 1582 Pls
932 1654 Pls
924 1692 Pls
922 1711 Pls
922 1721 Pls
921 1726 Pls
921 1728 Pls
922 1729 Pls
922 1730 Pls
921 1730 Pls
921 1730 Pls
921 1730 Pls
0.300 UL
LT3
2994 551 M
-23 -21 V
-159 46 V
5 -94 V
49 111 V
-52 -19 V
8 -81 V
-211 53 V
100 -14 V
3 40 V
-174 27 V
38 33 V
2435 771 L
-473 30 V
1559 913 L
-281 165 V
-155 260 V
-104 100 V
4 155 V
-77 -4 V
822 1548 L
27 138 V
-45 6 V
74 117 V
92 -34 V
-53 -29 V
798 1571 L
131 96 V
-28 169 V
42 -77 V
2773 551 M
-55 -53 V
103 2 V
66 49 V
121 -38 V
74 -31 V
44 49 V
2949 505 L
-197 0 V
-47 6 V
-155 45 V
68 -7 V
-255 2 V
-193 79 V
-306 18 V
1500 943 L
-347 79 V
-30 209 V
-118 209 V
905 1663 L
-24 74 V
58 35 V
20 -42 V
112 -55 V
2 62 V
936 1722 L
103 25 V
23 -36 V
-198 -5 V
40 39 V
2771 551 M
234 -25 V
-187 2 V
120 30 V
-83 17 V
7 11 V
68 -97 V
12 -11 V
2623 578 L
48 -7 V
-25 -26 V
2476 687 L
2281 672 L
-416 52 V
1706 878 L
-226 217 V
-264 42 V
-123 175 V
-58 233 V
-70 70 V
-158 72 V
124 14 V
33 138 V
938 1708 L
91 74 V
921 1709 L
-76 81 V
793 1663 L
170 58 V
2 -46 V
2954 551 M
61 13 V
-297 4 V
151 -22 V
77 -6 V
-210 49 V
148 -67 V
-107 61 V
-79 -26 V
220 -31 V
-86 -5 V
-363 69 V
-202 -4 V
2019 686 L
-334 64 V
-264 341 V
-301 174 V
-95 284 V
75 65 V
-126 66 V
-21 -77 V
-93 113 V
11 49 V
87 30 V
54 37 V
-25 -11 V
-36 -46 V
-55 -97 V
-46 15 V
62 -17 V
2855 551 M
-31 91 V
47 3 V
146 -78 V
-178 32 V
73 -37 V
-108 -2 V
-172 94 V
175 -45 V
59 -18 V
109 59 V
-505 2 V
-24 59 V
-567 27 V
1542 881 L
-279 252 V
-155 123 V
990 1442 L
872 1608 L
-42 35 V
58 270 V
-72 -63 V
86 -41 V
28 -63 V
-75 -1 V
63 -6 V
85 10 V
893 1692 L
-8 45 V
7 -33 V
2755 551 M
2652 492 L
143 77 V
102 -82 V
35 42 V
-53 17 V
-70 24 V
-64 -45 V
132 -38 V
-257 11 V
2415 487 L
2129 673 L
-143 18 V
1532 806 L
-59 237 V
-376 291 V
-90 208 V
-108 -1 V
27 187 V
9 36 V
65 -2 V
77 -94 V
-104 -5 V
20 30 V
-195 18 V
192 16 V
-69 54 V
136 48 V
979 1702 L
85 -10 V
2969 551 M
-211 13 V
6 37 V
205 0 V
20 -90 V
-18 103 V
2835 561 L
83 48 V
-105 2 V
194 -12 V
-82 35 V
15 28 V
2502 614 L
1942 742 L
1692 946 L
-391 130 V
-84 152 V
-155 128 V
872 1587 L
-38 46 V
-16 -1 V
-55 80 V
126 34 V
-45 -87 V
119 98 V
-93 45 V
-89 -19 V
38 -80 V
95 81 V
904 1670 L
2914 551 M
-11 5 V
-189 15 V
45 38 V
140 -40 V
2746 542 L
160 5 V
-78 12 V
11 -31 V
-42 1 V
-32 -22 V
-95 9 V
-276 75 V
2129 732 L
1694 909 L
-432 110 V
-189 326 V
33 -29 V
976 1681 L
-76 -90 V
118 200 V
897 1767 L
85 58 V
-13 -46 V
-99 32 V
86 -31 V
67 40 V
31 -23 V
-137 19 V
26 -69 V
2980 551 M
-107 53 V
25 -11 V
-33 -8 V
-79 -25 V
-26 25 V
-61 16 V
154 -20 V
278 -49 V
-45 71 V
2959 568 L
7 -8 V
2588 529 L
2162 704 L
-317 19 V
1413 996 L
-100 206 V
-230 208 V
-71 254 V
902 1767 L
-33 52 V
-5 -111 V
-90 111 V
74 -55 V
-19 -28 V
83 -128 V
799 1788 L
36 -6 V
57 -30 V
-22 -23 V
2943 551 M
-14 63 V
-2 -78 V
2817 488 L
83 46 V
28 -2 V
-206 0 V
28 13 V
-185 -5 V
66 -14 V
152 31 V
-485 25 V
-295 96 V
-82 130 V
-526 38 V
-176 356 V
-126 208 V
974 1613 L
-31 70 V
-54 -13 V
-42 19 V
195 6 V
-55 113 V
-12 -20 V
-66 -32 V
32 -49 V
-5 123 V
-32 -86 V
20 -38 V
-71 51 V
2853 551 M
-12 -31 V
67 38 V
-37 -37 V
-117 64 V
-35 -44 V
2666 439 L
-31 60 V
54 -41 V
-123 72 V
182 -31 V
-312 81 V
2 3 V
2160 691 L
-252 52 V
-429 352 V
-258 134 V
-150 194 V
-74 172 V
-72 -22 V
63 122 V
839 1659 L
18 -21 V
-47 161 V
111 -45 V
29 -14 V
18 -23 V
-128 34 V
67 75 V
16 -122 V
3027 551 M
-92 -34 V
-8 40 V
128 -18 V
2888 524 L
53 79 V
-243 -8 V
63 -29 V
-30 -27 V
-8 8 V
-322 53 V
-101 24 V
108 47 V
-476 63 V
1507 901 L
-329 386 V
954 1529 L
-2 119 V
-94 41 V
89 -31 V
-47 34 V
-108 20 V
195 -85 V
-44 58 V
837 1829 L
-61 -56 V
238 38 V
2 30 V
-60 -1 V
3 -27 V
3068 551 M
8 34 V
45 28 V
-4 16 V
45 -82 V
-88 53 V
-219 25 V
13 -61 V
-27 21 V
-32 -48 V
34 5 V
-4 -45 V
-222 66 V
2182 691 L
-100 52 V
-636 291 V
-281 278 V
878 1458 L
-6 164 V
31 12 V
105 98 V
-94 -63 V
-3 28 V
-63 -9 V
-21 64 V
-10 64 V
19 -107 V
22 107 V
58 40 V
16 -59 V
2982 551 M
-45 24 V
-40 -46 V
-29 -20 V
13 76 V
64 3 V
-3 -109 V
-9 31 V
-151 -1 V
27 -6 V
currentpoint stroke M
-24 19 V
-54 38 V
-308 54 V
1929 749 L
1542 957 L
-451 129 V
975 1380 L
-76 112 V
-71 160 V
-56 164 V
98 -128 V
4 -100 V
-50 153 V
149 108 V
-32 0 V
928 1666 L
-85 -45 V
127 77 V
948 1587 L
39 203 V
3141 551 M
3037 512 L
-57 49 V
85 4 V
-4 13 V
-92 -39 V
114 23 V
-163 15 V
2741 470 L
-10 107 V
-98 -4 V
-69 -10 V
-315 94 V
1996 792 L
1623 934 L
-379 224 V
996 1369 L
-65 183 V
-58 117 V
8 77 V
-11 -1 V
44 16 V
-74 33 V
960 1660 L
-61 2 V
-4 12 V
4 123 V
15 -40 V
814 1857 L
58 17 V
2956 551 M
2 -87 V
69 57 V
63 -19 V
-28 -48 V
-146 39 V
24 -68 V
-249 7 V
204 41 V
74 -30 V
-162 1 V
68 44 V
-86 44 V
2247 520 L
2032 730 L
1600 937 L
-147 339 V
-378 139 V
48 246 V
-31 59 V
-56 121 V
57 -19 V
-97 -5 V
968 1692 L
108 -116 V
-90 77 V
62 107 V
60 39 V
-99 -60 V
125 57 V
2975 551 M
139 5 V
-42 5 V
-51 20 V
-45 52 V
61 -77 V
2804 522 L
-79 -12 V
382 56 V
2909 514 L
-287 31 V
10 3 V
-451 57 V
1876 781 L
1498 918 L
-229 338 V
-262 305 V
52 -83 V
894 1721 L
49 121 V
-30 56 V
-29 -77 V
41 -11 V
15 13 V
-28 -23 V
41 51 V
27 -220 V
-105 71 V
135 57 V
-2 -81 V
2918 551 M
68 -76 V
7 54 V
209 29 V
2993 530 L
5 28 V
2808 492 L
132 19 V
-144 -4 V
83 48 V
-131 -9 V
-270 58 V
2134 704 L
1828 868 L
-157 82 V
-337 291 V
929 1402 L
22 206 V
-54 142 V
-39 52 V
82 -144 V
-31 37 V
17 -32 V
119 -18 V
-74 104 V
-4 -10 V
-40 -21 V
52 79 V
-17 -19 V
933 1655 L
2809 551 M
30 9 V
-27 -14 V
260 28 V
2866 548 L
213 62 V
2956 595 L
-44 -19 V
-142 36 V
-61 -49 V
-136 71 V
2348 607 L
-188 39 V
1834 885 L
-371 20 V
-232 258 V
-138 197 V
-100 88 V
855 1638 L
51 62 V
12 -42 V
-11 10 V
-131 67 V
121 34 V
0 11 V
18 -116 V
810 1849 L
93 39 V
838 1700 L
91 -156 V
2979 551 M
2857 522 L
285 -30 V
-86 15 V
-94 27 V
-54 -33 V
142 68 V
-125 -4 V
25 -10 V
-107 -7 V
-40 77 V
59 -76 V
-264 63 V
-540 12 V
1769 775 L
-303 248 V
-326 211 V
-58 117 V
9 146 V
-48 236 V
-164 13 V
13 -81 V
-76 48 V
204 70 V
-89 -88 V
139 45 V
945 1638 L
7 162 V
-5 -107 V
11 7 V
stroke
grestore
end
showpage
}}%
\put(1975,50){\makebox(0,0){$\mathrm{Re}\, \overline{l_{p}}$}}%
\put(100,1180){%
\makebox(0,0)[b]{\shortstack{$\mathrm{Im}\, \overline{l_{p}}$}}%
}%
\put(3450,200){\makebox(0,0){0.3}}%
\put(3029,200){\makebox(0,0){0.25}}%
\put(2607,200){\makebox(0,0){0.2}}%
\put(2186,200){\makebox(0,0){0.15}}%
\put(1764,200){\makebox(0,0){0.1}}%
\put(1343,200){\makebox(0,0){0.05}}%
\put(921,200){\makebox(0,0){0}}%
\put(500,200){\makebox(0,0){-0.05}}%
\put(450,2060){\makebox(0,0)[r]{0.3}}%
\put(450,1809){\makebox(0,0)[r]{0.25}}%
\put(450,1557){\makebox(0,0)[r]{0.2}}%
\put(450,1306){\makebox(0,0)[r]{0.15}}%
\put(450,1054){\makebox(0,0)[r]{0.1}}%
\put(450,803){\makebox(0,0)[r]{0.05}}%
\put(450,551){\makebox(0,0)[r]{0}}%
\put(450,300){\makebox(0,0)[r]{-0.05}}%
\end{picture}%
\endgroup

\end	{center}
\vskip 0.15in
\caption{The values in the complex plane taken by the Polyakov 
loop for $x_3\in [0,L-1]$ for a sample of  individual $k=1$ walls 
in SU(4) with $aT=0.25$ and $T\simeq 1.88 T_c$. Solid line
is the average.} 
\label{fig_profc3}
\end 	{figure}

\begin	{figure}[p]
\begin	{center}
\leavevmode
\begingroup%
  \makeatletter%
  \newcommand{\GNUPLOTspecial}{%
    \@sanitize\catcode`\%=14\relax\special}%
  \setlength{\unitlength}{0.1bp}%
{\GNUPLOTspecial{!
/gnudict 256 dict def
gnudict begin
/Color false def
/Solid false def
/gnulinewidth 5.000 def
/userlinewidth gnulinewidth def
/vshift -33 def
/dl {10 mul} def
/hpt_ 31.5 def
/vpt_ 31.5 def
/hpt hpt_ def
/vpt vpt_ def
/M {moveto} bind def
/L {lineto} bind def
/R {rmoveto} bind def
/V {rlineto} bind def
/vpt2 vpt 2 mul def
/hpt2 hpt 2 mul def
/Lshow { currentpoint stroke M
  0 vshift R show } def
/Rshow { currentpoint stroke M
  dup stringwidth pop neg vshift R show } def
/Cshow { currentpoint stroke M
  dup stringwidth pop -2 div vshift R show } def
/UP { dup vpt_ mul /vpt exch def hpt_ mul /hpt exch def
  /hpt2 hpt 2 mul def /vpt2 vpt 2 mul def } def
/DL { Color {setrgbcolor Solid {pop []} if 0 setdash }
 {pop pop pop Solid {pop []} if 0 setdash} ifelse } def
/BL { stroke userlinewidth 2 mul setlinewidth } def
/AL { stroke userlinewidth 2 div setlinewidth } def
/UL { dup gnulinewidth mul /userlinewidth exch def
      10 mul /udl exch def } def
/PL { stroke userlinewidth setlinewidth } def
/LTb { BL [] 0 0 0 DL } def
/LTa { AL [1 udl mul 2 udl mul] 0 setdash 0 0 0 setrgbcolor } def
/LT0 { PL [] 1 0 0 DL } def
/LT1 { PL [4 dl 2 dl] 0 1 0 DL } def
/LT2 { PL [2 dl 3 dl] 0 0 1 DL } def
/LT3 { PL [1 dl 1.5 dl] 1 0 1 DL } def
/LT4 { PL [5 dl 2 dl 1 dl 2 dl] 0 1 1 DL } def
/LT5 { PL [4 dl 3 dl 1 dl 3 dl] 1 1 0 DL } def
/LT6 { PL [2 dl 2 dl 2 dl 4 dl] 0 0 0 DL } def
/LT7 { PL [2 dl 2 dl 2 dl 2 dl 2 dl 4 dl] 1 0.3 0 DL } def
/LT8 { PL [2 dl 2 dl 2 dl 2 dl 2 dl 2 dl 2 dl 4 dl] 0.5 0.5 0.5 DL } def
/Pnt { stroke [] 0 setdash
   gsave 1 setlinecap M 0 0 V stroke grestore } def
/Dia { stroke [] 0 setdash 2 copy vpt add M
  hpt neg vpt neg V hpt vpt neg V
  hpt vpt V hpt neg vpt V closepath stroke
  Pnt } def
/Pls { stroke [] 0 setdash vpt sub M 0 vpt2 V
  currentpoint stroke M
  hpt neg vpt neg R hpt2 0 V stroke
  } def
/Box { stroke [] 0 setdash 2 copy exch hpt sub exch vpt add M
  0 vpt2 neg V hpt2 0 V 0 vpt2 V
  hpt2 neg 0 V closepath stroke
  Pnt } def
/Crs { stroke [] 0 setdash exch hpt sub exch vpt add M
  hpt2 vpt2 neg V currentpoint stroke M
  hpt2 neg 0 R hpt2 vpt2 V stroke } def
/TriU { stroke [] 0 setdash 2 copy vpt 1.12 mul add M
  hpt neg vpt -1.62 mul V
  hpt 2 mul 0 V
  hpt neg vpt 1.62 mul V closepath stroke
  Pnt  } def
/Star { 2 copy Pls Crs } def
/BoxF { stroke [] 0 setdash exch hpt sub exch vpt add M
  0 vpt2 neg V  hpt2 0 V  0 vpt2 V
  hpt2 neg 0 V  closepath fill } def
/TriUF { stroke [] 0 setdash vpt 1.12 mul add M
  hpt neg vpt -1.62 mul V
  hpt 2 mul 0 V
  hpt neg vpt 1.62 mul V closepath fill } def
/TriD { stroke [] 0 setdash 2 copy vpt 1.12 mul sub M
  hpt neg vpt 1.62 mul V
  hpt 2 mul 0 V
  hpt neg vpt -1.62 mul V closepath stroke
  Pnt  } def
/TriDF { stroke [] 0 setdash vpt 1.12 mul sub M
  hpt neg vpt 1.62 mul V
  hpt 2 mul 0 V
  hpt neg vpt -1.62 mul V closepath fill} def
/DiaF { stroke [] 0 setdash vpt add M
  hpt neg vpt neg V hpt vpt neg V
  hpt vpt V hpt neg vpt V closepath fill } def
/Pent { stroke [] 0 setdash 2 copy gsave
  translate 0 hpt M 4 {72 rotate 0 hpt L} repeat
  closepath stroke grestore Pnt } def
/PentF { stroke [] 0 setdash gsave
  translate 0 hpt M 4 {72 rotate 0 hpt L} repeat
  closepath fill grestore } def
/Circle { stroke [] 0 setdash 2 copy
  hpt 0 360 arc stroke Pnt } def
/CircleF { stroke [] 0 setdash hpt 0 360 arc fill } def
/C0 { BL [] 0 setdash 2 copy moveto vpt 90 450  arc } bind def
/C1 { BL [] 0 setdash 2 copy        moveto
       2 copy  vpt 0 90 arc closepath fill
               vpt 0 360 arc closepath } bind def
/C2 { BL [] 0 setdash 2 copy moveto
       2 copy  vpt 90 180 arc closepath fill
               vpt 0 360 arc closepath } bind def
/C3 { BL [] 0 setdash 2 copy moveto
       2 copy  vpt 0 180 arc closepath fill
               vpt 0 360 arc closepath } bind def
/C4 { BL [] 0 setdash 2 copy moveto
       2 copy  vpt 180 270 arc closepath fill
               vpt 0 360 arc closepath } bind def
/C5 { BL [] 0 setdash 2 copy moveto
       2 copy  vpt 0 90 arc
       2 copy moveto
       2 copy  vpt 180 270 arc closepath fill
               vpt 0 360 arc } bind def
/C6 { BL [] 0 setdash 2 copy moveto
      2 copy  vpt 90 270 arc closepath fill
              vpt 0 360 arc closepath } bind def
/C7 { BL [] 0 setdash 2 copy moveto
      2 copy  vpt 0 270 arc closepath fill
              vpt 0 360 arc closepath } bind def
/C8 { BL [] 0 setdash 2 copy moveto
      2 copy vpt 270 360 arc closepath fill
              vpt 0 360 arc closepath } bind def
/C9 { BL [] 0 setdash 2 copy moveto
      2 copy  vpt 270 450 arc closepath fill
              vpt 0 360 arc closepath } bind def
/C10 { BL [] 0 setdash 2 copy 2 copy moveto vpt 270 360 arc closepath fill
       2 copy moveto
       2 copy vpt 90 180 arc closepath fill
               vpt 0 360 arc closepath } bind def
/C11 { BL [] 0 setdash 2 copy moveto
       2 copy  vpt 0 180 arc closepath fill
       2 copy moveto
       2 copy  vpt 270 360 arc closepath fill
               vpt 0 360 arc closepath } bind def
/C12 { BL [] 0 setdash 2 copy moveto
       2 copy  vpt 180 360 arc closepath fill
               vpt 0 360 arc closepath } bind def
/C13 { BL [] 0 setdash  2 copy moveto
       2 copy  vpt 0 90 arc closepath fill
       2 copy moveto
       2 copy  vpt 180 360 arc closepath fill
               vpt 0 360 arc closepath } bind def
/C14 { BL [] 0 setdash 2 copy moveto
       2 copy  vpt 90 360 arc closepath fill
               vpt 0 360 arc } bind def
/C15 { BL [] 0 setdash 2 copy vpt 0 360 arc closepath fill
               vpt 0 360 arc closepath } bind def
/Rec   { newpath 4 2 roll moveto 1 index 0 rlineto 0 exch rlineto
       neg 0 rlineto closepath } bind def
/Square { dup Rec } bind def
/Bsquare { vpt sub exch vpt sub exch vpt2 Square } bind def
/S0 { BL [] 0 setdash 2 copy moveto 0 vpt rlineto BL Bsquare } bind def
/S1 { BL [] 0 setdash 2 copy vpt Square fill Bsquare } bind def
/S2 { BL [] 0 setdash 2 copy exch vpt sub exch vpt Square fill Bsquare } bind def
/S3 { BL [] 0 setdash 2 copy exch vpt sub exch vpt2 vpt Rec fill Bsquare } bind def
/S4 { BL [] 0 setdash 2 copy exch vpt sub exch vpt sub vpt Square fill Bsquare } bind def
/S5 { BL [] 0 setdash 2 copy 2 copy vpt Square fill
       exch vpt sub exch vpt sub vpt Square fill Bsquare } bind def
/S6 { BL [] 0 setdash 2 copy exch vpt sub exch vpt sub vpt vpt2 Rec fill Bsquare } bind def
/S7 { BL [] 0 setdash 2 copy exch vpt sub exch vpt sub vpt vpt2 Rec fill
       2 copy vpt Square fill
       Bsquare } bind def
/S8 { BL [] 0 setdash 2 copy vpt sub vpt Square fill Bsquare } bind def
/S9 { BL [] 0 setdash 2 copy vpt sub vpt vpt2 Rec fill Bsquare } bind def
/S10 { BL [] 0 setdash 2 copy vpt sub vpt Square fill 2 copy exch vpt sub exch vpt Square fill
       Bsquare } bind def
/S11 { BL [] 0 setdash 2 copy vpt sub vpt Square fill 2 copy exch vpt sub exch vpt2 vpt Rec fill
       Bsquare } bind def
/S12 { BL [] 0 setdash 2 copy exch vpt sub exch vpt sub vpt2 vpt Rec fill Bsquare } bind def
/S13 { BL [] 0 setdash 2 copy exch vpt sub exch vpt sub vpt2 vpt Rec fill
       2 copy vpt Square fill Bsquare } bind def
/S14 { BL [] 0 setdash 2 copy exch vpt sub exch vpt sub vpt2 vpt Rec fill
       2 copy exch vpt sub exch vpt Square fill Bsquare } bind def
/S15 { BL [] 0 setdash 2 copy Bsquare fill Bsquare } bind def
/D0 { gsave translate 45 rotate 0 0 S0 stroke grestore } bind def
/D1 { gsave translate 45 rotate 0 0 S1 stroke grestore } bind def
/D2 { gsave translate 45 rotate 0 0 S2 stroke grestore } bind def
/D3 { gsave translate 45 rotate 0 0 S3 stroke grestore } bind def
/D4 { gsave translate 45 rotate 0 0 S4 stroke grestore } bind def
/D5 { gsave translate 45 rotate 0 0 S5 stroke grestore } bind def
/D6 { gsave translate 45 rotate 0 0 S6 stroke grestore } bind def
/D7 { gsave translate 45 rotate 0 0 S7 stroke grestore } bind def
/D8 { gsave translate 45 rotate 0 0 S8 stroke grestore } bind def
/D9 { gsave translate 45 rotate 0 0 S9 stroke grestore } bind def
/D10 { gsave translate 45 rotate 0 0 S10 stroke grestore } bind def
/D11 { gsave translate 45 rotate 0 0 S11 stroke grestore } bind def
/D12 { gsave translate 45 rotate 0 0 S12 stroke grestore } bind def
/D13 { gsave translate 45 rotate 0 0 S13 stroke grestore } bind def
/D14 { gsave translate 45 rotate 0 0 S14 stroke grestore } bind def
/D15 { gsave translate 45 rotate 0 0 S15 stroke grestore } bind def
/DiaE { stroke [] 0 setdash vpt add M
  hpt neg vpt neg V hpt vpt neg V
  hpt vpt V hpt neg vpt V closepath stroke } def
/BoxE { stroke [] 0 setdash exch hpt sub exch vpt add M
  0 vpt2 neg V hpt2 0 V 0 vpt2 V
  hpt2 neg 0 V closepath stroke } def
/TriUE { stroke [] 0 setdash vpt 1.12 mul add M
  hpt neg vpt -1.62 mul V
  hpt 2 mul 0 V
  hpt neg vpt 1.62 mul V closepath stroke } def
/TriDE { stroke [] 0 setdash vpt 1.12 mul sub M
  hpt neg vpt 1.62 mul V
  hpt 2 mul 0 V
  hpt neg vpt -1.62 mul V closepath stroke } def
/PentE { stroke [] 0 setdash gsave
  translate 0 hpt M 4 {72 rotate 0 hpt L} repeat
  closepath stroke grestore } def
/CircE { stroke [] 0 setdash 
  hpt 0 360 arc stroke } def
/Opaque { gsave closepath 1 setgray fill grestore 0 setgray closepath } def
/DiaW { stroke [] 0 setdash vpt add M
  hpt neg vpt neg V hpt vpt neg V
  hpt vpt V hpt neg vpt V Opaque stroke } def
/BoxW { stroke [] 0 setdash exch hpt sub exch vpt add M
  0 vpt2 neg V hpt2 0 V 0 vpt2 V
  hpt2 neg 0 V Opaque stroke } def
/TriUW { stroke [] 0 setdash vpt 1.12 mul add M
  hpt neg vpt -1.62 mul V
  hpt 2 mul 0 V
  hpt neg vpt 1.62 mul V Opaque stroke } def
/TriDW { stroke [] 0 setdash vpt 1.12 mul sub M
  hpt neg vpt 1.62 mul V
  hpt 2 mul 0 V
  hpt neg vpt -1.62 mul V Opaque stroke } def
/PentW { stroke [] 0 setdash gsave
  translate 0 hpt M 4 {72 rotate 0 hpt L} repeat
  Opaque stroke grestore } def
/CircW { stroke [] 0 setdash 
  hpt 0 360 arc Opaque stroke } def
/BoxFill { gsave Rec 1 setgray fill grestore } def
end
}}%
\begin{picture}(3600,2160)(0,0)%
{\GNUPLOTspecial{"
gnudict begin
gsave
0 0 translate
0.100 0.100 scale
0 setgray
newpath
1.000 UL
LTb
500 300 M
63 0 V
2887 0 R
-63 0 V
500 460 M
63 0 V
2887 0 R
-63 0 V
500 620 M
63 0 V
2887 0 R
-63 0 V
500 780 M
63 0 V
2887 0 R
-63 0 V
500 940 M
63 0 V
2887 0 R
-63 0 V
500 1100 M
63 0 V
2887 0 R
-63 0 V
500 1260 M
63 0 V
2887 0 R
-63 0 V
500 1420 M
63 0 V
2887 0 R
-63 0 V
500 1580 M
63 0 V
2887 0 R
-63 0 V
500 1740 M
63 0 V
2887 0 R
-63 0 V
500 1900 M
63 0 V
2887 0 R
-63 0 V
500 2060 M
63 0 V
2887 0 R
-63 0 V
500 300 M
0 63 V
0 1697 R
0 -63 V
768 300 M
0 63 V
0 1697 R
0 -63 V
1036 300 M
0 63 V
0 1697 R
0 -63 V
1305 300 M
0 63 V
0 1697 R
0 -63 V
1573 300 M
0 63 V
0 1697 R
0 -63 V
1841 300 M
0 63 V
0 1697 R
0 -63 V
2109 300 M
0 63 V
0 1697 R
0 -63 V
2377 300 M
0 63 V
0 1697 R
0 -63 V
2645 300 M
0 63 V
0 1697 R
0 -63 V
2914 300 M
0 63 V
0 1697 R
0 -63 V
3182 300 M
0 63 V
0 1697 R
0 -63 V
3450 300 M
0 63 V
0 1697 R
0 -63 V
1.000 UL
LTb
500 300 M
2950 0 V
0 1760 V
-2950 0 V
500 300 L
1.000 UP
1.500 UL
LT0
2895 620 M
-6 1 V
-6 0 V
-12 0 V
-23 0 V
-32 -2 V
-42 1 V
-69 0 V
-97 2 V
-129 2 V
-162 8 V
-191 13 V
-206 24 V
-209 39 V
-190 61 V
1283 906 L
-106 108 V
-69 119 V
-40 118 V
-18 110 V
-10 95 V
-2 79 V
-2 61 V
-1 45 V
0 32 V
0 19 V
-1 14 V
2 9 V
-1 5 V
1 4 V
2895 620 Pls
2889 621 Pls
2883 621 Pls
2871 621 Pls
2848 621 Pls
2816 619 Pls
2774 620 Pls
2705 620 Pls
2608 622 Pls
2479 624 Pls
2317 632 Pls
2126 645 Pls
1920 669 Pls
1711 708 Pls
1521 769 Pls
1283 906 Pls
1177 1014 Pls
1108 1133 Pls
1068 1251 Pls
1050 1361 Pls
1040 1456 Pls
1038 1535 Pls
1036 1596 Pls
1035 1641 Pls
1035 1673 Pls
1035 1692 Pls
1034 1706 Pls
1036 1715 Pls
1035 1720 Pls
1036 1724 Pls
0.300 UL
LT3
3033 620 M
2713 554 L
2929 411 L
212 -19 V
-85 204 V
2858 536 L
7 9 V
235 71 V
2946 534 L
31 85 V
2319 487 L
-191 53 V
-316 89 V
1513 482 L
-192 77 V
42 318 V
888 1023 L
247 312 V
167 105 V
-111 -38 V
-115 322 V
168 -64 V
-213 -49 V
56 86 V
38 167 V
100 -14 V
-71 73 V
-21 -141 V
-83 -2 V
194 -88 V
2861 620 M
38 -97 V
2645 670 L
260 23 V
138 3 V
-1 -49 V
-83 2 V
110 -45 V
2671 499 L
2461 795 L
-346 33 V
1881 694 L
-38 22 V
1536 878 L
-284 32 V
-313 4 V
-5 389 V
23 141 V
22 54 V
-56 127 V
67 49 V
-22 166 V
324 -64 V
-161 54 V
-8 -92 V
3 155 V
-19 70 V
-176 27 V
771 1756 L
97 -40 V
2858 620 M
3026 519 L
-51 -33 V
-37 240 V
3039 619 L
92 -143 V
-228 8 V
129 155 V
-508 84 V
2481 528 L
54 155 V
-546 22 V
149 -58 V
-332 48 V
-82 100 V
1478 956 L
-266 98 V
-88 344 V
1023 1247 L
-13 217 V
229 144 V
194 65 V
-330 -40 V
117 3 V
-147 24 V
24 95 V
139 -101 V
-3 151 V
-118 30 V
-21 -48 V
2930 620 M
-128 68 V
2631 603 L
-43 123 V
2963 566 L
87 15 V
2778 693 L
2534 535 L
126 101 V
108 -43 V
2552 564 L
-272 8 V
1898 522 L
129 212 V
-727 52 V
-170 259 V
-30 -35 V
145 214 V
945 1365 L
139 115 V
85 217 V
14 54 V
-154 -31 V
918 1689 L
240 24 V
-249 73 V
218 -101 V
-130 3 V
10 178 V
-171 -9 V
2829 620 M
2548 600 L
492 52 V
97 -69 V
-253 75 V
147 -16 V
2784 619 L
-246 78 V
1 4 V
2386 531 L
199 265 V
2397 667 L
-422 83 V
1472 688 L
-41 46 V
1238 962 L
-189 -2 V
22 221 V
907 1441 L
143 -60 V
-116 70 V
94 124 V
-34 48 V
8 -5 V
17 230 V
-46 -65 V
859 1721 L
151 -92 V
93 206 V
824 1696 L
2957 620 M
2938 473 L
2767 459 L
166 9 V
253 -31 V
-138 -1 V
-210 33 V
-60 55 V
19 14 V
-81 -63 V
-69 166 V
-326 54 V
-5 -104 V
1722 737 L
-137 -2 V
-391 82 V
92 63 V
-212 229 V
-6 187 V
146 197 V
105 117 V
-111 -31 V
-121 4 V
16 206 V
61 -214 V
47 27 V
114 114 V
999 1652 L
261 -87 V
139 267 V
2768 620 M
10 101 V
77 107 V
42 -68 V
160 -34 V
-8 54 V
2767 702 L
19 -86 V
2584 472 L
153 247 V
2222 708 L
5 -134 V
-156 1 V
1643 725 L
1327 704 L
1182 872 L
-60 180 V
177 4 V
952 1303 L
34 155 V
-74 43 V
126 147 V
-226 7 V
10 24 V
-7 -57 V
108 19 V
-12 5 V
-24 10 V
362 -175 V
-147 88 V
3090 620 M
-533 16 V
261 76 V
-87 -28 V
316 61 V
-316 10 V
2543 596 L
201 80 V
2376 573 L
447 105 V
2532 602 L
-344 15 V
-365 55 V
-432 93 V
-78 63 V
-310 245 V
-2 134 V
113 -24 V
-23 188 V
72 176 V
-119 -91 V
75 61 V
-26 14 V
-337 6 V
203 -65 V
29 212 V
755 1585 L
334 -119 V
-150 53 V
161 77 V
2694 620 M
115 69 V
35 93 V
255 5 V
2701 683 L
203 38 V
-78 12 V
2581 586 L
-26 9 V
-180 88 V
255 30 V
-69 -63 V
2125 817 L
-194 1 V
1584 739 L
1285 919 L
-242 145 V
246 -43 V
-242 367 V
185 146 V
-54 -180 V
-142 77 V
-14 41 V
899 1459 L
-27 4 V
4 75 V
8 -6 V
-6 -24 V
28 17 V
660 1771 L
2724 620 M
62 -18 V
-102 -5 V
318 34 V
-77 49 V
2783 661 L
251 -74 V
-519 33 V
175 50 V
261 23 V
-454 4 V
-457 22 V
18 146 V
1527 807 L
52 -30 V
1128 919 L
110 39 V
-206 333 V
97 48 V
94 72 V
-394 71 V
24 -41 V
-69 126 V
151 77 V
701 1814 L
244 -87 V
761 1705 L
655 1637 L
163 62 V
218 -82 V
2526 620 M
-129 50 V
530 35 V
2727 681 L
189 26 V
2753 500 L
234 37 V
2802 652 L
44 18 V
2527 541 L
49 65 V
-48 -41 V
2185 729 L
-48 -61 V
1855 658 L
1418 908 L
1168 887 L
5 35 V
-58 235 V
68 27 V
-63 143 V
19 76 V
1015 1281 L
914 1566 L
180 -135 V
11 62 V
-10 147 V
205 27 V
-82 50 V
970 1558 L
2719 620 M
77 -12 V
162 -73 V
-153 8 V
86 20 V
175 39 V
2825 506 L
-125 15 V
-206 10 V
-277 70 V
76 -76 V
2134 681 L
1586 606 L
135 68 V
-243 45 V
-72 240 V
-260 163 V
155 142 V
-30 96 V
-230 29 V
46 107 V
98 -89 V
38 -13 V
-77 210 V
149 16 V
-112 -64 V
37 -108 V
-1 64 V
-181 -12 V
17 44 V
2616 620 M
70 79 V
-20 25 V
-66 30 V
106 -60 V
406 117 V
-502 2 V
25 44 V
189 -42 V
-4 -16 V
2647 673 L
2091 599 L
-58 152 V
49 37 V
1842 758 L
1530 902 L
-266 109 V
-274 16 V
108 20 V
876 1266 L
147 19 V
905 1400 L
167 118 V
896 1417 L
-24 89 V
131 75 V
-108 7 V
154 -38 V
-90 208 V
-119 14 V
2439 620 M
270 86 V
153 -71 V
-59 112 V
55 -68 V
54 -52 V
-57 35 V
2833 562 L
2375 682 L
49 -81 V
currentpoint stroke M
2241 762 L
-27 98 V
2157 553 L
1664 724 L
-164 64 V
1250 965 L
-52 135 V
990 1226 L
92 91 V
785 1267 L
390 121 V
-94 -2 V
-114 41 V
105 74 V
143 293 V
1092 1540 L
-103 79 V
-7 151 V
90 -179 V
927 1508 L
2637 620 M
332 91 V
-217 48 V
155 -94 V
127 27 V
-100 -5 V
-33 44 V
2857 627 L
-415 58 V
-107 44 V
2060 567 L
1929 542 L
1787 644 L
1564 805 L
1306 968 L
854 942 L
57 114 V
109 33 V
-84 328 V
911 1197 L
32 30 V
342 219 V
-269 110 V
143 13 V
-92 83 V
628 1553 L
477 89 V
-258 23 V
29 -71 V
348 -104 V
2900 620 M
2773 739 L
2621 676 L
-143 38 V
26 9 V
317 21 V
-26 -80 V
232 -70 V
2795 719 L
2371 847 L
-21 -8 V
185 -78 V
2170 747 L
2158 631 L
-392 41 V
1380 799 L
-41 206 V
1109 990 L
936 971 L
-15 177 V
200 160 V
-139 44 V
741 1243 L
36 87 V
-93 45 V
72 129 V
81 117 V
155 25 V
-3 -132 V
839 1688 L
2759 620 M
-44 -61 V
90 36 V
-59 42 V
-99 -30 V
149 30 V
19 -37 V
-36 54 V
2655 581 L
-169 20 V
2100 706 L
1822 595 L
165 34 V
-628 38 V
108 1 V
1105 786 L
-16 119 V
-59 158 V
890 1177 L
119 -25 V
887 1302 L
669 1402 L
302 -32 V
-32 48 V
125 158 V
111 -4 V
145 94 V
-86 -66 V
991 1705 L
82 -70 V
2487 620 M
2805 455 L
-97 91 V
140 -44 V
2796 399 L
-487 0 V
2183 550 L
1876 523 L
1751 697 L
-172 2 V
-77 -22 V
1352 665 L
-86 -13 V
51 84 V
-8 196 V
-295 183 V
829 1090 L
341 33 V
-92 227 V
212 87 V
-24 292 V
125 127 V
-175 14 V
26 85 V
124 -193 V
2 -24 V
-147 -90 V
171 -67 V
-265 159 V
197 -120 V
2647 620 M
25 22 V
2532 610 L
78 -2 V
53 90 V
55 -179 V
-25 87 V
-175 35 V
-397 -8 V
-10 61 V
1965 548 L
1784 695 L
-252 66 V
119 -73 V
-106 55 V
-461 60 V
141 141 V
-225 171 V
-145 99 V
186 54 V
186 194 V
-212 208 V
-70 228 V
101 34 V
190 26 V
-95 -172 V
-96 -62 V
303 45 V
-251 -46 V
32 -40 V
2600 620 M
54 -42 V
-47 169 V
8 -106 V
2515 608 L
90 32 V
210 -19 V
67 -55 V
-436 43 V
-210 35 V
1821 586 L
256 80 V
-132 17 V
1819 623 L
1505 731 L
1272 981 L
1565 862 L
-269 213 V
971 919 L
857 1021 L
23 219 V
54 360 V
76 21 V
-14 128 V
26 -4 V
41 66 V
-9 14 V
-18 26 V
-270 72 V
281 -88 V
stroke
grestore
end
showpage
}}%
\put(1975,50){\makebox(0,0){$\mathrm{Re}\, \overline{l_{p}}$}}%
\put(100,1180){%
\makebox(0,0)[b]{\shortstack{$\mathrm{Im}\, \overline{l_{p}}$}}%
}%
\put(3450,200){\makebox(0,0){0.18}}%
\put(3182,200){\makebox(0,0){0.16}}%
\put(2914,200){\makebox(0,0){0.14}}%
\put(2645,200){\makebox(0,0){0.12}}%
\put(2377,200){\makebox(0,0){0.1}}%
\put(2109,200){\makebox(0,0){0.08}}%
\put(1841,200){\makebox(0,0){0.06}}%
\put(1573,200){\makebox(0,0){0.04}}%
\put(1305,200){\makebox(0,0){0.02}}%
\put(1036,200){\makebox(0,0){0}}%
\put(768,200){\makebox(0,0){-0.02}}%
\put(500,200){\makebox(0,0){-0.04}}%
\put(450,2060){\makebox(0,0)[r]{0.18}}%
\put(450,1900){\makebox(0,0)[r]{0.16}}%
\put(450,1740){\makebox(0,0)[r]{0.14}}%
\put(450,1580){\makebox(0,0)[r]{0.12}}%
\put(450,1420){\makebox(0,0)[r]{0.1}}%
\put(450,1260){\makebox(0,0)[r]{0.08}}%
\put(450,1100){\makebox(0,0)[r]{0.06}}%
\put(450,940){\makebox(0,0)[r]{0.04}}%
\put(450,780){\makebox(0,0)[r]{0.02}}%
\put(450,620){\makebox(0,0)[r]{0}}%
\put(450,460){\makebox(0,0)[r]{-0.02}}%
\put(450,300){\makebox(0,0)[r]{-0.04}}%
\end{picture}%
\endgroup

\end	{center}
\vskip 0.15in
\caption{The values in the complex plane taken by the Polyakov 
loop for $x_3\in [0,L-1]$ for a sample of  individual $k=1$ walls 
in SU(4) with $aT=0.25$ and $T\simeq 1.02 T_c$. Solid line 
is the average.} 
\label{fig_profc4}
\end 	{figure}

\begin	{figure}[p]
\begin	{center}
\leavevmode
\begingroup%
  \makeatletter%
  \newcommand{\GNUPLOTspecial}{%
    \@sanitize\catcode`\%=14\relax\special}%
  \setlength{\unitlength}{0.1bp}%
{\GNUPLOTspecial{!
/gnudict 256 dict def
gnudict begin
/Color false def
/Solid false def
/gnulinewidth 5.000 def
/userlinewidth gnulinewidth def
/vshift -33 def
/dl {10 mul} def
/hpt_ 31.5 def
/vpt_ 31.5 def
/hpt hpt_ def
/vpt vpt_ def
/M {moveto} bind def
/L {lineto} bind def
/R {rmoveto} bind def
/V {rlineto} bind def
/vpt2 vpt 2 mul def
/hpt2 hpt 2 mul def
/Lshow { currentpoint stroke M
  0 vshift R show } def
/Rshow { currentpoint stroke M
  dup stringwidth pop neg vshift R show } def
/Cshow { currentpoint stroke M
  dup stringwidth pop -2 div vshift R show } def
/UP { dup vpt_ mul /vpt exch def hpt_ mul /hpt exch def
  /hpt2 hpt 2 mul def /vpt2 vpt 2 mul def } def
/DL { Color {setrgbcolor Solid {pop []} if 0 setdash }
 {pop pop pop Solid {pop []} if 0 setdash} ifelse } def
/BL { stroke userlinewidth 2 mul setlinewidth } def
/AL { stroke userlinewidth 2 div setlinewidth } def
/UL { dup gnulinewidth mul /userlinewidth exch def
      10 mul /udl exch def } def
/PL { stroke userlinewidth setlinewidth } def
/LTb { BL [] 0 0 0 DL } def
/LTa { AL [1 udl mul 2 udl mul] 0 setdash 0 0 0 setrgbcolor } def
/LT0 { PL [] 1 0 0 DL } def
/LT1 { PL [4 dl 2 dl] 0 1 0 DL } def
/LT2 { PL [2 dl 3 dl] 0 0 1 DL } def
/LT3 { PL [1 dl 1.5 dl] 1 0 1 DL } def
/LT4 { PL [5 dl 2 dl 1 dl 2 dl] 0 1 1 DL } def
/LT5 { PL [4 dl 3 dl 1 dl 3 dl] 1 1 0 DL } def
/LT6 { PL [2 dl 2 dl 2 dl 4 dl] 0 0 0 DL } def
/LT7 { PL [2 dl 2 dl 2 dl 2 dl 2 dl 4 dl] 1 0.3 0 DL } def
/LT8 { PL [2 dl 2 dl 2 dl 2 dl 2 dl 2 dl 2 dl 4 dl] 0.5 0.5 0.5 DL } def
/Pnt { stroke [] 0 setdash
   gsave 1 setlinecap M 0 0 V stroke grestore } def
/Dia { stroke [] 0 setdash 2 copy vpt add M
  hpt neg vpt neg V hpt vpt neg V
  hpt vpt V hpt neg vpt V closepath stroke
  Pnt } def
/Pls { stroke [] 0 setdash vpt sub M 0 vpt2 V
  currentpoint stroke M
  hpt neg vpt neg R hpt2 0 V stroke
  } def
/Box { stroke [] 0 setdash 2 copy exch hpt sub exch vpt add M
  0 vpt2 neg V hpt2 0 V 0 vpt2 V
  hpt2 neg 0 V closepath stroke
  Pnt } def
/Crs { stroke [] 0 setdash exch hpt sub exch vpt add M
  hpt2 vpt2 neg V currentpoint stroke M
  hpt2 neg 0 R hpt2 vpt2 V stroke } def
/TriU { stroke [] 0 setdash 2 copy vpt 1.12 mul add M
  hpt neg vpt -1.62 mul V
  hpt 2 mul 0 V
  hpt neg vpt 1.62 mul V closepath stroke
  Pnt  } def
/Star { 2 copy Pls Crs } def
/BoxF { stroke [] 0 setdash exch hpt sub exch vpt add M
  0 vpt2 neg V  hpt2 0 V  0 vpt2 V
  hpt2 neg 0 V  closepath fill } def
/TriUF { stroke [] 0 setdash vpt 1.12 mul add M
  hpt neg vpt -1.62 mul V
  hpt 2 mul 0 V
  hpt neg vpt 1.62 mul V closepath fill } def
/TriD { stroke [] 0 setdash 2 copy vpt 1.12 mul sub M
  hpt neg vpt 1.62 mul V
  hpt 2 mul 0 V
  hpt neg vpt -1.62 mul V closepath stroke
  Pnt  } def
/TriDF { stroke [] 0 setdash vpt 1.12 mul sub M
  hpt neg vpt 1.62 mul V
  hpt 2 mul 0 V
  hpt neg vpt -1.62 mul V closepath fill} def
/DiaF { stroke [] 0 setdash vpt add M
  hpt neg vpt neg V hpt vpt neg V
  hpt vpt V hpt neg vpt V closepath fill } def
/Pent { stroke [] 0 setdash 2 copy gsave
  translate 0 hpt M 4 {72 rotate 0 hpt L} repeat
  closepath stroke grestore Pnt } def
/PentF { stroke [] 0 setdash gsave
  translate 0 hpt M 4 {72 rotate 0 hpt L} repeat
  closepath fill grestore } def
/Circle { stroke [] 0 setdash 2 copy
  hpt 0 360 arc stroke Pnt } def
/CircleF { stroke [] 0 setdash hpt 0 360 arc fill } def
/C0 { BL [] 0 setdash 2 copy moveto vpt 90 450  arc } bind def
/C1 { BL [] 0 setdash 2 copy        moveto
       2 copy  vpt 0 90 arc closepath fill
               vpt 0 360 arc closepath } bind def
/C2 { BL [] 0 setdash 2 copy moveto
       2 copy  vpt 90 180 arc closepath fill
               vpt 0 360 arc closepath } bind def
/C3 { BL [] 0 setdash 2 copy moveto
       2 copy  vpt 0 180 arc closepath fill
               vpt 0 360 arc closepath } bind def
/C4 { BL [] 0 setdash 2 copy moveto
       2 copy  vpt 180 270 arc closepath fill
               vpt 0 360 arc closepath } bind def
/C5 { BL [] 0 setdash 2 copy moveto
       2 copy  vpt 0 90 arc
       2 copy moveto
       2 copy  vpt 180 270 arc closepath fill
               vpt 0 360 arc } bind def
/C6 { BL [] 0 setdash 2 copy moveto
      2 copy  vpt 90 270 arc closepath fill
              vpt 0 360 arc closepath } bind def
/C7 { BL [] 0 setdash 2 copy moveto
      2 copy  vpt 0 270 arc closepath fill
              vpt 0 360 arc closepath } bind def
/C8 { BL [] 0 setdash 2 copy moveto
      2 copy vpt 270 360 arc closepath fill
              vpt 0 360 arc closepath } bind def
/C9 { BL [] 0 setdash 2 copy moveto
      2 copy  vpt 270 450 arc closepath fill
              vpt 0 360 arc closepath } bind def
/C10 { BL [] 0 setdash 2 copy 2 copy moveto vpt 270 360 arc closepath fill
       2 copy moveto
       2 copy vpt 90 180 arc closepath fill
               vpt 0 360 arc closepath } bind def
/C11 { BL [] 0 setdash 2 copy moveto
       2 copy  vpt 0 180 arc closepath fill
       2 copy moveto
       2 copy  vpt 270 360 arc closepath fill
               vpt 0 360 arc closepath } bind def
/C12 { BL [] 0 setdash 2 copy moveto
       2 copy  vpt 180 360 arc closepath fill
               vpt 0 360 arc closepath } bind def
/C13 { BL [] 0 setdash  2 copy moveto
       2 copy  vpt 0 90 arc closepath fill
       2 copy moveto
       2 copy  vpt 180 360 arc closepath fill
               vpt 0 360 arc closepath } bind def
/C14 { BL [] 0 setdash 2 copy moveto
       2 copy  vpt 90 360 arc closepath fill
               vpt 0 360 arc } bind def
/C15 { BL [] 0 setdash 2 copy vpt 0 360 arc closepath fill
               vpt 0 360 arc closepath } bind def
/Rec   { newpath 4 2 roll moveto 1 index 0 rlineto 0 exch rlineto
       neg 0 rlineto closepath } bind def
/Square { dup Rec } bind def
/Bsquare { vpt sub exch vpt sub exch vpt2 Square } bind def
/S0 { BL [] 0 setdash 2 copy moveto 0 vpt rlineto BL Bsquare } bind def
/S1 { BL [] 0 setdash 2 copy vpt Square fill Bsquare } bind def
/S2 { BL [] 0 setdash 2 copy exch vpt sub exch vpt Square fill Bsquare } bind def
/S3 { BL [] 0 setdash 2 copy exch vpt sub exch vpt2 vpt Rec fill Bsquare } bind def
/S4 { BL [] 0 setdash 2 copy exch vpt sub exch vpt sub vpt Square fill Bsquare } bind def
/S5 { BL [] 0 setdash 2 copy 2 copy vpt Square fill
       exch vpt sub exch vpt sub vpt Square fill Bsquare } bind def
/S6 { BL [] 0 setdash 2 copy exch vpt sub exch vpt sub vpt vpt2 Rec fill Bsquare } bind def
/S7 { BL [] 0 setdash 2 copy exch vpt sub exch vpt sub vpt vpt2 Rec fill
       2 copy vpt Square fill
       Bsquare } bind def
/S8 { BL [] 0 setdash 2 copy vpt sub vpt Square fill Bsquare } bind def
/S9 { BL [] 0 setdash 2 copy vpt sub vpt vpt2 Rec fill Bsquare } bind def
/S10 { BL [] 0 setdash 2 copy vpt sub vpt Square fill 2 copy exch vpt sub exch vpt Square fill
       Bsquare } bind def
/S11 { BL [] 0 setdash 2 copy vpt sub vpt Square fill 2 copy exch vpt sub exch vpt2 vpt Rec fill
       Bsquare } bind def
/S12 { BL [] 0 setdash 2 copy exch vpt sub exch vpt sub vpt2 vpt Rec fill Bsquare } bind def
/S13 { BL [] 0 setdash 2 copy exch vpt sub exch vpt sub vpt2 vpt Rec fill
       2 copy vpt Square fill Bsquare } bind def
/S14 { BL [] 0 setdash 2 copy exch vpt sub exch vpt sub vpt2 vpt Rec fill
       2 copy exch vpt sub exch vpt Square fill Bsquare } bind def
/S15 { BL [] 0 setdash 2 copy Bsquare fill Bsquare } bind def
/D0 { gsave translate 45 rotate 0 0 S0 stroke grestore } bind def
/D1 { gsave translate 45 rotate 0 0 S1 stroke grestore } bind def
/D2 { gsave translate 45 rotate 0 0 S2 stroke grestore } bind def
/D3 { gsave translate 45 rotate 0 0 S3 stroke grestore } bind def
/D4 { gsave translate 45 rotate 0 0 S4 stroke grestore } bind def
/D5 { gsave translate 45 rotate 0 0 S5 stroke grestore } bind def
/D6 { gsave translate 45 rotate 0 0 S6 stroke grestore } bind def
/D7 { gsave translate 45 rotate 0 0 S7 stroke grestore } bind def
/D8 { gsave translate 45 rotate 0 0 S8 stroke grestore } bind def
/D9 { gsave translate 45 rotate 0 0 S9 stroke grestore } bind def
/D10 { gsave translate 45 rotate 0 0 S10 stroke grestore } bind def
/D11 { gsave translate 45 rotate 0 0 S11 stroke grestore } bind def
/D12 { gsave translate 45 rotate 0 0 S12 stroke grestore } bind def
/D13 { gsave translate 45 rotate 0 0 S13 stroke grestore } bind def
/D14 { gsave translate 45 rotate 0 0 S14 stroke grestore } bind def
/D15 { gsave translate 45 rotate 0 0 S15 stroke grestore } bind def
/DiaE { stroke [] 0 setdash vpt add M
  hpt neg vpt neg V hpt vpt neg V
  hpt vpt V hpt neg vpt V closepath stroke } def
/BoxE { stroke [] 0 setdash exch hpt sub exch vpt add M
  0 vpt2 neg V hpt2 0 V 0 vpt2 V
  hpt2 neg 0 V closepath stroke } def
/TriUE { stroke [] 0 setdash vpt 1.12 mul add M
  hpt neg vpt -1.62 mul V
  hpt 2 mul 0 V
  hpt neg vpt 1.62 mul V closepath stroke } def
/TriDE { stroke [] 0 setdash vpt 1.12 mul sub M
  hpt neg vpt 1.62 mul V
  hpt 2 mul 0 V
  hpt neg vpt -1.62 mul V closepath stroke } def
/PentE { stroke [] 0 setdash gsave
  translate 0 hpt M 4 {72 rotate 0 hpt L} repeat
  closepath stroke grestore } def
/CircE { stroke [] 0 setdash 
  hpt 0 360 arc stroke } def
/Opaque { gsave closepath 1 setgray fill grestore 0 setgray closepath } def
/DiaW { stroke [] 0 setdash vpt add M
  hpt neg vpt neg V hpt vpt neg V
  hpt vpt V hpt neg vpt V Opaque stroke } def
/BoxW { stroke [] 0 setdash exch hpt sub exch vpt add M
  0 vpt2 neg V hpt2 0 V 0 vpt2 V
  hpt2 neg 0 V Opaque stroke } def
/TriUW { stroke [] 0 setdash vpt 1.12 mul add M
  hpt neg vpt -1.62 mul V
  hpt 2 mul 0 V
  hpt neg vpt 1.62 mul V Opaque stroke } def
/TriDW { stroke [] 0 setdash vpt 1.12 mul sub M
  hpt neg vpt 1.62 mul V
  hpt 2 mul 0 V
  hpt neg vpt -1.62 mul V Opaque stroke } def
/PentW { stroke [] 0 setdash gsave
  translate 0 hpt M 4 {72 rotate 0 hpt L} repeat
  Opaque stroke grestore } def
/CircW { stroke [] 0 setdash 
  hpt 0 360 arc Opaque stroke } def
/BoxFill { gsave Rec 1 setgray fill grestore } def
end
}}%
\begin{picture}(3600,2160)(0,0)%
{\GNUPLOTspecial{"
gnudict begin
gsave
0 0 translate
0.100 0.100 scale
0 setgray
newpath
1.000 UL
LTb
450 301 M
63 0 V
2937 0 R
-63 0 V
450 653 M
63 0 V
2937 0 R
-63 0 V
450 1004 M
63 0 V
2937 0 R
-63 0 V
450 1356 M
63 0 V
2937 0 R
-63 0 V
450 1708 M
63 0 V
2937 0 R
-63 0 V
450 2060 M
63 0 V
2937 0 R
-63 0 V
450 300 M
0 63 V
0 1697 R
0 -63 V
950 300 M
0 63 V
0 1697 R
0 -63 V
1450 300 M
0 63 V
0 1697 R
0 -63 V
1950 300 M
0 63 V
0 1697 R
0 -63 V
2450 300 M
0 63 V
0 1697 R
0 -63 V
2950 300 M
0 63 V
0 1697 R
0 -63 V
3450 300 M
0 63 V
0 1697 R
0 -63 V
1.000 UL
LTb
450 300 M
3000 0 V
0 1760 V
-3000 0 V
450 300 L
1.000 UP
1.000 UL
LT0
450 1951 M
100 -1 V
100 0 V
100 -1 V
100 0 V
100 -1 V
100 -3 V
100 -7 V
100 -13 V
100 -28 V
100 -54 V
100 -98 V
100 -171 V
100 -252 V
100 -306 V
1950 704 L
2050 498 L
2150 381 L
100 -52 V
100 -19 V
100 -7 V
100 -2 V
100 0 V
100 0 V
100 0 V
100 0 V
100 0 V
100 0 V
100 0 V
100 0 V
450 1951 Pls
550 1950 Pls
650 1950 Pls
750 1949 Pls
850 1949 Pls
950 1948 Pls
1050 1945 Pls
1150 1938 Pls
1250 1925 Pls
1350 1897 Pls
1450 1843 Pls
1550 1745 Pls
1650 1574 Pls
1750 1322 Pls
1850 1016 Pls
1950 704 Pls
2050 498 Pls
2150 381 Pls
2250 329 Pls
2350 310 Pls
2450 303 Pls
2550 301 Pls
2650 301 Pls
2750 301 Pls
2850 301 Pls
2950 301 Pls
3050 301 Pls
3150 301 Pls
3250 301 Pls
3350 301 Pls
1.000 UL
LT1
450 1947 M
29 -1 V
30 0 V
29 -1 V
29 -1 V
29 -1 V
30 -1 V
29 -2 V
29 -1 V
30 -2 V
29 -2 V
29 -3 V
30 -2 V
29 -3 V
29 -4 V
29 -4 V
30 -5 V
29 -5 V
29 -6 V
30 -7 V
29 -7 V
29 -9 V
29 -10 V
30 -11 V
29 -12 V
29 -14 V
30 -15 V
29 -17 V
29 -19 V
29 -21 V
30 -23 V
29 -25 V
29 -28 V
30 -29 V
29 -33 V
29 -35 V
30 -37 V
29 -40 V
29 -42 V
29 -45 V
30 -46 V
29 -49 V
29 -50 V
30 -52 V
29 -52 V
29 -53 V
29 -54 V
30 -53 V
29 -53 V
29 -52 V
30 -51 V
29 -49 V
29 -47 V
30 -45 V
29 -43 V
29 -39 V
29 -38 V
30 -34 V
29 -31 V
29 -29 V
30 -26 V
29 -23 V
29 -21 V
29 -19 V
30 -16 V
29 -14 V
29 -13 V
30 -11 V
29 -9 V
29 -8 V
30 -7 V
29 -6 V
29 -5 V
29 -4 V
30 -4 V
29 -3 V
29 -2 V
30 -2 V
29 -2 V
29 -2 V
29 -1 V
30 -1 V
29 -1 V
29 0 V
30 -1 V
29 0 V
29 -1 V
29 0 V
30 0 V
29 0 V
29 -1 V
30 0 V
29 0 V
29 0 V
30 0 V
29 0 V
29 0 V
29 0 V
30 0 V
29 0 V
1.000 UL
LT2
450 1950 M
29 0 V
30 0 V
29 0 V
29 0 V
29 0 V
30 -1 V
29 0 V
29 0 V
30 -1 V
29 0 V
29 -1 V
30 -1 V
29 -1 V
29 -1 V
29 -1 V
30 -2 V
29 -2 V
29 -2 V
30 -3 V
29 -3 V
29 -4 V
29 -5 V
30 -5 V
29 -7 V
29 -7 V
30 -9 V
29 -10 V
29 -13 V
29 -14 V
30 -16 V
29 -19 V
29 -21 V
30 -25 V
29 -28 V
29 -32 V
30 -35 V
29 -39 V
29 -44 V
29 -47 V
30 -52 V
29 -56 V
29 -59 V
30 -62 V
29 -65 V
29 -67 V
29 -69 V
30 -68 V
29 -69 V
29 -67 V
30 -65 V
29 -63 V
29 -58 V
30 -55 V
29 -51 V
29 -46 V
29 -42 V
30 -37 V
29 -32 V
29 -28 V
30 -25 V
29 -20 V
29 -18 V
29 -14 V
30 -12 V
29 -10 V
29 -8 V
30 -7 V
29 -5 V
29 -5 V
30 -3 V
29 -3 V
29 -2 V
29 -1 V
30 -2 V
29 -1 V
29 -1 V
30 0 V
29 -1 V
29 0 V
29 0 V
30 -1 V
29 0 V
29 0 V
30 0 V
29 0 V
29 0 V
29 0 V
30 0 V
29 0 V
29 0 V
30 0 V
29 0 V
29 0 V
30 0 V
29 0 V
29 0 V
29 0 V
30 0 V
29 0 V
1.000 UL
LT3
450 1951 M
29 0 V
30 0 V
29 0 V
29 0 V
29 0 V
30 0 V
29 0 V
29 0 V
30 0 V
29 0 V
29 0 V
30 0 V
29 0 V
29 0 V
29 -1 V
30 0 V
29 0 V
29 0 V
30 0 V
29 -1 V
29 0 V
29 -1 V
30 -1 V
29 -1 V
29 -2 V
30 -2 V
29 -2 V
29 -4 V
29 -4 V
30 -5 V
29 -7 V
29 -9 V
30 -12 V
29 -14 V
29 -18 V
30 -23 V
29 -28 V
29 -33 V
29 -41 V
30 -48 V
29 -57 V
29 -65 V
30 -74 V
29 -83 V
29 -89 V
29 -96 V
30 -99 V
29 -100 V
29 -99 V
30 -95 V
29 -88 V
29 -81 V
30 -72 V
29 -61 V
29 -52 V
29 -43 V
30 -34 V
29 -27 V
29 -21 V
30 -16 V
29 -11 V
29 -9 V
29 -6 V
30 -5 V
29 -3 V
29 -2 V
30 -2 V
29 -1 V
29 -1 V
30 0 V
29 -1 V
29 0 V
29 0 V
30 0 V
29 0 V
29 0 V
30 0 V
29 0 V
29 0 V
29 0 V
30 0 V
29 0 V
29 0 V
30 0 V
29 0 V
29 0 V
29 0 V
30 0 V
29 0 V
29 0 V
30 0 V
29 0 V
29 0 V
30 0 V
29 0 V
29 0 V
29 0 V
30 0 V
29 0 V
1.000 UL
LT4
450 1951 M
29 0 V
30 0 V
29 0 V
29 0 V
29 0 V
30 0 V
29 0 V
29 0 V
30 0 V
29 0 V
29 0 V
30 -1 V
29 0 V
29 0 V
29 0 V
30 0 V
29 -1 V
29 0 V
30 -1 V
29 -1 V
29 -1 V
29 -1 V
30 -2 V
29 -2 V
29 -3 V
30 -3 V
29 -4 V
29 -5 V
29 -7 V
30 -8 V
29 -10 V
29 -12 V
30 -15 V
29 -19 V
29 -22 V
30 -27 V
29 -32 V
29 -38 V
29 -45 V
30 -51 V
29 -58 V
29 -65 V
30 -72 V
29 -78 V
29 -83 V
29 -88 V
30 -90 V
29 -90 V
29 -89 V
30 -85 V
29 -81 V
29 -74 V
30 -67 V
29 -59 V
29 -51 V
29 -43 V
30 -36 V
29 -30 V
29 -23 V
30 -19 V
29 -14 V
29 -12 V
29 -8 V
30 -7 V
29 -5 V
29 -3 V
30 -3 V
29 -2 V
29 -1 V
30 -1 V
29 -1 V
29 0 V
29 0 V
30 -1 V
29 0 V
29 0 V
30 0 V
29 0 V
29 0 V
29 0 V
30 0 V
29 0 V
29 0 V
30 0 V
29 0 V
29 0 V
29 0 V
30 0 V
29 0 V
29 0 V
30 0 V
29 0 V
29 0 V
30 0 V
29 0 V
29 0 V
29 0 V
30 0 V
29 0 V
stroke
grestore
end
showpage
}}%
\put(1950,50){\makebox(0,0){z}}%
\put(100,1180){%
\makebox(0,0)[b]{\shortstack{$\mathrm{Re}\, \overline{l_{p}}$}}%
}%
\put(3450,200){\makebox(0,0){30}}%
\put(2950,200){\makebox(0,0){25}}%
\put(2450,200){\makebox(0,0){20}}%
\put(1950,200){\makebox(0,0){15}}%
\put(1450,200){\makebox(0,0){10}}%
\put(950,200){\makebox(0,0){5}}%
\put(450,200){\makebox(0,0){0}}%
\put(400,2060){\makebox(0,0)[r]{0.25}}%
\put(400,1708){\makebox(0,0)[r]{0.2}}%
\put(400,1356){\makebox(0,0)[r]{0.15}}%
\put(400,1004){\makebox(0,0)[r]{0.1}}%
\put(400,653){\makebox(0,0)[r]{0.05}}%
\put(400,301){\makebox(0,0)[r]{0}}%
\end{picture}%
\endgroup

\end	{center}
\vskip 0.15in
\caption{Solid line: the average 
of the real part of the Polyakov loop as a function of $z$
for the $L_0=4$ SU(4) calculation at $T\simeq 1.88T_c$.
One loop perturbative predictions using for $g^2(T)$ the
lattice bare coupling (long dash), the mean field improved
lattice coupling (short dash), the Schrodinger functional coupling
(dot) and just a best fit (dash-dot).} 
\label{fig_profz1}
\end 	{figure}
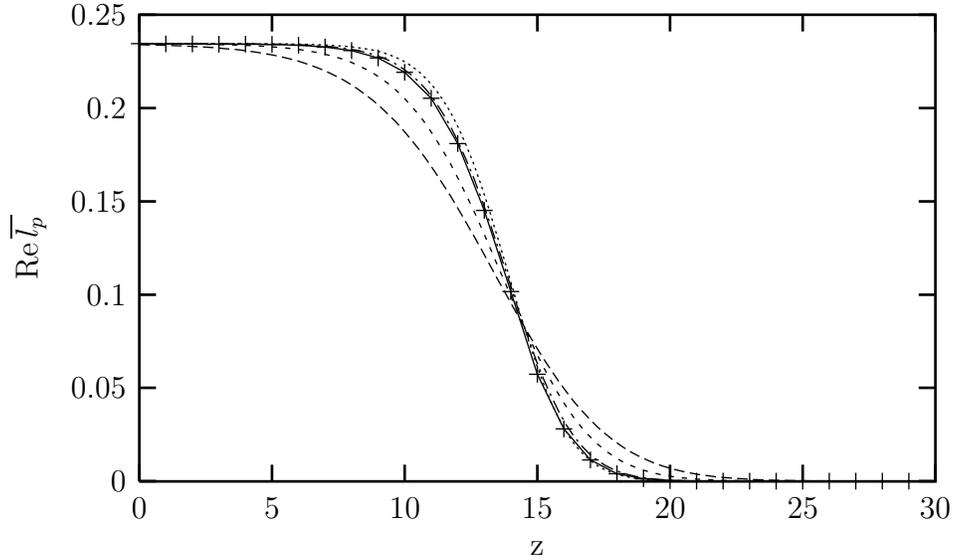

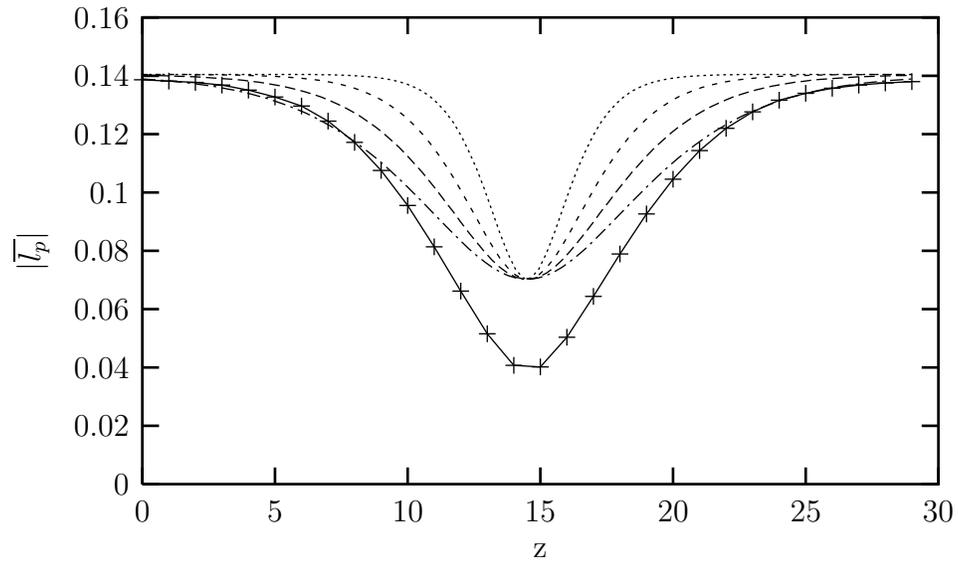
\begin	{figure}[p]
\begin	{center}
\leavevmode
\begingroup%
  \makeatletter%
  \newcommand{\GNUPLOTspecial}{%
    \@sanitize\catcode`\%=14\relax\special}%
  \setlength{\unitlength}{0.1bp}%
{\GNUPLOTspecial{!
/gnudict 256 dict def
gnudict begin
/Color false def
/Solid false def
/gnulinewidth 5.000 def
/userlinewidth gnulinewidth def
/vshift -33 def
/dl {10 mul} def
/hpt_ 31.5 def
/vpt_ 31.5 def
/hpt hpt_ def
/vpt vpt_ def
/M {moveto} bind def
/L {lineto} bind def
/R {rmoveto} bind def
/V {rlineto} bind def
/vpt2 vpt 2 mul def
/hpt2 hpt 2 mul def
/Lshow { currentpoint stroke M
  0 vshift R show } def
/Rshow { currentpoint stroke M
  dup stringwidth pop neg vshift R show } def
/Cshow { currentpoint stroke M
  dup stringwidth pop -2 div vshift R show } def
/UP { dup vpt_ mul /vpt exch def hpt_ mul /hpt exch def
  /hpt2 hpt 2 mul def /vpt2 vpt 2 mul def } def
/DL { Color {setrgbcolor Solid {pop []} if 0 setdash }
 {pop pop pop Solid {pop []} if 0 setdash} ifelse } def
/BL { stroke userlinewidth 2 mul setlinewidth } def
/AL { stroke userlinewidth 2 div setlinewidth } def
/UL { dup gnulinewidth mul /userlinewidth exch def
      10 mul /udl exch def } def
/PL { stroke userlinewidth setlinewidth } def
/LTb { BL [] 0 0 0 DL } def
/LTa { AL [1 udl mul 2 udl mul] 0 setdash 0 0 0 setrgbcolor } def
/LT0 { PL [] 1 0 0 DL } def
/LT1 { PL [4 dl 2 dl] 0 1 0 DL } def
/LT2 { PL [2 dl 3 dl] 0 0 1 DL } def
/LT3 { PL [1 dl 1.5 dl] 1 0 1 DL } def
/LT4 { PL [5 dl 2 dl 1 dl 2 dl] 0 1 1 DL } def
/LT5 { PL [4 dl 3 dl 1 dl 3 dl] 1 1 0 DL } def
/LT6 { PL [2 dl 2 dl 2 dl 4 dl] 0 0 0 DL } def
/LT7 { PL [2 dl 2 dl 2 dl 2 dl 2 dl 4 dl] 1 0.3 0 DL } def
/LT8 { PL [2 dl 2 dl 2 dl 2 dl 2 dl 2 dl 2 dl 4 dl] 0.5 0.5 0.5 DL } def
/Pnt { stroke [] 0 setdash
   gsave 1 setlinecap M 0 0 V stroke grestore } def
/Dia { stroke [] 0 setdash 2 copy vpt add M
  hpt neg vpt neg V hpt vpt neg V
  hpt vpt V hpt neg vpt V closepath stroke
  Pnt } def
/Pls { stroke [] 0 setdash vpt sub M 0 vpt2 V
  currentpoint stroke M
  hpt neg vpt neg R hpt2 0 V stroke
  } def
/Box { stroke [] 0 setdash 2 copy exch hpt sub exch vpt add M
  0 vpt2 neg V hpt2 0 V 0 vpt2 V
  hpt2 neg 0 V closepath stroke
  Pnt } def
/Crs { stroke [] 0 setdash exch hpt sub exch vpt add M
  hpt2 vpt2 neg V currentpoint stroke M
  hpt2 neg 0 R hpt2 vpt2 V stroke } def
/TriU { stroke [] 0 setdash 2 copy vpt 1.12 mul add M
  hpt neg vpt -1.62 mul V
  hpt 2 mul 0 V
  hpt neg vpt 1.62 mul V closepath stroke
  Pnt  } def
/Star { 2 copy Pls Crs } def
/BoxF { stroke [] 0 setdash exch hpt sub exch vpt add M
  0 vpt2 neg V  hpt2 0 V  0 vpt2 V
  hpt2 neg 0 V  closepath fill } def
/TriUF { stroke [] 0 setdash vpt 1.12 mul add M
  hpt neg vpt -1.62 mul V
  hpt 2 mul 0 V
  hpt neg vpt 1.62 mul V closepath fill } def
/TriD { stroke [] 0 setdash 2 copy vpt 1.12 mul sub M
  hpt neg vpt 1.62 mul V
  hpt 2 mul 0 V
  hpt neg vpt -1.62 mul V closepath stroke
  Pnt  } def
/TriDF { stroke [] 0 setdash vpt 1.12 mul sub M
  hpt neg vpt 1.62 mul V
  hpt 2 mul 0 V
  hpt neg vpt -1.62 mul V closepath fill} def
/DiaF { stroke [] 0 setdash vpt add M
  hpt neg vpt neg V hpt vpt neg V
  hpt vpt V hpt neg vpt V closepath fill } def
/Pent { stroke [] 0 setdash 2 copy gsave
  translate 0 hpt M 4 {72 rotate 0 hpt L} repeat
  closepath stroke grestore Pnt } def
/PentF { stroke [] 0 setdash gsave
  translate 0 hpt M 4 {72 rotate 0 hpt L} repeat
  closepath fill grestore } def
/Circle { stroke [] 0 setdash 2 copy
  hpt 0 360 arc stroke Pnt } def
/CircleF { stroke [] 0 setdash hpt 0 360 arc fill } def
/C0 { BL [] 0 setdash 2 copy moveto vpt 90 450  arc } bind def
/C1 { BL [] 0 setdash 2 copy        moveto
       2 copy  vpt 0 90 arc closepath fill
               vpt 0 360 arc closepath } bind def
/C2 { BL [] 0 setdash 2 copy moveto
       2 copy  vpt 90 180 arc closepath fill
               vpt 0 360 arc closepath } bind def
/C3 { BL [] 0 setdash 2 copy moveto
       2 copy  vpt 0 180 arc closepath fill
               vpt 0 360 arc closepath } bind def
/C4 { BL [] 0 setdash 2 copy moveto
       2 copy  vpt 180 270 arc closepath fill
               vpt 0 360 arc closepath } bind def
/C5 { BL [] 0 setdash 2 copy moveto
       2 copy  vpt 0 90 arc
       2 copy moveto
       2 copy  vpt 180 270 arc closepath fill
               vpt 0 360 arc } bind def
/C6 { BL [] 0 setdash 2 copy moveto
      2 copy  vpt 90 270 arc closepath fill
              vpt 0 360 arc closepath } bind def
/C7 { BL [] 0 setdash 2 copy moveto
      2 copy  vpt 0 270 arc closepath fill
              vpt 0 360 arc closepath } bind def
/C8 { BL [] 0 setdash 2 copy moveto
      2 copy vpt 270 360 arc closepath fill
              vpt 0 360 arc closepath } bind def
/C9 { BL [] 0 setdash 2 copy moveto
      2 copy  vpt 270 450 arc closepath fill
              vpt 0 360 arc closepath } bind def
/C10 { BL [] 0 setdash 2 copy 2 copy moveto vpt 270 360 arc closepath fill
       2 copy moveto
       2 copy vpt 90 180 arc closepath fill
               vpt 0 360 arc closepath } bind def
/C11 { BL [] 0 setdash 2 copy moveto
       2 copy  vpt 0 180 arc closepath fill
       2 copy moveto
       2 copy  vpt 270 360 arc closepath fill
               vpt 0 360 arc closepath } bind def
/C12 { BL [] 0 setdash 2 copy moveto
       2 copy  vpt 180 360 arc closepath fill
               vpt 0 360 arc closepath } bind def
/C13 { BL [] 0 setdash  2 copy moveto
       2 copy  vpt 0 90 arc closepath fill
       2 copy moveto
       2 copy  vpt 180 360 arc closepath fill
               vpt 0 360 arc closepath } bind def
/C14 { BL [] 0 setdash 2 copy moveto
       2 copy  vpt 90 360 arc closepath fill
               vpt 0 360 arc } bind def
/C15 { BL [] 0 setdash 2 copy vpt 0 360 arc closepath fill
               vpt 0 360 arc closepath } bind def
/Rec   { newpath 4 2 roll moveto 1 index 0 rlineto 0 exch rlineto
       neg 0 rlineto closepath } bind def
/Square { dup Rec } bind def
/Bsquare { vpt sub exch vpt sub exch vpt2 Square } bind def
/S0 { BL [] 0 setdash 2 copy moveto 0 vpt rlineto BL Bsquare } bind def
/S1 { BL [] 0 setdash 2 copy vpt Square fill Bsquare } bind def
/S2 { BL [] 0 setdash 2 copy exch vpt sub exch vpt Square fill Bsquare } bind def
/S3 { BL [] 0 setdash 2 copy exch vpt sub exch vpt2 vpt Rec fill Bsquare } bind def
/S4 { BL [] 0 setdash 2 copy exch vpt sub exch vpt sub vpt Square fill Bsquare } bind def
/S5 { BL [] 0 setdash 2 copy 2 copy vpt Square fill
       exch vpt sub exch vpt sub vpt Square fill Bsquare } bind def
/S6 { BL [] 0 setdash 2 copy exch vpt sub exch vpt sub vpt vpt2 Rec fill Bsquare } bind def
/S7 { BL [] 0 setdash 2 copy exch vpt sub exch vpt sub vpt vpt2 Rec fill
       2 copy vpt Square fill
       Bsquare } bind def
/S8 { BL [] 0 setdash 2 copy vpt sub vpt Square fill Bsquare } bind def
/S9 { BL [] 0 setdash 2 copy vpt sub vpt vpt2 Rec fill Bsquare } bind def
/S10 { BL [] 0 setdash 2 copy vpt sub vpt Square fill 2 copy exch vpt sub exch vpt Square fill
       Bsquare } bind def
/S11 { BL [] 0 setdash 2 copy vpt sub vpt Square fill 2 copy exch vpt sub exch vpt2 vpt Rec fill
       Bsquare } bind def
/S12 { BL [] 0 setdash 2 copy exch vpt sub exch vpt sub vpt2 vpt Rec fill Bsquare } bind def
/S13 { BL [] 0 setdash 2 copy exch vpt sub exch vpt sub vpt2 vpt Rec fill
       2 copy vpt Square fill Bsquare } bind def
/S14 { BL [] 0 setdash 2 copy exch vpt sub exch vpt sub vpt2 vpt Rec fill
       2 copy exch vpt sub exch vpt Square fill Bsquare } bind def
/S15 { BL [] 0 setdash 2 copy Bsquare fill Bsquare } bind def
/D0 { gsave translate 45 rotate 0 0 S0 stroke grestore } bind def
/D1 { gsave translate 45 rotate 0 0 S1 stroke grestore } bind def
/D2 { gsave translate 45 rotate 0 0 S2 stroke grestore } bind def
/D3 { gsave translate 45 rotate 0 0 S3 stroke grestore } bind def
/D4 { gsave translate 45 rotate 0 0 S4 stroke grestore } bind def
/D5 { gsave translate 45 rotate 0 0 S5 stroke grestore } bind def
/D6 { gsave translate 45 rotate 0 0 S6 stroke grestore } bind def
/D7 { gsave translate 45 rotate 0 0 S7 stroke grestore } bind def
/D8 { gsave translate 45 rotate 0 0 S8 stroke grestore } bind def
/D9 { gsave translate 45 rotate 0 0 S9 stroke grestore } bind def
/D10 { gsave translate 45 rotate 0 0 S10 stroke grestore } bind def
/D11 { gsave translate 45 rotate 0 0 S11 stroke grestore } bind def
/D12 { gsave translate 45 rotate 0 0 S12 stroke grestore } bind def
/D13 { gsave translate 45 rotate 0 0 S13 stroke grestore } bind def
/D14 { gsave translate 45 rotate 0 0 S14 stroke grestore } bind def
/D15 { gsave translate 45 rotate 0 0 S15 stroke grestore } bind def
/DiaE { stroke [] 0 setdash vpt add M
  hpt neg vpt neg V hpt vpt neg V
  hpt vpt V hpt neg vpt V closepath stroke } def
/BoxE { stroke [] 0 setdash exch hpt sub exch vpt add M
  0 vpt2 neg V hpt2 0 V 0 vpt2 V
  hpt2 neg 0 V closepath stroke } def
/TriUE { stroke [] 0 setdash vpt 1.12 mul add M
  hpt neg vpt -1.62 mul V
  hpt 2 mul 0 V
  hpt neg vpt 1.62 mul V closepath stroke } def
/TriDE { stroke [] 0 setdash vpt 1.12 mul sub M
  hpt neg vpt 1.62 mul V
  hpt 2 mul 0 V
  hpt neg vpt -1.62 mul V closepath stroke } def
/PentE { stroke [] 0 setdash gsave
  translate 0 hpt M 4 {72 rotate 0 hpt L} repeat
  closepath stroke grestore } def
/CircE { stroke [] 0 setdash 
  hpt 0 360 arc stroke } def
/Opaque { gsave closepath 1 setgray fill grestore 0 setgray closepath } def
/DiaW { stroke [] 0 setdash vpt add M
  hpt neg vpt neg V hpt vpt neg V
  hpt vpt V hpt neg vpt V Opaque stroke } def
/BoxW { stroke [] 0 setdash exch hpt sub exch vpt add M
  0 vpt2 neg V hpt2 0 V 0 vpt2 V
  hpt2 neg 0 V Opaque stroke } def
/TriUW { stroke [] 0 setdash vpt 1.12 mul add M
  hpt neg vpt -1.62 mul V
  hpt 2 mul 0 V
  hpt neg vpt 1.62 mul V Opaque stroke } def
/TriDW { stroke [] 0 setdash vpt 1.12 mul sub M
  hpt neg vpt 1.62 mul V
  hpt 2 mul 0 V
  hpt neg vpt -1.62 mul V Opaque stroke } def
/PentW { stroke [] 0 setdash gsave
  translate 0 hpt M 4 {72 rotate 0 hpt L} repeat
  Opaque stroke grestore } def
/CircW { stroke [] 0 setdash 
  hpt 0 360 arc Opaque stroke } def
/BoxFill { gsave Rec 1 setgray fill grestore } def
end
}}%
\begin{picture}(3600,2160)(0,0)%
{\GNUPLOTspecial{"
gnudict begin
gsave
0 0 translate
0.100 0.100 scale
0 setgray
newpath
1.000 UL
LTb
450 300 M
63 0 V
2937 0 R
-63 0 V
450 520 M
63 0 V
2937 0 R
-63 0 V
450 740 M
63 0 V
2937 0 R
-63 0 V
450 960 M
63 0 V
2937 0 R
-63 0 V
450 1180 M
63 0 V
2937 0 R
-63 0 V
450 1400 M
63 0 V
2937 0 R
-63 0 V
450 1620 M
63 0 V
2937 0 R
-63 0 V
450 1840 M
63 0 V
2937 0 R
-63 0 V
450 2060 M
63 0 V
2937 0 R
-63 0 V
450 300 M
0 63 V
0 1697 R
0 -63 V
950 300 M
0 63 V
0 1697 R
0 -63 V
1450 300 M
0 63 V
0 1697 R
0 -63 V
1950 300 M
0 63 V
0 1697 R
0 -63 V
2450 300 M
0 63 V
0 1697 R
0 -63 V
2950 300 M
0 63 V
0 1697 R
0 -63 V
3450 300 M
0 63 V
0 1697 R
0 -63 V
1.000 UL
LTb
450 300 M
3000 0 V
0 1760 V
-3000 0 V
450 300 L
1.000 UP
1.000 UL
LT0
450 1825 M
100 -5 V
100 -5 V
100 -10 V
100 -19 V
100 -26 V
100 -34 V
100 -57 V
100 -80 V
100 -106 V
100 -132 V
100 -156 V
100 -167 V
1750 867 L
1850 748 L
100 -6 V
100 112 V
100 154 V
100 160 V
100 151 V
100 131 V
100 108 V
100 84 V
100 62 V
100 44 V
100 26 V
100 19 V
100 12 V
100 8 V
100 5 V
450 1825 Pls
550 1820 Pls
650 1815 Pls
750 1805 Pls
850 1786 Pls
950 1760 Pls
1050 1726 Pls
1150 1669 Pls
1250 1589 Pls
1350 1483 Pls
1450 1351 Pls
1550 1195 Pls
1650 1028 Pls
1750 867 Pls
1850 748 Pls
1950 742 Pls
2050 854 Pls
2150 1008 Pls
2250 1168 Pls
2350 1319 Pls
2450 1450 Pls
2550 1558 Pls
2650 1642 Pls
2750 1704 Pls
2850 1748 Pls
2950 1774 Pls
3050 1793 Pls
3150 1805 Pls
3250 1813 Pls
3350 1818 Pls
1.000 UL
LT1
450 1842 M
29 -1 V
30 0 V
29 -1 V
29 -1 V
29 -1 V
30 -1 V
29 -1 V
29 -2 V
30 -2 V
29 -2 V
29 -2 V
30 -2 V
29 -3 V
29 -4 V
29 -3 V
30 -5 V
29 -5 V
29 -5 V
30 -7 V
29 -7 V
29 -8 V
29 -9 V
30 -10 V
29 -12 V
29 -13 V
30 -14 V
29 -16 V
29 -17 V
29 -20 V
30 -21 V
29 -23 V
29 -26 V
30 -27 V
29 -30 V
29 -32 V
30 -33 V
29 -36 V
29 -36 V
29 -39 V
30 -38 V
29 -40 V
29 -38 V
30 -37 V
29 -35 V
29 -32 V
29 -27 V
30 -22 V
29 -15 V
29 -7 V
30 0 V
29 7 V
29 15 V
30 22 V
29 27 V
29 32 V
29 35 V
30 37 V
29 38 V
29 40 V
30 38 V
29 39 V
29 36 V
29 36 V
30 33 V
29 32 V
29 30 V
30 27 V
29 26 V
29 23 V
30 21 V
29 20 V
29 17 V
29 16 V
30 14 V
29 13 V
29 12 V
30 10 V
29 9 V
29 8 V
29 7 V
30 7 V
29 5 V
29 5 V
30 5 V
29 3 V
29 4 V
29 3 V
30 2 V
29 2 V
29 2 V
30 2 V
29 2 V
29 1 V
30 1 V
29 1 V
29 1 V
29 1 V
30 0 V
29 1 V
1.000 UL
LT2
450 1845 M
29 0 V
30 0 V
29 0 V
29 0 V
29 0 V
30 0 V
29 -1 V
29 0 V
30 0 V
29 0 V
29 -1 V
30 0 V
29 -1 V
29 -1 V
29 -1 V
30 -1 V
29 -1 V
29 -2 V
30 -2 V
29 -2 V
29 -3 V
29 -3 V
30 -4 V
29 -5 V
29 -6 V
30 -7 V
29 -8 V
29 -10 V
29 -12 V
30 -13 V
29 -16 V
29 -19 V
30 -21 V
29 -24 V
29 -28 V
30 -32 V
29 -35 V
29 -39 V
29 -43 V
30 -47 V
29 -50 V
29 -52 V
30 -53 V
29 -53 V
29 -50 V
29 -46 V
30 -37 V
29 -27 V
29 -14 V
30 0 V
29 14 V
29 27 V
30 37 V
29 46 V
29 50 V
29 53 V
30 53 V
29 52 V
29 50 V
30 47 V
29 43 V
29 39 V
29 35 V
30 32 V
29 28 V
29 24 V
30 21 V
29 19 V
29 16 V
30 13 V
29 12 V
29 10 V
29 8 V
30 7 V
29 6 V
29 5 V
30 4 V
29 3 V
29 3 V
29 2 V
30 2 V
29 2 V
29 1 V
30 1 V
29 1 V
29 1 V
29 1 V
30 0 V
29 1 V
29 0 V
30 0 V
29 0 V
29 1 V
30 0 V
29 0 V
29 0 V
29 0 V
30 0 V
29 0 V
1.000 UL
LT3
450 1845 M
29 0 V
30 0 V
29 0 V
29 0 V
29 0 V
30 0 V
29 0 V
29 0 V
30 0 V
29 0 V
29 0 V
30 0 V
29 0 V
29 0 V
29 0 V
30 0 V
29 0 V
29 0 V
30 0 V
29 0 V
29 0 V
29 0 V
30 -1 V
29 0 V
29 -1 V
30 -1 V
29 -1 V
29 -2 V
29 -2 V
30 -3 V
29 -4 V
29 -5 V
30 -7 V
29 -9 V
29 -12 V
30 -15 V
29 -20 V
29 -26 V
29 -32 V
30 -40 V
29 -49 V
29 -59 V
30 -67 V
29 -76 V
29 -82 V
29 -83 V
30 -77 V
29 -60 V
29 -34 V
30 0 V
29 34 V
29 60 V
30 77 V
29 83 V
29 82 V
29 76 V
30 67 V
29 59 V
29 49 V
30 40 V
29 32 V
29 26 V
29 20 V
30 15 V
29 12 V
29 9 V
30 7 V
29 5 V
29 4 V
30 3 V
29 2 V
29 2 V
29 1 V
30 1 V
29 1 V
29 0 V
30 1 V
29 0 V
29 0 V
29 0 V
30 0 V
29 0 V
29 0 V
30 0 V
29 0 V
29 0 V
29 0 V
30 0 V
29 0 V
29 0 V
30 0 V
29 0 V
29 0 V
30 0 V
29 0 V
29 0 V
29 0 V
30 0 V
29 0 V
1.000 UL
LT4
450 1828 M
29 -2 V
30 -2 V
29 -2 V
29 -3 V
29 -3 V
30 -3 V
29 -4 V
29 -4 V
30 -4 V
29 -5 V
29 -6 V
30 -5 V
29 -7 V
29 -7 V
29 -8 V
30 -8 V
29 -10 V
29 -10 V
30 -11 V
29 -12 V
29 -12 V
29 -14 V
30 -16 V
29 -16 V
29 -17 V
30 -19 V
29 -20 V
29 -21 V
29 -22 V
30 -24 V
29 -25 V
29 -26 V
30 -28 V
29 -28 V
29 -29 V
30 -30 V
29 -30 V
29 -30 V
29 -30 V
30 -30 V
29 -29 V
29 -28 V
30 -26 V
29 -23 V
29 -21 V
29 -17 V
30 -14 V
29 -9 V
29 -5 V
30 0 V
29 5 V
29 9 V
30 14 V
29 17 V
29 21 V
29 23 V
30 26 V
29 28 V
29 29 V
30 30 V
29 30 V
29 30 V
29 30 V
30 30 V
29 29 V
29 28 V
30 28 V
29 26 V
29 25 V
30 24 V
29 22 V
29 21 V
29 20 V
30 19 V
29 17 V
29 16 V
30 16 V
29 14 V
29 12 V
29 12 V
30 11 V
29 10 V
29 10 V
30 8 V
29 8 V
29 7 V
29 7 V
30 5 V
29 6 V
29 5 V
30 4 V
29 4 V
29 4 V
30 3 V
29 3 V
29 3 V
29 2 V
30 2 V
29 2 V
stroke
grestore
end
showpage
}}%
\put(1950,50){\makebox(0,0){z}}%
\put(100,1180){%
\makebox(0,0)[b]{\shortstack{$|\overline{l_{p}}|$}}%
}%
\put(3450,200){\makebox(0,0){30}}%
\put(2950,200){\makebox(0,0){25}}%
\put(2450,200){\makebox(0,0){20}}%
\put(1950,200){\makebox(0,0){15}}%
\put(1450,200){\makebox(0,0){10}}%
\put(950,200){\makebox(0,0){5}}%
\put(450,200){\makebox(0,0){0}}%
\put(400,2060){\makebox(0,0)[r]{0.16}}%
\put(400,1840){\makebox(0,0)[r]{0.14}}%
\put(400,1620){\makebox(0,0)[r]{0.12}}%
\put(400,1400){\makebox(0,0)[r]{0.1}}%
\put(400,1180){\makebox(0,0)[r]{0.08}}%
\put(400,960){\makebox(0,0)[r]{0.06}}%
\put(400,740){\makebox(0,0)[r]{0.04}}%
\put(400,520){\makebox(0,0)[r]{0.02}}%
\put(400,300){\makebox(0,0)[r]{0}}%
\end{picture}%
\endgroup

\end	{center}
\vskip 0.15in
\caption{ The average 
of the modulus of the Polyakov loop as a function of $z$
for the $L_0=4$ SU(4) calculation at $T\simeq 1.02T_c$.
One loop perturbative predictions using for $g^2(T)$ the
lattice bare coupling (long dash), the mean field improved
lattice coupling (short dash), the Schrodinger functional coupling
(dot) and just a best fit (dash-dot).} 
\label{fig_profz2}
\end 	{figure}

\clearpage

\begin	{figure}[p]
\begin	{center}
\leavevmode
\begingroup%
  \makeatletter%
  \newcommand{\GNUPLOTspecial}{%
    \@sanitize\catcode`\%=14\relax\special}%
  \setlength{\unitlength}{0.1bp}%
{\GNUPLOTspecial{!
/gnudict 256 dict def
gnudict begin
/Color false def
/Solid false def
/gnulinewidth 5.000 def
/userlinewidth gnulinewidth def
/vshift -33 def
/dl {10 mul} def
/hpt_ 31.5 def
/vpt_ 31.5 def
/hpt hpt_ def
/vpt vpt_ def
/M {moveto} bind def
/L {lineto} bind def
/R {rmoveto} bind def
/V {rlineto} bind def
/vpt2 vpt 2 mul def
/hpt2 hpt 2 mul def
/Lshow { currentpoint stroke M
  0 vshift R show } def
/Rshow { currentpoint stroke M
  dup stringwidth pop neg vshift R show } def
/Cshow { currentpoint stroke M
  dup stringwidth pop -2 div vshift R show } def
/UP { dup vpt_ mul /vpt exch def hpt_ mul /hpt exch def
  /hpt2 hpt 2 mul def /vpt2 vpt 2 mul def } def
/DL { Color {setrgbcolor Solid {pop []} if 0 setdash }
 {pop pop pop Solid {pop []} if 0 setdash} ifelse } def
/BL { stroke userlinewidth 2 mul setlinewidth } def
/AL { stroke userlinewidth 2 div setlinewidth } def
/UL { dup gnulinewidth mul /userlinewidth exch def
      10 mul /udl exch def } def
/PL { stroke userlinewidth setlinewidth } def
/LTb { BL [] 0 0 0 DL } def
/LTa { AL [1 udl mul 2 udl mul] 0 setdash 0 0 0 setrgbcolor } def
/LT0 { PL [] 1 0 0 DL } def
/LT1 { PL [4 dl 2 dl] 0 1 0 DL } def
/LT2 { PL [2 dl 3 dl] 0 0 1 DL } def
/LT3 { PL [1 dl 1.5 dl] 1 0 1 DL } def
/LT4 { PL [5 dl 2 dl 1 dl 2 dl] 0 1 1 DL } def
/LT5 { PL [4 dl 3 dl 1 dl 3 dl] 1 1 0 DL } def
/LT6 { PL [2 dl 2 dl 2 dl 4 dl] 0 0 0 DL } def
/LT7 { PL [2 dl 2 dl 2 dl 2 dl 2 dl 4 dl] 1 0.3 0 DL } def
/LT8 { PL [2 dl 2 dl 2 dl 2 dl 2 dl 2 dl 2 dl 4 dl] 0.5 0.5 0.5 DL } def
/Pnt { stroke [] 0 setdash
   gsave 1 setlinecap M 0 0 V stroke grestore } def
/Dia { stroke [] 0 setdash 2 copy vpt add M
  hpt neg vpt neg V hpt vpt neg V
  hpt vpt V hpt neg vpt V closepath stroke
  Pnt } def
/Pls { stroke [] 0 setdash vpt sub M 0 vpt2 V
  currentpoint stroke M
  hpt neg vpt neg R hpt2 0 V stroke
  } def
/Box { stroke [] 0 setdash 2 copy exch hpt sub exch vpt add M
  0 vpt2 neg V hpt2 0 V 0 vpt2 V
  hpt2 neg 0 V closepath stroke
  Pnt } def
/Crs { stroke [] 0 setdash exch hpt sub exch vpt add M
  hpt2 vpt2 neg V currentpoint stroke M
  hpt2 neg 0 R hpt2 vpt2 V stroke } def
/TriU { stroke [] 0 setdash 2 copy vpt 1.12 mul add M
  hpt neg vpt -1.62 mul V
  hpt 2 mul 0 V
  hpt neg vpt 1.62 mul V closepath stroke
  Pnt  } def
/Star { 2 copy Pls Crs } def
/BoxF { stroke [] 0 setdash exch hpt sub exch vpt add M
  0 vpt2 neg V  hpt2 0 V  0 vpt2 V
  hpt2 neg 0 V  closepath fill } def
/TriUF { stroke [] 0 setdash vpt 1.12 mul add M
  hpt neg vpt -1.62 mul V
  hpt 2 mul 0 V
  hpt neg vpt 1.62 mul V closepath fill } def
/TriD { stroke [] 0 setdash 2 copy vpt 1.12 mul sub M
  hpt neg vpt 1.62 mul V
  hpt 2 mul 0 V
  hpt neg vpt -1.62 mul V closepath stroke
  Pnt  } def
/TriDF { stroke [] 0 setdash vpt 1.12 mul sub M
  hpt neg vpt 1.62 mul V
  hpt 2 mul 0 V
  hpt neg vpt -1.62 mul V closepath fill} def
/DiaF { stroke [] 0 setdash vpt add M
  hpt neg vpt neg V hpt vpt neg V
  hpt vpt V hpt neg vpt V closepath fill } def
/Pent { stroke [] 0 setdash 2 copy gsave
  translate 0 hpt M 4 {72 rotate 0 hpt L} repeat
  closepath stroke grestore Pnt } def
/PentF { stroke [] 0 setdash gsave
  translate 0 hpt M 4 {72 rotate 0 hpt L} repeat
  closepath fill grestore } def
/Circle { stroke [] 0 setdash 2 copy
  hpt 0 360 arc stroke Pnt } def
/CircleF { stroke [] 0 setdash hpt 0 360 arc fill } def
/C0 { BL [] 0 setdash 2 copy moveto vpt 90 450  arc } bind def
/C1 { BL [] 0 setdash 2 copy        moveto
       2 copy  vpt 0 90 arc closepath fill
               vpt 0 360 arc closepath } bind def
/C2 { BL [] 0 setdash 2 copy moveto
       2 copy  vpt 90 180 arc closepath fill
               vpt 0 360 arc closepath } bind def
/C3 { BL [] 0 setdash 2 copy moveto
       2 copy  vpt 0 180 arc closepath fill
               vpt 0 360 arc closepath } bind def
/C4 { BL [] 0 setdash 2 copy moveto
       2 copy  vpt 180 270 arc closepath fill
               vpt 0 360 arc closepath } bind def
/C5 { BL [] 0 setdash 2 copy moveto
       2 copy  vpt 0 90 arc
       2 copy moveto
       2 copy  vpt 180 270 arc closepath fill
               vpt 0 360 arc } bind def
/C6 { BL [] 0 setdash 2 copy moveto
      2 copy  vpt 90 270 arc closepath fill
              vpt 0 360 arc closepath } bind def
/C7 { BL [] 0 setdash 2 copy moveto
      2 copy  vpt 0 270 arc closepath fill
              vpt 0 360 arc closepath } bind def
/C8 { BL [] 0 setdash 2 copy moveto
      2 copy vpt 270 360 arc closepath fill
              vpt 0 360 arc closepath } bind def
/C9 { BL [] 0 setdash 2 copy moveto
      2 copy  vpt 270 450 arc closepath fill
              vpt 0 360 arc closepath } bind def
/C10 { BL [] 0 setdash 2 copy 2 copy moveto vpt 270 360 arc closepath fill
       2 copy moveto
       2 copy vpt 90 180 arc closepath fill
               vpt 0 360 arc closepath } bind def
/C11 { BL [] 0 setdash 2 copy moveto
       2 copy  vpt 0 180 arc closepath fill
       2 copy moveto
       2 copy  vpt 270 360 arc closepath fill
               vpt 0 360 arc closepath } bind def
/C12 { BL [] 0 setdash 2 copy moveto
       2 copy  vpt 180 360 arc closepath fill
               vpt 0 360 arc closepath } bind def
/C13 { BL [] 0 setdash  2 copy moveto
       2 copy  vpt 0 90 arc closepath fill
       2 copy moveto
       2 copy  vpt 180 360 arc closepath fill
               vpt 0 360 arc closepath } bind def
/C14 { BL [] 0 setdash 2 copy moveto
       2 copy  vpt 90 360 arc closepath fill
               vpt 0 360 arc } bind def
/C15 { BL [] 0 setdash 2 copy vpt 0 360 arc closepath fill
               vpt 0 360 arc closepath } bind def
/Rec   { newpath 4 2 roll moveto 1 index 0 rlineto 0 exch rlineto
       neg 0 rlineto closepath } bind def
/Square { dup Rec } bind def
/Bsquare { vpt sub exch vpt sub exch vpt2 Square } bind def
/S0 { BL [] 0 setdash 2 copy moveto 0 vpt rlineto BL Bsquare } bind def
/S1 { BL [] 0 setdash 2 copy vpt Square fill Bsquare } bind def
/S2 { BL [] 0 setdash 2 copy exch vpt sub exch vpt Square fill Bsquare } bind def
/S3 { BL [] 0 setdash 2 copy exch vpt sub exch vpt2 vpt Rec fill Bsquare } bind def
/S4 { BL [] 0 setdash 2 copy exch vpt sub exch vpt sub vpt Square fill Bsquare } bind def
/S5 { BL [] 0 setdash 2 copy 2 copy vpt Square fill
       exch vpt sub exch vpt sub vpt Square fill Bsquare } bind def
/S6 { BL [] 0 setdash 2 copy exch vpt sub exch vpt sub vpt vpt2 Rec fill Bsquare } bind def
/S7 { BL [] 0 setdash 2 copy exch vpt sub exch vpt sub vpt vpt2 Rec fill
       2 copy vpt Square fill
       Bsquare } bind def
/S8 { BL [] 0 setdash 2 copy vpt sub vpt Square fill Bsquare } bind def
/S9 { BL [] 0 setdash 2 copy vpt sub vpt vpt2 Rec fill Bsquare } bind def
/S10 { BL [] 0 setdash 2 copy vpt sub vpt Square fill 2 copy exch vpt sub exch vpt Square fill
       Bsquare } bind def
/S11 { BL [] 0 setdash 2 copy vpt sub vpt Square fill 2 copy exch vpt sub exch vpt2 vpt Rec fill
       Bsquare } bind def
/S12 { BL [] 0 setdash 2 copy exch vpt sub exch vpt sub vpt2 vpt Rec fill Bsquare } bind def
/S13 { BL [] 0 setdash 2 copy exch vpt sub exch vpt sub vpt2 vpt Rec fill
       2 copy vpt Square fill Bsquare } bind def
/S14 { BL [] 0 setdash 2 copy exch vpt sub exch vpt sub vpt2 vpt Rec fill
       2 copy exch vpt sub exch vpt Square fill Bsquare } bind def
/S15 { BL [] 0 setdash 2 copy Bsquare fill Bsquare } bind def
/D0 { gsave translate 45 rotate 0 0 S0 stroke grestore } bind def
/D1 { gsave translate 45 rotate 0 0 S1 stroke grestore } bind def
/D2 { gsave translate 45 rotate 0 0 S2 stroke grestore } bind def
/D3 { gsave translate 45 rotate 0 0 S3 stroke grestore } bind def
/D4 { gsave translate 45 rotate 0 0 S4 stroke grestore } bind def
/D5 { gsave translate 45 rotate 0 0 S5 stroke grestore } bind def
/D6 { gsave translate 45 rotate 0 0 S6 stroke grestore } bind def
/D7 { gsave translate 45 rotate 0 0 S7 stroke grestore } bind def
/D8 { gsave translate 45 rotate 0 0 S8 stroke grestore } bind def
/D9 { gsave translate 45 rotate 0 0 S9 stroke grestore } bind def
/D10 { gsave translate 45 rotate 0 0 S10 stroke grestore } bind def
/D11 { gsave translate 45 rotate 0 0 S11 stroke grestore } bind def
/D12 { gsave translate 45 rotate 0 0 S12 stroke grestore } bind def
/D13 { gsave translate 45 rotate 0 0 S13 stroke grestore } bind def
/D14 { gsave translate 45 rotate 0 0 S14 stroke grestore } bind def
/D15 { gsave translate 45 rotate 0 0 S15 stroke grestore } bind def
/DiaE { stroke [] 0 setdash vpt add M
  hpt neg vpt neg V hpt vpt neg V
  hpt vpt V hpt neg vpt V closepath stroke } def
/BoxE { stroke [] 0 setdash exch hpt sub exch vpt add M
  0 vpt2 neg V hpt2 0 V 0 vpt2 V
  hpt2 neg 0 V closepath stroke } def
/TriUE { stroke [] 0 setdash vpt 1.12 mul add M
  hpt neg vpt -1.62 mul V
  hpt 2 mul 0 V
  hpt neg vpt 1.62 mul V closepath stroke } def
/TriDE { stroke [] 0 setdash vpt 1.12 mul sub M
  hpt neg vpt 1.62 mul V
  hpt 2 mul 0 V
  hpt neg vpt -1.62 mul V closepath stroke } def
/PentE { stroke [] 0 setdash gsave
  translate 0 hpt M 4 {72 rotate 0 hpt L} repeat
  closepath stroke grestore } def
/CircE { stroke [] 0 setdash 
  hpt 0 360 arc stroke } def
/Opaque { gsave closepath 1 setgray fill grestore 0 setgray closepath } def
/DiaW { stroke [] 0 setdash vpt add M
  hpt neg vpt neg V hpt vpt neg V
  hpt vpt V hpt neg vpt V Opaque stroke } def
/BoxW { stroke [] 0 setdash exch hpt sub exch vpt add M
  0 vpt2 neg V hpt2 0 V 0 vpt2 V
  hpt2 neg 0 V Opaque stroke } def
/TriUW { stroke [] 0 setdash vpt 1.12 mul add M
  hpt neg vpt -1.62 mul V
  hpt 2 mul 0 V
  hpt neg vpt 1.62 mul V Opaque stroke } def
/TriDW { stroke [] 0 setdash vpt 1.12 mul sub M
  hpt neg vpt 1.62 mul V
  hpt 2 mul 0 V
  hpt neg vpt -1.62 mul V Opaque stroke } def
/PentW { stroke [] 0 setdash gsave
  translate 0 hpt M 4 {72 rotate 0 hpt L} repeat
  Opaque stroke grestore } def
/CircW { stroke [] 0 setdash 
  hpt 0 360 arc Opaque stroke } def
/BoxFill { gsave Rec 1 setgray fill grestore } def
end
}}%
\begin{picture}(3600,2160)(0,0)%
{\GNUPLOTspecial{"
gnudict begin
gsave
0 0 translate
0.100 0.100 scale
0 setgray
newpath
1.000 UL
LTb
450 300 M
63 0 V
2937 0 R
-63 0 V
450 551 M
63 0 V
2937 0 R
-63 0 V
450 803 M
63 0 V
2937 0 R
-63 0 V
450 1054 M
63 0 V
2937 0 R
-63 0 V
450 1306 M
63 0 V
2937 0 R
-63 0 V
450 1557 M
63 0 V
2937 0 R
-63 0 V
450 1809 M
63 0 V
2937 0 R
-63 0 V
450 2060 M
63 0 V
2937 0 R
-63 0 V
450 300 M
0 63 V
0 1697 R
0 -63 V
950 300 M
0 63 V
0 1697 R
0 -63 V
1450 300 M
0 63 V
0 1697 R
0 -63 V
1950 300 M
0 63 V
0 1697 R
0 -63 V
2450 300 M
0 63 V
0 1697 R
0 -63 V
2950 300 M
0 63 V
0 1697 R
0 -63 V
3450 300 M
0 63 V
0 1697 R
0 -63 V
1.000 UL
LTb
450 300 M
3000 0 V
0 1760 V
-3000 0 V
450 300 L
1.000 UP
1.000 UL
LT0
450 1990 M
100 -1 V
100 0 V
100 -2 V
100 0 V
100 -2 V
100 -6 V
100 -12 V
100 -23 V
100 -49 V
100 -96 V
100 -175 V
100 -301 V
1750 901 L
1850 510 L
100 -2 V
100 390 V
100 422 V
100 300 V
100 179 V
100 96 V
100 47 V
100 25 V
100 11 V
100 7 V
100 3 V
100 1 V
100 1 V
100 -1 V
100 -1 V
450 1990 Pls
550 1989 Pls
650 1989 Pls
750 1987 Pls
850 1987 Pls
950 1985 Pls
1050 1979 Pls
1150 1967 Pls
1250 1944 Pls
1350 1895 Pls
1450 1799 Pls
1550 1624 Pls
1650 1323 Pls
1750 901 Pls
1850 510 Pls
1950 508 Pls
2050 898 Pls
2150 1320 Pls
2250 1620 Pls
2350 1799 Pls
2450 1895 Pls
2550 1942 Pls
2650 1967 Pls
2750 1978 Pls
2850 1985 Pls
2950 1988 Pls
3050 1989 Pls
3150 1990 Pls
3250 1989 Pls
3350 1988 Pls
1.000 UL
LT1
450 1984 M
29 -1 V
30 -1 V
29 -2 V
29 -1 V
29 -2 V
30 -3 V
29 -2 V
29 -3 V
30 -3 V
29 -4 V
29 -4 V
30 -5 V
29 -6 V
29 -6 V
29 -7 V
30 -9 V
29 -9 V
29 -11 V
30 -12 V
29 -14 V
29 -15 V
29 -17 V
30 -20 V
29 -22 V
29 -24 V
30 -28 V
29 -30 V
29 -33 V
29 -37 V
30 -41 V
29 -45 V
29 -48 V
30 -52 V
29 -57 V
29 -60 V
30 -64 V
29 -68 V
29 -70 V
29 -73 V
30 -74 V
29 -75 V
29 -73 V
30 -71 V
29 -67 V
29 -60 V
29 -52 V
30 -41 V
29 -29 V
29 -14 V
30 0 V
29 14 V
29 29 V
30 41 V
29 52 V
29 60 V
29 67 V
30 71 V
29 73 V
29 75 V
30 74 V
29 73 V
29 70 V
29 68 V
30 64 V
29 60 V
29 57 V
30 52 V
29 48 V
29 45 V
30 41 V
29 37 V
29 33 V
29 30 V
30 28 V
29 24 V
29 22 V
30 20 V
29 17 V
29 15 V
29 14 V
30 12 V
29 11 V
29 9 V
30 9 V
29 7 V
29 6 V
29 6 V
30 5 V
29 4 V
29 4 V
30 3 V
29 3 V
29 2 V
30 3 V
29 2 V
29 1 V
29 2 V
30 1 V
29 1 V
1.000 UL
LT2
450 1990 M
29 0 V
30 0 V
29 -1 V
29 0 V
29 0 V
30 -1 V
29 0 V
29 -1 V
30 -1 V
29 -1 V
29 -1 V
30 -2 V
29 -1 V
29 -2 V
29 -3 V
30 -3 V
29 -3 V
29 -4 V
30 -5 V
29 -6 V
29 -7 V
29 -8 V
30 -10 V
29 -11 V
29 -14 V
30 -16 V
29 -18 V
29 -22 V
29 -25 V
30 -29 V
29 -34 V
29 -38 V
30 -44 V
29 -50 V
29 -56 V
30 -62 V
29 -69 V
29 -75 V
29 -82 V
30 -87 V
29 -92 V
29 -95 V
30 -96 V
29 -93 V
29 -89 V
29 -78 V
30 -65 V
29 -46 V
29 -24 V
30 0 V
29 24 V
29 46 V
30 65 V
29 78 V
29 89 V
29 93 V
30 96 V
29 95 V
29 92 V
30 87 V
29 82 V
29 75 V
29 69 V
30 62 V
29 56 V
29 50 V
30 44 V
29 38 V
29 34 V
30 29 V
29 25 V
29 22 V
29 18 V
30 16 V
29 14 V
29 11 V
30 10 V
29 8 V
29 7 V
29 6 V
30 5 V
29 4 V
29 3 V
30 3 V
29 3 V
29 2 V
29 1 V
30 2 V
29 1 V
29 1 V
30 1 V
29 1 V
29 0 V
30 1 V
29 0 V
29 0 V
29 1 V
30 0 V
29 0 V
1.000 UL
LT3
450 1991 M
29 0 V
30 0 V
29 0 V
29 0 V
29 0 V
30 0 V
29 0 V
29 0 V
30 0 V
29 0 V
29 0 V
30 0 V
29 -1 V
29 0 V
29 0 V
30 0 V
29 0 V
29 -1 V
30 0 V
29 -1 V
29 -1 V
29 -1 V
30 -2 V
29 -2 V
29 -3 V
30 -3 V
29 -5 V
29 -6 V
29 -8 V
30 -9 V
29 -13 V
29 -16 V
30 -21 V
29 -25 V
29 -33 V
30 -40 V
29 -49 V
29 -60 V
29 -72 V
30 -85 V
29 -100 V
29 -113 V
30 -125 V
29 -136 V
29 -139 V
29 -135 V
30 -121 V
29 -92 V
29 -50 V
30 0 V
29 50 V
29 92 V
30 121 V
29 135 V
29 139 V
29 136 V
30 125 V
29 113 V
29 100 V
30 85 V
29 72 V
29 60 V
29 49 V
30 40 V
29 33 V
29 25 V
30 21 V
29 16 V
29 13 V
30 9 V
29 8 V
29 6 V
29 5 V
30 3 V
29 3 V
29 2 V
30 2 V
29 1 V
29 1 V
29 1 V
30 0 V
29 1 V
29 0 V
30 0 V
29 0 V
29 0 V
29 1 V
30 0 V
29 0 V
29 0 V
30 0 V
29 0 V
29 0 V
30 0 V
29 0 V
29 0 V
29 0 V
30 0 V
29 0 V
1.000 UL
LT4
450 1991 M
29 0 V
30 0 V
29 0 V
29 0 V
29 0 V
30 0 V
29 0 V
29 0 V
30 -1 V
29 0 V
29 0 V
30 0 V
29 0 V
29 -1 V
29 0 V
30 -1 V
29 0 V
29 -1 V
30 -1 V
29 -2 V
29 -2 V
29 -2 V
30 -3 V
29 -4 V
29 -4 V
30 -6 V
29 -8 V
29 -9 V
29 -12 V
30 -14 V
29 -18 V
29 -22 V
30 -27 V
29 -33 V
29 -40 V
30 -48 V
29 -57 V
29 -68 V
29 -77 V
30 -90 V
29 -100 V
29 -111 V
30 -120 V
29 -125 V
29 -125 V
29 -118 V
30 -102 V
29 -76 V
29 -41 V
30 0 V
29 41 V
29 76 V
30 102 V
29 118 V
29 125 V
29 125 V
30 120 V
29 111 V
29 100 V
30 90 V
29 77 V
29 68 V
29 57 V
30 48 V
29 40 V
29 33 V
30 27 V
29 22 V
29 18 V
30 14 V
29 12 V
29 9 V
29 8 V
30 6 V
29 4 V
29 4 V
30 3 V
29 2 V
29 2 V
29 2 V
30 1 V
29 1 V
29 0 V
30 1 V
29 0 V
29 1 V
29 0 V
30 0 V
29 0 V
29 0 V
30 1 V
29 0 V
29 0 V
30 0 V
29 0 V
29 0 V
29 0 V
30 0 V
29 0 V
stroke
grestore
end
showpage
}}%
\put(1950,50){\makebox(0,0){z}}%
\put(100,1180){%
\makebox(0,0)[b]{\shortstack{$|\overline{l_{p}}|$}}%
}%
\put(3450,200){\makebox(0,0){30}}%
\put(2950,200){\makebox(0,0){25}}%
\put(2450,200){\makebox(0,0){20}}%
\put(1950,200){\makebox(0,0){15}}%
\put(1450,200){\makebox(0,0){10}}%
\put(950,200){\makebox(0,0){5}}%
\put(450,200){\makebox(0,0){0}}%
\put(400,2060){\makebox(0,0)[r]{0.24}}%
\put(400,1809){\makebox(0,0)[r]{0.22}}%
\put(400,1557){\makebox(0,0)[r]{0.2}}%
\put(400,1306){\makebox(0,0)[r]{0.18}}%
\put(400,1054){\makebox(0,0)[r]{0.16}}%
\put(400,803){\makebox(0,0)[r]{0.14}}%
\put(400,551){\makebox(0,0)[r]{0.12}}%
\put(400,300){\makebox(0,0)[r]{0.1}}%
\end{picture}%
\endgroup

\end	{center}
\vskip 0.15in
\caption{ The average 
of the modulus of the Polyakov loop as a function of $z$
for the $L_0=4$ SU(4) calculation at $T\simeq 1.88T_c$.
One loop perturbative predictions using for $g^2(T)$ the
lattice bare coupling (long dash), the mean field improved
lattice coupling (short dash), the Schrodinger functional coupling
(dot) and just a best fit (dash-dot).} 
\label{fig_profz3}
\end 	{figure}
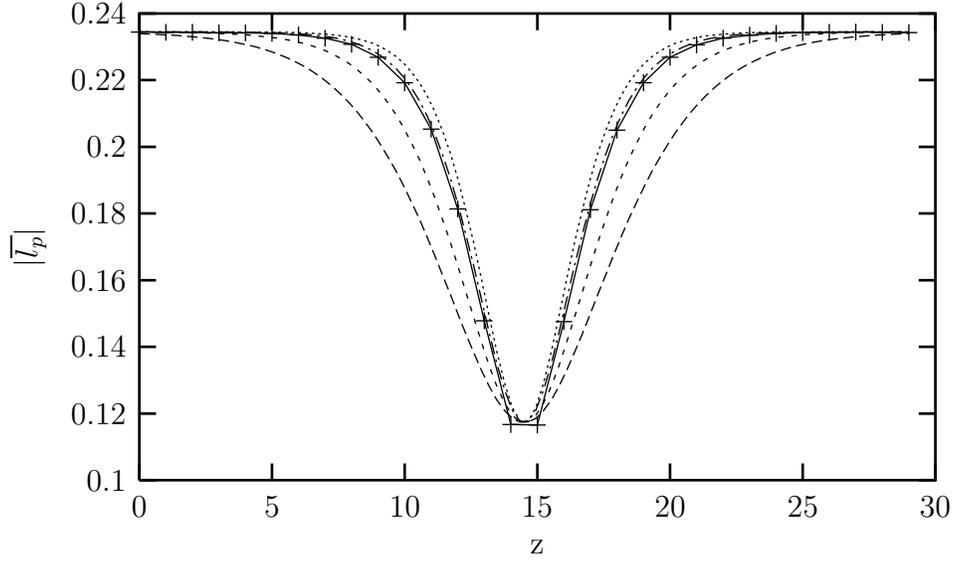

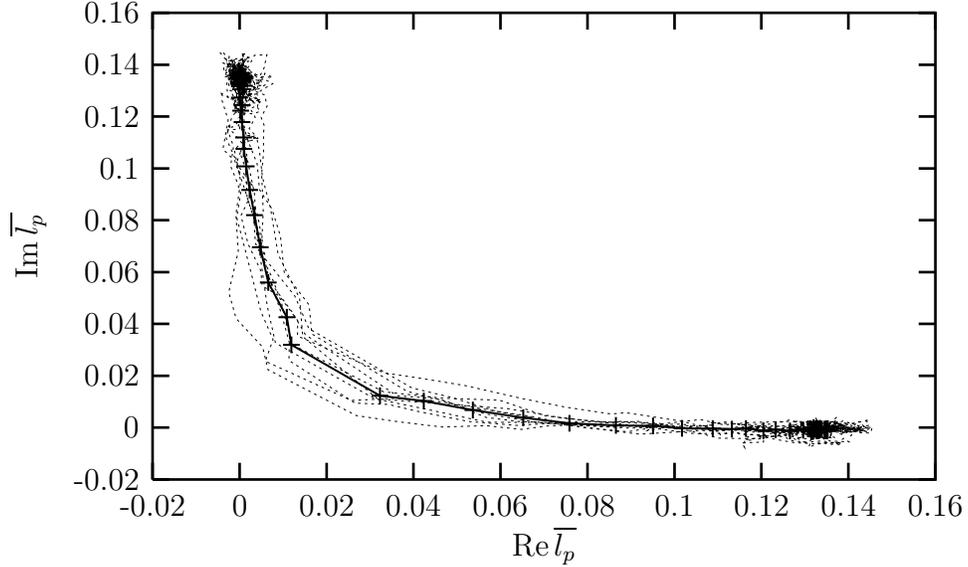
\begin	{figure}[p]
\begin	{center}
\leavevmode
\begingroup%
  \makeatletter%
  \newcommand{\GNUPLOTspecial}{%
    \@sanitize\catcode`\%=14\relax\special}%
  \setlength{\unitlength}{0.1bp}%
{\GNUPLOTspecial{!
/gnudict 256 dict def
gnudict begin
/Color false def
/Solid false def
/gnulinewidth 5.000 def
/userlinewidth gnulinewidth def
/vshift -33 def
/dl {10 mul} def
/hpt_ 31.5 def
/vpt_ 31.5 def
/hpt hpt_ def
/vpt vpt_ def
/M {moveto} bind def
/L {lineto} bind def
/R {rmoveto} bind def
/V {rlineto} bind def
/vpt2 vpt 2 mul def
/hpt2 hpt 2 mul def
/Lshow { currentpoint stroke M
  0 vshift R show } def
/Rshow { currentpoint stroke M
  dup stringwidth pop neg vshift R show } def
/Cshow { currentpoint stroke M
  dup stringwidth pop -2 div vshift R show } def
/UP { dup vpt_ mul /vpt exch def hpt_ mul /hpt exch def
  /hpt2 hpt 2 mul def /vpt2 vpt 2 mul def } def
/DL { Color {setrgbcolor Solid {pop []} if 0 setdash }
 {pop pop pop Solid {pop []} if 0 setdash} ifelse } def
/BL { stroke userlinewidth 2 mul setlinewidth } def
/AL { stroke userlinewidth 2 div setlinewidth } def
/UL { dup gnulinewidth mul /userlinewidth exch def
      10 mul /udl exch def } def
/PL { stroke userlinewidth setlinewidth } def
/LTb { BL [] 0 0 0 DL } def
/LTa { AL [1 udl mul 2 udl mul] 0 setdash 0 0 0 setrgbcolor } def
/LT0 { PL [] 1 0 0 DL } def
/LT1 { PL [4 dl 2 dl] 0 1 0 DL } def
/LT2 { PL [2 dl 3 dl] 0 0 1 DL } def
/LT3 { PL [1 dl 1.5 dl] 1 0 1 DL } def
/LT4 { PL [5 dl 2 dl 1 dl 2 dl] 0 1 1 DL } def
/LT5 { PL [4 dl 3 dl 1 dl 3 dl] 1 1 0 DL } def
/LT6 { PL [2 dl 2 dl 2 dl 4 dl] 0 0 0 DL } def
/LT7 { PL [2 dl 2 dl 2 dl 2 dl 2 dl 4 dl] 1 0.3 0 DL } def
/LT8 { PL [2 dl 2 dl 2 dl 2 dl 2 dl 2 dl 2 dl 4 dl] 0.5 0.5 0.5 DL } def
/Pnt { stroke [] 0 setdash
   gsave 1 setlinecap M 0 0 V stroke grestore } def
/Dia { stroke [] 0 setdash 2 copy vpt add M
  hpt neg vpt neg V hpt vpt neg V
  hpt vpt V hpt neg vpt V closepath stroke
  Pnt } def
/Pls { stroke [] 0 setdash vpt sub M 0 vpt2 V
  currentpoint stroke M
  hpt neg vpt neg R hpt2 0 V stroke
  } def
/Box { stroke [] 0 setdash 2 copy exch hpt sub exch vpt add M
  0 vpt2 neg V hpt2 0 V 0 vpt2 V
  hpt2 neg 0 V closepath stroke
  Pnt } def
/Crs { stroke [] 0 setdash exch hpt sub exch vpt add M
  hpt2 vpt2 neg V currentpoint stroke M
  hpt2 neg 0 R hpt2 vpt2 V stroke } def
/TriU { stroke [] 0 setdash 2 copy vpt 1.12 mul add M
  hpt neg vpt -1.62 mul V
  hpt 2 mul 0 V
  hpt neg vpt 1.62 mul V closepath stroke
  Pnt  } def
/Star { 2 copy Pls Crs } def
/BoxF { stroke [] 0 setdash exch hpt sub exch vpt add M
  0 vpt2 neg V  hpt2 0 V  0 vpt2 V
  hpt2 neg 0 V  closepath fill } def
/TriUF { stroke [] 0 setdash vpt 1.12 mul add M
  hpt neg vpt -1.62 mul V
  hpt 2 mul 0 V
  hpt neg vpt 1.62 mul V closepath fill } def
/TriD { stroke [] 0 setdash 2 copy vpt 1.12 mul sub M
  hpt neg vpt 1.62 mul V
  hpt 2 mul 0 V
  hpt neg vpt -1.62 mul V closepath stroke
  Pnt  } def
/TriDF { stroke [] 0 setdash vpt 1.12 mul sub M
  hpt neg vpt 1.62 mul V
  hpt 2 mul 0 V
  hpt neg vpt -1.62 mul V closepath fill} def
/DiaF { stroke [] 0 setdash vpt add M
  hpt neg vpt neg V hpt vpt neg V
  hpt vpt V hpt neg vpt V closepath fill } def
/Pent { stroke [] 0 setdash 2 copy gsave
  translate 0 hpt M 4 {72 rotate 0 hpt L} repeat
  closepath stroke grestore Pnt } def
/PentF { stroke [] 0 setdash gsave
  translate 0 hpt M 4 {72 rotate 0 hpt L} repeat
  closepath fill grestore } def
/Circle { stroke [] 0 setdash 2 copy
  hpt 0 360 arc stroke Pnt } def
/CircleF { stroke [] 0 setdash hpt 0 360 arc fill } def
/C0 { BL [] 0 setdash 2 copy moveto vpt 90 450  arc } bind def
/C1 { BL [] 0 setdash 2 copy        moveto
       2 copy  vpt 0 90 arc closepath fill
               vpt 0 360 arc closepath } bind def
/C2 { BL [] 0 setdash 2 copy moveto
       2 copy  vpt 90 180 arc closepath fill
               vpt 0 360 arc closepath } bind def
/C3 { BL [] 0 setdash 2 copy moveto
       2 copy  vpt 0 180 arc closepath fill
               vpt 0 360 arc closepath } bind def
/C4 { BL [] 0 setdash 2 copy moveto
       2 copy  vpt 180 270 arc closepath fill
               vpt 0 360 arc closepath } bind def
/C5 { BL [] 0 setdash 2 copy moveto
       2 copy  vpt 0 90 arc
       2 copy moveto
       2 copy  vpt 180 270 arc closepath fill
               vpt 0 360 arc } bind def
/C6 { BL [] 0 setdash 2 copy moveto
      2 copy  vpt 90 270 arc closepath fill
              vpt 0 360 arc closepath } bind def
/C7 { BL [] 0 setdash 2 copy moveto
      2 copy  vpt 0 270 arc closepath fill
              vpt 0 360 arc closepath } bind def
/C8 { BL [] 0 setdash 2 copy moveto
      2 copy vpt 270 360 arc closepath fill
              vpt 0 360 arc closepath } bind def
/C9 { BL [] 0 setdash 2 copy moveto
      2 copy  vpt 270 450 arc closepath fill
              vpt 0 360 arc closepath } bind def
/C10 { BL [] 0 setdash 2 copy 2 copy moveto vpt 270 360 arc closepath fill
       2 copy moveto
       2 copy vpt 90 180 arc closepath fill
               vpt 0 360 arc closepath } bind def
/C11 { BL [] 0 setdash 2 copy moveto
       2 copy  vpt 0 180 arc closepath fill
       2 copy moveto
       2 copy  vpt 270 360 arc closepath fill
               vpt 0 360 arc closepath } bind def
/C12 { BL [] 0 setdash 2 copy moveto
       2 copy  vpt 180 360 arc closepath fill
               vpt 0 360 arc closepath } bind def
/C13 { BL [] 0 setdash  2 copy moveto
       2 copy  vpt 0 90 arc closepath fill
       2 copy moveto
       2 copy  vpt 180 360 arc closepath fill
               vpt 0 360 arc closepath } bind def
/C14 { BL [] 0 setdash 2 copy moveto
       2 copy  vpt 90 360 arc closepath fill
               vpt 0 360 arc } bind def
/C15 { BL [] 0 setdash 2 copy vpt 0 360 arc closepath fill
               vpt 0 360 arc closepath } bind def
/Rec   { newpath 4 2 roll moveto 1 index 0 rlineto 0 exch rlineto
       neg 0 rlineto closepath } bind def
/Square { dup Rec } bind def
/Bsquare { vpt sub exch vpt sub exch vpt2 Square } bind def
/S0 { BL [] 0 setdash 2 copy moveto 0 vpt rlineto BL Bsquare } bind def
/S1 { BL [] 0 setdash 2 copy vpt Square fill Bsquare } bind def
/S2 { BL [] 0 setdash 2 copy exch vpt sub exch vpt Square fill Bsquare } bind def
/S3 { BL [] 0 setdash 2 copy exch vpt sub exch vpt2 vpt Rec fill Bsquare } bind def
/S4 { BL [] 0 setdash 2 copy exch vpt sub exch vpt sub vpt Square fill Bsquare } bind def
/S5 { BL [] 0 setdash 2 copy 2 copy vpt Square fill
       exch vpt sub exch vpt sub vpt Square fill Bsquare } bind def
/S6 { BL [] 0 setdash 2 copy exch vpt sub exch vpt sub vpt vpt2 Rec fill Bsquare } bind def
/S7 { BL [] 0 setdash 2 copy exch vpt sub exch vpt sub vpt vpt2 Rec fill
       2 copy vpt Square fill
       Bsquare } bind def
/S8 { BL [] 0 setdash 2 copy vpt sub vpt Square fill Bsquare } bind def
/S9 { BL [] 0 setdash 2 copy vpt sub vpt vpt2 Rec fill Bsquare } bind def
/S10 { BL [] 0 setdash 2 copy vpt sub vpt Square fill 2 copy exch vpt sub exch vpt Square fill
       Bsquare } bind def
/S11 { BL [] 0 setdash 2 copy vpt sub vpt Square fill 2 copy exch vpt sub exch vpt2 vpt Rec fill
       Bsquare } bind def
/S12 { BL [] 0 setdash 2 copy exch vpt sub exch vpt sub vpt2 vpt Rec fill Bsquare } bind def
/S13 { BL [] 0 setdash 2 copy exch vpt sub exch vpt sub vpt2 vpt Rec fill
       2 copy vpt Square fill Bsquare } bind def
/S14 { BL [] 0 setdash 2 copy exch vpt sub exch vpt sub vpt2 vpt Rec fill
       2 copy exch vpt sub exch vpt Square fill Bsquare } bind def
/S15 { BL [] 0 setdash 2 copy Bsquare fill Bsquare } bind def
/D0 { gsave translate 45 rotate 0 0 S0 stroke grestore } bind def
/D1 { gsave translate 45 rotate 0 0 S1 stroke grestore } bind def
/D2 { gsave translate 45 rotate 0 0 S2 stroke grestore } bind def
/D3 { gsave translate 45 rotate 0 0 S3 stroke grestore } bind def
/D4 { gsave translate 45 rotate 0 0 S4 stroke grestore } bind def
/D5 { gsave translate 45 rotate 0 0 S5 stroke grestore } bind def
/D6 { gsave translate 45 rotate 0 0 S6 stroke grestore } bind def
/D7 { gsave translate 45 rotate 0 0 S7 stroke grestore } bind def
/D8 { gsave translate 45 rotate 0 0 S8 stroke grestore } bind def
/D9 { gsave translate 45 rotate 0 0 S9 stroke grestore } bind def
/D10 { gsave translate 45 rotate 0 0 S10 stroke grestore } bind def
/D11 { gsave translate 45 rotate 0 0 S11 stroke grestore } bind def
/D12 { gsave translate 45 rotate 0 0 S12 stroke grestore } bind def
/D13 { gsave translate 45 rotate 0 0 S13 stroke grestore } bind def
/D14 { gsave translate 45 rotate 0 0 S14 stroke grestore } bind def
/D15 { gsave translate 45 rotate 0 0 S15 stroke grestore } bind def
/DiaE { stroke [] 0 setdash vpt add M
  hpt neg vpt neg V hpt vpt neg V
  hpt vpt V hpt neg vpt V closepath stroke } def
/BoxE { stroke [] 0 setdash exch hpt sub exch vpt add M
  0 vpt2 neg V hpt2 0 V 0 vpt2 V
  hpt2 neg 0 V closepath stroke } def
/TriUE { stroke [] 0 setdash vpt 1.12 mul add M
  hpt neg vpt -1.62 mul V
  hpt 2 mul 0 V
  hpt neg vpt 1.62 mul V closepath stroke } def
/TriDE { stroke [] 0 setdash vpt 1.12 mul sub M
  hpt neg vpt 1.62 mul V
  hpt 2 mul 0 V
  hpt neg vpt -1.62 mul V closepath stroke } def
/PentE { stroke [] 0 setdash gsave
  translate 0 hpt M 4 {72 rotate 0 hpt L} repeat
  closepath stroke grestore } def
/CircE { stroke [] 0 setdash 
  hpt 0 360 arc stroke } def
/Opaque { gsave closepath 1 setgray fill grestore 0 setgray closepath } def
/DiaW { stroke [] 0 setdash vpt add M
  hpt neg vpt neg V hpt vpt neg V
  hpt vpt V hpt neg vpt V Opaque stroke } def
/BoxW { stroke [] 0 setdash exch hpt sub exch vpt add M
  0 vpt2 neg V hpt2 0 V 0 vpt2 V
  hpt2 neg 0 V Opaque stroke } def
/TriUW { stroke [] 0 setdash vpt 1.12 mul add M
  hpt neg vpt -1.62 mul V
  hpt 2 mul 0 V
  hpt neg vpt 1.62 mul V Opaque stroke } def
/TriDW { stroke [] 0 setdash vpt 1.12 mul sub M
  hpt neg vpt 1.62 mul V
  hpt 2 mul 0 V
  hpt neg vpt -1.62 mul V Opaque stroke } def
/PentW { stroke [] 0 setdash gsave
  translate 0 hpt M 4 {72 rotate 0 hpt L} repeat
  Opaque stroke grestore } def
/CircW { stroke [] 0 setdash 
  hpt 0 360 arc Opaque stroke } def
/BoxFill { gsave Rec 1 setgray fill grestore } def
end
}}%
\begin{picture}(3600,2160)(0,0)%
{\GNUPLOTspecial{"
gnudict begin
gsave
0 0 translate
0.100 0.100 scale
0 setgray
newpath
1.000 UL
LTb
500 300 M
63 0 V
2887 0 R
-63 0 V
500 496 M
63 0 V
2887 0 R
-63 0 V
500 691 M
63 0 V
2887 0 R
-63 0 V
500 887 M
63 0 V
2887 0 R
-63 0 V
500 1082 M
63 0 V
2887 0 R
-63 0 V
500 1278 M
63 0 V
2887 0 R
-63 0 V
500 1473 M
63 0 V
2887 0 R
-63 0 V
500 1669 M
63 0 V
2887 0 R
-63 0 V
500 1864 M
63 0 V
2887 0 R
-63 0 V
500 2060 M
63 0 V
2887 0 R
-63 0 V
500 300 M
0 63 V
0 1697 R
0 -63 V
828 300 M
0 63 V
0 1697 R
0 -63 V
1156 300 M
0 63 V
0 1697 R
0 -63 V
1483 300 M
0 63 V
0 1697 R
0 -63 V
1811 300 M
0 63 V
0 1697 R
0 -63 V
2139 300 M
0 63 V
0 1697 R
0 -63 V
2467 300 M
0 63 V
0 1697 R
0 -63 V
2794 300 M
0 63 V
0 1697 R
0 -63 V
3122 300 M
0 63 V
0 1697 R
0 -63 V
3450 300 M
0 63 V
0 1697 R
0 -63 V
1.000 UL
LTb
500 300 M
2950 0 V
0 1760 V
-2950 0 V
500 300 L
1.000 UP
1.500 UL
LT0
2991 496 M
-5 -7 V
-7 3 V
-9 -1 V
7 -1 V
5 -5 V
4 -4 V
2 3 V
24 5 V
7 -1 V
-3 5 V
-8 -1 V
-5 -1 V
8 -5 V
-8 6 V
-3 1 V
11 -5 V
-9 -1 V
11 0 V
-3 1 V
-22 -4 V
1 -2 V
0 2 V
-2 3 V
8 2 V
27 1 V
18 -5 V
2 -2 V
-9 6 V
1 -1 V
0 -4 V
10 8 V
-6 -2 V
2 0 V
-38 0 V
-50 -4 V
-50 -3 V
-54 5 V
-46 -4 V
-66 8 V
-53 -2 V
-72 2 V
-116 1 V
-109 9 V
-140 2 V
-175 8 V
-174 21 V
-190 29 V
-185 34 V
-165 20 V
1023 808 L
-17 104 V
-70 131 V
-30 133 V
-22 121 V
-18 95 V
-13 89 V
-9 66 V
-1 43 V
-6 58 V
-5 43 V
6 21 V
-11 28 V
14 32 V
5 13 V
-8 18 V
3 -2 V
1 9 V
0 6 V
6 1 V
1 1 V
-12 -14 V
4 0 V
-17 -6 V
4 -8 V
-4 2 V
2 5 V
1 11 V
-6 5 V
-7 10 V
18 9 V
-4 -3 V
2 3 V
-5 -7 V
5 -5 V
2 -7 V
-3 -5 V
4 3 V
8 8 V
-3 12 V
-6 2 V
3 -3 V
8 -7 V
-10 -4 V
8 -2 V
-5 -4 V
5 -8 V
-1 -3 V
-12 -19 V
-1 1 V
2991 496 Pls
2986 489 Pls
2979 492 Pls
2970 491 Pls
2977 490 Pls
2982 485 Pls
2986 481 Pls
2988 484 Pls
3012 489 Pls
3019 488 Pls
3016 493 Pls
3008 492 Pls
3003 491 Pls
3011 486 Pls
3003 492 Pls
3000 493 Pls
3011 488 Pls
3002 487 Pls
3013 487 Pls
3010 488 Pls
2988 484 Pls
2989 482 Pls
2989 484 Pls
2987 487 Pls
2995 489 Pls
3022 490 Pls
3040 485 Pls
3042 483 Pls
3033 489 Pls
3034 488 Pls
3034 484 Pls
3044 492 Pls
3038 490 Pls
3040 490 Pls
3002 490 Pls
2952 486 Pls
2902 483 Pls
2848 488 Pls
2802 484 Pls
2736 492 Pls
2683 490 Pls
2611 492 Pls
2495 493 Pls
2386 502 Pls
2246 504 Pls
2071 512 Pls
1897 533 Pls
1707 562 Pls
1522 596 Pls
1357 616 Pls
1023 808 Pls
1006 912 Pls
936 1043 Pls
906 1176 Pls
884 1297 Pls
866 1392 Pls
853 1481 Pls
844 1547 Pls
843 1590 Pls
837 1648 Pls
832 1691 Pls
838 1712 Pls
827 1740 Pls
841 1772 Pls
846 1785 Pls
838 1803 Pls
841 1801 Pls
842 1810 Pls
842 1816 Pls
848 1817 Pls
849 1818 Pls
837 1804 Pls
841 1804 Pls
824 1798 Pls
828 1790 Pls
824 1792 Pls
826 1797 Pls
827 1808 Pls
821 1813 Pls
814 1823 Pls
832 1832 Pls
828 1829 Pls
830 1832 Pls
825 1825 Pls
830 1820 Pls
832 1813 Pls
829 1808 Pls
833 1811 Pls
841 1819 Pls
838 1831 Pls
832 1833 Pls
835 1830 Pls
843 1823 Pls
833 1819 Pls
841 1817 Pls
836 1813 Pls
841 1805 Pls
840 1802 Pls
828 1783 Pls
827 1784 Pls
0.300 UL
LT3
2819 496 M
-33 6 V
7 -6 V
-52 -5 V
79 -5 V
60 8 V
50 2 V
25 -26 V
46 12 V
-22 6 V
-7 6 V
-27 -3 V
-34 -2 V
35 10 V
-19 0 V
0 -8 V
86 4 V
-16 9 V
42 -13 V
8 -1 V
-6 -9 V
68 7 V
-9 -13 V
43 11 V
-1 9 V
69 -1 V
-5 -19 V
-53 2 V
-19 12 V
-22 -1 V
-5 1 V
-27 12 V
16 -1 V
-21 -6 V
-77 2 V
-69 2 V
-98 4 V
-66 -1 V
-83 9 V
-76 4 V
-27 -16 V
-86 22 V
-79 12 V
-132 20 V
-53 -5 V
-173 26 V
-158 31 V
-219 50 V
-183 30 V
-123 18 V
1046 841 L
1 58 V
-69 93 V
-65 38 V
-5 57 V
-48 101 V
-27 107 V
50 104 V
-58 80 V
-48 109 V
20 73 V
17 28 V
0 48 V
9 36 V
0 18 V
-26 27 V
36 -15 V
13 -22 V
5 6 V
7 -1 V
-6 13 V
-5 9 V
-6 23 V
-27 13 V
12 27 V
-4 -17 V
-13 -8 V
-14 8 V
18 -2 V
-19 17 V
29 -16 V
-2 -10 V
2 -9 V
-5 -21 V
11 -14 V
-12 -26 V
22 13 V
2 22 V
-8 10 V
2 2 V
8 -10 V
-4 -20 V
20 -6 V
-39 -4 V
36 11 V
-8 -25 V
-10 29 V
7 -30 V
-23 -27 V
-9 -41 V
2993 496 M
12 -12 V
-36 13 V
-22 10 V
-2 -4 V
-3 -5 V
-27 0 V
-16 -8 V
7 12 V
35 -13 V
2 18 V
2 -6 V
-13 -1 V
40 -8 V
46 -1 V
81 -9 V
39 7 V
-23 -5 V
-8 10 V
-26 5 V
-36 3 V
35 -15 V
-7 14 V
-27 0 V
-2 3 V
32 29 V
-1 -16 V
-38 -21 V
-7 24 V
13 0 V
31 -2 V
18 -11 V
-61 1 V
-24 0 V
-43 1 V
-66 15 V
-58 -15 V
-115 -2 V
-55 -21 V
-60 -3 V
-46 2 V
-93 -6 V
-137 -8 V
-161 20 V
-106 13 V
-86 11 V
-103 4 V
-121 15 V
-152 62 V
-145 36 V
1100 849 L
-10 109 V
-94 124 V
-14 135 V
-32 103 V
-47 73 V
-9 98 V
-38 96 V
-16 65 V
8 50 V
-32 43 V
19 25 V
-7 35 V
37 16 V
-15 18 V
-20 -1 V
5 -37 V
12 25 V
2 -2 V
-9 24 V
0 -11 V
-26 -15 V
9 -5 V
-20 -25 V
8 -8 V
-15 6 V
14 15 V
9 47 V
-28 -28 V
17 -24 V
-6 18 V
10 -7 V
3 10 V
15 2 V
1 11 V
-10 19 V
-36 6 V
34 -14 V
-13 -35 V
-3 15 V
-33 -6 V
14 0 V
-5 17 V
-8 -6 V
21 47 V
21 6 V
18 -18 V
4 -38 V
-14 -60 V
-8 -3 V
3025 496 M
47 6 V
-20 4 V
-22 0 V
16 -8 V
-4 16 V
-52 -26 V
-55 6 V
-30 10 V
-5 -2 V
28 18 V
30 -6 V
-2 5 V
29 -10 V
15 22 V
18 -33 V
27 -12 V
65 20 V
14 8 V
-23 7 V
-3 -2 V
-54 -15 V
-12 15 V
44 -2 V
-4 -19 V
75 -13 V
21 6 V
-16 20 V
-65 0 V
-1 -5 V
-29 -15 V
11 25 V
-10 2 V
54 -19 V
-50 14 V
-74 -4 V
-56 1 V
-96 -30 V
-66 22 V
-112 15 V
2529 489 L
-127 24 V
-143 6 V
-88 -10 V
-183 -7 V
-99 -13 V
-132 14 V
-167 -5 V
-158 19 V
-164 24 V
935 714 L
-20 85 V
815 907 L
-28 99 V
36 154 V
-9 140 V
18 125 V
-24 119 V
28 26 V
-43 50 V
-7 18 V
13 -10 V
15 10 V
16 28 V
-16 18 V
-11 53 V
-20 4 V
-3 47 V
1 22 V
30 22 V
1 38 V
-34 -16 V
5 0 V
-1 13 V
10 -21 V
-1 3 V
-6 -7 V
10 -17 V
5 16 V
-11 -1 V
48 34 V
-32 -15 V
-12 -3 V
8 0 V
20 8 V
18 -30 V
-28 6 V
-13 -3 V
19 50 V
-36 -11 V
-2 4 V
-26 32 V
14 -3 V
7 -8 V
19 -78 V
-8 -23 V
12 2 V
-5 -6 V
30 5 V
13 7 V
3116 496 M
-17 -3 V
-44 -4 V
-35 -7 V
-4 14 V
13 3 V
-14 -15 V
14 25 V
44 -20 V
12 0 V
-9 0 V
24 2 V
8 6 V
-74 -18 V
-46 10 V
-6 0 V
-107 -3 V
-42 6 V
-57 -15 V
-5 13 V
-34 -6 V
15 2 V
60 -4 V
-10 13 V
85 8 V
40 -7 V
35 12 V
99 -10 V
-5 4 V
13 -2 V
22 -6 V
-10 2 V
17 -3 V
3 1 V
-50 2 V
40 -6 V
38 -8 V
-51 36 V
42 -22 V
-39 -10 V
-57 4 V
2907 477 L
-148 7 V
-224 6 V
-228 11 V
-258 9 V
-202 38 V
-250 44 V
-184 14 V
-164 -8 V
925 746 L
29 82 V
908 934 L
-33 131 V
-33 128 V
-25 95 V
15 92 V
-15 46 V
-23 56 V
11 74 V
-22 63 V
45 2 V
1 58 V
-21 40 V
21 34 V
-37 14 V
-7 34 V
22 28 V
45 30 V
-7 48 V
0 -11 V
-26 -33 V
23 -13 V
-30 -23 V
4 25 V
-27 -14 V
-1 5 V
28 -6 V
-13 11 V
-5 -4 V
16 20 V
24 -15 V
-26 17 V
1 4 V
-13 18 V
-15 -34 V
15 -33 V
-2 9 V
35 -19 V
-4 21 V
-29 -11 V
46 14 V
-14 -47 V
-12 26 V
22 12 V
-10 -16 V
4 21 V
-18 -2 V
-24 14 V
2 14 V
currentpoint stroke M
2933 496 M
-24 -29 V
77 2 V
-4 -15 V
-1 -6 V
-8 -27 V
42 19 V
55 -9 V
34 24 V
20 -13 V
-71 3 V
-40 21 V
-9 -9 V
-45 -3 V
-81 13 V
-64 -8 V
-34 4 V
-2 3 V
71 -3 V
36 -1 V
13 -14 V
22 -3 V
46 -5 V
58 6 V
69 -16 V
-22 11 V
29 -5 V
-42 4 V
-49 -6 V
-91 5 V
-74 13 V
-38 9 V
-72 -11 V
23 -2 V
-14 14 V
-15 -24 V
3 -17 V
-10 17 V
50 2 V
31 1 V
1 12 V
-49 1 V
-121 2 V
-74 19 V
-134 31 V
-210 -7 V
-249 16 V
-206 1 V
-228 49 V
-196 40 V
963 815 L
932 955 L
-24 169 V
9 188 V
-6 140 V
7 79 V
0 74 V
3 29 V
-12 32 V
23 43 V
-10 41 V
-21 10 V
-22 4 V
17 39 V
20 7 V
7 22 V
30 -16 V
-44 3 V
-19 8 V
13 -8 V
50 -26 V
-35 -20 V
-28 -19 V
-47 -11 V
18 5 V
12 7 V
21 20 V
-11 6 V
-13 3 V
-28 -17 V
35 -27 V
0 -15 V
34 -1 V
-52 14 V
11 28 V
25 9 V
9 46 V
20 39 V
0 11 V
7 31 V
-47 -2 V
-26 -50 V
66 -62 V
-32 -19 V
5 -37 V
-12 15 V
13 -21 V
4 6 V
-15 -5 V
-27 3 V
2986 496 M
-14 -13 V
-67 7 V
-8 -22 V
8 -6 V
-7 5 V
26 -17 V
49 23 V
51 -2 V
-6 11 V
4 16 V
-32 8 V
-37 -11 V
-98 -22 V
-84 15 V
-45 12 V
0 -19 V
28 -2 V
70 1 V
84 -11 V
46 -28 V
59 16 V
60 1 V
-22 6 V
12 15 V
-11 -20 V
-11 1 V
-3 16 V
-20 -3 V
-28 12 V
-22 -11 V
3 4 V
-10 -13 V
5 10 V
21 -16 V
-15 5 V
53 -13 V
30 9 V
-25 -5 V
-37 10 V
-42 15 V
-20 -2 V
-86 -16 V
-143 13 V
-161 -2 V
-246 33 V
-249 26 V
-207 35 V
-269 20 V
-222 17 V
1066 839 L
-12 144 V
-81 167 V
-29 175 V
-30 115 V
-11 97 V
-11 55 V
-36 -19 V
41 14 V
-33 36 V
23 43 V
-21 36 V
-1 60 V
35 34 V
-10 -8 V
16 -2 V
-53 -28 V
29 -18 V
22 -15 V
-9 -20 V
3 14 V
-28 2 V
-3 18 V
-23 30 V
18 -13 V
9 -3 V
-5 8 V
1 3 V
-6 -16 V
-17 42 V
39 12 V
-4 2 V
5 -10 V
-8 -4 V
-39 5 V
32 16 V
-14 -11 V
1 -2 V
-10 8 V
12 9 V
51 -13 V
-24 -15 V
-3 24 V
-22 0 V
27 18 V
-2 -7 V
-17 -9 V
-8 -15 V
-10 -26 V
-22 3 V
2937 496 M
-33 -9 V
-52 -21 V
-29 25 V
4 3 V
-53 -1 V
-52 -13 V
46 24 V
40 -19 V
90 4 V
41 0 V
-13 -19 V
2 9 V
36 5 V
5 -8 V
12 5 V
80 8 V
-11 -9 V
24 10 V
-48 -15 V
-75 -6 V
-93 11 V
-99 13 V
-42 3 V
-4 0 V
38 4 V
46 -2 V
53 -6 V
22 -5 V
131 -1 V
69 -3 V
67 -3 V
48 13 V
-28 -10 V
-17 -4 V
-51 0 V
-103 -5 V
-36 -1 V
-106 8 V
-94 9 V
1 5 V
-89 -8 V
-101 17 V
-91 -4 V
-176 -6 V
-192 10 V
-138 9 V
-214 35 V
-184 57 V
-147 11 V
1076 846 L
-49 84 V
-92 126 V
-42 154 V
-9 168 V
-27 131 V
-16 109 V
-22 102 V
30 49 V
8 53 V
-19 14 V
3 25 V
-20 -3 V
31 18 V
-29 3 V
40 7 V
5 6 V
-29 -5 V
-10 -14 V
-5 -8 V
0 -7 V
26 -35 V
-6 -11 V
3 -26 V
-17 -10 V
13 5 V
32 41 V
-9 22 V
-29 22 V
6 -1 V
-1 0 V
-42 -16 V
2 -11 V
11 -21 V
1 -52 V
-7 -29 V
0 -23 V
3 -12 V
37 -8 V
6 18 V
19 -11 V
18 9 V
16 -20 V
-26 -10 V
-45 3 V
7 41 V
15 1 V
12 15 V
-17 4 V
-23 13 V
3008 496 M
17 5 V
62 6 V
76 -17 V
-27 -2 V
21 -4 V
-36 9 V
-75 -6 V
-30 25 V
-55 -7 V
-12 -2 V
40 -9 V
27 -6 V
39 -1 V
17 -6 V
26 13 V
49 -12 V
-17 -4 V
25 10 V
-5 1 V
-67 15 V
-28 -28 V
-16 8 V
-15 -2 V
-16 24 V
13 -7 V
19 -20 V
12 6 V
22 2 V
-5 -8 V
-8 -2 V
-2 2 V
-43 9 V
-16 -4 V
-75 7 V
2795 467 L
-83 4 V
-16 11 V
-55 -14 V
-37 37 V
-70 -6 V
-42 -3 V
-129 15 V
-29 8 V
-117 -3 V
-89 7 V
-147 17 V
-174 59 V
-220 8 V
-217 24 V
1016 814 L
984 930 L
-51 151 V
-32 150 V
-45 134 V
-5 107 V
-21 58 V
16 50 V
35 15 V
-19 2 V
23 -8 V
-14 10 V
-48 36 V
17 54 V
13 52 V
-24 41 V
19 12 V
-17 13 V
-5 13 V
23 -27 V
-24 -13 V
4 -38 V
23 8 V
-2 -28 V
-14 -49 V
-6 -4 V
6 7 V
-25 18 V
-1 41 V
-13 25 V
14 35 V
12 39 V
25 11 V
-16 -22 V
40 -25 V
-27 -4 V
22 11 V
-4 1 V
-3 41 V
-21 32 V
6 2 V
-8 31 V
4 -15 V
17 -7 V
2 3 V
-21 8 V
8 -32 V
-12 -11 V
-22 -30 V
24 -4 V
currentpoint stroke M
2964 496 M
-18 -18 V
0 12 V
-16 13 V
6 -2 V
6 -16 V
79 -1 V
38 8 V
69 3 V
47 -12 V
22 -3 V
-40 10 V
-29 -1 V
41 -7 V
33 9 V
-37 14 V
-39 -22 V
-26 -14 V
-10 4 V
6 14 V
-5 -17 V
24 15 V
-3 6 V
-40 1 V
-58 -16 V
27 10 V
76 -7 V
15 -14 V
21 26 V
2 -1 V
15 -15 V
-16 23 V
15 -12 V
-18 4 V
-70 8 V
-97 -24 V
-119 14 V
-107 11 V
2620 481 L
-123 28 V
-54 -1 V
-18 -15 V
-60 -8 V
18 11 V
-92 -16 V
-172 17 V
-179 39 V
-148 22 V
-167 52 V
-147 35 V
1089 869 L
-53 123 V
926 1130 L
-30 167 V
-31 90 V
-13 38 V
-34 48 V
3 35 V
18 46 V
5 51 V
2 75 V
12 67 V
-10 18 V
-8 51 V
28 16 V
-22 15 V
-4 25 V
20 6 V
-26 -13 V
0 -27 V
3 8 V
-13 -12 V
13 -27 V
6 -2 V
3 -9 V
-10 -10 V
-42 -20 V
25 28 V
-9 6 V
25 30 V
-5 7 V
6 13 V
3 25 V
-3 11 V
6 -11 V
-6 17 V
-35 -48 V
12 -22 V
20 3 V
14 7 V
-26 38 V
24 -9 V
-12 9 V
14 -10 V
14 -42 V
-18 -19 V
-1 -35 V
-6 28 V
-13 -33 V
19 3 V
3124 496 M
19 2 V
2 9 V
20 8 V
-5 4 V
24 -21 V
24 -2 V
-57 -2 V
8 0 V
-51 19 V
-27 -13 V
-27 -3 V
41 -1 V
80 2 V
26 7 V
-14 20 V
11 -1 V
-48 -10 V
-57 -12 V
-56 2 V
-54 13 V
-40 -8 V
-21 -15 V
-7 -1 V
-2 12 V
8 -1 V
-29 3 V
-1 -14 V
8 3 V
-1 -5 V
-2 -1 V
92 13 V
44 -4 V
41 22 V
-5 -22 V
-19 13 V
-81 6 V
-74 1 V
-26 -1 V
-115 -3 V
-100 -9 V
-88 20 V
2412 509 L
-160 7 V
-158 2 V
1874 504 L
-184 15 V
-196 39 V
-104 25 V
-127 5 V
1014 749 L
2 94 V
984 973 L
-38 98 V
-59 121 V
-4 88 V
-38 121 V
-35 101 V
-45 44 V
20 109 V
-24 77 V
2 16 V
-8 15 V
3 -6 V
39 -21 V
-3 -2 V
22 3 V
1 13 V
-10 24 V
14 8 V
-22 6 V
15 18 V
11 24 V
-27 -5 V
-4 -25 V
-6 40 V
9 -8 V
1 6 V
12 -4 V
-20 35 V
8 -2 V
-12 -6 V
-20 10 V
4 -35 V
13 -17 V
20 -13 V
18 -10 V
-12 10 V
4 14 V
-13 -4 V
-4 32 V
16 -25 V
-8 38 V
6 -2 V
-21 36 V
-3 -15 V
10 -22 V
10 28 V
-9 -29 V
20 7 V
stroke
grestore
end
showpage
}}%
\put(1975,50){\makebox(0,0){$\mathrm{Re}\, \overline{l_{p}}$}}%
\put(100,1180){%
\makebox(0,0)[b]{\shortstack{$\mathrm{Im}\, \overline{l_{p}}$}}%
}%
\put(3450,200){\makebox(0,0){0.16}}%
\put(3122,200){\makebox(0,0){0.14}}%
\put(2794,200){\makebox(0,0){0.12}}%
\put(2467,200){\makebox(0,0){0.1}}%
\put(2139,200){\makebox(0,0){0.08}}%
\put(1811,200){\makebox(0,0){0.06}}%
\put(1483,200){\makebox(0,0){0.04}}%
\put(1156,200){\makebox(0,0){0.02}}%
\put(828,200){\makebox(0,0){0}}%
\put(500,200){\makebox(0,0){-0.02}}%
\put(450,2060){\makebox(0,0)[r]{0.16}}%
\put(450,1864){\makebox(0,0)[r]{0.14}}%
\put(450,1669){\makebox(0,0)[r]{0.12}}%
\put(450,1473){\makebox(0,0)[r]{0.1}}%
\put(450,1278){\makebox(0,0)[r]{0.08}}%
\put(450,1082){\makebox(0,0)[r]{0.06}}%
\put(450,887){\makebox(0,0)[r]{0.04}}%
\put(450,691){\makebox(0,0)[r]{0.02}}%
\put(450,496){\makebox(0,0)[r]{0}}%
\put(450,300){\makebox(0,0)[r]{-0.02}}%
\end{picture}%
\endgroup

\end	{center}
\vskip 0.15in
\caption{The values in the complex plane taken by the Polyakov 
loop for $x_3\in [0,L-1]$ for a sample of  $k=1$ walls 
in SU(4) at $\beta_c=10.491$, i.e $T\simeq 1.006T_c$. The
lines represent averages over subsequences of 100 sweeps,
taken over the first 1000 sweeps.}
\label{fig_profc5}
\end 	{figure}

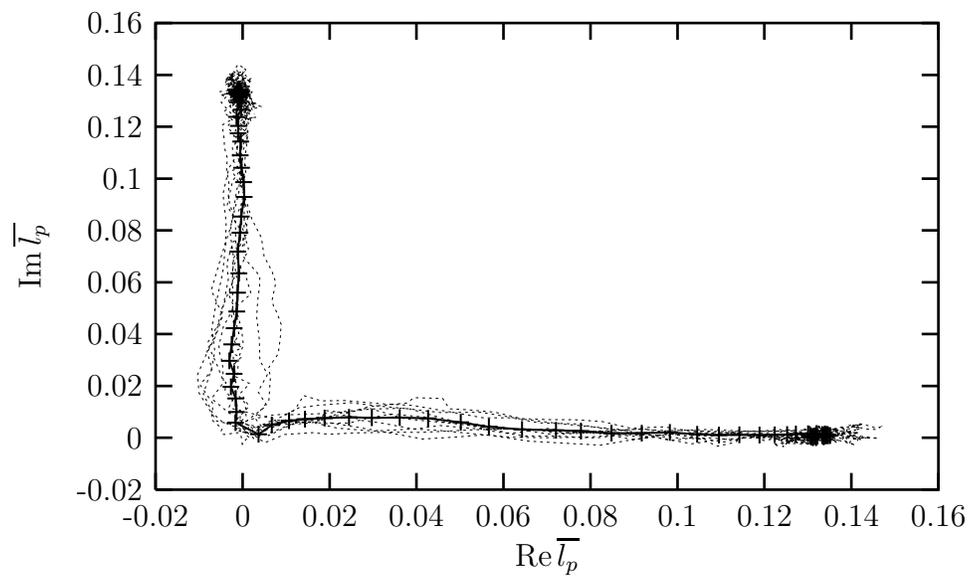
\begin	{figure}[p]
\begin	{center}
\leavevmode
\begingroup%
  \makeatletter%
  \newcommand{\GNUPLOTspecial}{%
    \@sanitize\catcode`\%=14\relax\special}%
  \setlength{\unitlength}{0.1bp}%
{\GNUPLOTspecial{!
/gnudict 256 dict def
gnudict begin
/Color false def
/Solid false def
/gnulinewidth 5.000 def
/userlinewidth gnulinewidth def
/vshift -33 def
/dl {10 mul} def
/hpt_ 31.5 def
/vpt_ 31.5 def
/hpt hpt_ def
/vpt vpt_ def
/M {moveto} bind def
/L {lineto} bind def
/R {rmoveto} bind def
/V {rlineto} bind def
/vpt2 vpt 2 mul def
/hpt2 hpt 2 mul def
/Lshow { currentpoint stroke M
  0 vshift R show } def
/Rshow { currentpoint stroke M
  dup stringwidth pop neg vshift R show } def
/Cshow { currentpoint stroke M
  dup stringwidth pop -2 div vshift R show } def
/UP { dup vpt_ mul /vpt exch def hpt_ mul /hpt exch def
  /hpt2 hpt 2 mul def /vpt2 vpt 2 mul def } def
/DL { Color {setrgbcolor Solid {pop []} if 0 setdash }
 {pop pop pop Solid {pop []} if 0 setdash} ifelse } def
/BL { stroke userlinewidth 2 mul setlinewidth } def
/AL { stroke userlinewidth 2 div setlinewidth } def
/UL { dup gnulinewidth mul /userlinewidth exch def
      10 mul /udl exch def } def
/PL { stroke userlinewidth setlinewidth } def
/LTb { BL [] 0 0 0 DL } def
/LTa { AL [1 udl mul 2 udl mul] 0 setdash 0 0 0 setrgbcolor } def
/LT0 { PL [] 1 0 0 DL } def
/LT1 { PL [4 dl 2 dl] 0 1 0 DL } def
/LT2 { PL [2 dl 3 dl] 0 0 1 DL } def
/LT3 { PL [1 dl 1.5 dl] 1 0 1 DL } def
/LT4 { PL [5 dl 2 dl 1 dl 2 dl] 0 1 1 DL } def
/LT5 { PL [4 dl 3 dl 1 dl 3 dl] 1 1 0 DL } def
/LT6 { PL [2 dl 2 dl 2 dl 4 dl] 0 0 0 DL } def
/LT7 { PL [2 dl 2 dl 2 dl 2 dl 2 dl 4 dl] 1 0.3 0 DL } def
/LT8 { PL [2 dl 2 dl 2 dl 2 dl 2 dl 2 dl 2 dl 4 dl] 0.5 0.5 0.5 DL } def
/Pnt { stroke [] 0 setdash
   gsave 1 setlinecap M 0 0 V stroke grestore } def
/Dia { stroke [] 0 setdash 2 copy vpt add M
  hpt neg vpt neg V hpt vpt neg V
  hpt vpt V hpt neg vpt V closepath stroke
  Pnt } def
/Pls { stroke [] 0 setdash vpt sub M 0 vpt2 V
  currentpoint stroke M
  hpt neg vpt neg R hpt2 0 V stroke
  } def
/Box { stroke [] 0 setdash 2 copy exch hpt sub exch vpt add M
  0 vpt2 neg V hpt2 0 V 0 vpt2 V
  hpt2 neg 0 V closepath stroke
  Pnt } def
/Crs { stroke [] 0 setdash exch hpt sub exch vpt add M
  hpt2 vpt2 neg V currentpoint stroke M
  hpt2 neg 0 R hpt2 vpt2 V stroke } def
/TriU { stroke [] 0 setdash 2 copy vpt 1.12 mul add M
  hpt neg vpt -1.62 mul V
  hpt 2 mul 0 V
  hpt neg vpt 1.62 mul V closepath stroke
  Pnt  } def
/Star { 2 copy Pls Crs } def
/BoxF { stroke [] 0 setdash exch hpt sub exch vpt add M
  0 vpt2 neg V  hpt2 0 V  0 vpt2 V
  hpt2 neg 0 V  closepath fill } def
/TriUF { stroke [] 0 setdash vpt 1.12 mul add M
  hpt neg vpt -1.62 mul V
  hpt 2 mul 0 V
  hpt neg vpt 1.62 mul V closepath fill } def
/TriD { stroke [] 0 setdash 2 copy vpt 1.12 mul sub M
  hpt neg vpt 1.62 mul V
  hpt 2 mul 0 V
  hpt neg vpt -1.62 mul V closepath stroke
  Pnt  } def
/TriDF { stroke [] 0 setdash vpt 1.12 mul sub M
  hpt neg vpt 1.62 mul V
  hpt 2 mul 0 V
  hpt neg vpt -1.62 mul V closepath fill} def
/DiaF { stroke [] 0 setdash vpt add M
  hpt neg vpt neg V hpt vpt neg V
  hpt vpt V hpt neg vpt V closepath fill } def
/Pent { stroke [] 0 setdash 2 copy gsave
  translate 0 hpt M 4 {72 rotate 0 hpt L} repeat
  closepath stroke grestore Pnt } def
/PentF { stroke [] 0 setdash gsave
  translate 0 hpt M 4 {72 rotate 0 hpt L} repeat
  closepath fill grestore } def
/Circle { stroke [] 0 setdash 2 copy
  hpt 0 360 arc stroke Pnt } def
/CircleF { stroke [] 0 setdash hpt 0 360 arc fill } def
/C0 { BL [] 0 setdash 2 copy moveto vpt 90 450  arc } bind def
/C1 { BL [] 0 setdash 2 copy        moveto
       2 copy  vpt 0 90 arc closepath fill
               vpt 0 360 arc closepath } bind def
/C2 { BL [] 0 setdash 2 copy moveto
       2 copy  vpt 90 180 arc closepath fill
               vpt 0 360 arc closepath } bind def
/C3 { BL [] 0 setdash 2 copy moveto
       2 copy  vpt 0 180 arc closepath fill
               vpt 0 360 arc closepath } bind def
/C4 { BL [] 0 setdash 2 copy moveto
       2 copy  vpt 180 270 arc closepath fill
               vpt 0 360 arc closepath } bind def
/C5 { BL [] 0 setdash 2 copy moveto
       2 copy  vpt 0 90 arc
       2 copy moveto
       2 copy  vpt 180 270 arc closepath fill
               vpt 0 360 arc } bind def
/C6 { BL [] 0 setdash 2 copy moveto
      2 copy  vpt 90 270 arc closepath fill
              vpt 0 360 arc closepath } bind def
/C7 { BL [] 0 setdash 2 copy moveto
      2 copy  vpt 0 270 arc closepath fill
              vpt 0 360 arc closepath } bind def
/C8 { BL [] 0 setdash 2 copy moveto
      2 copy vpt 270 360 arc closepath fill
              vpt 0 360 arc closepath } bind def
/C9 { BL [] 0 setdash 2 copy moveto
      2 copy  vpt 270 450 arc closepath fill
              vpt 0 360 arc closepath } bind def
/C10 { BL [] 0 setdash 2 copy 2 copy moveto vpt 270 360 arc closepath fill
       2 copy moveto
       2 copy vpt 90 180 arc closepath fill
               vpt 0 360 arc closepath } bind def
/C11 { BL [] 0 setdash 2 copy moveto
       2 copy  vpt 0 180 arc closepath fill
       2 copy moveto
       2 copy  vpt 270 360 arc closepath fill
               vpt 0 360 arc closepath } bind def
/C12 { BL [] 0 setdash 2 copy moveto
       2 copy  vpt 180 360 arc closepath fill
               vpt 0 360 arc closepath } bind def
/C13 { BL [] 0 setdash  2 copy moveto
       2 copy  vpt 0 90 arc closepath fill
       2 copy moveto
       2 copy  vpt 180 360 arc closepath fill
               vpt 0 360 arc closepath } bind def
/C14 { BL [] 0 setdash 2 copy moveto
       2 copy  vpt 90 360 arc closepath fill
               vpt 0 360 arc } bind def
/C15 { BL [] 0 setdash 2 copy vpt 0 360 arc closepath fill
               vpt 0 360 arc closepath } bind def
/Rec   { newpath 4 2 roll moveto 1 index 0 rlineto 0 exch rlineto
       neg 0 rlineto closepath } bind def
/Square { dup Rec } bind def
/Bsquare { vpt sub exch vpt sub exch vpt2 Square } bind def
/S0 { BL [] 0 setdash 2 copy moveto 0 vpt rlineto BL Bsquare } bind def
/S1 { BL [] 0 setdash 2 copy vpt Square fill Bsquare } bind def
/S2 { BL [] 0 setdash 2 copy exch vpt sub exch vpt Square fill Bsquare } bind def
/S3 { BL [] 0 setdash 2 copy exch vpt sub exch vpt2 vpt Rec fill Bsquare } bind def
/S4 { BL [] 0 setdash 2 copy exch vpt sub exch vpt sub vpt Square fill Bsquare } bind def
/S5 { BL [] 0 setdash 2 copy 2 copy vpt Square fill
       exch vpt sub exch vpt sub vpt Square fill Bsquare } bind def
/S6 { BL [] 0 setdash 2 copy exch vpt sub exch vpt sub vpt vpt2 Rec fill Bsquare } bind def
/S7 { BL [] 0 setdash 2 copy exch vpt sub exch vpt sub vpt vpt2 Rec fill
       2 copy vpt Square fill
       Bsquare } bind def
/S8 { BL [] 0 setdash 2 copy vpt sub vpt Square fill Bsquare } bind def
/S9 { BL [] 0 setdash 2 copy vpt sub vpt vpt2 Rec fill Bsquare } bind def
/S10 { BL [] 0 setdash 2 copy vpt sub vpt Square fill 2 copy exch vpt sub exch vpt Square fill
       Bsquare } bind def
/S11 { BL [] 0 setdash 2 copy vpt sub vpt Square fill 2 copy exch vpt sub exch vpt2 vpt Rec fill
       Bsquare } bind def
/S12 { BL [] 0 setdash 2 copy exch vpt sub exch vpt sub vpt2 vpt Rec fill Bsquare } bind def
/S13 { BL [] 0 setdash 2 copy exch vpt sub exch vpt sub vpt2 vpt Rec fill
       2 copy vpt Square fill Bsquare } bind def
/S14 { BL [] 0 setdash 2 copy exch vpt sub exch vpt sub vpt2 vpt Rec fill
       2 copy exch vpt sub exch vpt Square fill Bsquare } bind def
/S15 { BL [] 0 setdash 2 copy Bsquare fill Bsquare } bind def
/D0 { gsave translate 45 rotate 0 0 S0 stroke grestore } bind def
/D1 { gsave translate 45 rotate 0 0 S1 stroke grestore } bind def
/D2 { gsave translate 45 rotate 0 0 S2 stroke grestore } bind def
/D3 { gsave translate 45 rotate 0 0 S3 stroke grestore } bind def
/D4 { gsave translate 45 rotate 0 0 S4 stroke grestore } bind def
/D5 { gsave translate 45 rotate 0 0 S5 stroke grestore } bind def
/D6 { gsave translate 45 rotate 0 0 S6 stroke grestore } bind def
/D7 { gsave translate 45 rotate 0 0 S7 stroke grestore } bind def
/D8 { gsave translate 45 rotate 0 0 S8 stroke grestore } bind def
/D9 { gsave translate 45 rotate 0 0 S9 stroke grestore } bind def
/D10 { gsave translate 45 rotate 0 0 S10 stroke grestore } bind def
/D11 { gsave translate 45 rotate 0 0 S11 stroke grestore } bind def
/D12 { gsave translate 45 rotate 0 0 S12 stroke grestore } bind def
/D13 { gsave translate 45 rotate 0 0 S13 stroke grestore } bind def
/D14 { gsave translate 45 rotate 0 0 S14 stroke grestore } bind def
/D15 { gsave translate 45 rotate 0 0 S15 stroke grestore } bind def
/DiaE { stroke [] 0 setdash vpt add M
  hpt neg vpt neg V hpt vpt neg V
  hpt vpt V hpt neg vpt V closepath stroke } def
/BoxE { stroke [] 0 setdash exch hpt sub exch vpt add M
  0 vpt2 neg V hpt2 0 V 0 vpt2 V
  hpt2 neg 0 V closepath stroke } def
/TriUE { stroke [] 0 setdash vpt 1.12 mul add M
  hpt neg vpt -1.62 mul V
  hpt 2 mul 0 V
  hpt neg vpt 1.62 mul V closepath stroke } def
/TriDE { stroke [] 0 setdash vpt 1.12 mul sub M
  hpt neg vpt 1.62 mul V
  hpt 2 mul 0 V
  hpt neg vpt -1.62 mul V closepath stroke } def
/PentE { stroke [] 0 setdash gsave
  translate 0 hpt M 4 {72 rotate 0 hpt L} repeat
  closepath stroke grestore } def
/CircE { stroke [] 0 setdash 
  hpt 0 360 arc stroke } def
/Opaque { gsave closepath 1 setgray fill grestore 0 setgray closepath } def
/DiaW { stroke [] 0 setdash vpt add M
  hpt neg vpt neg V hpt vpt neg V
  hpt vpt V hpt neg vpt V Opaque stroke } def
/BoxW { stroke [] 0 setdash exch hpt sub exch vpt add M
  0 vpt2 neg V hpt2 0 V 0 vpt2 V
  hpt2 neg 0 V Opaque stroke } def
/TriUW { stroke [] 0 setdash vpt 1.12 mul add M
  hpt neg vpt -1.62 mul V
  hpt 2 mul 0 V
  hpt neg vpt 1.62 mul V Opaque stroke } def
/TriDW { stroke [] 0 setdash vpt 1.12 mul sub M
  hpt neg vpt 1.62 mul V
  hpt 2 mul 0 V
  hpt neg vpt -1.62 mul V Opaque stroke } def
/PentW { stroke [] 0 setdash gsave
  translate 0 hpt M 4 {72 rotate 0 hpt L} repeat
  Opaque stroke grestore } def
/CircW { stroke [] 0 setdash 
  hpt 0 360 arc Opaque stroke } def
/BoxFill { gsave Rec 1 setgray fill grestore } def
end
}}%
\begin{picture}(3600,2160)(0,0)%
{\GNUPLOTspecial{"
gnudict begin
gsave
0 0 translate
0.100 0.100 scale
0 setgray
newpath
1.000 UL
LTb
500 300 M
63 0 V
2887 0 R
-63 0 V
500 496 M
63 0 V
2887 0 R
-63 0 V
500 691 M
63 0 V
2887 0 R
-63 0 V
500 887 M
63 0 V
2887 0 R
-63 0 V
500 1082 M
63 0 V
2887 0 R
-63 0 V
500 1278 M
63 0 V
2887 0 R
-63 0 V
500 1473 M
63 0 V
2887 0 R
-63 0 V
500 1669 M
63 0 V
2887 0 R
-63 0 V
500 1864 M
63 0 V
2887 0 R
-63 0 V
500 2060 M
63 0 V
2887 0 R
-63 0 V
500 300 M
0 63 V
0 1697 R
0 -63 V
828 300 M
0 63 V
0 1697 R
0 -63 V
1156 300 M
0 63 V
0 1697 R
0 -63 V
1483 300 M
0 63 V
0 1697 R
0 -63 V
1811 300 M
0 63 V
0 1697 R
0 -63 V
2139 300 M
0 63 V
0 1697 R
0 -63 V
2467 300 M
0 63 V
0 1697 R
0 -63 V
2794 300 M
0 63 V
0 1697 R
0 -63 V
3122 300 M
0 63 V
0 1697 R
0 -63 V
3450 300 M
0 63 V
0 1697 R
0 -63 V
1.000 UL
LTb
500 300 M
2950 0 V
0 1760 V
-2950 0 V
500 300 L
1.000 UP
1.500 UL
LT0
2989 496 M
-12 0 V
0 3 V
-3 2 V
-7 3 V
3 -5 V
14 2 V
10 0 V
20 3 V
4 -5 V
16 4 V
8 1 V
-10 4 V
-8 -2 V
6 -1 V
3 2 V
-8 2 V
-10 -4 V
0 -3 V
7 2 V
0 6 V
3 -6 V
0 1 V
-13 3 V
-21 4 V
-32 -1 V
-46 -1 V
-33 -2 V
-47 -1 V
-57 -1 V
-76 1 V
-77 -1 V
-82 3 V
-103 7 V
-106 -3 V
-114 0 V
-115 8 V
-94 3 V
-128 3 V
-125 10 V
-106 17 V
-123 15 V
-107 3 V
-105 -1 V
-86 2 V
-92 -3 V
-74 -4 V
-60 -6 V
938 544 L
888 509 L
-87 44 V
4 40 V
-3 51 V
-18 44 V
12 49 V
-18 49 V
9 62 V
9 61 V
10 63 V
3 71 V
5 73 V
-4 82 V
8 71 V
5 61 V
11 74 V
-2 56 V
-8 54 V
-5 47 V
2 52 V
-10 31 V
1 28 V
-7 34 V
18 27 V
-14 28 V
3 8 V
2 15 V
3 12 V
-11 4 V
6 4 V
-2 4 V
7 1 V
8 -5 V
-10 -2 V
-4 -6 V
5 -4 V
1 -6 V
-2 -5 V
-17 13 V
16 3 V
4 -8 V
-1 -2 V
4 5 V
-11 10 V
12 -3 V
-1 3 V
0 -4 V
-13 -7 V
5 -1 V
11 -1 V
3 -4 V
2989 496 Pls
2977 496 Pls
2977 499 Pls
2974 501 Pls
2967 504 Pls
2970 499 Pls
2984 501 Pls
2994 501 Pls
3014 504 Pls
3018 499 Pls
3034 503 Pls
3042 504 Pls
3032 508 Pls
3024 506 Pls
3030 505 Pls
3033 507 Pls
3025 509 Pls
3015 505 Pls
3015 502 Pls
3022 504 Pls
3022 510 Pls
3025 504 Pls
3025 505 Pls
3012 508 Pls
2991 512 Pls
2959 511 Pls
2913 510 Pls
2880 508 Pls
2833 507 Pls
2776 506 Pls
2700 507 Pls
2623 506 Pls
2541 509 Pls
2438 516 Pls
2332 513 Pls
2218 513 Pls
2103 521 Pls
2009 524 Pls
1881 527 Pls
1756 537 Pls
1650 554 Pls
1527 569 Pls
1420 572 Pls
1315 571 Pls
1229 573 Pls
1137 570 Pls
1063 566 Pls
1003 560 Pls
938 544 Pls
888 509 Pls
801 553 Pls
805 593 Pls
802 644 Pls
784 688 Pls
796 737 Pls
778 786 Pls
787 848 Pls
796 909 Pls
806 972 Pls
809 1043 Pls
814 1116 Pls
810 1198 Pls
818 1269 Pls
823 1330 Pls
834 1404 Pls
832 1460 Pls
824 1514 Pls
819 1561 Pls
821 1613 Pls
811 1644 Pls
812 1672 Pls
805 1706 Pls
823 1733 Pls
809 1761 Pls
812 1769 Pls
814 1784 Pls
817 1796 Pls
806 1800 Pls
812 1804 Pls
810 1808 Pls
817 1809 Pls
825 1804 Pls
815 1802 Pls
811 1796 Pls
816 1792 Pls
817 1786 Pls
815 1781 Pls
798 1794 Pls
814 1797 Pls
818 1789 Pls
817 1787 Pls
821 1792 Pls
810 1802 Pls
822 1799 Pls
821 1802 Pls
821 1798 Pls
808 1791 Pls
813 1790 Pls
824 1789 Pls
827 1785 Pls
0.300 UL
LT3
3044 496 M
-11 2 V
21 1 V
-13 -3 V
-17 -6 V
40 -10 V
-7 9 V
-2 -4 V
-37 7 V
9 -17 V
-34 15 V
-43 -15 V
-11 11 V
19 11 V
-29 -1 V
22 1 V
-11 -3 V
-35 9 V
-26 1 V
9 0 V
-1 4 V
14 -16 V
53 5 V
-11 -12 V
-35 19 V
-44 4 V
-37 -7 V
-41 -2 V
-50 -2 V
16 16 V
-55 -24 V
-12 -10 V
-67 -15 V
-61 27 V
-47 3 V
-67 -9 V
-63 13 V
-76 -8 V
2178 468 L
-95 5 V
-176 10 V
-163 24 V
-106 -8 V
-131 -8 V
-157 29 V
-93 16 V
-87 14 V
-95 17 V
983 551 L
908 522 L
-89 33 V
0 44 V
-57 76 V
3 29 V
-11 9 V
-23 59 V
-36 80 V
59 103 V
-44 97 V
39 133 V
33 103 V
32 134 V
5 104 V
-17 54 V
48 99 V
-11 47 V
2 45 V
6 26 V
-2 25 V
6 -2 V
-14 -23 V
2 42 V
-16 14 V
5 16 V
0 -20 V
4 4 V
21 -12 V
-21 -16 V
30 -6 V
-43 7 V
36 -12 V
-28 -27 V
23 -18 V
-6 -10 V
-5 -11 V
-20 0 V
5 -28 V
24 -16 V
31 -3 V
-3 4 V
-26 -7 V
10 -12 V
-10 48 V
8 -6 V
-15 11 V
10 21 V
-37 32 V
4 4 V
30 31 V
-37 -7 V
2965 496 M
-64 21 V
-22 -21 V
-30 -10 V
-10 19 V
15 -16 V
10 -12 V
23 16 V
16 1 V
0 14 V
9 -2 V
30 3 V
5 -15 V
-62 2 V
3 -7 V
25 19 V
18 -6 V
4 -4 V
40 -15 V
7 17 V
34 1 V
5 0 V
40 0 V
26 11 V
2 -6 V
-68 -4 V
-58 7 V
-72 -15 V
-12 9 V
-64 -10 V
-94 3 V
-78 -16 V
-68 10 V
-142 21 V
2302 488 L
-108 22 V
-135 8 V
-79 13 V
1834 512 L
-94 24 V
-113 13 V
-187 18 V
-138 41 V
-57 -10 V
1130 587 L
-112 -4 V
-67 0 V
855 565 L
820 547 L
16 -24 V
-85 33 V
-37 34 V
-34 71 V
8 24 V
25 79 V
10 59 V
-11 86 V
16 94 V
25 115 V
10 97 V
3 92 V
-12 108 V
22 97 V
39 64 V
35 61 V
-11 76 V
-6 54 V
0 5 V
7 62 V
-19 16 V
10 -20 V
-1 21 V
-10 -6 V
-12 -17 V
9 -24 V
-6 -5 V
2 -19 V
-25 -22 V
25 -21 V
0 -26 V
16 20 V
-1 2 V
-24 13 V
-5 -5 V
14 21 V
-1 9 V
-11 8 V
-2 35 V
-3 0 V
20 -2 V
-1 11 V
-5 16 V
-13 -23 V
5 -1 V
2 -17 V
6 -6 V
-2 -10 V
16 29 V
3 -3 V
-4 -38 V
3106 496 M
39 16 V
-45 9 V
-33 -1 V
-25 11 V
-4 -11 V
-19 2 V
-17 16 V
-16 -19 V
10 -8 V
-30 16 V
38 -6 V
-50 22 V
15 -16 V
49 -8 V
-38 2 V
4 20 V
45 -29 V
-18 -7 V
1 4 V
3 1 V
18 1 V
-36 26 V
-19 -6 V
25 10 V
-2 -24 V
-2 3 V
47 4 V
-4 11 V
-32 -12 V
0 3 V
-71 -19 V
-41 15 V
-131 -3 V
-188 8 V
-156 -5 V
2258 507 L
-99 -12 V
-147 7 V
-119 0 V
-131 20 V
-112 30 V
-75 19 V
-114 33 V
-125 12 V
-102 12 V
1112 618 L
-72 -9 V
964 570 L
880 522 L
776 590 L
-65 53 V
45 61 V
-4 75 V
-12 41 V
-1 23 V
16 71 V
35 64 V
70 61 V
-21 69 V
-28 45 V
-9 81 V
-1 78 V
-31 61 V
20 91 V
36 51 V
-15 52 V
-12 60 V
-5 9 V
-22 34 V
17 25 V
-32 29 V
52 51 V
1 7 V
0 32 V
-38 -4 V
30 0 V
-25 34 V
30 11 V
22 16 V
-39 -44 V
12 8 V
-12 -11 V
-19 -3 V
6 -19 V
34 -35 V
-9 -2 V
0 23 V
16 28 V
-4 8 V
0 -4 V
-8 13 V
-36 17 V
28 6 V
-21 19 V
4 -10 V
13 -21 V
-6 3 V
35 5 V
9 8 V
2953 496 M
22 6 V
-11 -4 V
22 25 V
-51 4 V
30 0 V
78 -12 V
11 6 V
38 10 V
11 -14 V
7 25 V
38 2 V
-30 -4 V
47 6 V
-40 -8 V
-1 -2 V
35 0 V
-10 -7 V
-8 -8 V
7 29 V
28 -6 V
38 -11 V
8 0 V
8 3 V
-12 -9 V
-42 2 V
-21 5 V
-27 3 V
-55 -2 V
-14 -5 V
-63 -9 V
-47 2 V
-57 3 V
-50 -6 V
-43 0 V
-149 -1 V
-93 2 V
2437 505 L
-125 1 V
-166 9 V
-113 -5 V
-129 20 V
-205 -4 V
1500 516 L
-111 1 V
-142 -3 V
-93 -5 V
-64 6 V
986 508 L
911 477 L
-97 70 V
27 31 V
-3 48 V
-47 53 V
19 66 V
2 50 V
19 51 V
-5 41 V
-12 58 V
-12 38 V
8 94 V
-13 73 V
36 81 V
-7 40 V
-27 42 V
-25 86 V
3 53 V
-10 62 V
4 34 V
-12 44 V
0 28 V
-8 55 V
47 28 V
-51 58 V
29 32 V
-1 37 V
-6 17 V
-7 -2 V
2 -12 V
7 24 V
15 -6 V
1 -3 V
-7 7 V
1 -32 V
-3 -11 V
-4 -1 V
12 -4 V
-46 -6 V
15 -16 V
6 1 V
8 8 V
-22 -5 V
24 3 V
23 16 V
-11 -3 V
27 -7 V
-22 9 V
12 -30 V
-1 -8 V
10 -12 V
currentpoint stroke M
2781 496 M
-27 1 V
2 10 V
0 -5 V
37 -17 V
27 5 V
46 1 V
33 4 V
27 8 V
-21 -12 V
67 -8 V
-8 2 V
-10 1 V
-53 1 V
58 10 V
2 -1 V
-71 -14 V
-16 -10 V
19 15 V
25 8 V
34 13 V
4 6 V
19 -31 V
17 16 V
-12 -9 V
56 7 V
3 -6 V
29 -6 V
-26 1 V
-80 1 V
-56 1 V
-27 -5 V
-48 12 V
-77 14 V
-64 -22 V
-95 8 V
-110 -1 V
-96 7 V
-147 27 V
-201 19 V
-167 20 V
-149 25 V
-153 -1 V
-138 16 V
-129 -3 V
1191 588 L
-83 -33 V
-86 -4 V
943 519 L
870 493 L
-45 43 V
-20 28 V
47 50 V
-31 51 V
34 35 V
-53 26 V
-9 67 V
-2 47 V
23 49 V
-23 87 V
-1 117 V
4 82 V
-13 94 V
13 87 V
13 80 V
34 51 V
-5 44 V
2 29 V
7 54 V
1 31 V
-3 31 V
-11 57 V
38 32 V
-43 43 V
8 5 V
0 19 V
23 18 V
-23 2 V
22 -11 V
-28 -13 V
20 -16 V
-33 -12 V
7 9 V
28 -15 V
-43 -1 V
14 4 V
-28 13 V
10 15 V
-8 -37 V
24 -9 V
-1 -10 V
15 -14 V
-29 -23 V
15 -11 V
32 -27 V
-21 11 V
2 -9 V
0 -7 V
11 -11 V
-23 6 V
2892 496 M
26 -19 V
32 6 V
-17 -2 V
-61 1 V
-27 -11 V
-18 2 V
15 -7 V
44 15 V
44 -4 V
66 -4 V
3 28 V
-2 -16 V
-36 -4 V
18 6 V
-1 -3 V
40 11 V
-49 -17 V
-16 12 V
24 -19 V
-93 4 V
18 0 V
-7 12 V
22 8 V
-5 2 V
-63 -5 V
-51 10 V
-61 -4 V
-33 -4 V
-84 -6 V
-99 -15 V
-130 24 V
-159 9 V
-105 -8 V
-143 -2 V
-143 2 V
-156 14 V
-112 11 V
-101 6 V
-125 23 V
-70 9 V
-55 21 V
-91 -7 V
-47 -6 V
-13 -5 V
-78 0 V
936 546 L
898 535 L
-44 9 V
843 519 L
757 508 L
-12 9 V
6 13 V
-41 35 V
7 58 V
-49 72 V
45 58 V
-19 68 V
32 62 V
24 100 V
39 96 V
3 90 V
41 80 V
1 92 V
1 89 V
-10 64 V
-8 49 V
10 59 V
8 49 V
-27 28 V
11 8 V
13 -2 V
24 17 V
-13 31 V
7 -2 V
47 14 V
-37 15 V
-10 -11 V
-29 -5 V
28 -12 V
-20 5 V
34 7 V
-17 21 V
18 -20 V
16 -4 V
-31 23 V
27 -7 V
-31 32 V
8 15 V
-15 -8 V
2 -28 V
-8 6 V
9 15 V
13 -6 V
10 19 V
-9 -16 V
2 -19 V
-20 -5 V
11 4 V
24 -49 V
3149 496 M
-30 9 V
16 -14 V
16 4 V
73 -1 V
-57 6 V
-17 2 V
4 -13 V
31 9 V
-35 -3 V
-7 -13 V
12 19 V
-59 2 V
20 -26 V
16 35 V
-10 0 V
-30 0 V
-16 -5 V
35 -11 V
-18 19 V
-32 -8 V
-11 -19 V
-14 -6 V
-41 22 V
-85 -1 V
-73 3 V
-67 -11 V
-92 21 V
-52 -14 V
-131 2 V
-156 25 V
-156 -7 V
2066 512 L
-84 13 V
-130 3 V
-136 1 V
-145 13 V
-83 -12 V
-167 -9 V
-136 11 V
-103 24 V
-87 -1 V
-64 3 V
885 537 L
-14 -9 V
840 515 L
-6 -3 V
21 -17 V
-14 -5 V
827 471 L
-9 61 V
-13 10 V
-75 17 V
-26 41 V
-40 64 V
-6 69 V
32 75 V
37 72 V
18 53 V
13 76 V
16 83 V
-5 62 V
23 81 V
56 29 V
-15 68 V
-10 15 V
0 32 V
11 46 V
-47 78 V
51 19 V
-23 43 V
4 39 V
27 35 V
-21 29 V
-10 39 V
-24 3 V
2 38 V
20 36 V
-16 32 V
16 15 V
22 1 V
-5 -23 V
-25 -32 V
10 7 V
2 12 V
10 7 V
-14 7 V
-32 1 V
28 25 V
46 0 V
-17 29 V
1 20 V
-4 3 V
3 -2 V
-26 18 V
-3 -2 V
-14 -26 V
27 -14 V
-14 16 V
-4 13 V
3054 496 M
-36 5 V
-14 3 V
22 0 V
-14 16 V
9 -9 V
80 13 V
-22 -18 V
48 16 V
-26 -7 V
36 10 V
-19 -12 V
5 15 V
-47 7 V
-46 -26 V
-12 -4 V
-17 13 V
15 15 V
6 -16 V
-16 -4 V
60 19 V
-17 -9 V
-9 -10 V
-67 -1 V
7 -6 V
-38 10 V
-108 1 V
-64 -5 V
-98 0 V
-77 5 V
-70 7 V
-89 -2 V
-117 6 V
-131 -8 V
-119 -4 V
-89 -6 V
-84 13 V
-91 13 V
-126 14 V
-106 6 V
-58 18 V
-126 11 V
-89 -15 V
-71 -19 V
-49 0 V
-33 -2 V
-49 5 V
-81 -8 V
-90 3 V
895 505 L
-95 20 V
74 58 V
45 34 V
-33 53 V
36 65 V
-29 52 V
-2 79 V
-10 86 V
4 63 V
7 52 V
1 39 V
-59 68 V
-4 20 V
10 67 V
8 79 V
-13 54 V
-5 69 V
-23 61 V
36 89 V
-46 21 V
14 69 V
-31 44 V
-6 12 V
8 27 V
2 -19 V
32 26 V
-28 34 V
-21 1 V
22 14 V
-16 11 V
20 0 V
38 -8 V
-23 25 V
-32 -24 V
17 31 V
-5 -30 V
44 -9 V
-64 12 V
26 18 V
-19 -41 V
17 -9 V
14 25 V
-21 17 V
-7 3 V
29 10 V
19 2 V
-37 -15 V
-4 -4 V
28 -47 V
20 10 V
currentpoint stroke M
3039 496 M
5 -3 V
-20 7 V
0 12 V
49 -8 V
3 7 V
-20 9 V
-1 2 V
10 4 V
12 -6 V
1 -7 V
-16 -8 V
31 10 V
-14 -1 V
33 -1 V
41 2 V
-60 1 V
2 -5 V
-6 25 V
-33 -21 V
14 9 V
-36 -10 V
-14 -1 V
-69 2 V
-49 8 V
-58 4 V
-69 -7 V
-35 1 V
-87 -9 V
-51 -4 V
-98 15 V
-89 14 V
-126 -5 V
-106 25 V
-105 7 V
-89 0 V
-111 4 V
-98 24 V
-99 9 V
-74 4 V
-25 39 V
-109 3 V
-25 -20 V
1336 616 L
-30 -2 V
-82 -25 V
-80 9 V
-25 -34 V
-90 -10 V
926 504 L
-53 68 V
54 65 V
6 90 V
-8 31 V
37 54 V
8 59 V
2 35 V
2 16 V
-16 44 V
-14 28 V
-5 20 V
14 67 V
-14 42 V
-30 57 V
-10 70 V
-36 70 V
-17 56 V
-22 63 V
-8 66 V
-22 34 V
-5 58 V
11 7 V
11 62 V
-15 48 V
-6 30 V
6 21 V
-4 26 V
-1 2 V
-6 7 V
-15 2 V
15 30 V
12 2 V
-8 -17 V
-6 31 V
33 -40 V
-22 -16 V
20 -24 V
-26 29 V
20 -4 V
-6 -18 V
-13 9 V
36 -5 V
-11 28 V
8 4 V
-5 1 V
-17 -9 V
-13 2 V
18 8 V
6 -25 V
13 23 V
2905 496 M
-42 -34 V
37 32 V
6 2 V
-52 5 V
-7 -10 V
9 6 V
61 -4 V
34 -16 V
33 9 V
51 7 V
48 -9 V
15 18 V
29 1 V
4 -12 V
-3 1 V
13 0 V
-43 14 V
-23 -20 V
63 -18 V
-50 20 V
0 3 V
-38 14 V
9 -12 V
-51 26 V
12 -5 V
-55 -2 V
-9 -14 V
-56 -10 V
-54 11 V
-61 3 V
-76 9 V
-23 7 V
-135 -7 V
-92 1 V
2350 500 L
-94 24 V
-78 8 V
-94 16 V
-139 4 V
-103 19 V
-107 0 V
-135 24 V
-133 23 V
-118 6 V
-133 12 V
-92 -1 V
-66 20 V
-27 -44 V
983 557 L
776 609 L
34 72 V
-16 51 V
1 40 V
27 27 V
-37 17 V
40 19 V
-27 22 V
-1 26 V
4 28 V
-15 43 V
5 46 V
-16 37 V
20 60 V
34 61 V
21 50 V
-23 80 V
-15 61 V
25 59 V
-16 82 V
8 63 V
-23 40 V
19 25 V
-3 39 V
-11 8 V
2 43 V
28 -7 V
2 18 V
-14 30 V
3 18 V
-16 32 V
48 2 V
-8 -9 V
-30 0 V
11 -7 V
38 -30 V
-70 4 V
-1 1 V
28 6 V
-3 -15 V
18 -18 V
4 8 V
-22 13 V
28 -29 V
-7 -10 V
-19 -16 V
-19 -15 V
0 6 V
3 25 V
27 11 V
stroke
grestore
end
showpage
}}%
\put(1975,50){\makebox(0,0){$\mathrm{Re}\, \overline{l_{p}}$}}%
\put(100,1180){%
\makebox(0,0)[b]{\shortstack{$\mathrm{Im}\, \overline{l_{p}}$}}%
}%
\put(3450,200){\makebox(0,0){0.16}}%
\put(3122,200){\makebox(0,0){0.14}}%
\put(2794,200){\makebox(0,0){0.12}}%
\put(2467,200){\makebox(0,0){0.1}}%
\put(2139,200){\makebox(0,0){0.08}}%
\put(1811,200){\makebox(0,0){0.06}}%
\put(1483,200){\makebox(0,0){0.04}}%
\put(1156,200){\makebox(0,0){0.02}}%
\put(828,200){\makebox(0,0){0}}%
\put(500,200){\makebox(0,0){-0.02}}%
\put(450,2060){\makebox(0,0)[r]{0.16}}%
\put(450,1864){\makebox(0,0)[r]{0.14}}%
\put(450,1669){\makebox(0,0)[r]{0.12}}%
\put(450,1473){\makebox(0,0)[r]{0.1}}%
\put(450,1278){\makebox(0,0)[r]{0.08}}%
\put(450,1082){\makebox(0,0)[r]{0.06}}%
\put(450,887){\makebox(0,0)[r]{0.04}}%
\put(450,691){\makebox(0,0)[r]{0.02}}%
\put(450,496){\makebox(0,0)[r]{0}}%
\put(450,300){\makebox(0,0)[r]{-0.02}}%
\end{picture}%
\endgroup

\end	{center}
\vskip 0.15in
\caption{As in the previous figure, but for a later sequence of
1000 sweeps.}
\label{fig_profc6}
\end 	{figure}

\end{document}